\documentclass[11pt,a4paper]{article}
\usepackage{amssymb} 
\usepackage{amsmath}
\usepackage{graphicx}
\usepackage{epsfig,latexsym}
\usepackage{tabularx}
\def\bef{\begin{figure}}
\def\eef{\end{figure}}

\newcommand{\be}[1]{\begin{equation}\label{#1}}
\newcommand{\beq}{\begin{equation}}
\newcommand{\ee}{\end{equation}}
\newcommand{\beqn}[1]{\begin{eqnarray}\label{#1}}
\newcommand{\eeqn}{\end{eqnarray}}
\newcommand{\bd}{\begin{displaymath}}
\newcommand{\ed}{\end{displaymath}}

\def\simlt{\mathrel{\lower2.5pt\vbox{\lineskip=0pt\baselineskip=0pt
           \hbox{$<$}\hbox{$\sim$}}}}
\def\simgt{\mathrel{\lower2.5pt\vbox{\lineskip=0pt\baselineskip=0pt
           \hbox{$>$}\hbox{$\sim$}}}}
\def\unity{{\hbox{1\kern-.8mm l}}}

\newcommand{\vect}[1]{\mbox{\boldmath$#1$}}

\def\be{\begin{equation}}
\def\ee{\end{equation}}
\def\bea{\begin{eqnarray}}
\def\eea{\end{eqnarray}}
\newcommand{\lsim}{\mathrel{\mathop{\kern 0pt \rlap
  {\raise.2ex\hbox{$<$}}} \lower.9ex\hbox{\kern-.190em $\sim$}}}
\newcommand{\gsim}{\mathrel{\mathop{\kern 0pt \rlap
  {\raise.2ex\hbox{$>$}}}
  \lower.9ex\hbox{\kern-.190em $\sim$}}}

\normalsize
\begin{document}

\baselineskip=0.65cm

\Large{\bf Improved model-dependent corollary analyses after the first six annual cycles of 
DAMA/LIBRA--phase2}
\rm

\large
\vspace{0.4cm}

\begin{center}

R.\,Bernabei$^{a,b}$,~P.\,Belli$^{a,b}$,~F.\,Cappella$^{c,d}$,
V.\,Caracciolo$^{e}$,~R.\,Cerulli$^{a,b}$,~C.J.\,Dai$^{f}$,~A.\,d'Angelo$^{c,d}$, \\
A. Di Marco$^{b}$,~H.L.\,He$^{f}$, A.\,Incicchitti$^{c,d}$,
~X.H.\,Ma$^{f}$,~V.\,Merlo$^{a,b}$, \\
F.\,Montecchia$^{b,g}$,~X.D.\,Sheng$^{f}$,~Z.P.\,Ye$^{f,h}$

\normalsize
\vspace{0.3cm}
\end{center}

$^{a}${\it Dip. di Fisica, Universit\`a di Roma ``Tor Vergata'', Rome, Italy}\
\vspace{1mm}

$^{b}${\it INFN, sez. Roma ``Tor Vergata'', Rome, Italy}\
\vspace{1mm}

$^{c}${\it Dip. di Fisica, Universit\`a di Roma ``La Sapienza'', Rome, Italy}\
\vspace{1mm}

$^{d}${\it INFN, sez. Roma, Rome, Italy}\
\vspace{1mm}

$^{e}${\it INFN Laboratori Nazionali del Gran Sasso, Assergi, Italy}\
\vspace{1mm}

$^{f}${\it Key Laboratory of Particle Astrophysics, Institute of High Energy 
Physics, Chinese Academy of Sciences, Beijing, P.R. China}\
\vspace{1mm}

$^{g}${\it Dip. Ingegneria Civile e Ingegneria Informatica, Universit\`a di Roma 
``Tor Vergata'', Rome, Italy}\
\vspace{1mm}

$^{h}${\it University of Jinggangshan, Ji'an, Jiangxi, P.R. China}\

\normalsize

\begin{abstract}
Several of the many proposed Dark Matter candidate particles, 
already investigated with lower exposure and a higher software energy threshold,
are further analyzed including the first DAMA/LIBRA--phase2 
data release, with an exposure of 1.13 ton $\times$ yr and a lower software energy threshold (1 keV).
The cumulative exposure above 2 keV considering also DAMA/NaI and DAMA/LIBRA--phase1 results is now 
2.46 ton $\times$ yr. 
The analysis permits to constraint the parameters' space of the considered candidates restricting their values 
-- with respect to previous analyses -- thanks to the increase of the exposure and to the lower energy threshold.
\end{abstract}

\vspace{5.0mm}

%
%

\section{Introduction}

The model-independent results of the first six full annual cycles measured by DAMA/LIBRA--phase2 
with software energy threshold lowered down to 1 keV\footnote{Throughout this paper keV means keV electron equivalent, where not
otherwise specified.} \cite{pmts,review}
have been released \cite{uni18,bled18,npae18,npp19,mg15}. On the basis of the exploited Dark Matter (DM) annual modulation signature,
the model-independent evidence for the presence of DM particles in 
the galactic halo has been further confirmed after the previous DAMA/LIBRA--phase1 
\cite{perflibra,modlibra,modlibra2,modlibra3,pmts,review,mu,norole,daissue}
and the former DAMA/NaI 
\cite{RNC,ijmd};
the cumulative C.L. is increased from the previous 9.3 $\sigma$ (data from 14 independent 
annual cycles: cumulative exposure 1.33 ton $\times$ yr) to 12.9 $\sigma$ (data from 
20 independent annual cycles: cumulative exposure 2.46 ton $\times$ yr)\footnote{Throughout this paper 
ton means metric ton (1000 kg).}. 

The expected differential counting rate of DM particles depends on the Earth's velocity 
in the galactic frame, which depends on the time:
$v_E(t) = v_{\odot} + v_{\oplus} cos\gamma cos\omega(t-t_0)$.
Here $v_{\odot}$ is the Sun velocity with respect
to the galactic halo ($v_{\odot} \simeq v_0 + 12$ km/s and $v_0$ is
the local velocity),
$v_{\oplus} \simeq$ 30 km/s is the Earth's orbital
velocity around the Sun on a plane with inclination
$\gamma$ = 60$^o$ with respect to the galactic plane. Furthermore,
$\omega$= 2$\pi/T$ with $T=1$ year and roughly $t_0 \simeq $ June 2$^{nd}$
(when the Earth's speed in the galactic halo is at maximum). 
Hence, the expected counting rate averaged in a given energy interval
can be conveniently worked out through a first order Taylor expansion:
\begin{equation}
        \mathcal{S} (t) = \mathcal{S}_0 + \mathcal{S}_m cos\omega(t-t_0),
\label{eq:sm}
\end{equation}
with the contribution from the highest order terms being less than $0.1\%$;
$\mathcal{S}_{m}$ and $\mathcal{S}_{0}$ are the modulation amplitude and the un-modulated
part of the expected differential counting rate, respectively.

Since in DAMA experiments the model-independent DM annual modulation signature is exploited, 
the experimental observable is the modulation amplitude, $\mathcal{S}_m$, as a function of the energy,
and the identification of the constant part of the signal, $\mathcal{S}_0$, is not required 
as opposed to other methods.
This approach has several advantages; in particular,
the only background of interest is the one able to mimic the
signature, i.e. able to account for the whole observed modulation amplitude and to
simultaneously satisfy all the many specific peculiarities of this signature (see e.g. Ref. \cite{npae18}). 
No background of this sort has been found or suggested by anyone over some decades, see Refs. 
\cite{review,uni18,bled18,npae18,npp19,mg15,perflibra,modlibra,modlibra2,modlibra3,mu,norole};
in particular, in the latter two references the case of neutrons, muons, and solar neutrinos has further been
addressed in details\footnote{
Concerning recently appeared remarks, let us comment that
any hypothetical effect due to environmental Helium diffusion inside the photomultipliers (PMTs) can
be excluded, even considering the following simple arguments \cite{mg15}:
i) the PMTs are kept in high purity (HP; 5.5 grade) Nitrogen atmosphere, the gas being continuously flushed
through the apparatus ($\simeq$ 250 liters / hour) \cite{perflibra}; thus no Helium accumulation
process can take place (typical characteristic time for accumulation of Helium in PMTs
through glass -- considering the permeability of the materials -- is $\simeq 1$ year). An 
estimate of Helium concentration within the DAMA shield is less than $5 \times 10^{-11}$ ppm.
Thus, any hypothetical effect due to He-correlated events is negligible;
ii) any migration of Helium into PMTs would cause their irreversible
degradation which has not been 
observed: e.g. the dark noise of the PMTs ranges from 40 to 500 Hz \cite{pmts} and 
improves over time on the contrary of what it is expected by any hypothetical He 
migration inside PMTs; iii) the PMTs used in DAMA/LIBRA--phase2 and DAMA/LIBRA--phase1 are 
different and with different voltage values on the first dynode; in the He erroneous conjecture 
this should have produced different modulation amplitudes in the two phases, which has not been observed;
iv) this conjecture needs that the He is modulated and with the same phase and period as
the dark matter, but getting the right phase and period over 20 
annual cycles is practically excluded.}.
Noteworthy, as already pointed out 
in Refs. \cite{Freese1,Freese2}, this signature acts as an efficient background rejection 
procedure and does not require any identification of $\mathcal{S}_0$ from the total counting rate
in order to establish the presence of DM particles in the galactic halo.
Therefore, the DM annual modulation signature allows
one to overcome -- in the identification of the existence of a signal -- the large uncertainties associated to: i) 
many statistical data selections/subtractions/discrimination
procedures; ii) strongly uncertain modeling of the background in particular in keV region; 
iii) {\it a priori} assumption on the nature, interaction type, etc. of the DM particle(s). 
On the other hand, it requires an uncontested stability at level of less than 1\% of the operational experimental parameters.

In Table \ref{T1} the experimental modulation amplitudes, $\mathcal{S}_m$, measured by DAMA/NaI,
DAMA/LIBRA--phase1 and DAMA/LIBRA--phase2 are shown;
the data below 2 keV refer, instead, only to the DAMA/LIBRA--phase2 \cite{npae18}.

\begin{table}[!ht]
        \caption{Experimental modulation amplitudes, $\mathcal{S}_{m}$, measured by DAMA/NaI, DAMA/LIBRA--phase1 and DAMA/LIBRA--phase2
                (total exposure 2.46 ton$\times$yr); data below 2 keV refer, instead, only to the DAMA/LIBRA--phase2 (exposure
                1.13 ton$\times$yr) \cite{npae18}.}
        \begin{center}
                \begin{tabular}{|cc|cc|}
                        \hline
                        Energy          &   $\mathcal{S}_m$ (cpd/kg/keV)   &   Energy          &   $\mathcal{S}_m$ (cpd/kg/keV)   \\
                        \hline
                        (1.0--1.5) keV  &   (0.0232$\pm$0.0052)  &   (6.5--7.0) keV  &   (0.0016$\pm$0.0018)  \\
                        (1.5--2.0) keV  &   (0.0164$\pm$0.0043)  &   (7.0--7.5) keV  &   (0.0007$\pm$0.0018)  \\
                        (2.0--2.5) keV  &   (0.0178$\pm$0.0028)  &   (7.5--8.0) keV  &   (0.0016$\pm$0.0018)  \\
                        (2.5--3.0) keV  &   (0.0190$\pm$0.0029)  &   (8.0--8.5) keV  &   (0.0014$\pm$0.0018)  \\
                        (3.0--3.5) keV  &   (0.0178$\pm$0.0028)  &   (8.5--9.0) keV  &   (0.0029$\pm$0.0018)  \\
                        (3.5--4.0) keV  &   (0.0109$\pm$0.0025)  &   (9.0--9.5) keV  &   (0.0014$\pm$0.0018)  \\
                        (4.0--4.5) keV  &   (0.0110$\pm$0.0022)  &  (9.5--10.0) keV  &  -(0.0029$\pm$0.0019)  \\
                        (4.5--5.0) keV  &   (0.0040$\pm$0.0020)  & (10.0--10.5) keV  &   (0.0035$\pm$0.0019)  \\
                        (5.0--5.5) keV  &   (0.0065$\pm$0.0020)  & (10.5--11.0) keV  &  -(0.0038$\pm$0.0019)  \\
                        (5.5--6.0) keV  &   (0.0066$\pm$0.0019)  & (11.0--11.5) keV  &  -(0.0013$\pm$0.0019)  \\
                        (6.0--6.5) keV  &   (0.0009$\pm$0.0018)  & (11.5--12.0) keV  &  -(0.0019$\pm$0.0019)  \\
                        \hline
                \end{tabular}
        \end{center}
        \label{T1}
\end{table}

The aim of the present paper is to update the implications on several models (of the many available in literature) 
we already investigated with lower exposure and higher software energy threshold
with the data previously collected with DAMA/NaI and DAMA/LIBRA-phase1. 

The plan of the paper is as follows. In Sect. \ref{data_analysis} the description of the data analysis
and the inclusion of the uncertainties used in the evaluation of the allowed regions are described.
In Sect. \ref{update} the results achieved for the considered scenarios of DM particles are reported:
i)   DM particles which elastically interact with target nuclei with Spin-Independent (SI) or Spin-Dependent (SD) or mixed coupling (Sect. \ref{DM1}); 
ii)  DM particles with preferred electron interaction (Sect. \ref{DM2}); 
iii) DM particles with preferred inelastic scattering (Sect. \ref{DM3}); 
iv)  Light DM (Sect. \ref{DM4});
v)   asymmetric and symmetric Mirror DM (Sect. \ref{DM5}). 
Moreover, some of the many other interesting scenarios 
available in literature are introduced for the first time.
Finally, Sect. \ref{concl} is devoted to our conclusions.

\section{Data analysis}
\label{data_analysis}

As mentioned the corollary analyses presented here are model-dependent; thus, it
is important to point out at least the main topics which enter
in the determination of results and the related uncertainties. 
These arguments have been already addressed at various extents in previous 
corollary model-dependent analyses. The DM candidates considered here have been previously discussed in the Refs.
\cite{review,RNC,ijmd,bot11,mirasim,mirsim,bot99,dama_inel,bel02,ijma,epj06,ijma07,chan,wimpele,ldm}.

As first, in order to derive the allowed regions of the parameter's space of the DM particles
in the considered scenarios, a specific phase-space
distribution function (DF) of each DM candidate in the
galactic halo has to be adopted. A large number of possibilities
is available in literature; these models are continuously in evolution thanks to the
new simulations and new astrophysical observations, as the recent GAIA ones (see e.g. Refs. \cite{gre17,eva18} and references therein).  
Thus, large uncertainties in the predicted theoretical rate are present. 

In this paper, to account at some extent for the uncertainties in halo models, 
we consider the same not-exhaustive set of halo models as in our previous published analyses \cite{RNC,ijmd,bel02}, for all the considered DM candidates.
This can give an idea of the role of some astrophysical 
uncertainties on DM model-dependent analyses and offers a direct impact on the increasing of 
the exposure and of the lowering of the software energy threshold, achieved with DAMA/LIBRA-phase2. 
Further analyses in other frameworks, such as e.g. 
exploiting recent GAIA data, can be possible in dedicated future papers.
The considered models are summarized in Table \ref{tb:halo}.
\begin{table}[!hbt]
\begin{center}
\vspace{-0.3cm}
\caption{
  Summary of the considered consistent halo models \cite{bel02,RNC}.
  The labels in the first column identify the models.
  In the third column the values of the related considered parameters are reported \cite{bel02,RNC};
  other choices are also possible as well as other halo models.
  The models of the Class C have also been considered including possible co--rotation and counter-rotation of the dark halo.
}
\vspace{0.3cm}
\begin{tabular}{|c|l|c|}
\hline\hline
\multicolumn{3}{|l|}{{\bf Class A:  spherical $\bf \rho_{dm}$,
isotropic velocity dispersion}} \\
\hline
A0 & {\rm ~Isothermal Sphere}   &     \\
A1 & {\rm ~Evans' logarithmic}  & $R_c=5$ kpc \\
A2 & {\rm ~Evans' power-law}  & $R_c=16$ kpc, $\beta=0.7$ \\
A3 & {\rm ~Evans' power-law}  & $R_c=2$ kpc, $\beta=-0.1$ \\
A4 & {\rm ~Jaffe}               & $\alpha=1$, $\beta=4$,
$\gamma=2$, $a=160$ kpc \\
A5 & {\rm ~NFW}                  & $\alpha=1$, $\beta=3$,
$\gamma=1$, $a=20$ kpc \\
A6 & {\rm ~Moore et al.}    & $\alpha=1.5$, $\beta=3$,
$\gamma=1.5$, $a=28$ kpc  \\
A7 & {\rm ~Kravtsov et al.} & $\alpha=2$, $\beta=3$,
$\gamma=0.4$, $a=10$ kpc   \\
\hline
\multicolumn{3}{|l|}{{\bf Class B: spherical $\bf \rho_{dm}$,
non--isotropic velocity dispersion    }} \\
\multicolumn{3}{|l|}{{\bf (Osipkov--Merrit, $\bf \beta_0=0.4$)}} \\
\hline
B1 & {\rm ~Evans' logarithmic} & $R_c=5$ kpc \\
B2 & {\rm ~Evans' power-law} & $R_c=16$ kpc, $\beta=0.7$  \\
B3 & {\rm ~Evans' power-law} & $R_c=2$ kpc, $\beta=-0.1$  \\
B4 & {\rm ~Jaffe}           & $\alpha=1$, $\beta=4$,
$\gamma=2$, $a=160$ kpc  \\
B5 & {\rm ~NFW}             & $\alpha=1$, $\beta=3$,
$\gamma=1$, $a=20$ kpc   \\
B6 & {\rm ~Moore et al.}    & $\alpha=1.5$, $\beta=3$,
$\gamma=1.5$, $a=28$ kpc   \\
B7 & {\rm ~Kravtsov et al.} &  $\alpha=2$, $\beta=3$,
$\gamma=0.4$, $a=10$ kpc    \\
\hline
\multicolumn{3}{|l|}{{\bf Class C:  Axisymmetric $\bf \rho_{dm}$}} \\
\hline
C1 & {\rm ~Evans' logarithmic} & $R_c=0$, $q=1/\sqrt{2}$ \\
C2 & {\rm ~Evans' logarithmic} & $R_c=5$ kpc, $q=1/\sqrt{2}$ \\
C3 & {\rm ~Evans' power-law} & $R_c=16$ kpc, $q=0.95$, $\beta=0.9$ \\
C4 & {\rm ~Evans' power-law} & $R_c=2$ kpc, $q=1/\sqrt{2}$, $\beta=-0.1$\\
\hline
\multicolumn{3}{|l|}{{\bf Class D: Triaxial $\bf \rho_{dm}$
  ($\bf q=0.8$, $\bf p=0.9$)}} \\
\hline
D1 & {\rm ~Earth on maj. axis, rad. anis.}    & $\delta=-1.78$  \\
D2 & {\rm ~Earth on maj. axis, tang. anis. }    &   $\delta=16$ \\
D3 & {\rm ~Earth on interm. axis, rad. anis.}  &  $\delta=-1.78$ \\
D4 & {\rm ~Earth on interm. axis, tang. anis.} & $\delta=16$ \\
\hline\hline
\end{tabular}
\label{tb:halo}
\end{center}
\vspace{-0.5cm}
\end{table}
In particular, the considered
classes of halo models correspond to: (1) spherically symmetric
matter density with isotropic velocity dispersion (Class A);
(2) spherically symmetric matter density with non-isotropic
velocity dispersion (Class B); (3) axisymmetric models (Class C); (4) triaxial
models (Class D); (5) moreover, in the case of axisymmetric
models it is possible to include either an halo co-rotation or
an halo counter-rotation.

In our analysis we also consider the physical ranges of the
main halo parameters: the local total DM density, $\rho_{0}$, and the
local velocity $v_0$ as discussed in Ref. \cite{bel02}. 
The range of the possible $v_0$ values is from 170 km/s
to 270 km/s. For $\rho_{0}$, its minimal,
$\rho^{min}_{0}$, and its maximal, $\rho^{max}_{0}$, values
are estimated imposing essentially two astrophysical constraints: 
one on the amount of non-halo components and the other on the flatness of the rotational curve
in the Galaxy; for a detailed procedure see Ref. \cite{bel02}.

The values for $\rho^{min}_{0}$ and $\rho^{max}_{0}$
are related to the DF and the considered $v_0$; they are reported in Table III of Ref. \cite{bel02}.
The halo density $\rho_{0}$ ranges from 0.17 to 0.67 GeV/cm$^{3}$ for $v_0$ = 170 km/s,
while $\rho_{0}$ ranges from 0.29 to 1.11 GeV/cm$^{3}$ for $v_0$ = 220 km/s, and 
$\rho_{0}$ ranges from 0.45 to 1.68 GeV/cm$^{3}$ for $v_0$ = 270 km/s, depending on
the halo model.

To take into account that the considered DM candidate can be just one of the components of the dark halo,
the $\xi$ parameter is introduced; it is defined as the fractional amount of local density in terms of the considered
DM candidate ($\xi\leq 1$).
Thus, the local density of the DM particles is $\rho_{DM} = \xi \rho_0$.

Finally, we consider the DM escape velocity, $v_{esc}$, from the galactic gravitational potential; 
actually, it is also affected by
significant uncertainty:
$(528^{+24}_{-25})$ km/s \cite{dea19},
$498 < v_{esc} < 608$ km/s (90\% C.L.), with a median likelihood of 544 km/s \cite{smi07},
$(533^{+54}_{-41})$ km/s (90\% C.L.) \cite{pif14},
$(521^{+46}_{-30})$ km/s \cite{belo17}, and
$(580 \pm 63)$ km/s \cite{mon18}. 
In the following analysis $v_{esc}$ = 550 km/s is adopted, as often considered in literature.
However, no sizable differences are observed
in the final results when $v_{esc}$ values ranging from 550 to 650 km/s are considered.
In particular, for low-mass DM particles scattering off nuclei, the Na contribution is dominant and has
a small dependence on the tail of the velocity distribution.

In addition, it is also possible the presence of non-virialized components,
as streams in the dark halo coming from external sources
with respect to our Galaxy \cite{epj06,Fre04_1,Fre04_2} or other scenarios as e.g. that of Ref. \cite{caus,Fre05,Gel01}; 
however, these latter possibilities are not included in the present analyses.

In conclusion, to properly evaluate the allowed regions in the parameters' space of particle DM scenarios
it is limiting only considering
an isothermal profile\footnote{It is also worth noting that the isothermal halo is an unphysical model;
for example, the mass would diverge and one has to adopt a by-hand
cut-off. Let us remark, however, that flat density profile for the Galaxy
within the radius of 10 kpc can be obtained if the DM particles have
self-interaction cross-section $\sigma/M$ $\simeq 10^{-24} - 10^{-23}$ cm$^2$/GeV \cite{sper2000,Wand2001}.} with 
local parameters $v_0$ = 220 km/s and $\rho_{0}$ = 0.3 GeV/cm$^{3}$ without taking 
in consideration at least some of the other existing possibilities in the distribution of
velocity and spatial coordinates permitted by astrophysical
observations.

In the interaction of DM particles in the NaI(Tl) detectors the detected energy, $E_{det}$, is a key quantity.
It is connected with the energy released by the products of the interaction, $E_{rel}$;
two possibilities exist: 1) the products of the interaction have electromagnetic nature (mainly electrons);
2) a nuclear recoil with $E_R$ kinetic energy is produced by the DM particle scattering either off Sodium or off Iodine nucleus.
Since, the detectors are calibrated by $\gamma$ sources, in the first case $E_{det} = E_{rel}$, while 
in the second case a quenching factor for each recoiling nucleus must be included: $E_{det} = q_{Na,I} \times E_{rel}$. 
For completeness, we also recall that the energy resolutions of each detector in the two configurations (phase1 and phase2) 
are shown e.g. in Ref. \cite{pmts}.

\subsection{The case of DM particles inducing nuclear recoils}
\label{intro_rec}

The quenching factors are a property of the specific detector and not general properties of any NaI(Tl),
particularly in the very low energy range. In fact, in NaI(Tl) they depend on the adopted growing procedures, on Tl
concentration and uniformity in the detector, 
on the specific additives always used by companies to
strengthen the performance of the detectors, on the monocrystalline or polycrystalline nature of the NaI(Tl) crystal, 
etc. Moreover, their measurements are difficult and always affected by significant 
experimental uncertainties. 
All these aspects are always relevant sources of uncertainties when comparing whatever
results in terms of DM candidates inducing nuclear recoils.
Naively summarizing, different quenching factors values
imply that the same energy in keV electron equivalent corresponds to different recoil energies in the different 
experiments.

Arguments on various quenching factors determinations have already been addressed by us e.g. in Refs.
\cite{RNC,bot11,mirasim,mirsim,ldm}. It is worth noting that recently Ref. \cite{KIMqf} gave 
quenching factors for a small COSINE-100 like detector; in particular: Na quenching factor ranging from 0.1 to 0.23 with a significant 
energy dependence, 
and I quenching factor ranging from 0.04 to 0.06 were reported; a very high precision is quoted. 
However, these values cannot be consistently considered for other detectors because of the above mentioned arguments; in particular, those
crystals have been grown by different technique and protocols  than
those of DAMA/LIBRA. For example, the energy of the internal $\alpha$'s in those detectors roughly 
ranges between 2.3 and 3.0 MeV electron equivalent \cite{epjc2018}, 
while in DAMA/LIBRA it ranges between 2.6 e 4.5 MeV electron equivalent \cite{perflibra}, indicating a lower 
quenching factor for $\alpha$'s 
in COSINE-100 like detectors; thus, a much lower quenching factors at keV region is implied as well.

In literature one can find a lot of measurements on the Na and I quenching factors that, owing to the above considerations, show a wide 
spread.
It is evident also in Fig. 10 of Ref. \cite{KIMqf}, where systematically poorer quenching factors are obtained for crystals recently developed 
(with different technology and materials) with respect to previous measurements\footnote{For example, Ref. \cite{Toy1998} reports Na 
quenching factor substantially constant with energy and significantly higher than Ref. \cite{KIMqf} with a very
good precision as well. Moreover, Ref. \cite{KIMqf} claims agreement with Ref. \cite{Xu2015}, but this
latter reference gives mean values systematically higher.}.

In the following the same procedures previously adopted in Refs. \cite{bot11,mirasim,mirsim,ldm} are considered. 
This also allows us to point out -- by direct comparison 
with
previously published results -- the effect of increasing the
exposure and decreasing the energy threshold. Three possible instances can be considered:
\begin{itemize}
\item
$(Q_I)$ quenching factors of Na and I ``constants'' with respect to the recoil energy
$E_R$: the adopted values are $q_{Na} = 0.3$ and $q_I = 0.09$, measured
with neutron source integrating the data over the 6.5 -- 97 keV and
the 22 -- 330 keV recoil energy range, respectively \cite{psd96};
\item $(Q_{II})$  quenching factors varying as a function of E$_R$ evaluated as in Ref. \cite{Tretyak};
\item $(Q_{III})$ quenching factors with the same behavior of Ref. \cite{Tretyak}, but normalized in order to have 
their
mean values consistent with $Q_I$ in the energy range considered there.
\end{itemize}
Moreover, to account for the uncertainties on the measured quenching factors some discrete cases of possibilities 
will also be introduced
at the end of this Section.
For long time in the field the quenching factors have been considered locally constant with energy. 
On the contrary, the $(Q_{II})$, $(Q_{III})$ instances use energy dependent quenching factors
following the phenomenological prescription of Ref. \cite{Tretyak}.

Another important effect is the channeling of low energy ions along axes and planes of the NaI(Tl) DAMA crystals. 
This effect can lead to a further important deviation, in addition to the uncertainties discussed in section II of 
Ref. \cite{bot11} and in Ref. \cite{mirasim}. In fact, the channeling effect in crystals implies that a fraction of nuclear recoils 
are channeled and experience much larger quenching factors than those derived from neutron calibration (see Refs. \cite{chan,bot11}
for a discussion of these aspects). Anyhow, the channeling effect in solid crystal detectors is not a well fixed issue.
There could be several uncertainties in the modeling. Moreover, the experimental approaches (as that in Ref. \cite{collar_qnai}) 
are rather difficult since the channeled nuclear recoils are -- even in the most optimistic model -- a very tiny fraction 
of the not-channeled ones. In particular, the modeling of the channeling effect described in Ref. \cite{chan},
where the recoiling nuclei are considered free in the lattice, 
is able to reproduce the recoil spectrum measured at neutron beam by some other groups \cite{chan}. 
For completeness, we mention: 
i) the alternative channeling model of Ref. \cite{Mat08}, where larger probabilities of the planar 
channeling are expected; 
ii) the analytic calculation of Ref. \cite{gelmini}, where it is claimed that the channeling effect holds for recoils coming from outside 
a crystal and not from recoils from lattice sites, due to the blocking effect. Nevertheless, although some  
amount of blocking effect could be present, the precise description of the crystal lattice with dopant and trace contaminants 
is quite difficult and analytical calculations require some simplifications, which can affect the result. Because of the 
difficulties of experimental measurements and of theoretical estimate of the channeling effect, in the following 
it will be either included using the procedure given in Ref. \cite{chan} or not in order to give idea on the related uncertainty.

In case of low mass DM particles giving rise to nuclear recoils,
the Migdal effect (discussed in details in Refs. \cite{ijma07,mirasim}, where the impact in some corollary analyses 
was discussed) can also be considered.

Finally, some discrete cases are considered in the following to cautiously account for
possible uncertainties on the quenching factors measured by DAMA in its detectors and on
the parameters used in the SI and SD nuclear form factors, as already
done in previous analyses. Three cases are considered:
\begin{itemize}
\item Set A considers the mean values of the parameters of the used nuclear form factors \cite{RNC} and of the quenching 
factors.
\item Set B adopts the same procedure as in Refs. \cite{sisd,dama_inel,ijmd}, by varying (i) the mean values of the 
$^{23}$Na
      and $^{127}$I quenching factors as measured in Ref. \cite{psd96} up to $+2$
      times the errors; (ii) the nuclear radius, $r_A$, and the nuclear
      surface thickness parameter, $s$, in the SI nuclear form factor from their
      central values down to $-20\%$; (iii) the $b$ parameter in the considered SD nuclear
      form factor from the given value down to $-20\%$.
\item Set C where the Iodine nucleus parameters are fixed at the values of set B,
      while for the Sodium nucleus one considers \cite{RNC}: (i) $^{23}$Na quenching factor at
      $q_{Na} = 0.25 $\footnote{This value offers backward compatibility with previous similar model-dependent 
      DAMA studies and
      a safe realistic representation of possible uncertainties in the $q_{Na}$ measured for the DAMA detectors.
      };
      (ii) the nuclear radius, $r_A$, and the nuclear surface thickness parameter,
      $s$, in the SI nuclear form factor from their central values up to $+20\%$;
      (iii) the $b$ parameter in the considered SD nuclear form factor from the given value up to $+20\%$.
\end{itemize}

\subsection{The analysis procedure}

Model-dependent corollary analyses 
through a maximum likelihood procedure, which also takes into account the
energy behavior of each detector, can be pursued.

In the following for each considered scenario, the allowed domains in the corresponding parameters' space 
will be obtained by marginalizing over the halo models of Table \ref{tb:halo}, 
over halo parameters ($v_{0}$ and $\rho_0$) and over the sets A, B, C\footnote{In particular, each allowed domain 
encloses all the allowed regions 
obtained for each chosen configuration of model and parameters.}.
This procedure shows the impact of the uncertainties in the astrophysical, nuclear and particle physics 
on the model-dependent analyses.

However, for simplicity the allowed regions in the parameters' space of each considered scenario
can also be derived by comparing -- for each $k$-$th$ energy bin of 1 keV -- the measured DM
annual modulation amplitude, $\mathcal{S}^{exp}_{m,k} \pm \sigma_k$ \footnote{
The distributions of the measured modulation amplitudes around their mean value
show a perfect Gaussian behaviors, justifying the use of a symmetric uncertainty \cite{modlibra,review,modlibra3,uni18,npae18}.},
with the theoretical expectation in each considered framework, $\mathcal{S}_{m,k}^{th}$. 
Of course, the $\mathcal{S}^{th}_{m,k}$ values depend on the free parameters of the model $\bar{\theta}$,
such as the DM particle mass, the cross section, etc., on the uncertainties accounted for, on the proper accounting for the 
detector's features, and on priors.

As mentioned in previous works (as e.g. recently in Refs. \cite{mirasim,mirsim}),
a cautious prior on $\mathcal{S}_{0,k}$ -- assuring safe and more realistic allowed regions/volumes -- can be worked out from
the measured counting rate in the cumulative
energy spectrum; the latter is given by the sum of the un-modulated
background contribution $b_k$ (whose existence is shown by the detailed analyses on residual radioactive contaminations in the detectors
\cite{perflibra}) and of the constant part of the
signal $\mathcal{S}_{0,k}$. By adopting a standard procedure, used in the past in several low background fields, 
one can derive lower limits on $b_k$ and, thus, upper limits on $\mathcal{S}_{0,k}$ ($\mathcal{S}_{0,k}^{max}$). In particular, in DAMA/LIBRA--phase2
is obtained: $\mathcal{S}_0 \lsim 0.80$ cpd/kg/keV in the
(1-2) keV energy interval; $\mathcal{S}_0 \lsim 0.24$ cpd/kg/keV in 
(2-3) keV, and $\mathcal{S}_0 \lsim 0.12$ cpd/kg/keV
in (3-4) keV \footnote{Disregarding this prior is one of the main critical issues in model-dependent analyses by other authors.}.

Thus, the following $\chi^2$ can be calculated for each considered model:

\begin{equation}
\chi^2(\bar{\theta}) = \sum_k \frac{\left(\mathcal{S}_{m,k}^{exp} - \mathcal{S}_{m,k}^{th}(\bar{\theta})\right)^2}{\sigma_k^2} + 
\sum_{k'} \frac{\left(\mathcal{S}_{0,k'}^{max} - \mathcal{S}_{0,k'}^{th}(\bar{\theta})\right)^2}{\sigma_{0,k'}^2} 
\Theta\left(\mathcal{S}_{0,k'}^{th}(\bar{\theta}) - \mathcal{S}_{0,k'}^{max}\right)
\label{eq:chi2}
\end{equation}

\noindent where the second term encodes the experimental bounds about the un-modulated part of the signal;
$\sigma_{0,k'} \simeq 10^{-3}$ cpd/kg/keV, 
$\Theta$ is the Heaviside function, and
$\mathcal{S}_{0,k'}^{th}$ is the average expected signal counting rate in the $k'$ energy bin.
The sum in the first term in eq. \ref{eq:chi2} runs here from 1 keV to 20 keV.

The $\chi^2$ defined in eq. (\ref{eq:chi2}) can be calculated in each considered framework and 
is function of the model parameters $\bar{\theta}$.
Thus, we can define:

\begin{equation}
\Delta \chi^2 (\bar{\theta}) = \chi^2 (\bar{\theta}) - \chi^2_0
\end{equation}

\noindent where $\chi^2_0$ is the $\chi^2$ for $\bar{\theta}$ values corresponding to absence of signal.
The $\Delta \chi^2$ is used to determine the allowed intervals of the model parameters $\bar{\theta}$
at 10 $\sigma$ from the {\it null signal hypothesis}.

It is worth noting that the results presented in the following
are, of course, not exhaustive of the many possible scenarios.
For example, the possible contribution of non-thermalized component in the Dark Halo, 
which would extend the allowed regions of the DM particle's parameters,
is not included in the present paper.

Moreover, the improvement in the energy threshold achieved by DAMA/LIBRA--phase2 
at 1 keV prevents to find configurations due to the Migdal effects, which were
instead present with the 2 keV energy threshold data.
This is an example of the relevance of lowering the energy threshold to 
disentangle at least among some of the possible scenarios.

Finally, we have verified that the $Q_{III}$ option for the quenching factors provides
results similar to the case of the $Q_{I}$ option; thus, to avoid the overloading of the figures
in the following the $Q_{III}$ case is not considered.

\section{Updated corollary model-dependent scenarios}
\label{update}

In the following we will present the updated results for DM candidates in the frameworks described above,
using the total exposure of 2.46 ton $\times$ yr for the data from 2 to 6 keV
and of 1.13 ton $\times$ yr for the data below 2 keV.

\subsection{DM particles elastically interacting with target nuclei}
\label{DM1}

A lot of candidates have been proposed in theory extending the Standard Model of particles 
that includes candidates for Dark Matter elastically scattering off target nuclei.

In the DM particle-nucleus elastic scattering,   
the differential energy distribution of the recoil nuclei
can be calculated 
by means of the differential cross section of the DM-nucleus elastic
process \cite{psd96,bot99,RNC,ijmd,chan}. 
The latter is given by the sum of two contributions: 
the SI and the SD one.

In the purely SI case, the nuclear parameters can be decoupled from the
particle parameters and the nuclear cross sections, which are derived quantities, are usually scaled to a defined
point-like SI DM particle-nucleon cross section, $\sigma_{SI}$.
In principle, this procedure could allow -- within a framework of several other assumptions (that in turn introduce uncertainties in final evaluations) --
a model-dependent comparison among different target nuclei, otherwise impossible.
In the following, the usually considered coherent scaling law for the nuclear cross sections is adopted:
\begin{equation}
\sigma_{SI}(A,Z) \propto m_{red}^2(A,DM) \left[f_pZ + f_n(A-Z)\right]^2,
\end{equation}
where $\sigma_{SI}(A,Z)$ is the point-like cross section of DM particles scattering off 
nuclei of mass number A and atomic number Z, 
$m_{red}(A,DM)$ is the reduced
mass of the system DM particle and nucleus, $f_p$ and $f_n$ are the effective DM particle couplings to
protons and neutrons, respectively.
The case of isospin violation $f_p \neq f_n$ will be discussed in Sect. \ref{sec:isovio}; now we assume $f_p = f_n$
and, thus, we can write\footnote{It was also  
proposed that two-nucleon currents from pion exchange in the nucleus  
can give different contribution for nuclei with different atomic number \cite{pre03}, as  
a consequence the cross-section for some nuclei can be enhanced with respect to  
others. Also similar arguments have a great relevance in the model-dependent  
comparisons.}:
\begin{equation}
\sigma_{SI}(A,Z) = \frac{m_{red}^2(A,DM)}{m_{red}^2(1,DM)} A^2 \sigma_{SI}.
\end{equation}

As for nuclear SI form factors, the Helm form factor \cite{Helm1,Helm2} has been adopted\footnote{It should be 
noted that the Helm form factor is the least favorable one
e.g. for Iodine and requires larger SI cross-sections for a given signal
rate; in case other form factor profiles, considered in the literature, would
be used, the allowed parameters' space would extend \cite{RNC}.}.
Details on the used form factors can also be found in Ref. \cite{RNC}. 
As described above, some uncertainties on the nuclear radius and on the nuclear surface thickness parameters 
in the Helm SI form factors have been included in the following analysis by considering three discrete cases, labeled as 
set A, B, and C in Sect. \ref{intro_rec}.

The purely SD case is even more uncertain
since the nuclear and particle physics degrees of freedom cannot be decoupled and a dependence
on the assumed nuclear potential exists. 
Also in the purely SD case all the nuclear cross sections are usually scaled to a defined
point-like SD Dark Matter particle-nucleon cross section, $\sigma_{SD}$ \cite{sisd,RNC}.
The adopted scaling law for this case profits of the proportionality of the 
SD nuclear cross section to the nuclear spin factor 
$\Lambda^2 J(J+1)$ and to the squared reduced mass. To take into account the finiteness of the nucleus,
a SD nuclear form factor is also used; for details of its parametrization used in the following
see Ref. \cite{RNC}. A further parameter must be introduced;
in fact, following the notations reported in Ref. \cite{sisd}:
$tan \theta = \frac{a_n}{a_p}$, 
where $a_{p,n}$ are the effective DM-nucleon coupling strengths for SD interactions.
The mixing angle $\theta$ is defined in the $\left[ 0, \pi \right)$ interval;
in particular, $\theta$ values in the second sector account for $a_p$
and $a_n$ with different signs. 
Therefore, further significant uncertainties in the evaluation of the SD interaction rate also arise
from the adopted spin factor for the single target-nucleus.
In fact, the available calculated values are well different in different models (and 
differently
vary for each nucleus) and, in addition,
at  fixed model they depend on $\theta$ \cite{sisd,RNC}. 

It is worth noting that for the SD part of the interaction 
not only the target nuclei should have spin different from zero
(for example, this is not the case of Ar isotopes, and most of the Ca, Ge, Te, Xe, W isotopes)
to be sensitive to DM particles with a SD component in the coupling, but also
well different sensitivities can be expected among
odd-nuclei having an unpaired proton (as e.g. $^{23}$Na and $^{127}$I, and $^{1}$H, $^{19}$F, $^{27}$Al, $^{133}$Cs)
and odd-nuclei having an unpaired neutron (as e.g. the odd Xe and Te isotopes and $^{29}$Si, $^{43}$Ca, $^{73}$Ge, $^{183}$W).

In conclusion, the free parameters, once fixed the assumptions for the model framework, are the DM particle mass, $m_{DM}$,
the $\xi\sigma_{SI}$ for the purely SI case,
the $\xi\sigma_{SD}$ and $\theta$ for the purely SD case.
Therefore, in the SI case the allowed regions are presented in the plane  $\xi\sigma_{SI}$ vs $m_{DM}$, while
in the SD case $\xi\sigma_{SD}$, $\theta$ and $m_{DM}$ give rise to 3-dimensional allowed volume 
of which generally only slices in the plane $\xi\sigma_{SD}$ vs $m_{DM}$ are depicted at 
fixed $\theta$ values.
Obviously the situation is even more complex when the SI and SD mixed case is considered and both the large uncertainties 
existing for the SI and SD interactions are present. 
In this general scenario the data give rise to an allowed volume in
the 4-dimensional space ($m_{DM}$, $\xi \sigma_{SI}$,
$\xi \sigma_{SD}$, $\theta$); practically just some slices of this 4-dimensional allowed volume
in the plane
$\xi \sigma_{SI}$ vs $\xi \sigma_{SD}$ for some of the possible
$\theta$ and $m_{DM}$ values in some of the possible model frameworks are depicted.

\subsubsection{Spin-Independent interaction}

\begin{figure}[!ht]
\begin{center}
\vspace{-0.2cm}
\includegraphics[width=10.cm] {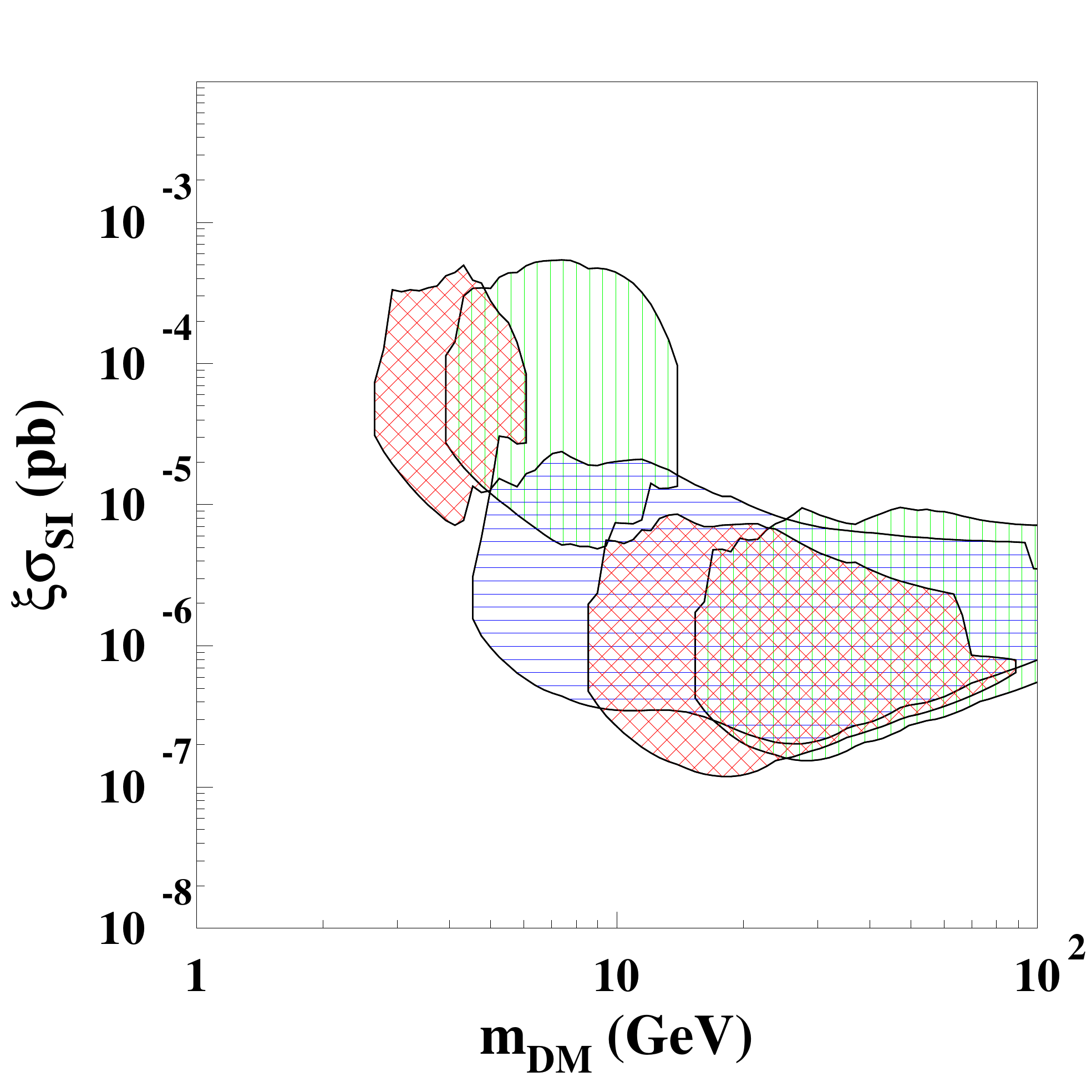}
\end{center}
\vspace{-0.8cm}
\caption{Regions in the nucleon cross-section vs DM particle mass plane allowed by DAMA experiments
in the case of a DM candidate elastically scattering off target nuclei and SI
interaction.
Three different instances for the Na and I quenching factors have been considered: 
(i) $Q_I$ case [(green on-line) vertically-hatched region],
(ii) with channeling effect [(blue on-line) horizontally-hatched region)] and 
(iii) $Q_{II}$ [(red on-line) cross-hatched region].
The regions have been obtained by marginalizing all the models for each considered scenario 
(see Sect. \ref{data_analysis}) and they represent the domain where the likelihood-function 
values differ more than 10 $\sigma$ from the {\it null hypothesis} (absence of modulation).}
\label{fg:si}
\end{figure}

Often the purely SI interaction with ordinary matter is assumed to be dominant.
In addition, most of the used target-nuclei are practically
not sensitive to SD interactions (on the contrary to
$^{23}$Na and $^{127}$I) and the theoretical calculations
and comparisons are even much more complex and uncertain.
Therefore, for the purely SI scenario in the considered model frameworks the allowed region in the plane
$m_{DM}$ and $\xi \sigma_{SI}$ have been calculated and shown in Fig. \ref{fg:si}.
Of course, best fit values of cross section and DM mass
span over a large range in the considered model frameworks.

The allowed domains in Fig. \ref{fg:si} are obtained 
by marginalizing all the models for each considered
scenario (see Sect. \ref{data_analysis}); they represent the domains where the likelihood-function values differ more than 10 $\sigma$ from 
absence of signal.
The three different instances described above for
the Na and I quenching factors have been considered: 
(i) $Q_I$ case, (ii) with channeling effect, and (iii) $Q_{II}$.

\begin{figure}[!ht]
\begin{center}
\vspace{-0.3cm}
\includegraphics[width=6.0cm] {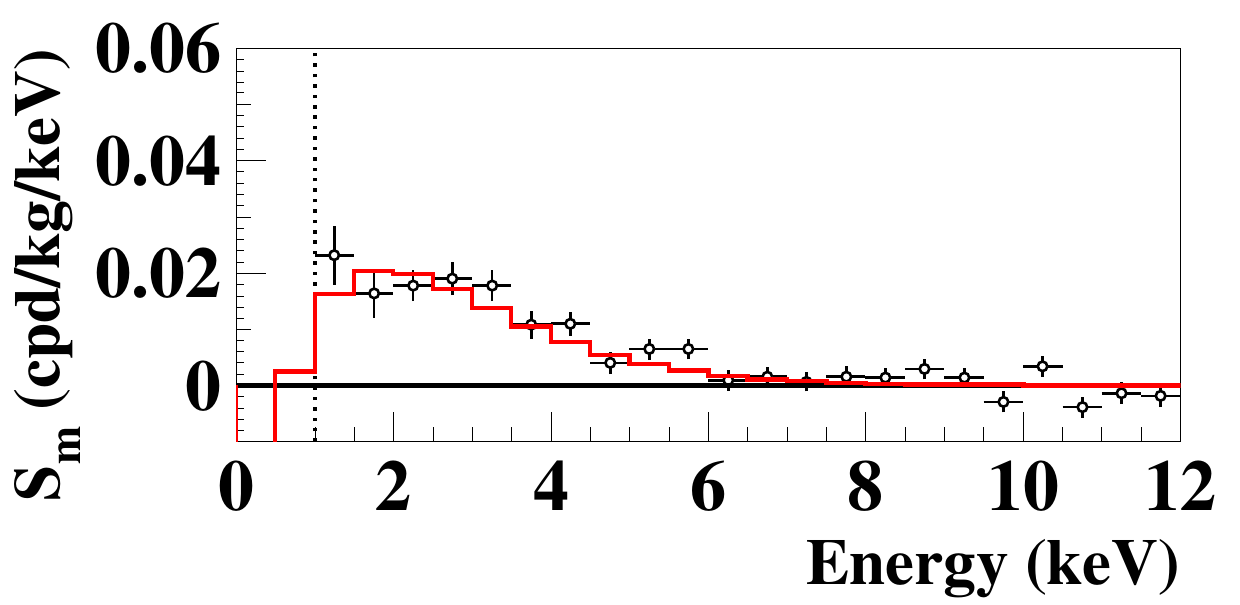}
\includegraphics[width=6.0cm] {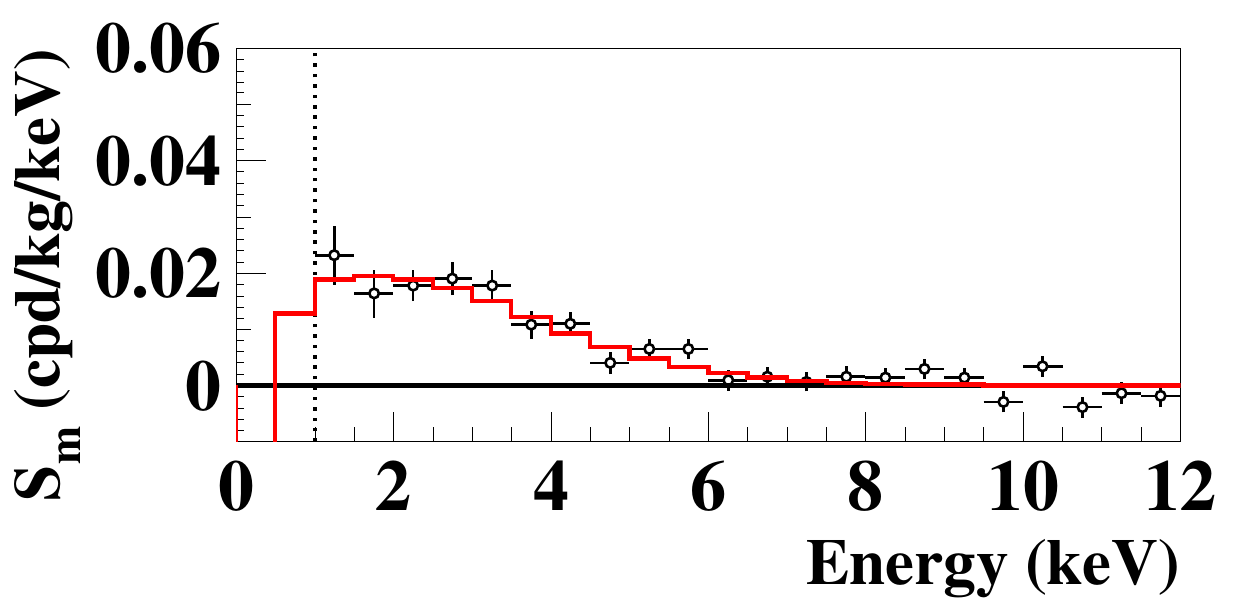}
\end{center}
\vspace{-0.5cm}
\caption{Examples of superposition of the measured $\mathcal{S}^{exp}_{m}$ vs energy (points 
with error bars) with theoretical expectations (solid histograms) for the SI case with 
quenching factors $Q_I$.
$Left:$ case of the A1 (Evans logarithmic) halo model with $\rho_0$ = 0.2 GeV/cm$^3$, 
$v_0$ = 170 km/s and set B of parameters values;
$Right:$ case of the C2 (Evans logarithmic) halo model with $\rho_0$ = 0.67 GeV/cm$^3$, 
$v_0$ = 170 km/s and set C of parameters values.
In both cases The mass of the DM particle is 60 GeV and $\xi\sigma_{SI}$ is equal to 3.9 
$\times$ 10$^{-6}$ pb and to 1.3 $\times$ 10$^{-6}$ pb, respectively.}
\label{fg:smvse_si_damaqf}
\end{figure}

\begin{figure}[!ht]
\begin{center}
\vspace{-0.3cm}
\includegraphics[width=4.3cm] {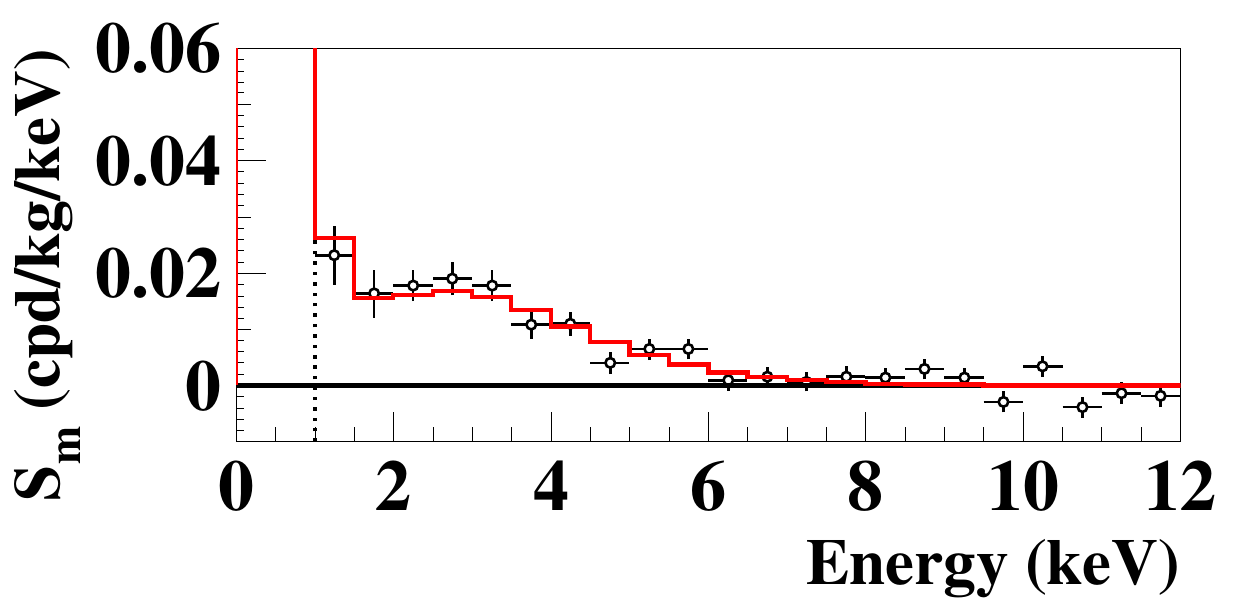}
\includegraphics[width=4.3cm] {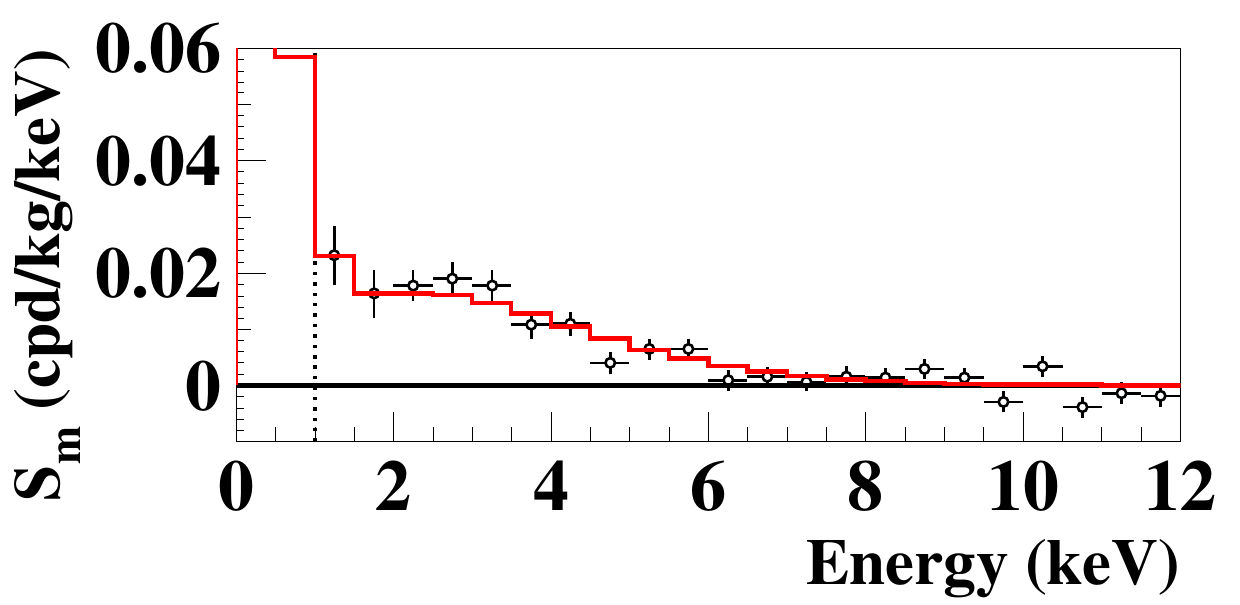}
\includegraphics[width=4.3cm] {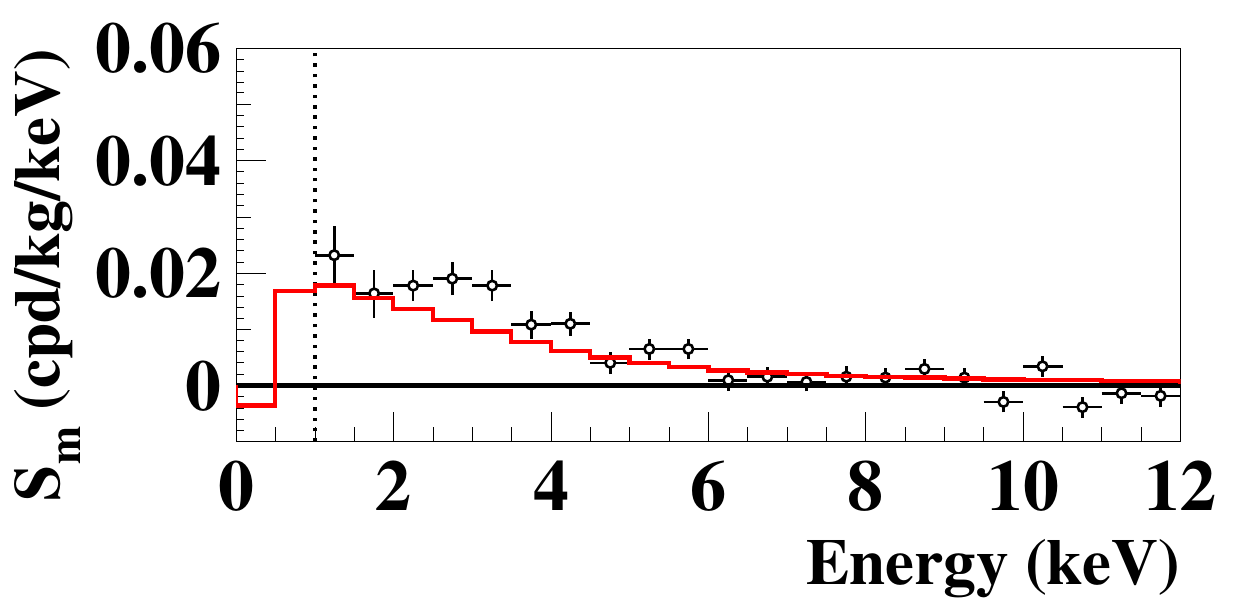}
\end{center}
\vspace{-0.4cm}
\caption{Examples of superposition of the measured $\mathcal{S}^{exp}_{m}$ vs energy (points
with error bars) with theoretical expectations (solid histograms) for the SI case, when the 
channeling effect is included \cite{chan}.
$Left:$ case of the C1 (Evans logarithmic) halo model with $\rho_0$ = 0.56 GeV/cm$^3$, 
$v_0$ = 170 km/s, set A of parameters values, DM particle mass 14 GeV and $\xi\sigma_{SI}$ 
equal to 5.5 $\times$ 10$^{-6}$ pb.
$Center:$ case of the C2 (Evans logarithmic) halo model with $\rho_0$ = 0.67 GeV/cm$^3$, 
$v_0$ = 170 km/s, set A of parameters values, DM particle mass 16 GeV and $\xi\sigma_{SI}$ 
equal to 2.6 $\times$ 10$^{-6}$ pb.  
$Right:$ case of the C2 (Evans logarithmic) halo model with $\rho_0$ = 0.67 GeV/cm$^3$, 
$v_0$ = 170 km/s, set C of parameters values, DM particle mass 50 GeV and $\xi\sigma_{SI}$ 
equal to 1.2 $\times$ 10$^{-6}$ pb.
Obviously many other possibilities are open for good agreement including cases of the 
isothermal halo model usually used by other experiments in the field to report their results.
}
\label{fg:smvse_si_channeling}
\end{figure}

When comparing with the previous results obtained only considering DAMA/NaI \cite{RNC} and DAMA/LIBRA--phase1 \cite{modlibra3} data,
one can derive that: 1) the C.L. associated to the regions allowed 
in the described frameworks is improved; 2) the allowed regions are restricted (i.e. several configurations for the specific 
considered frameworks are no more supported by the cumulative data at the given C.L.);
3) in the $Q_{I}$ and $Q_{II}$ cases the low and high mass regions, driven by the Na and I nuclei, respectively, are disconnected;
4) including the channeling effect the lower available mass is 4 GeV, instead of 2 GeV as in the previous analysis \cite{bot11,review}.

In Fig. \ref{fg:smvse_si_damaqf} few examples of superposition of the measured $\mathcal{S}^{exp}_{m}$ vs energy 
(points with error bars) with theoretical 
expectations (solid histograms) for SI case are shown. In general the comparison for most of considered scenarios is very stringent 
when the channeling effect is included
according to Ref. \cite{chan} as in the examples shown in Fig. \ref{fg:smvse_si_channeling}.
In these examples the Evans logarithmic halo model has been considered. We recall that the Evans model is an 
analytical solution giving the DF for particular families of logarithmic gravitational potentials related 
to the total matter distribution in the Galaxy.
Hereafter, the theoretical expectations are reported below 1 keV to show the importance
of further lowering the energy threshold to disentangle among the different models and scenarios.

In conclusion, the purely SI scenario is still supported by the data both for low and high mass candidates;
the inclusion of channeling effect also offers stringent agreement 
in many considered SI scenarios.

\subsubsection{Candidates with isospin violating SI coupling}
\label{sec:isovio}

To study the case of a DM candidate with SI isospin violating interaction, where $f_p \neq f_n$, 
a third parameter, namely the ratio $f_n/f_p$, must be considered together with
$\xi \sigma_{SI}$ and $m_{DM}$. Obviously the previous case of isospin conserving is restored whenever 
the ratio $f_n/f_p = 1$. 

\begin{figure}[!p]
\begin{center} 
\vspace{-0.5cm}
\includegraphics[width=4.2cm] {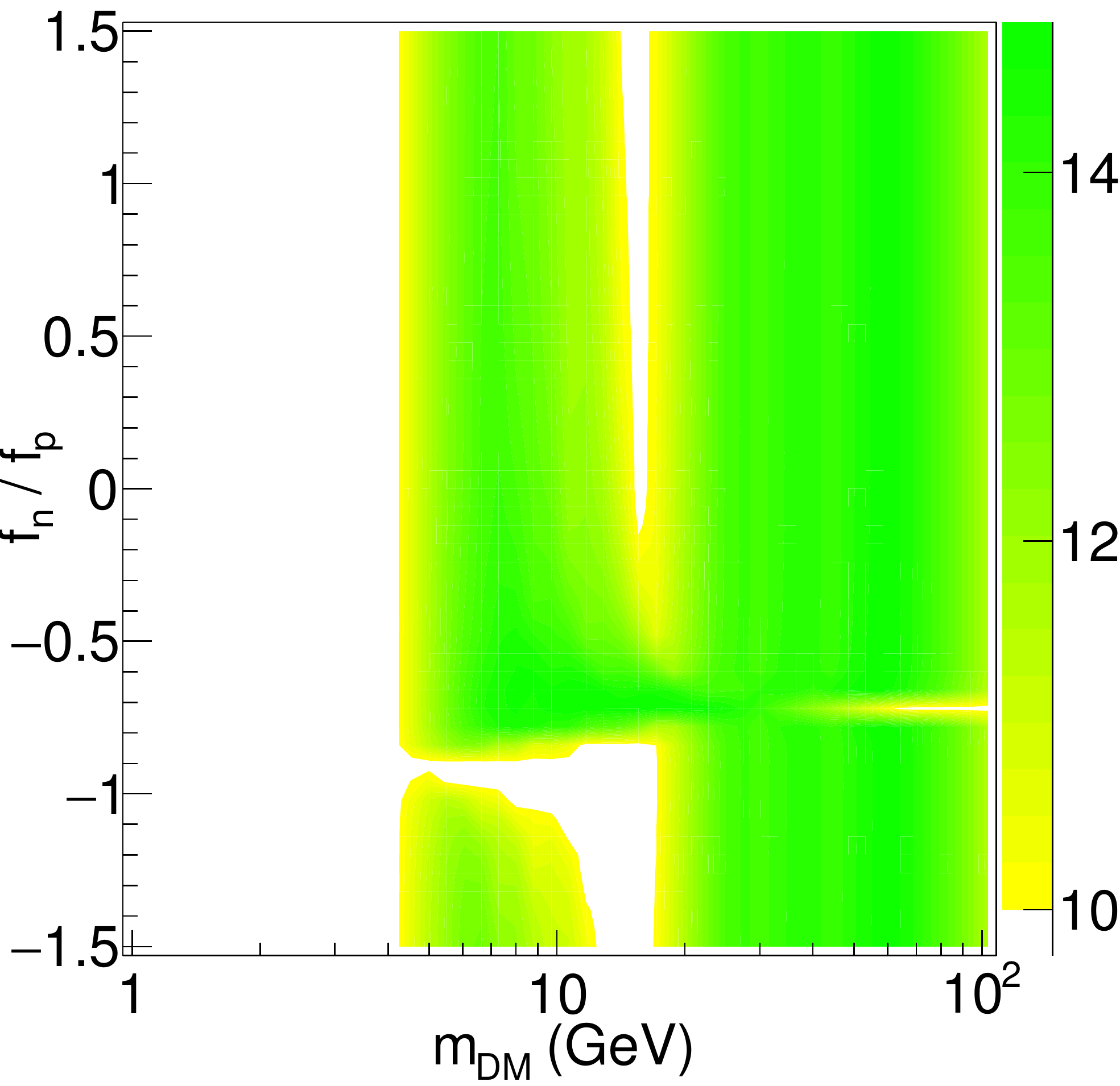}
\includegraphics[width=4.2cm] {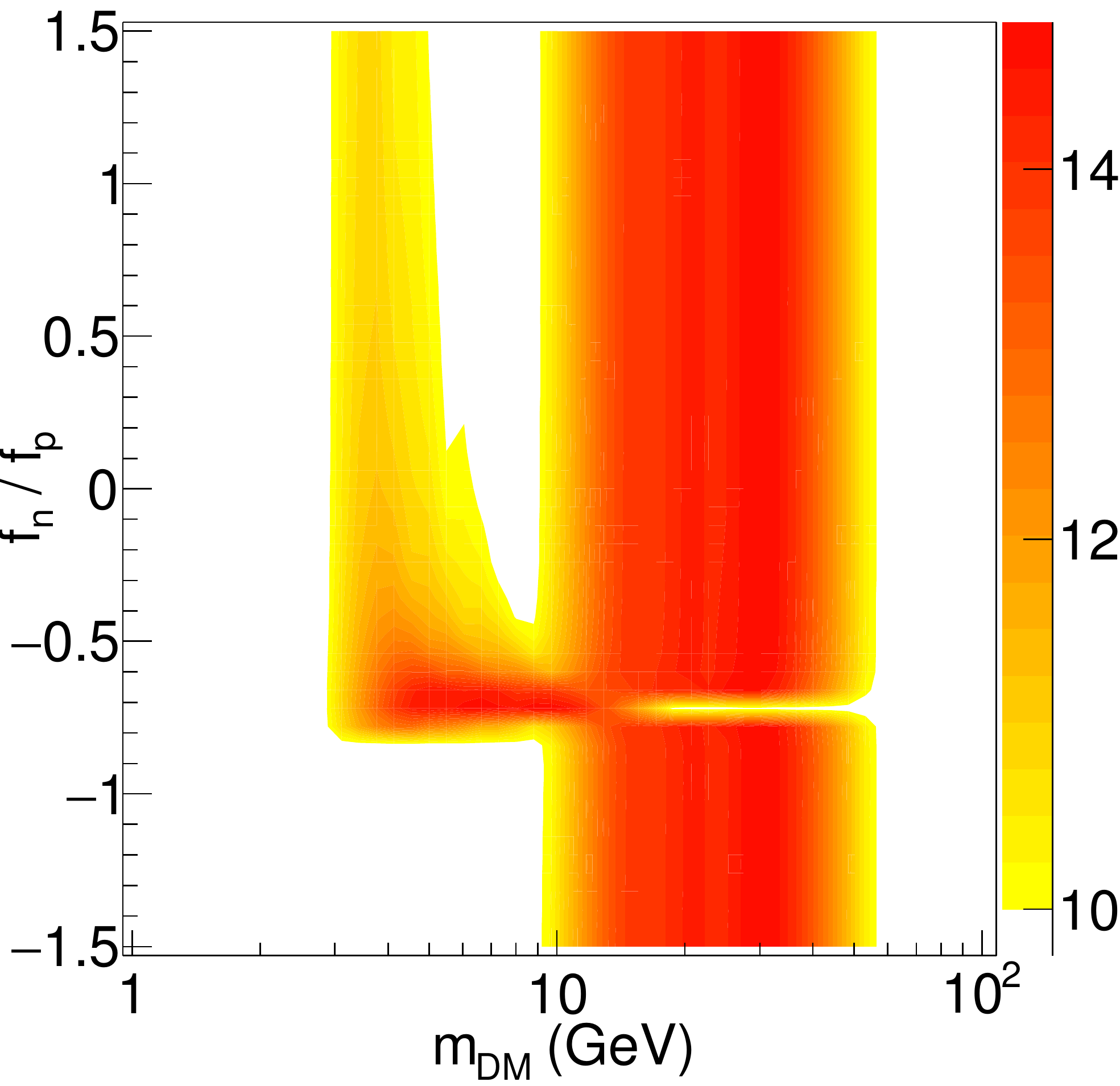}
\includegraphics[width=4.2cm] {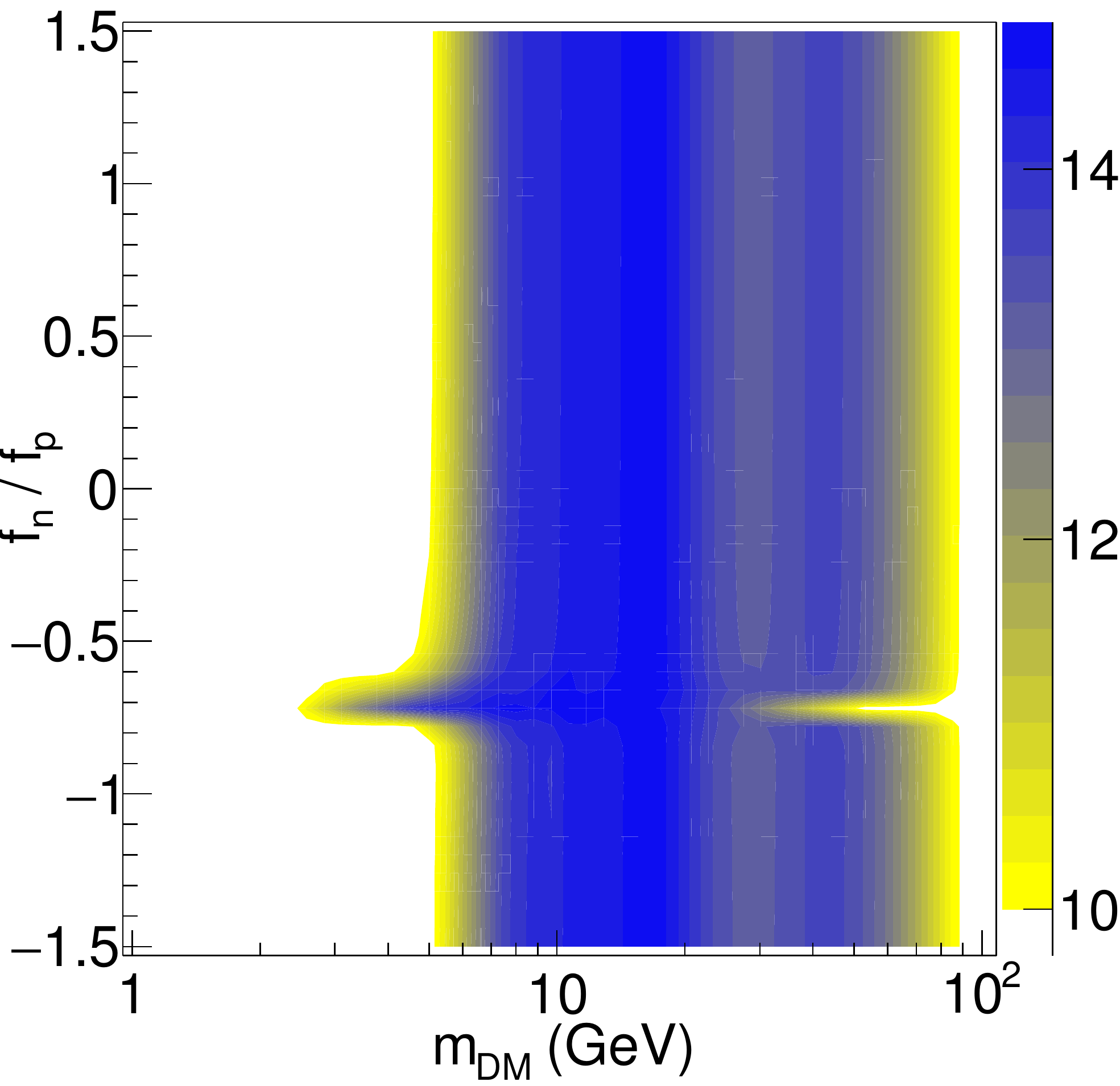}
\includegraphics[width=4.2cm] {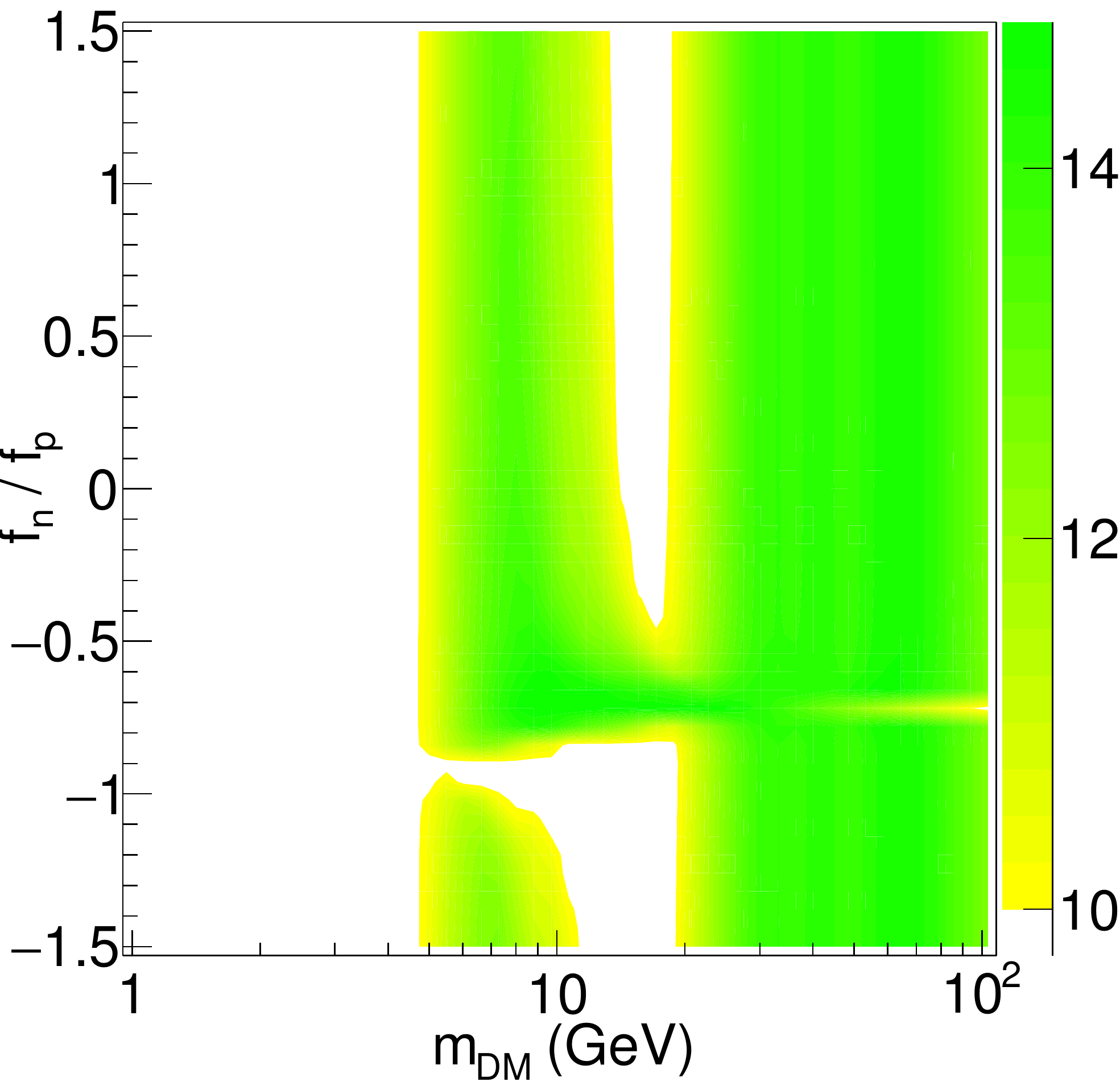}
\includegraphics[width=4.2cm] {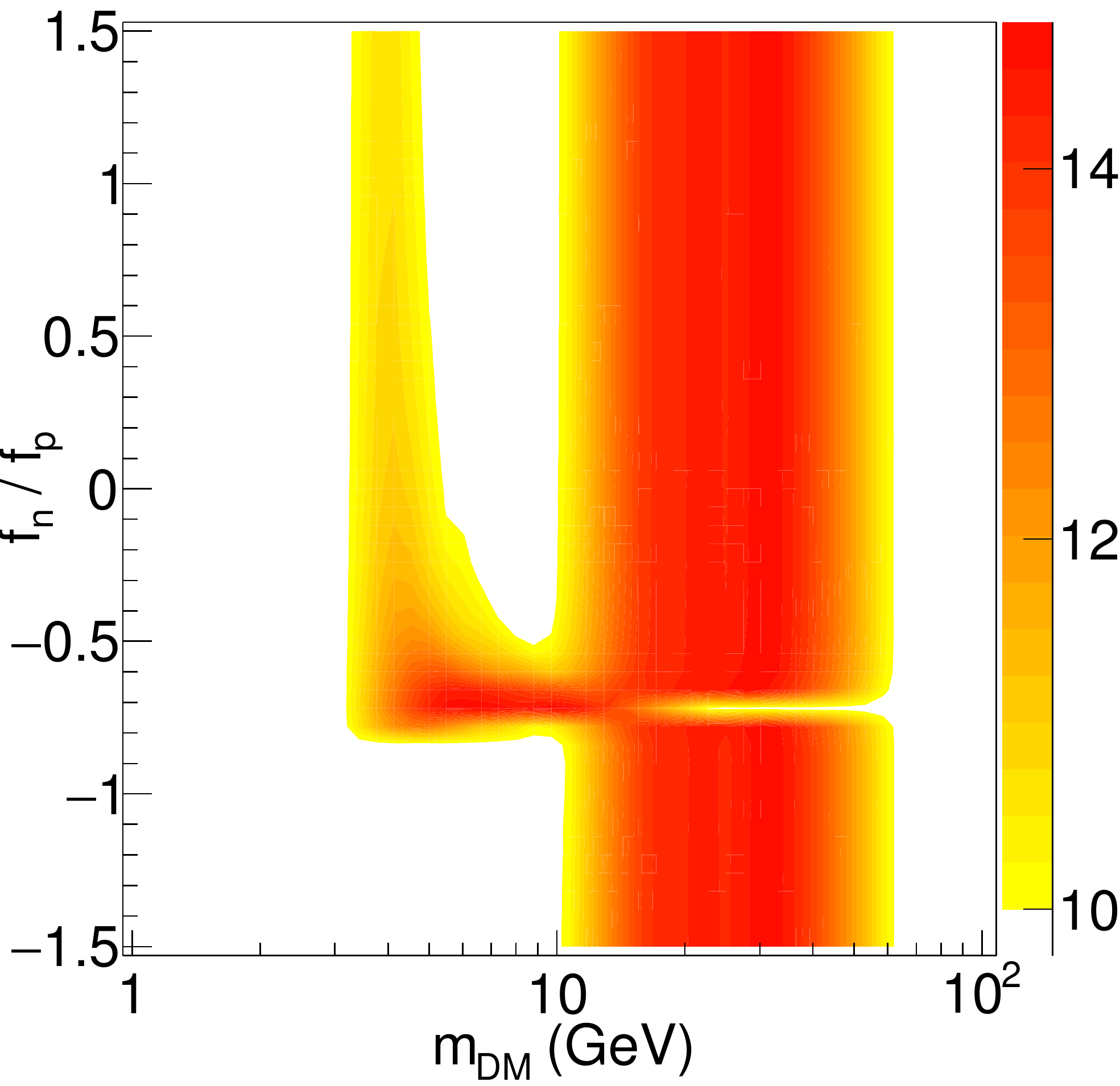}
\includegraphics[width=4.2cm] {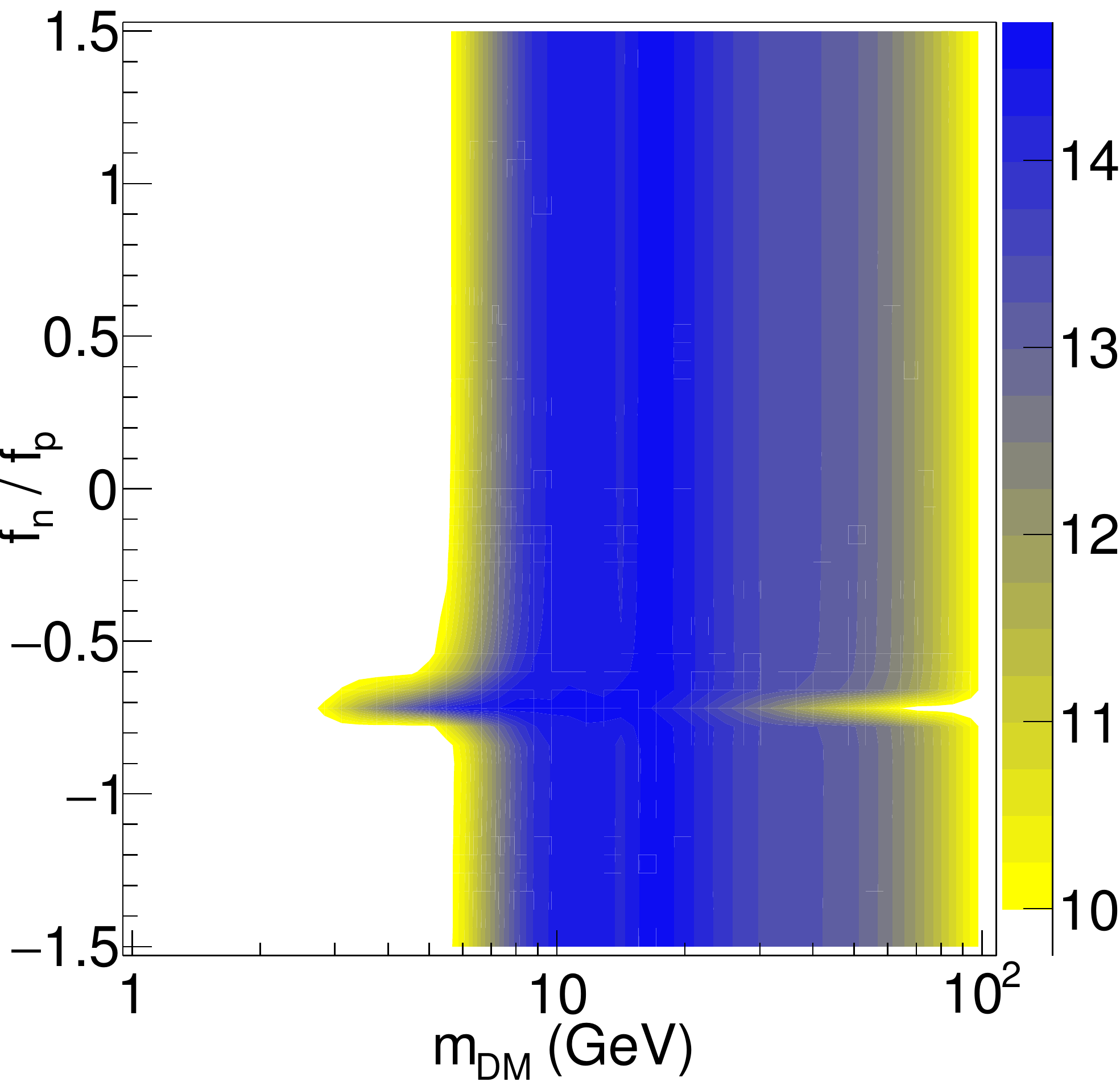}
\includegraphics[width=4.2cm] {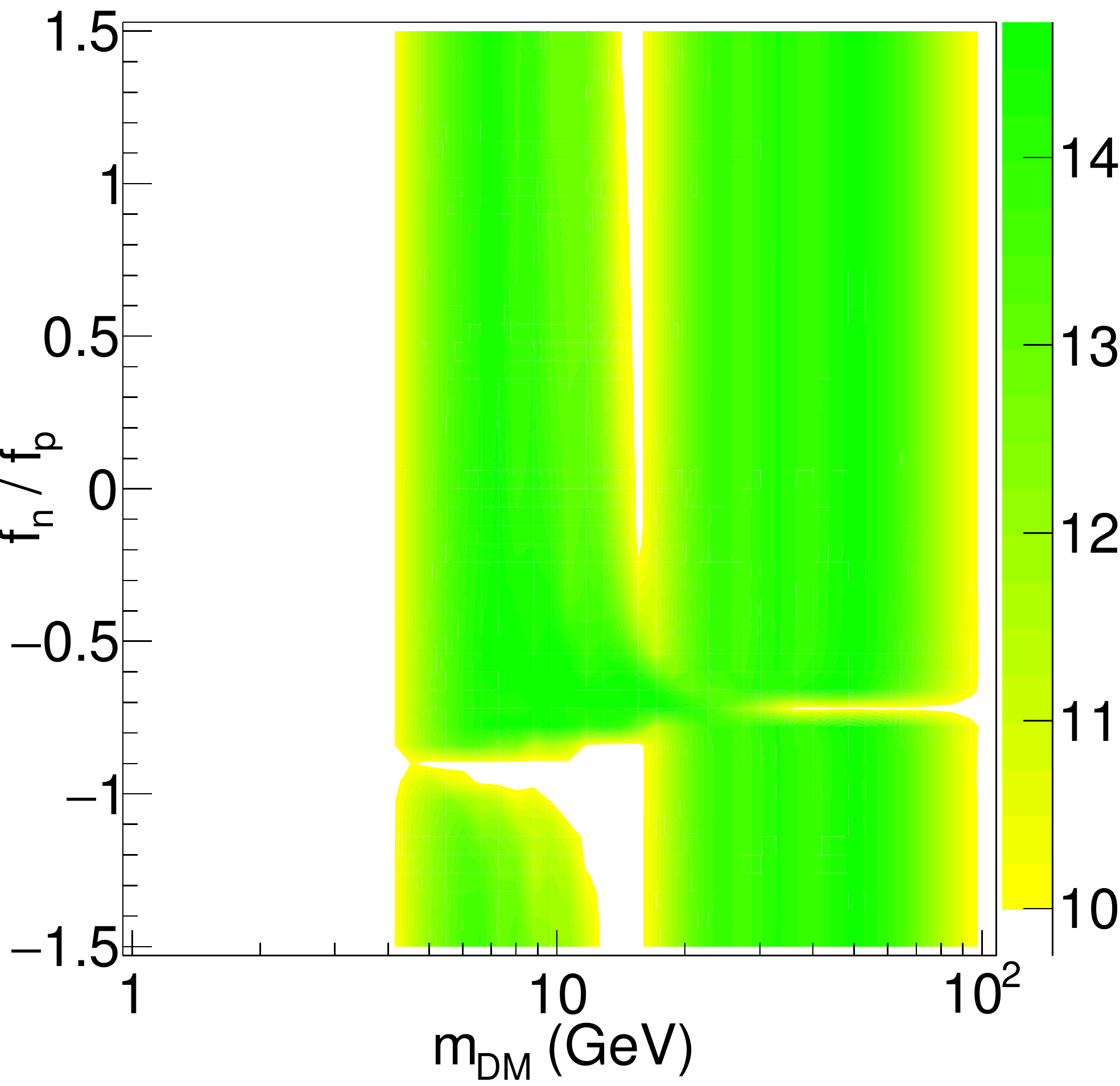}
\includegraphics[width=4.2cm] {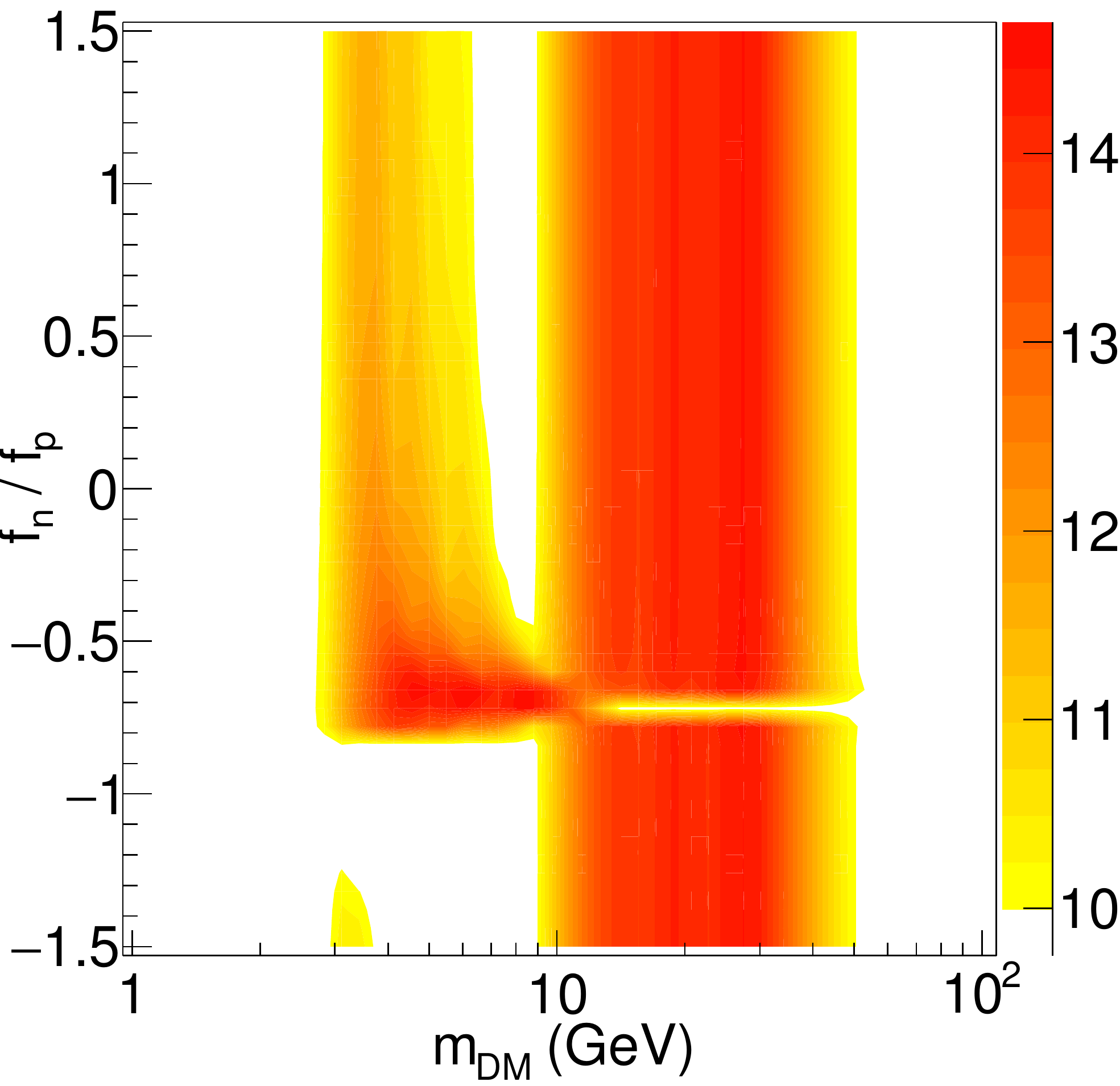}
\includegraphics[width=4.2cm] {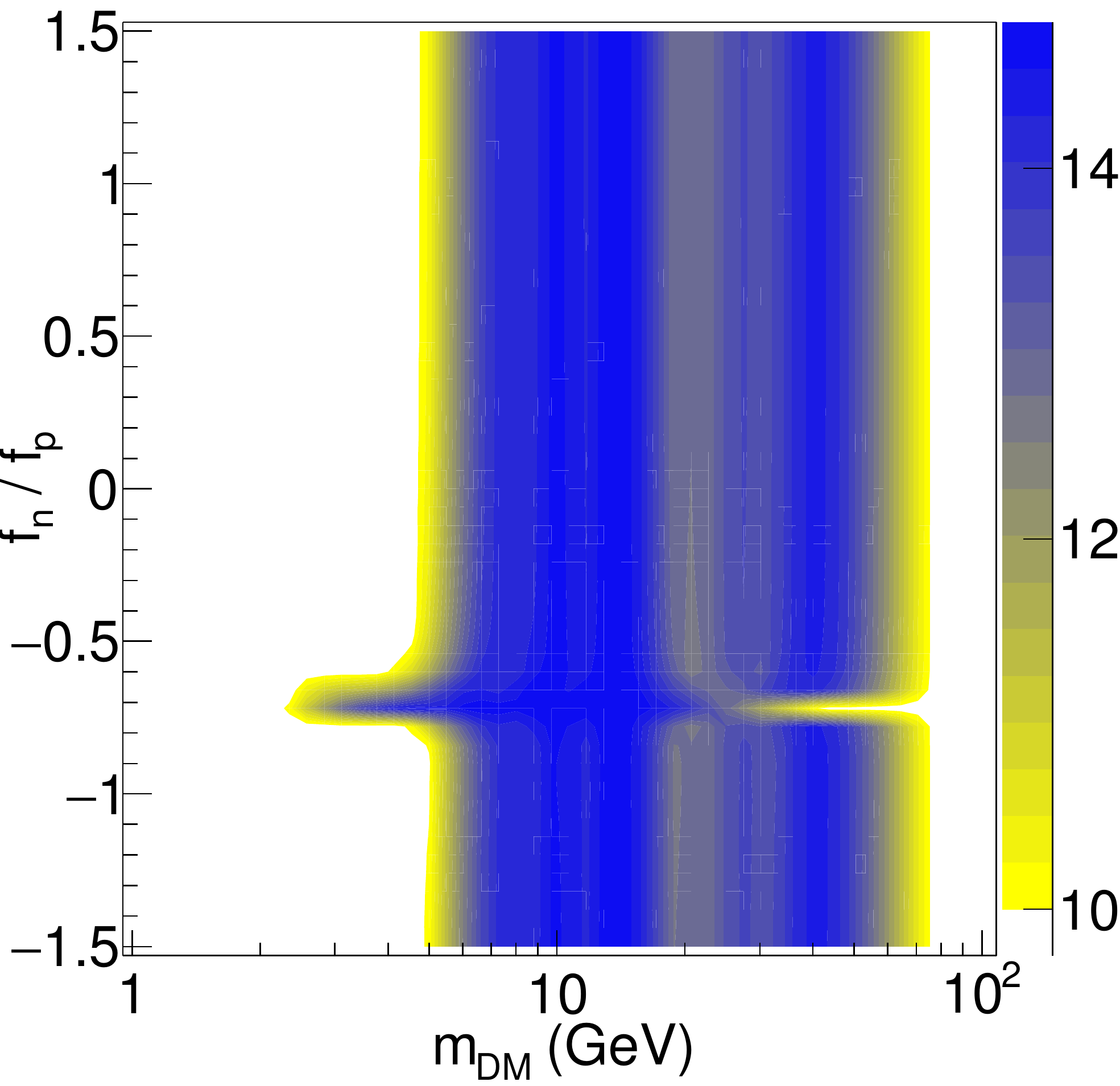}
\includegraphics[width=4.2cm] {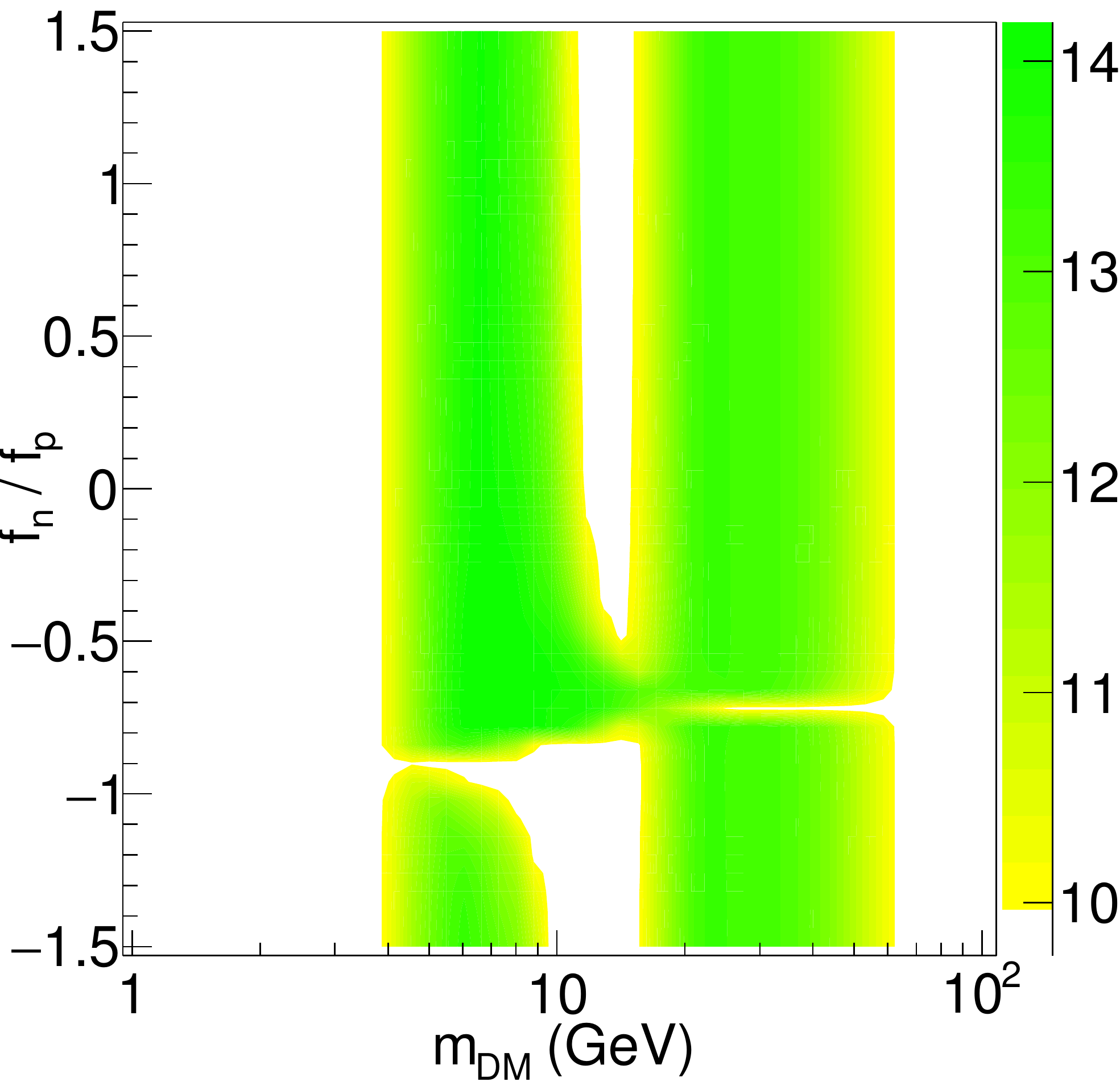}
\includegraphics[width=4.2cm] {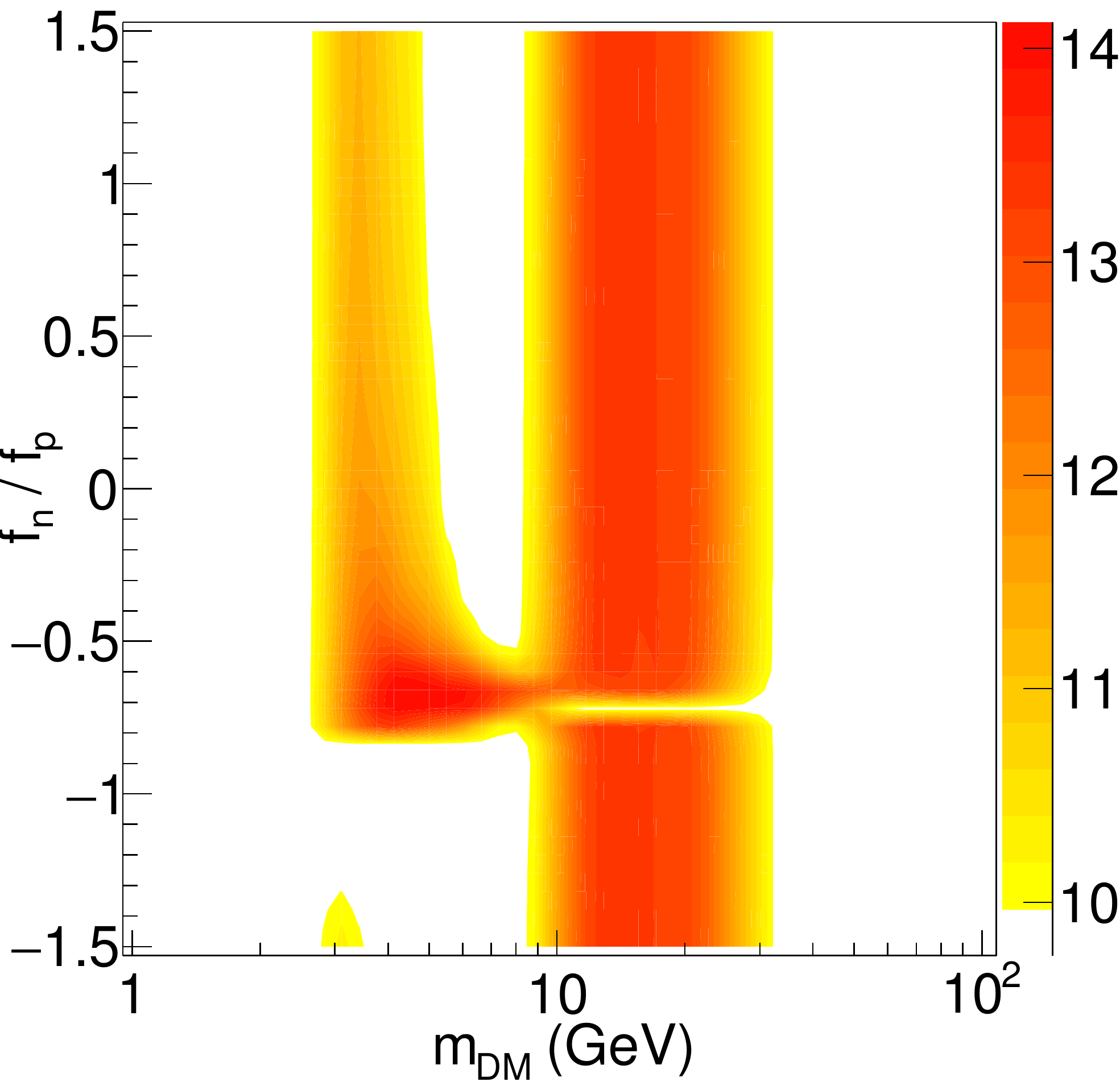}
\includegraphics[width=4.2cm] {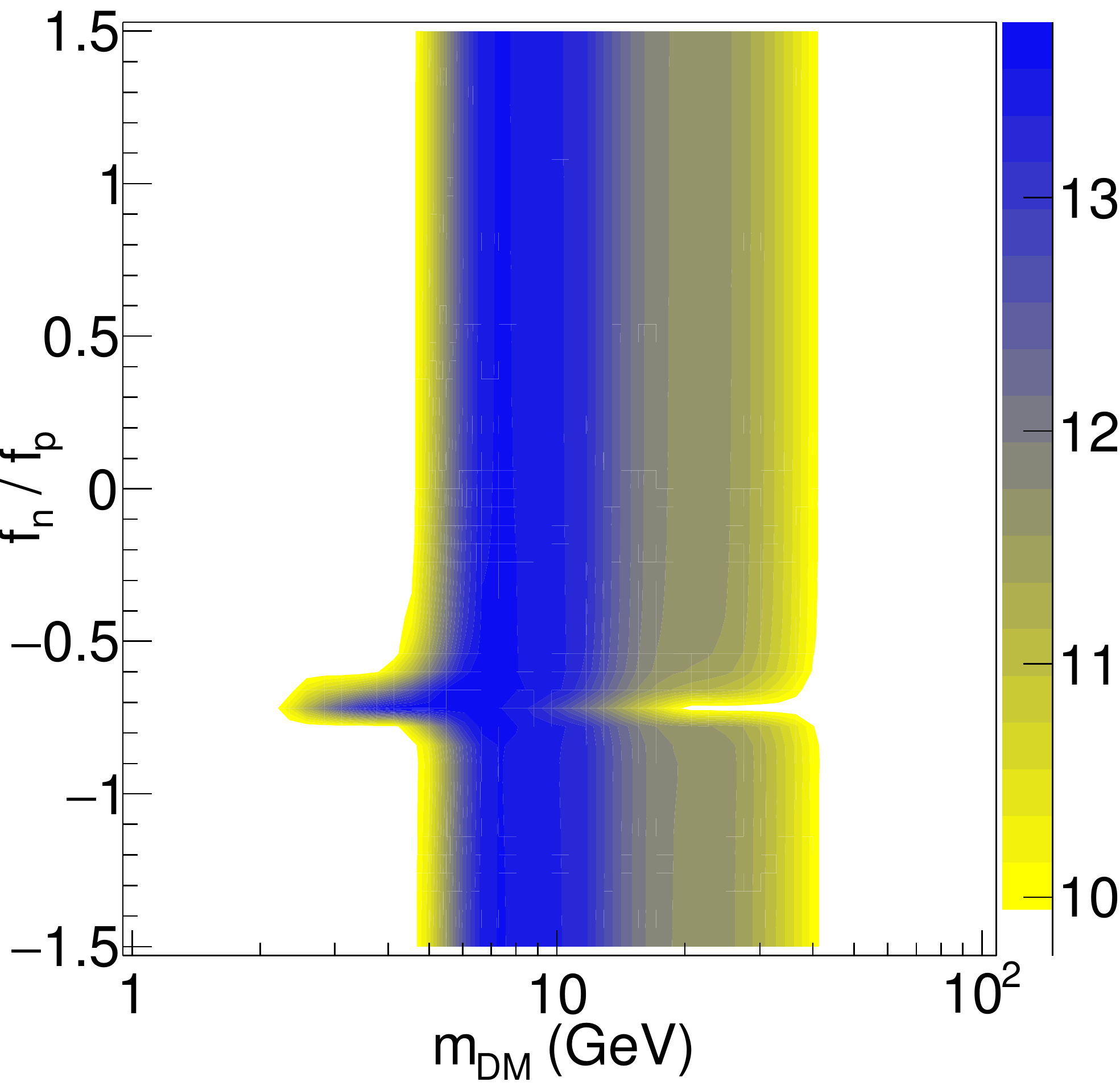}
\end{center}
\vspace{-0.5cm}
\caption{Regions in the $f_n/f_p$ vs $m_{DM}$ plane allowed by DAMA experiments in the case
of a Dark Matter candidate having isospin violating SI interaction.
The Na and I quenching factors are: 
$Q_I$ [$left$ (green on-line)], $Q_{II}$ [$center$ (red on-line)], and with channeling effect [$right$ (blue on-line)]. 
The considered halos (from top to bottom) are A0 (isothermal sphere), B1, C1, D3 with the $v_0$ and $\rho_0$ 
in the range of Table III of Ref. \cite{bel02}.
The three possible sets of parameters A, B 
and C are considered (see Sect. \ref{data_analysis}); the allowed regions represent the domain where the 
likelihood-function values differ more than 10 $\sigma$ from the {\it null hypothesis} (absence of 
modulation). 
The color scales give the confidence level in units of $\sigma$ from the {\it null hypothesis}.}
\label{fg:si-iv}
\end{figure}

The results of the analysis are reported in Fig. \ref{fg:si-iv}, where the allowed regions 
in the $f_n/f_p$ vs $m_{DM}$ plane are shown after marginalizing on $\xi \sigma_{SI}$.
For simplicity four halo models: A0 (isothermal sphere), B1, C1, D3 with the $v_0$ and $\rho_0$ 
in the range of Table III of Ref. \cite{bel02}, and three choices of the Na and I quenching factors:
$Q_I$, $Q_{II}$, and including the channeling effect are considered.

\begin{figure}[!ht]
\begin{center}
\includegraphics[width=7.0cm] {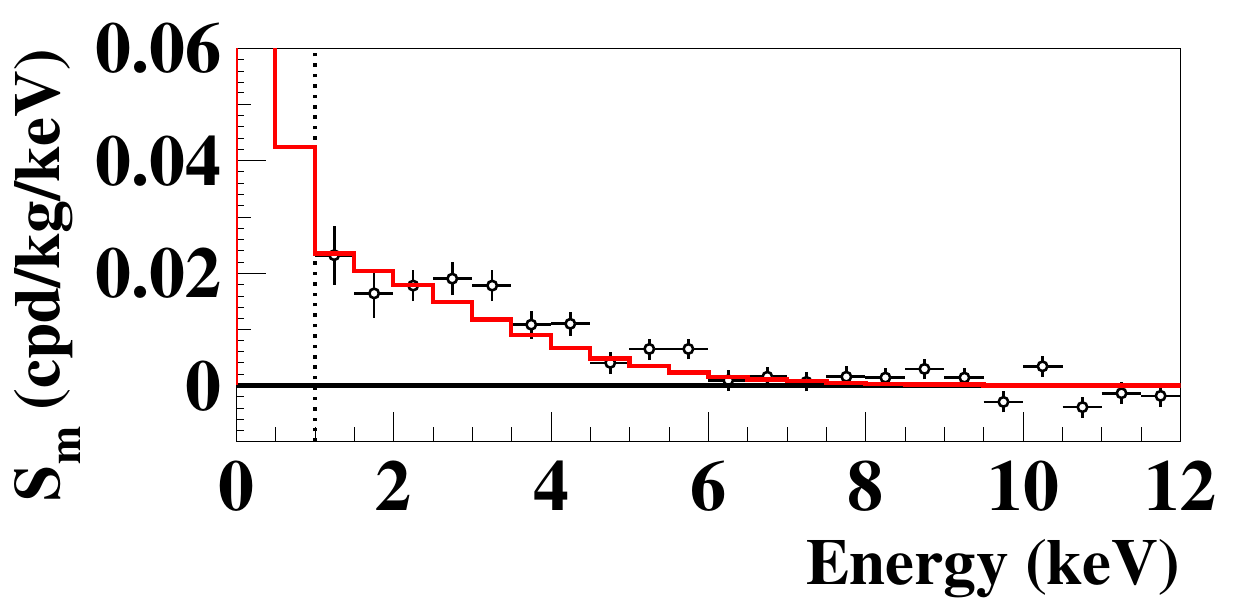}
\end{center}
\vspace{-0.6cm}
\caption{Example of superposition of the measured $\mathcal{S}^{exp}_{m}$ vs energy (points
with error bars) with theoretical expectations (solid histograms) for the SI case, when the
channeling effect is included \cite{chan}, for $m_{DM} = 11$ GeV and for the isothermal
sphere halo model A0, with $v_0$ = 220 km/s and $\rho_0$ = 0.3 GeV/cm$^3$. The
set A of parameters values and $\xi\sigma_{SI}$ equal to 7.0 $\times$ 10$^{-6}$ pb
are taken into account.}
\label{fg:smvse_si_iso}
\end{figure}

Typically, few considerations can be done:
\begin{itemize}
\item Two bands of $m_{DM}$ can be recognized, as expected: one at low mass and the other at higher mass. 
\item The low mass DM candidates have a good fit in correspondence of $f_n/f_p \simeq -53/74 = -0.72$, where 
      the $^{127}$I contribution vanishes and the signal is mostly due to $^{23}$Na recoils.
\item Similarly, at larger mass $f_n/f_p \simeq -0.72$ is instead disfavored.
\item The case of isospin-conserving $f_n/f_p = 1$ is well supported at different extent both at lower and larger mass.
\item When the channeling effect is included (panels on the $right$ of Fig. \ref{fg:si-iv}), the case of 
      $f_n/f_p = 1$ at low mass has even a stronger support, that is higher confidence level. 
      This argument is also supported by the agreement of the theoretical model and experimental data 
      shown in Fig. \ref{fg:smvse_si_channeling} and in Fig. \ref{fg:smvse_si_iso}. In particular, in Fig. \ref{fg:smvse_si_iso}
      the case of the isothermal sphere and $v_0$ = 220 km/s is considered.
\item Contrary to what was stated in Ref. \cite{bau19,Kang18,Kahl18} where the low mass DM candidates 
      were disfavored for $f_n/f_p = 1$ by DAMA data, the inclusion of the uncertainties related to 
      halo models, $v_0$ and $\rho_0$, quenching factors, channeling effect, nuclear form factors, etc., 
      and correctly accounting for other aspects, can also support
      low mass DM candidates either including or not the channeling effect. Some instances of this are reported in 
      Figs. \ref{fg:smvse_si_damaqf}, \ref{fg:smvse_si_channeling}, and \ref{fg:smvse_si_iso}. 
\end{itemize}

In conclusion, at present level of uncertainties the DAMA data, if interpreted in terms of DM particle inducing 
nuclear recoils through SI interaction, can account either for low and large DM particle mass and for a wide range
of the ratio $f_n/f_p$, even including the ``standard'' case $f_n/f_p = 1$.

\vspace{0.6cm}
\subsubsection{Spin-Dependent interaction}

The purely SD interaction, to which Na and I nuclei are fully sensitive, can also be considered. 
As mentioned above, any result and comparison in this case is even more 
uncertain considering the large uncertainties on spin factors and on form factor and the complementary 
sensitivities among different target nuclei depending on their unpaired nucleon \cite{sisd,RNC,ijmd}.

\begin{figure}[!ht]
\begin{center}
\vspace{-0.4cm}
\includegraphics[width=6.4cm] {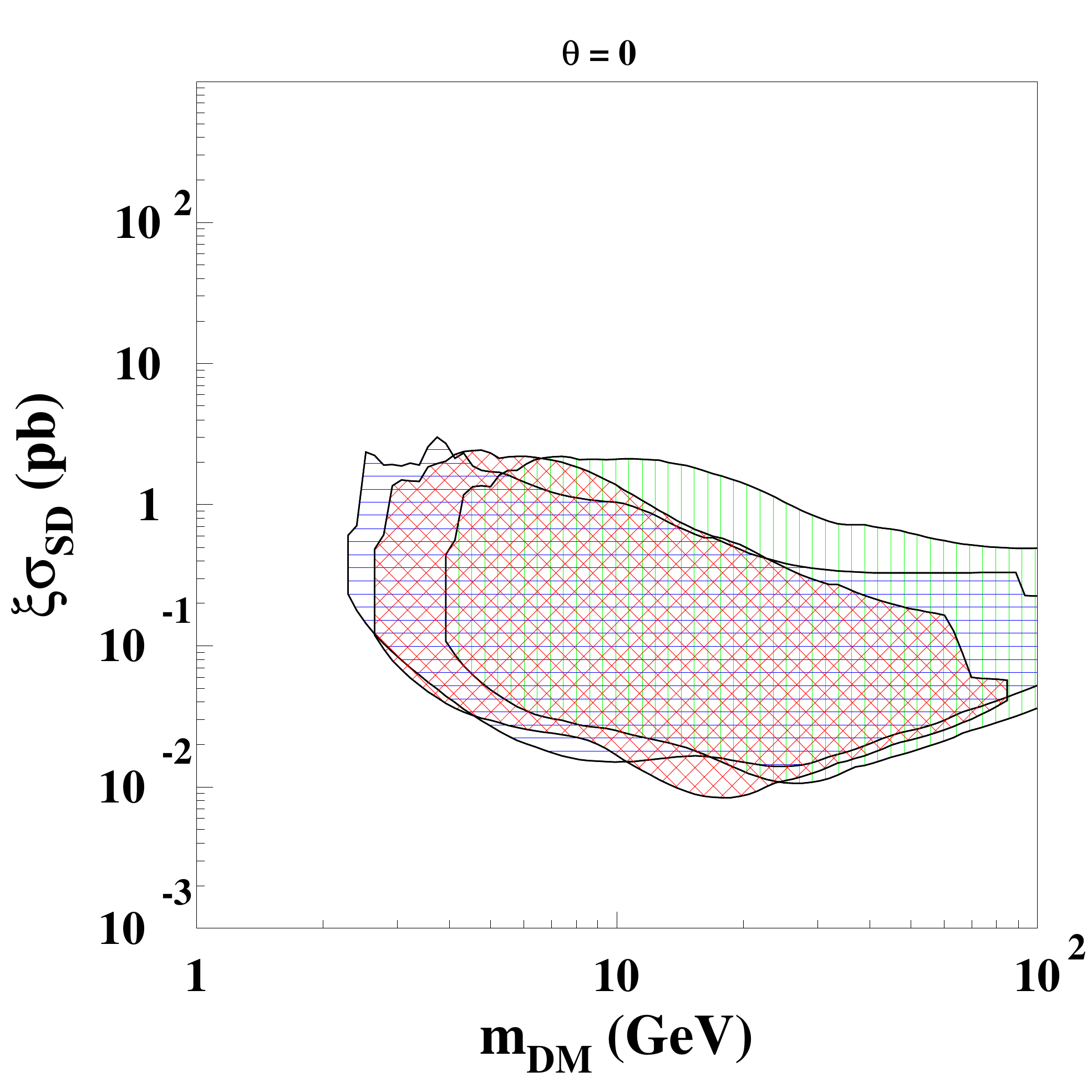}
\includegraphics[width=6.4cm] {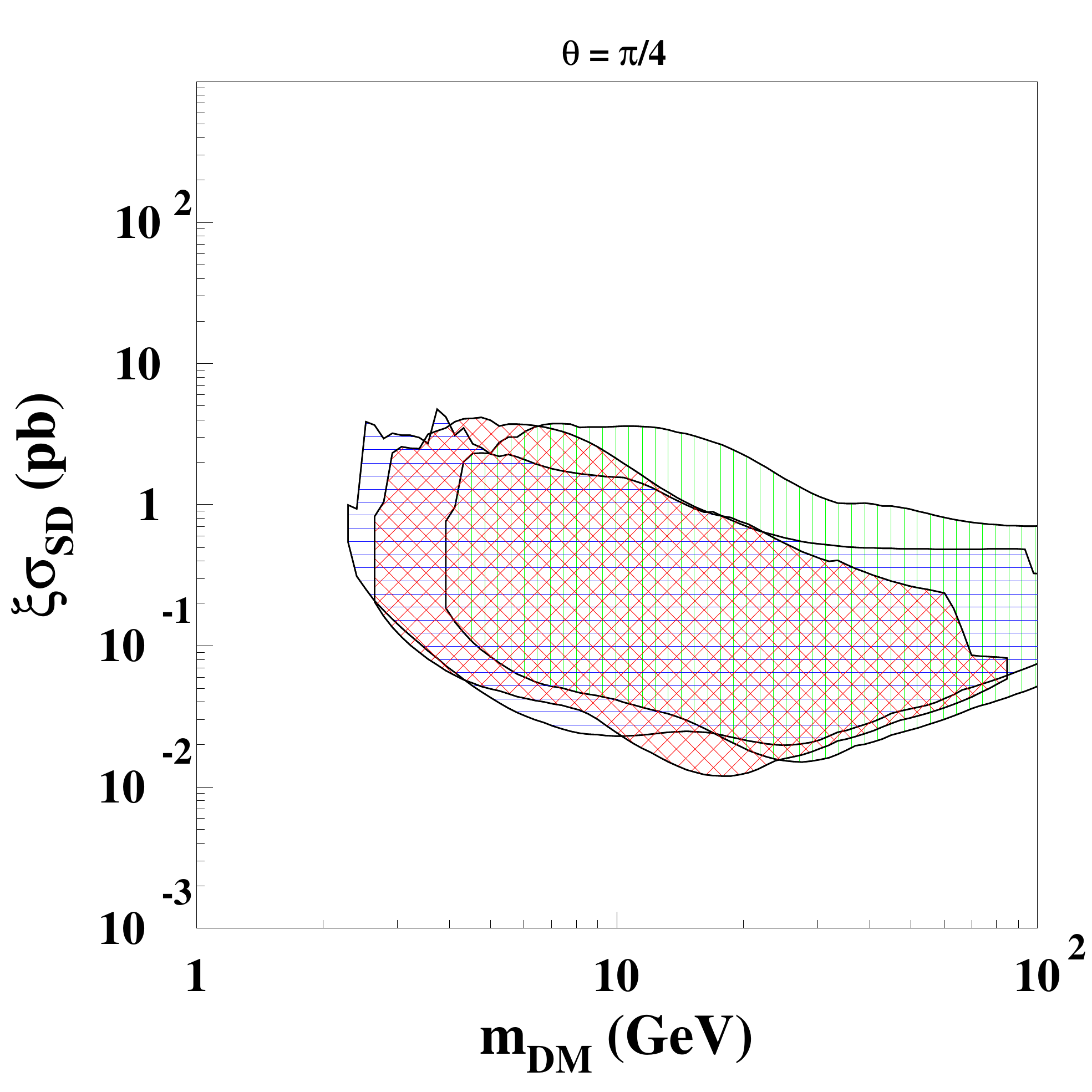}
\includegraphics[width=6.4cm] {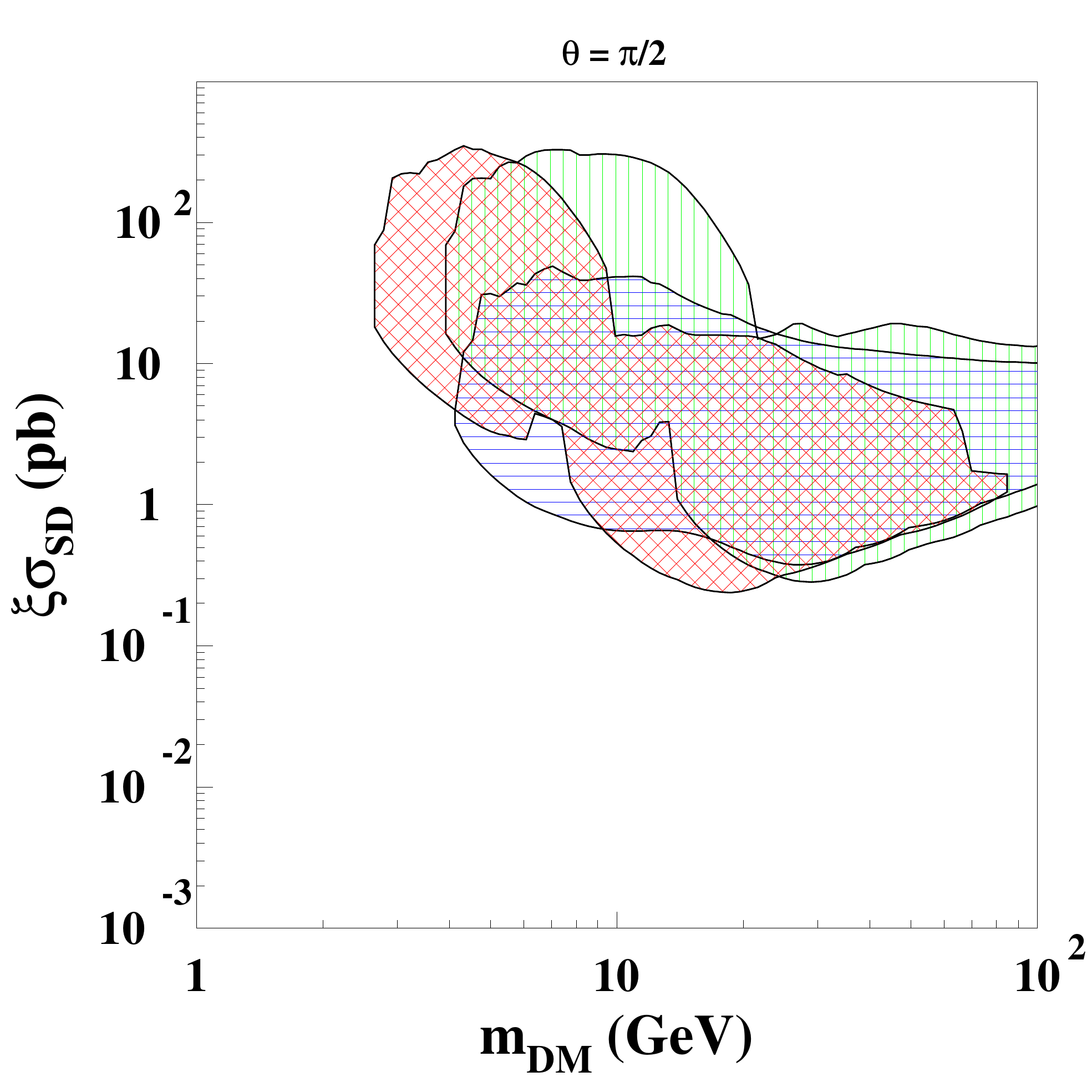}
\includegraphics[width=6.4cm] {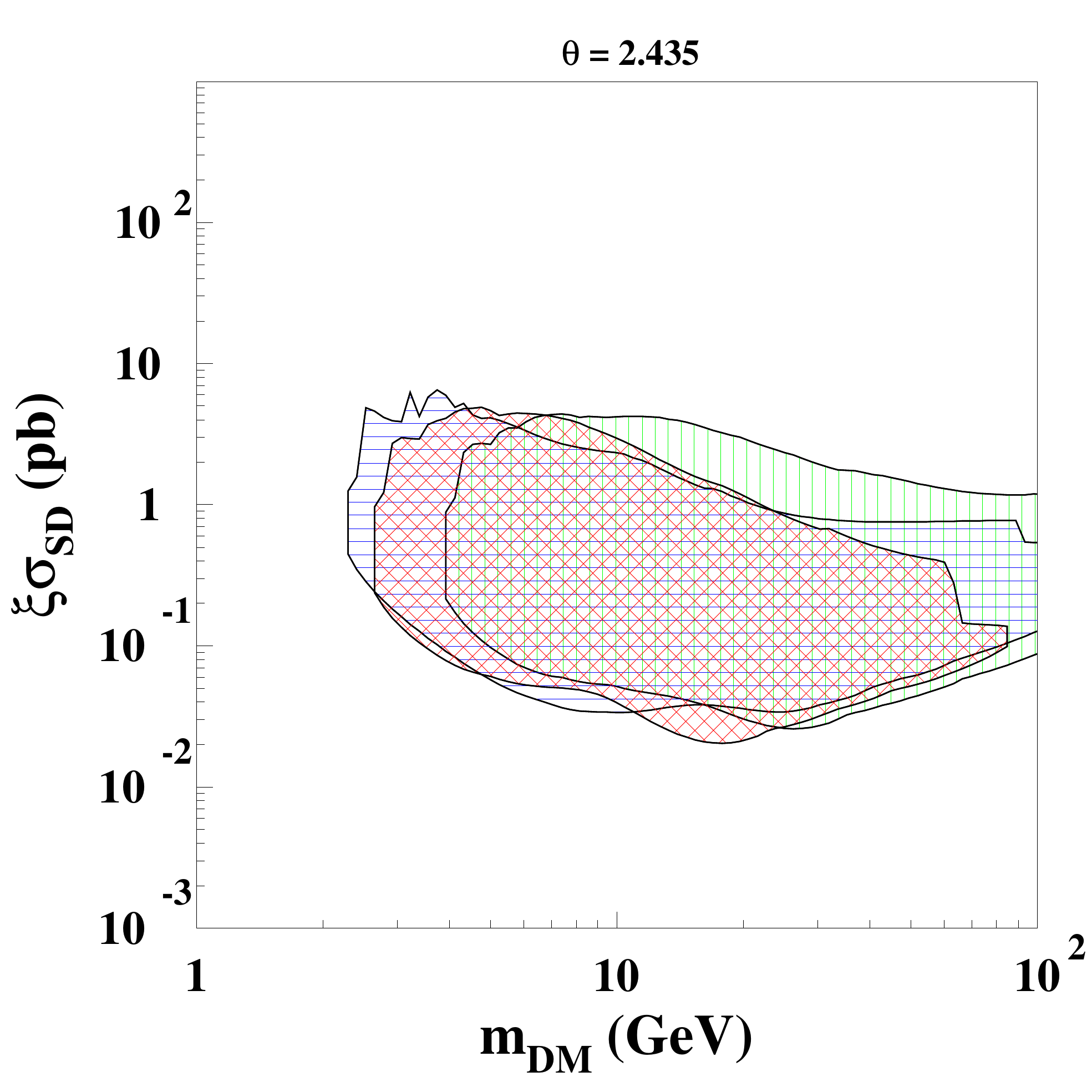}
\end{center}
\vspace{-0.4cm}
\caption{Slices of the 3-dimensional volume ($\xi \sigma_{SD}$, $m_{DM}$, $\theta$)
allowed by DAMA experiments in the case of a Dark Matter candidate elastically scattering off target 
nuclei and SD interaction.
Three different instances for the Na and I quenching factors have been considered: 
(i) $Q_I$ case [(green on-line) vertically-hatched region],
(ii) with channeling effect [(blue on-line) horizontally-hatched region)] and 
(iii) $Q_{II}$ [(red on-line) cross-hatched region].
The regions have been obtained by marginalizing all the models for each considered scenario 
(see Sect. \ref{data_analysis}) and they represent the domain where the likelihood-function 
values differ more than 10 $\sigma$ from the {\it null hypothesis} (absence of modulation).}
\label{fg:sd}
\end{figure}

\vspace{0.6cm}
The complete results would be described by a 3-dimensional volume: ($\xi\sigma_{SD}$, $m_{DM}$, $\theta$).
Thus, a very large number of possible configurations are
available; here for simplicity we show, as examples, the results obtained
only for 4 particular couplings, which correspond
to the following values of the mixing angle $\theta$:
(i)   $\theta = 0$ ($a_n = 0$ and $a_p \ne 0$ or $|a_p| \gg |a_n|$); 
(ii)  $\theta = \pi/4$ ($a_p = a_n$); 
(iii) $\theta = \pi/2$ ($a_n \ne 0$ and $a_p = 0$ or $|a_n| \gg |a_p|$);  
(iv)  $\theta = 2.435$ rad ($a_n/a_p = - 0.85$, pure $Z_0$ coupling). 
The case $a_p = -a_n$ is nearly similar to the case (iv). 

\vspace{0.6cm}
In Fig. \ref{fg:sd} slices ($\xi\sigma_{SD}$ vs $m_{DM}$) of the 3-dimensional allowed volume at the $\theta$ values given above
at 10 $\sigma$ from absence of signal are shown. For each configuration 3 regions are depicted accounting for the quenching 
factors uncertainties.

\begin{figure}[!ht]
\begin{center}
\vspace{-0.3cm}
\includegraphics[width=6.4cm] {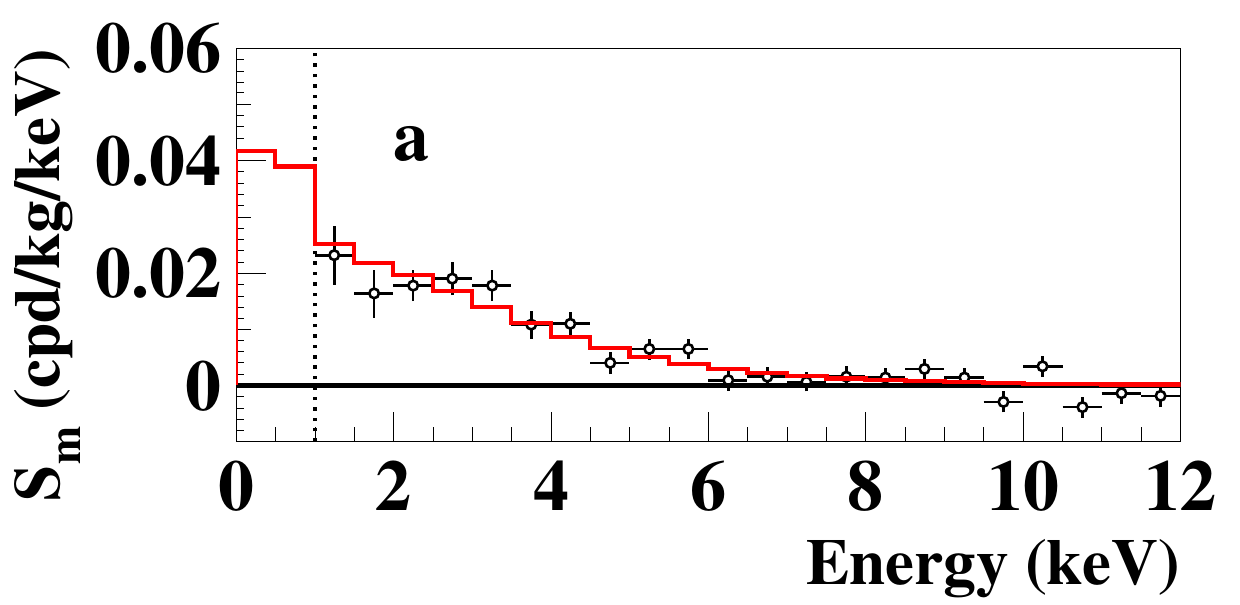}
\includegraphics[width=6.4cm] {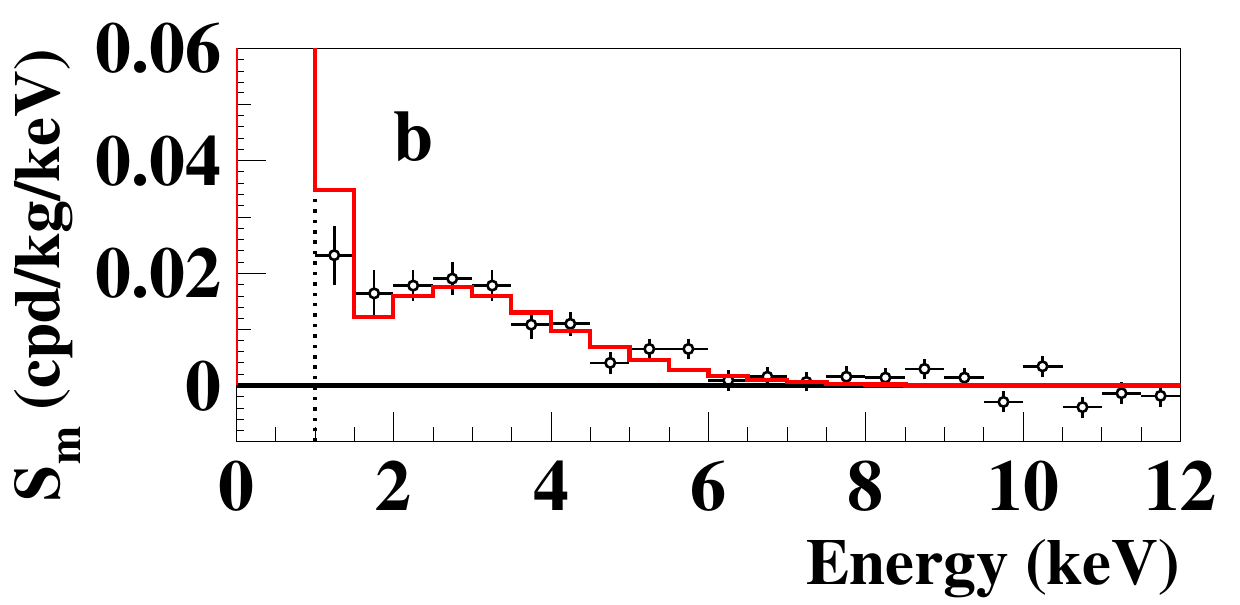}
\includegraphics[width=6.4cm] {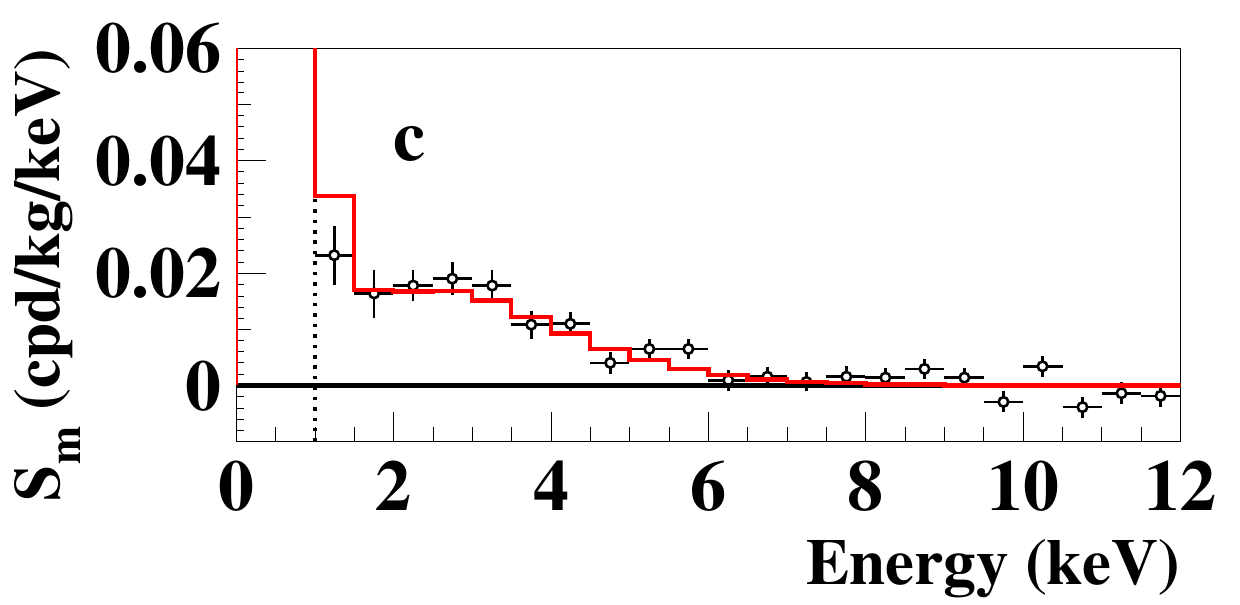}
\includegraphics[width=6.4cm] {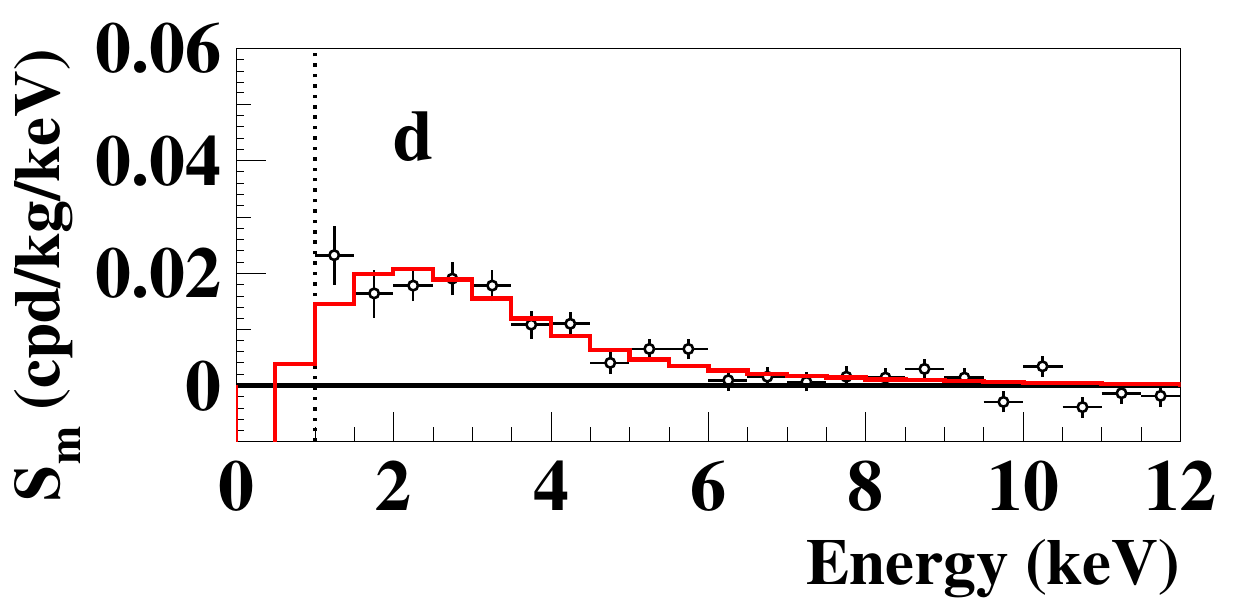}
\end{center}
\vspace{-0.4cm}
\caption{Examples of superposition of the measured $\mathcal{S}^{exp}_{m}$ vs energy (points
with error bars) with theoretical expectations (solid histograms) for the purely SD interaction.
(a) case of the A0 (isothermal sphere) halo model with $\rho_0$ = 0.18 GeV/cm$^3$,
    $v_0$ = 170 km/s, set B of parameters values, $\theta=0$, $m_{DM} = 15$ GeV, $\xi\sigma_{SD} = 0.47$ pb
    including channeling effect;
(b) case of the C1 (Evans logarithmic) halo model with $\rho_0$ = 0.94 GeV/cm$^3$,
    $v_0$ = 220 km/s, set A of parameters values, $\theta=\pi/2$, $m_{DM} = 10$ GeV, $\xi\sigma_{SD} = 23$ pb
    and quenching $Q_{I}$;
(c) case of the C4 (Evans power-law) halo model with $\rho_0$ = 0.65 GeV/cm$^3$,
    $v_0$ = 170 km/s, set A of parameters values, $\theta=2.435$, $m_{DM} = 8$ GeV, $\xi\sigma_{SD} = 0.49$ pb
    and quenching $Q_{II}$;
(d) case of the C2 (Evans logarithmic) halo model with $\rho_0$ = 0.67 GeV/cm$^3$,
    $v_0$ = 170 km/s, set C of parameters values, $\theta=\pi/4$, $m_{DM} = 52$ GeV, $\xi\sigma_{SD} = 0.10$ pb
    and quenching $Q_{I}$.}
\label{fg:smvsSD}
\end{figure}

\vspace{0.6cm}
In Fig. \ref{fg:smvsSD} the experimental $\mathcal{S}_m$ values are compared with some of the expectations in this scenario. As it can be seen, several 
configurations are in good agreement with the data. Obviously, much more can exist considering that only few configurations  
of the 3-dimensional volume are depicted here.

\begin{figure}[!p]
\begin{center}
\vspace{-0.5cm}
\includegraphics[width=4.2cm] {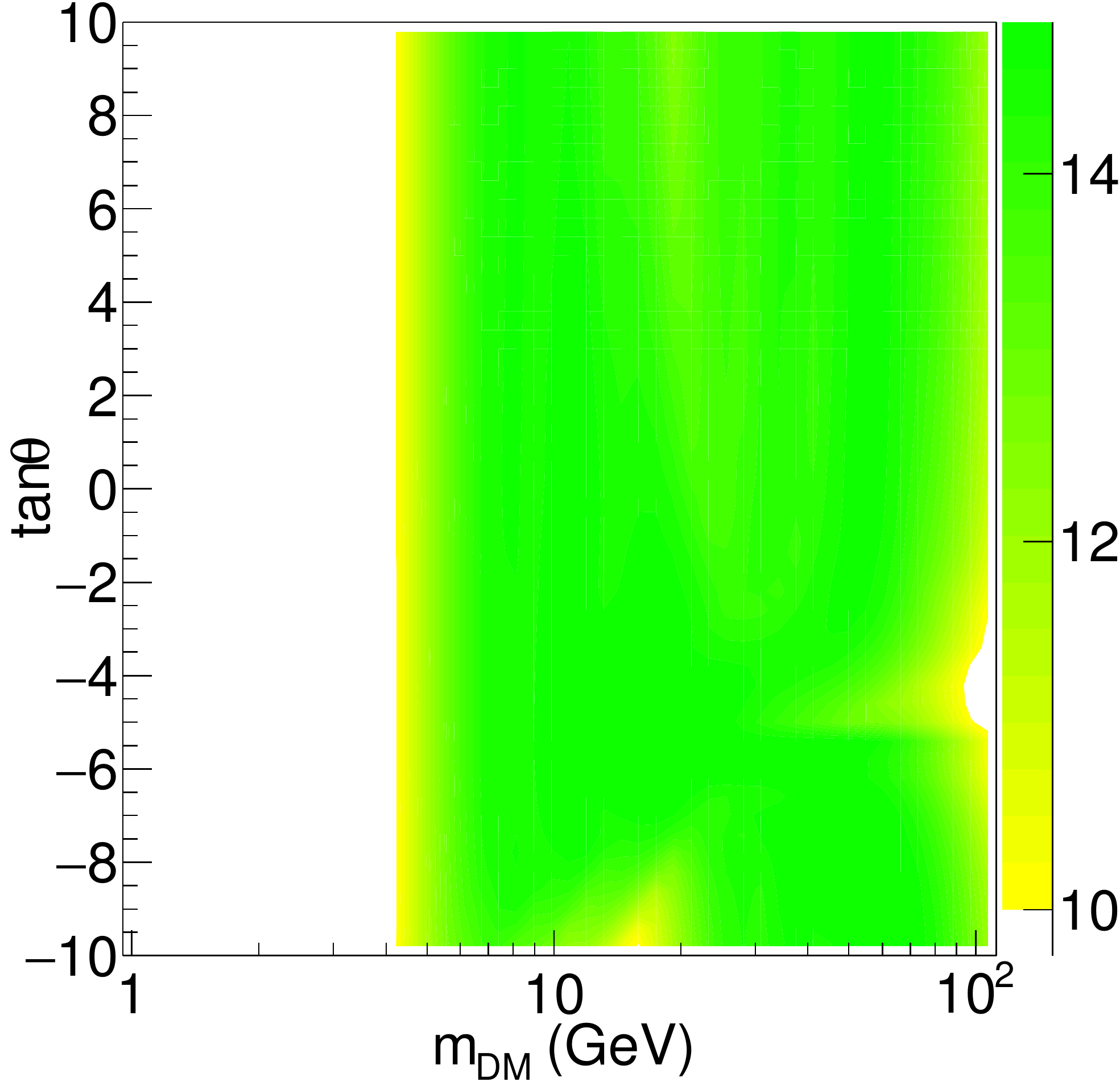}
\includegraphics[width=4.2cm] {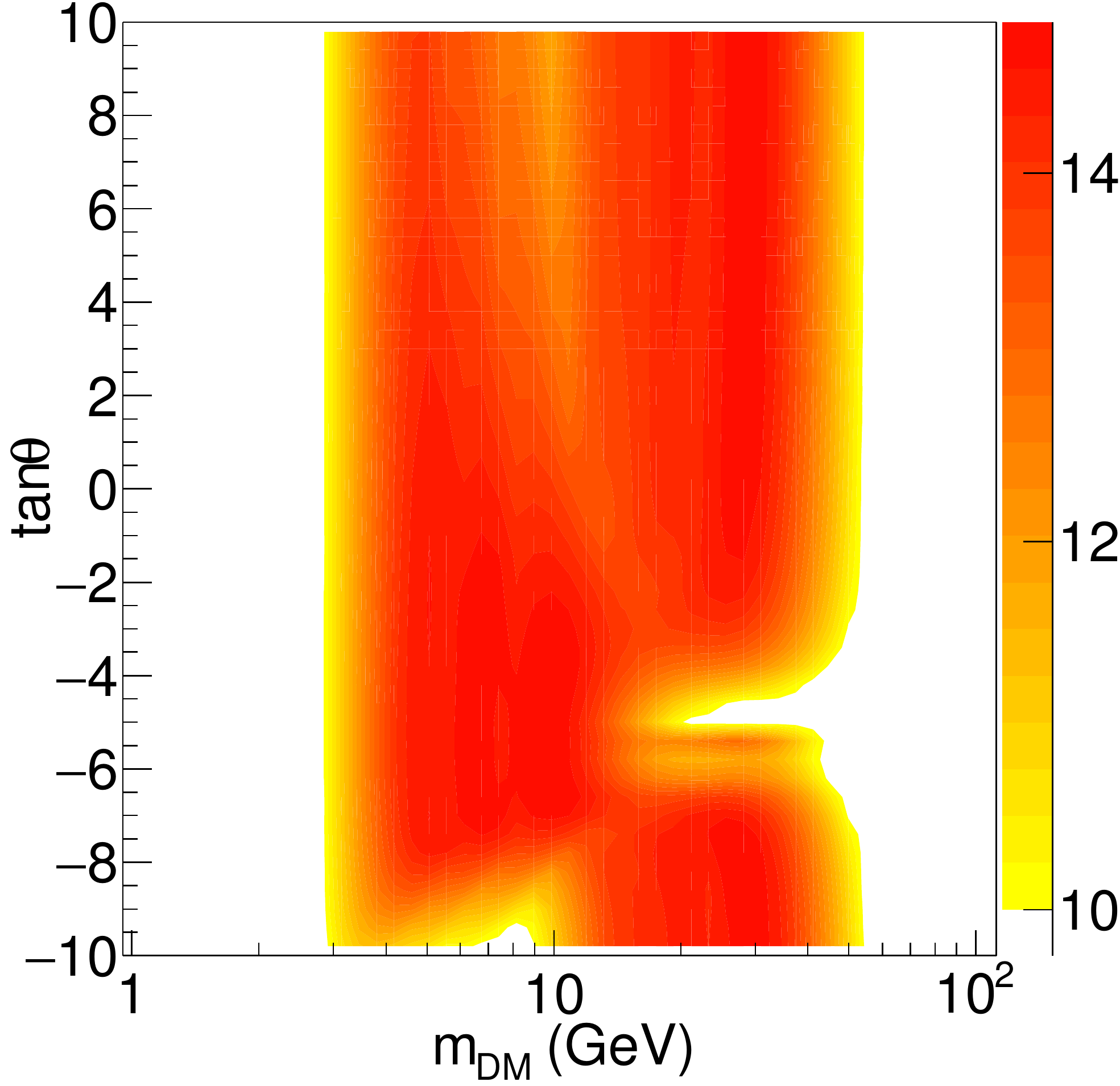}
\includegraphics[width=4.2cm] {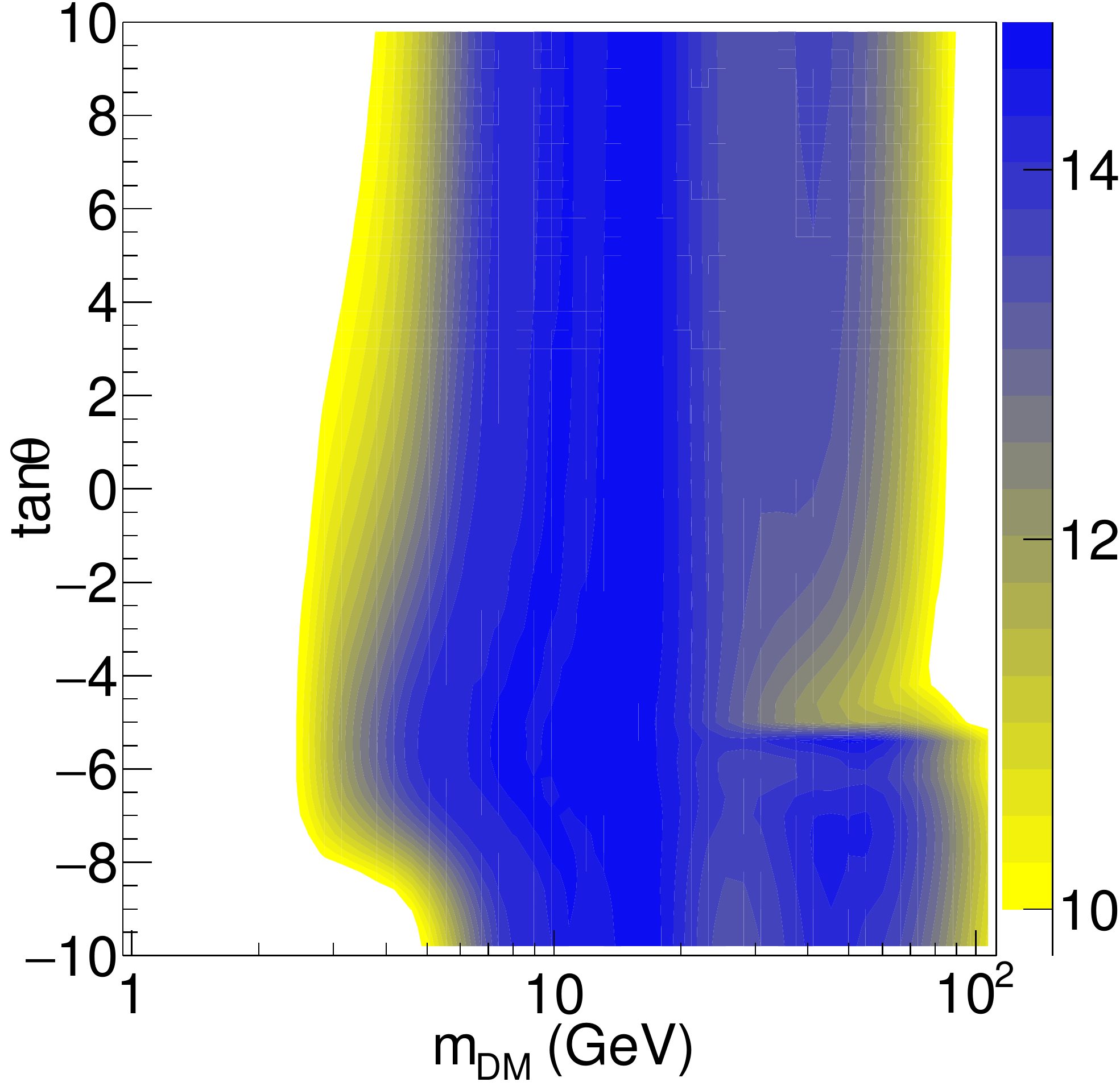}
\includegraphics[width=4.2cm] {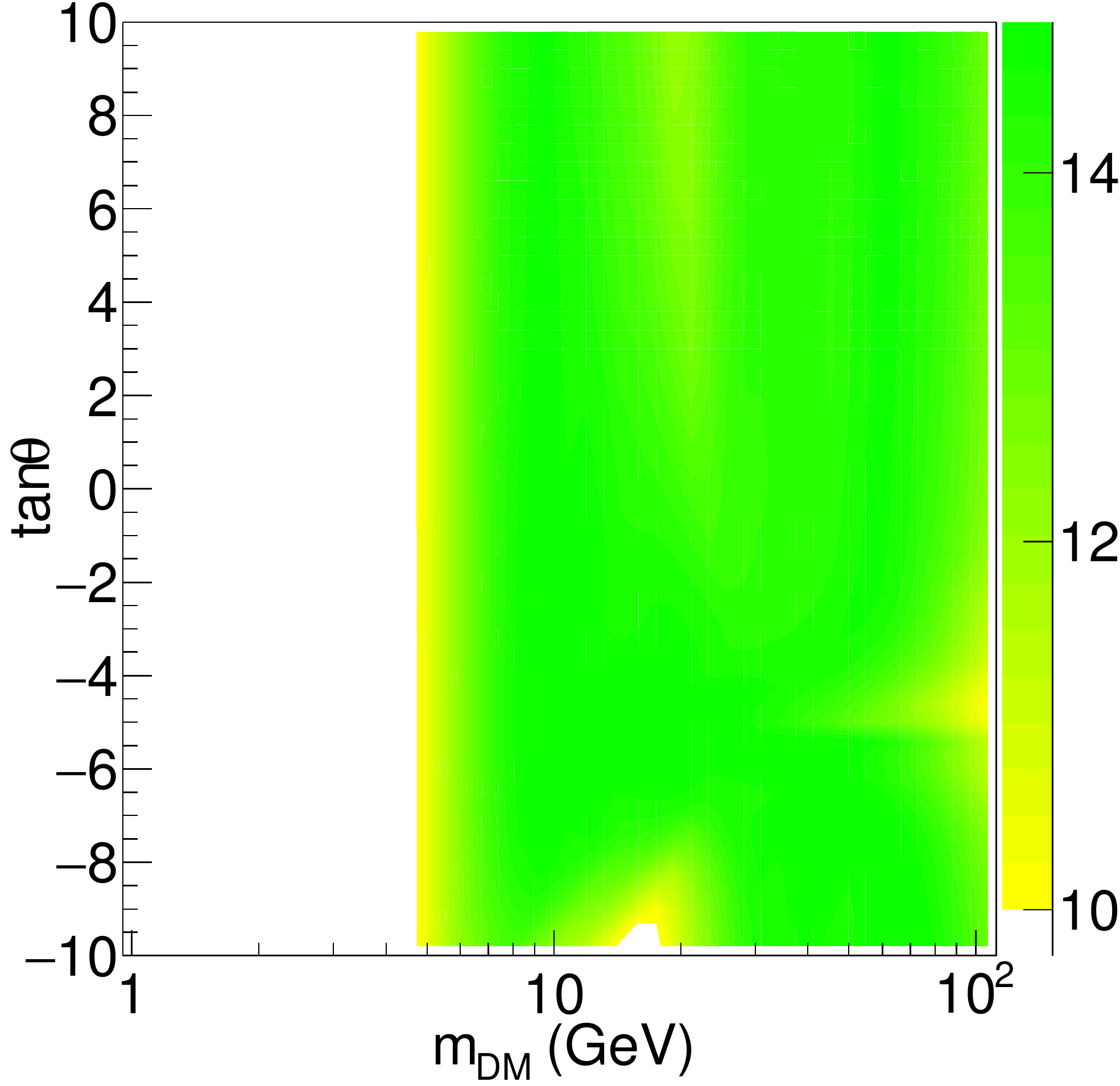}
\includegraphics[width=4.2cm] {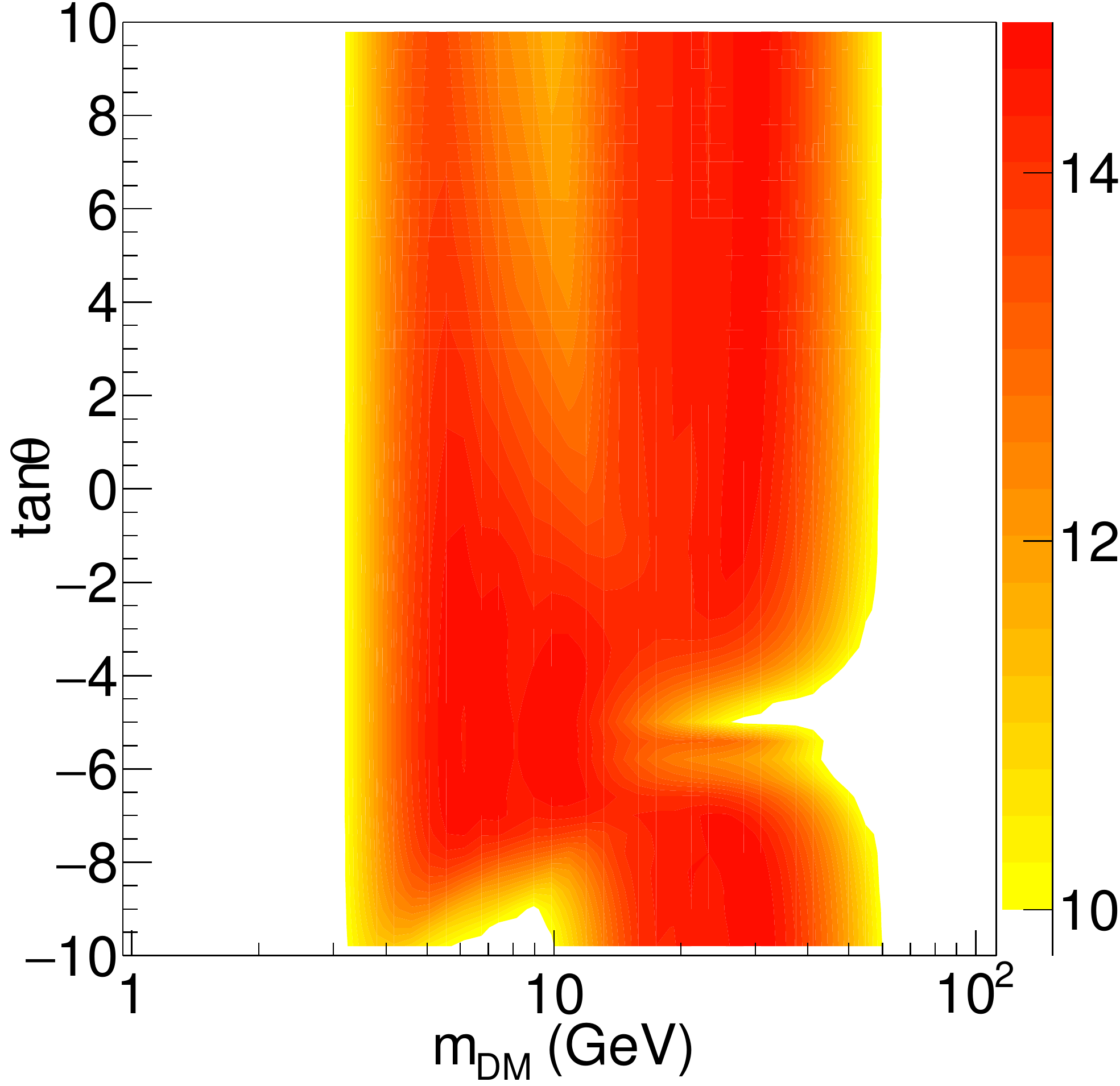}
\includegraphics[width=4.2cm] {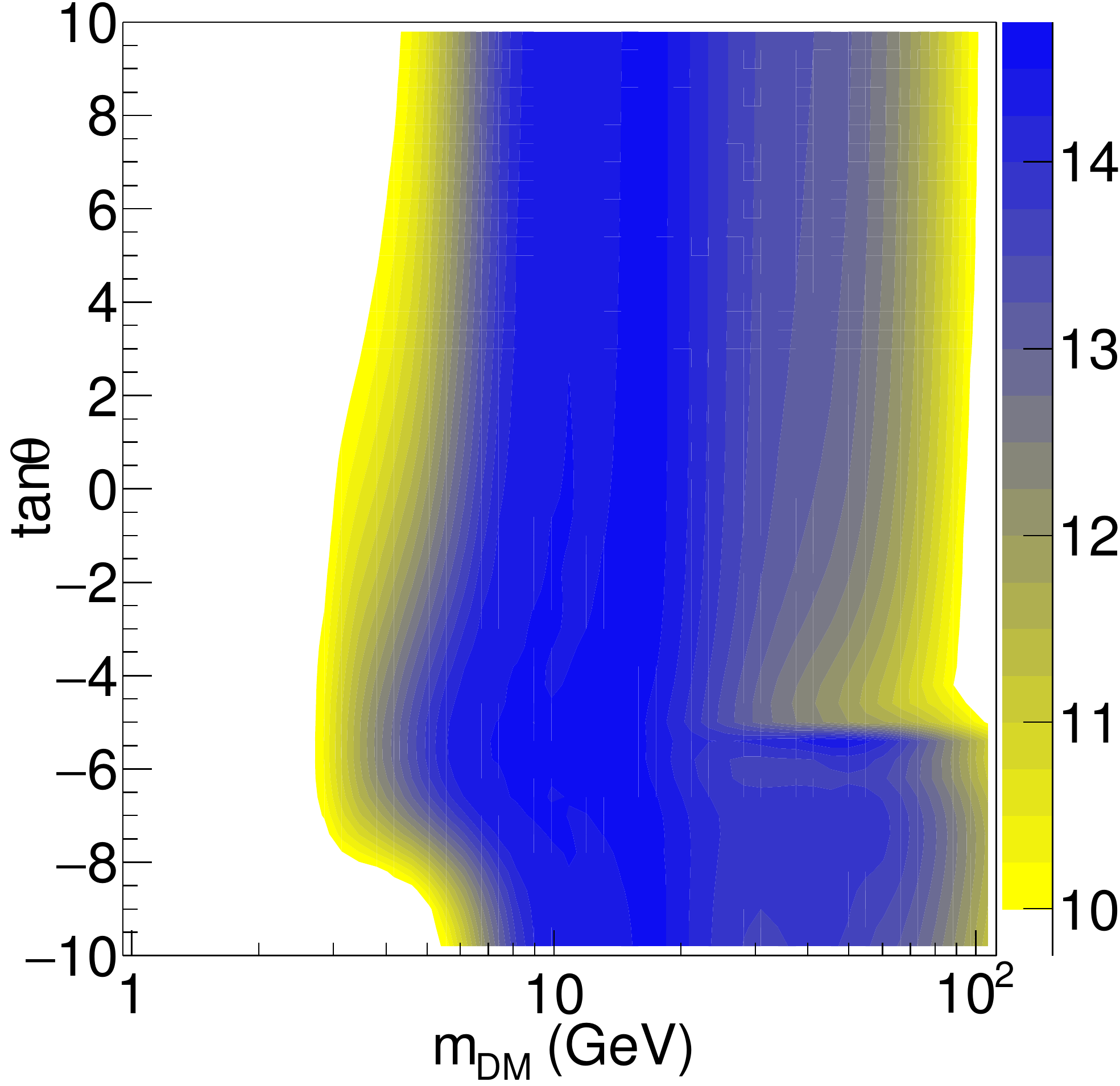}
\includegraphics[width=4.2cm] {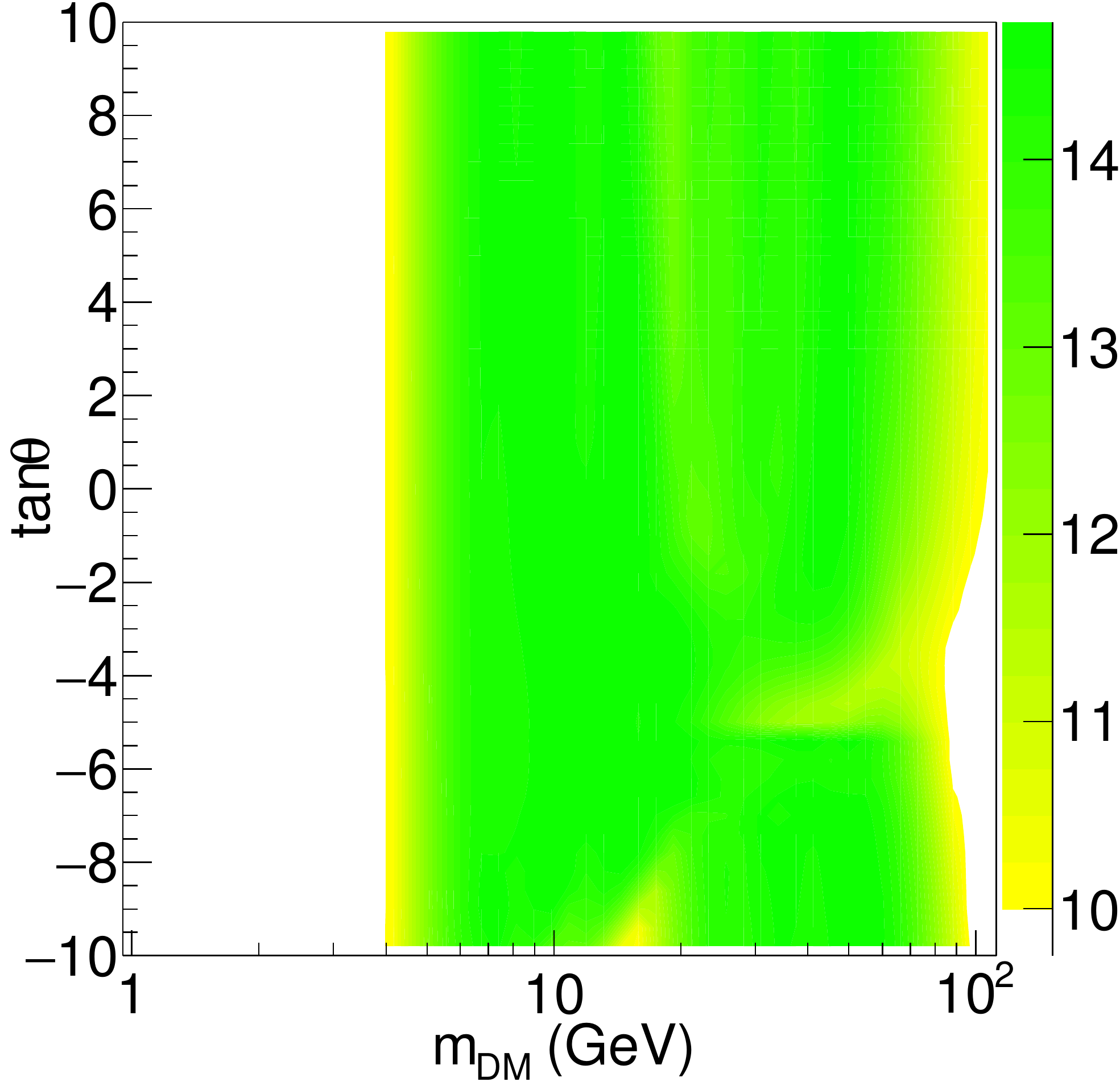}
\includegraphics[width=4.2cm] {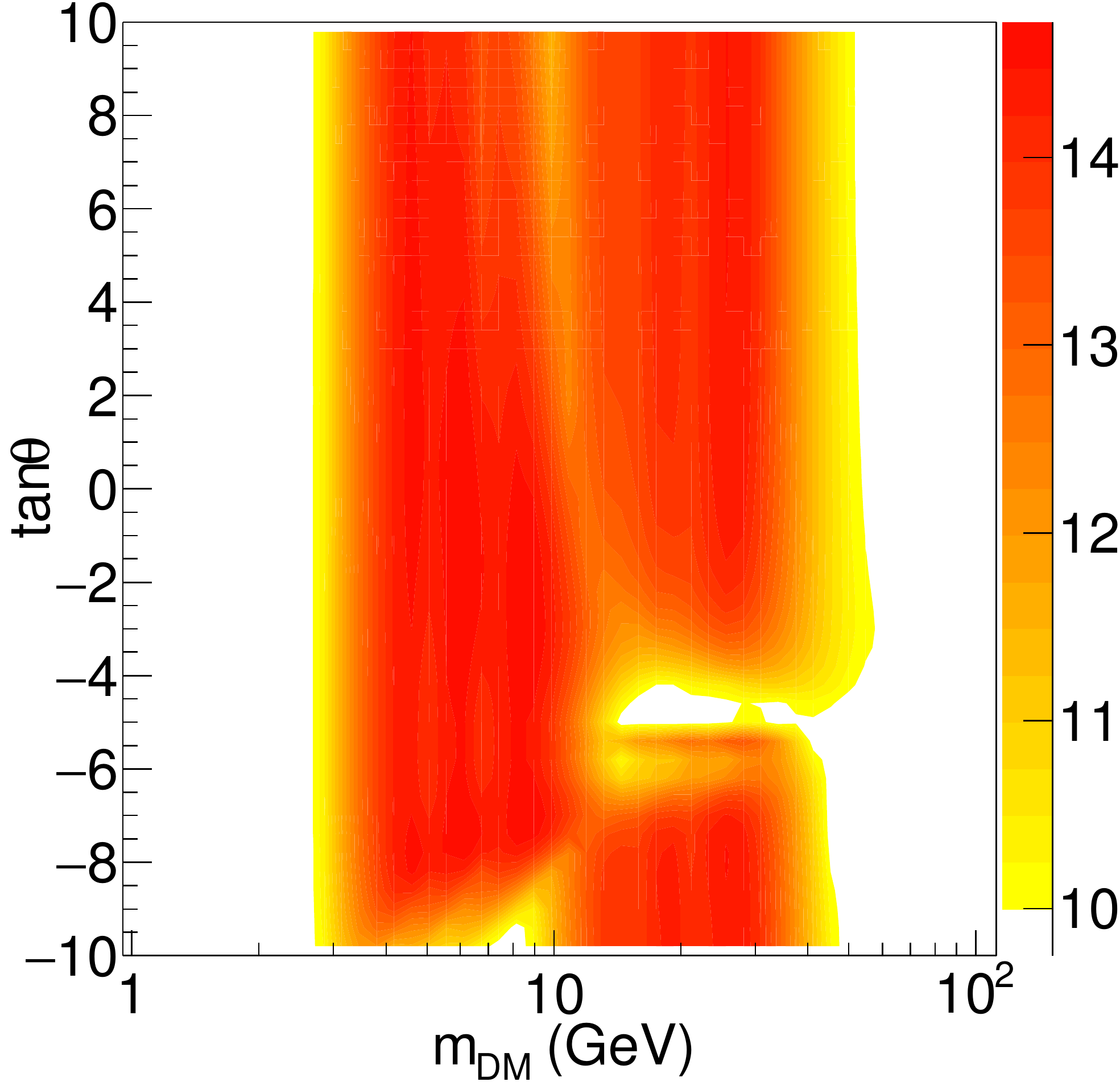}
\includegraphics[width=4.2cm] {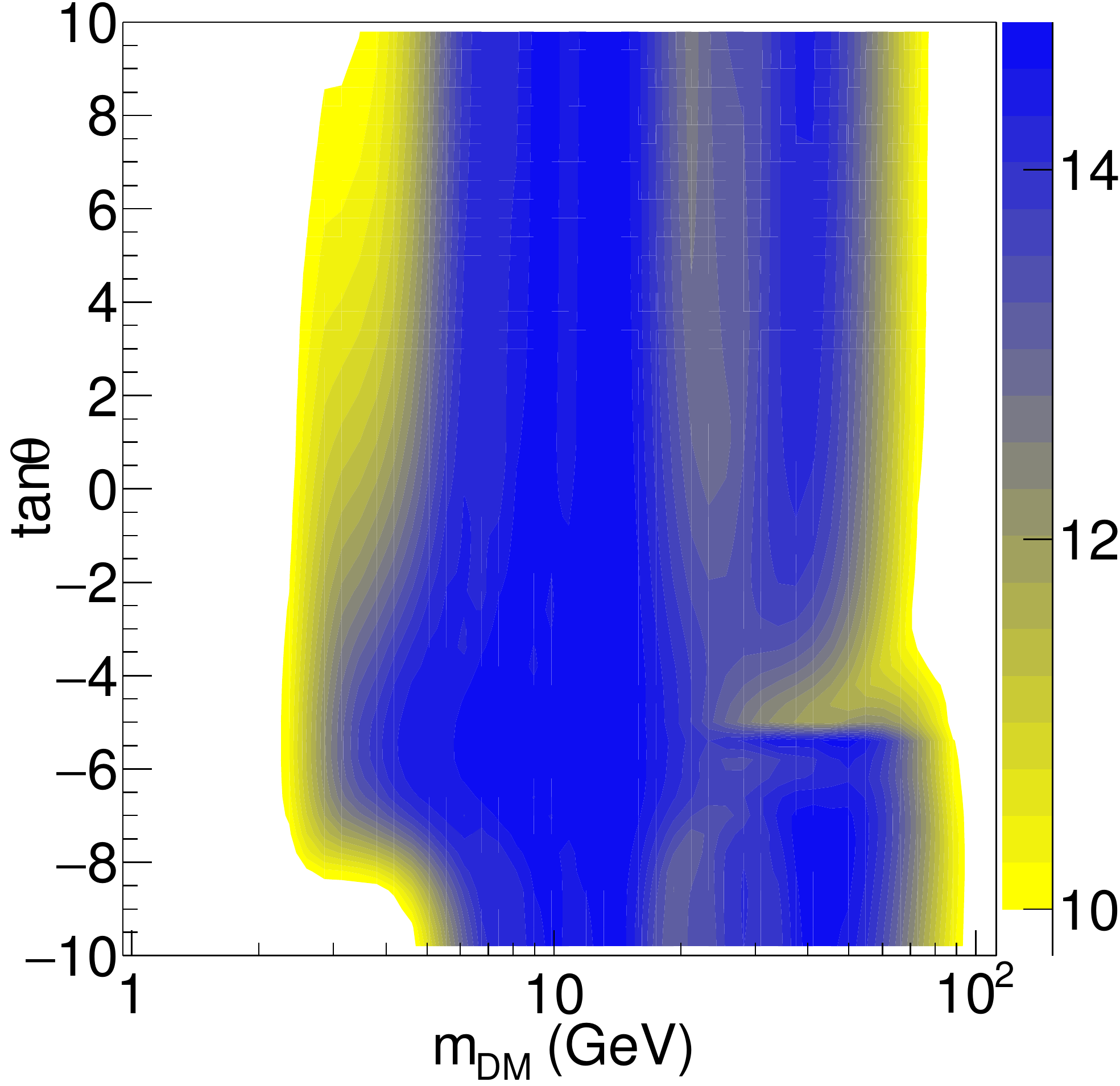}
\includegraphics[width=4.2cm] {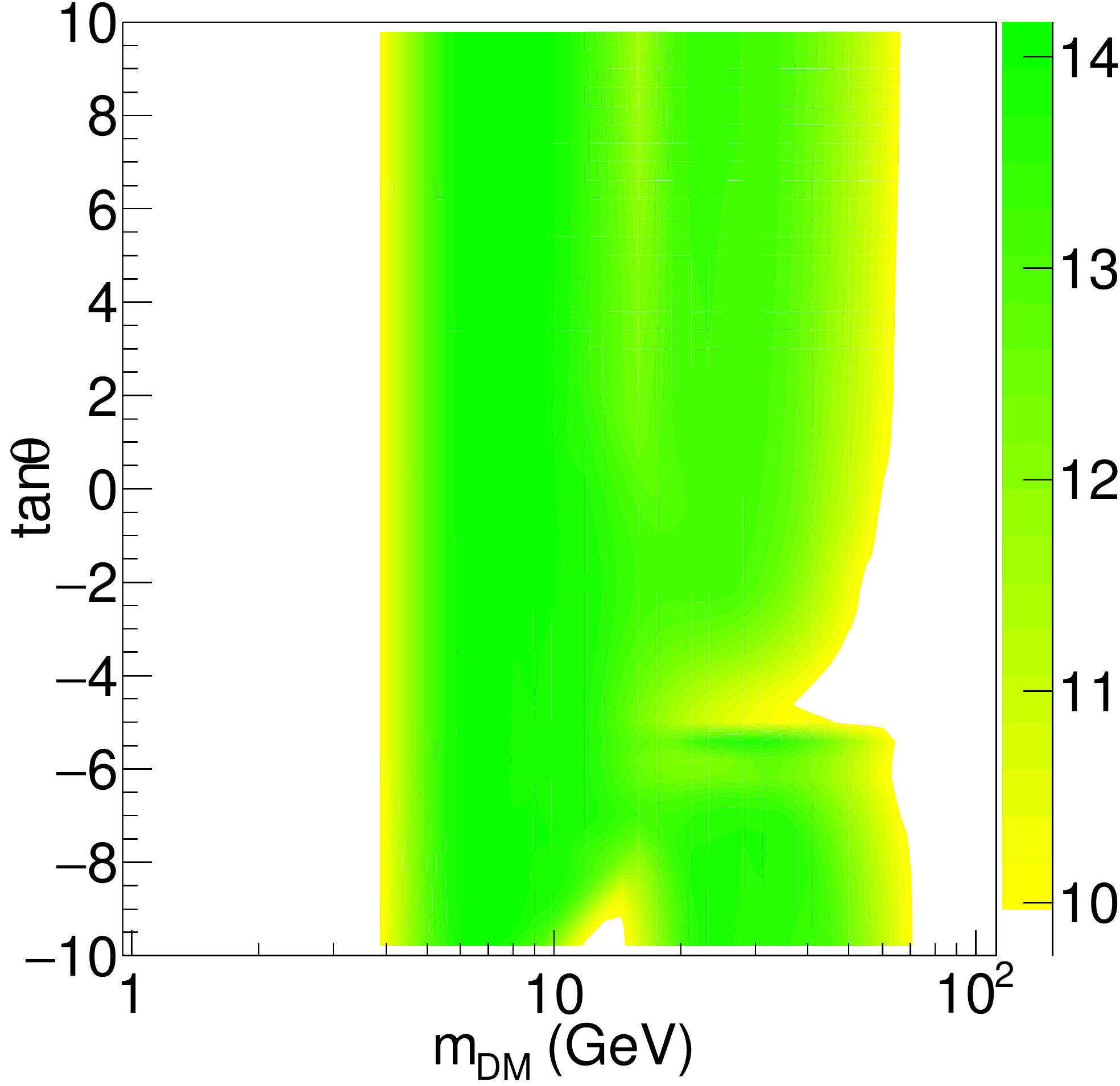}
\includegraphics[width=4.2cm] {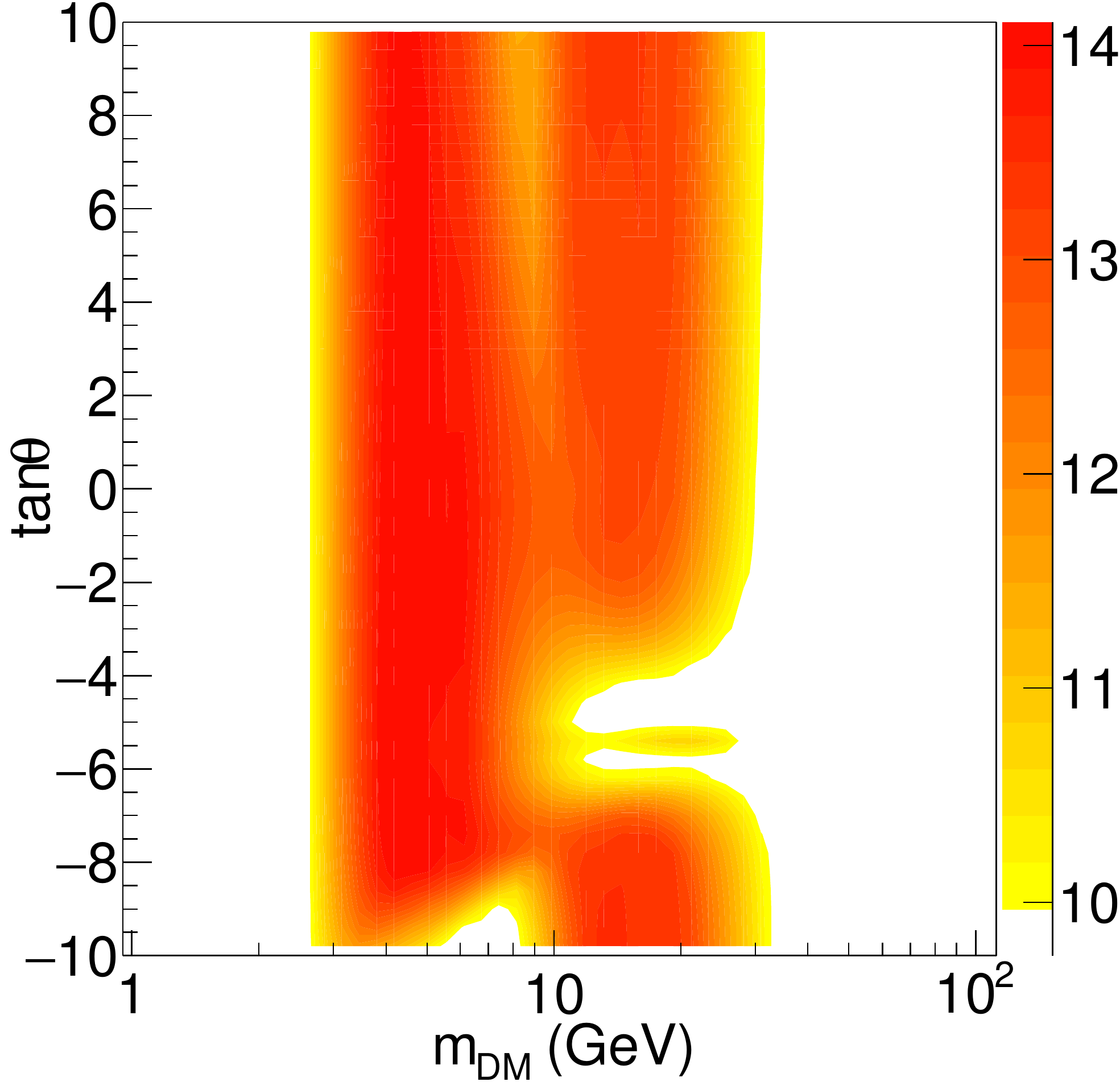}
\includegraphics[width=4.2cm] {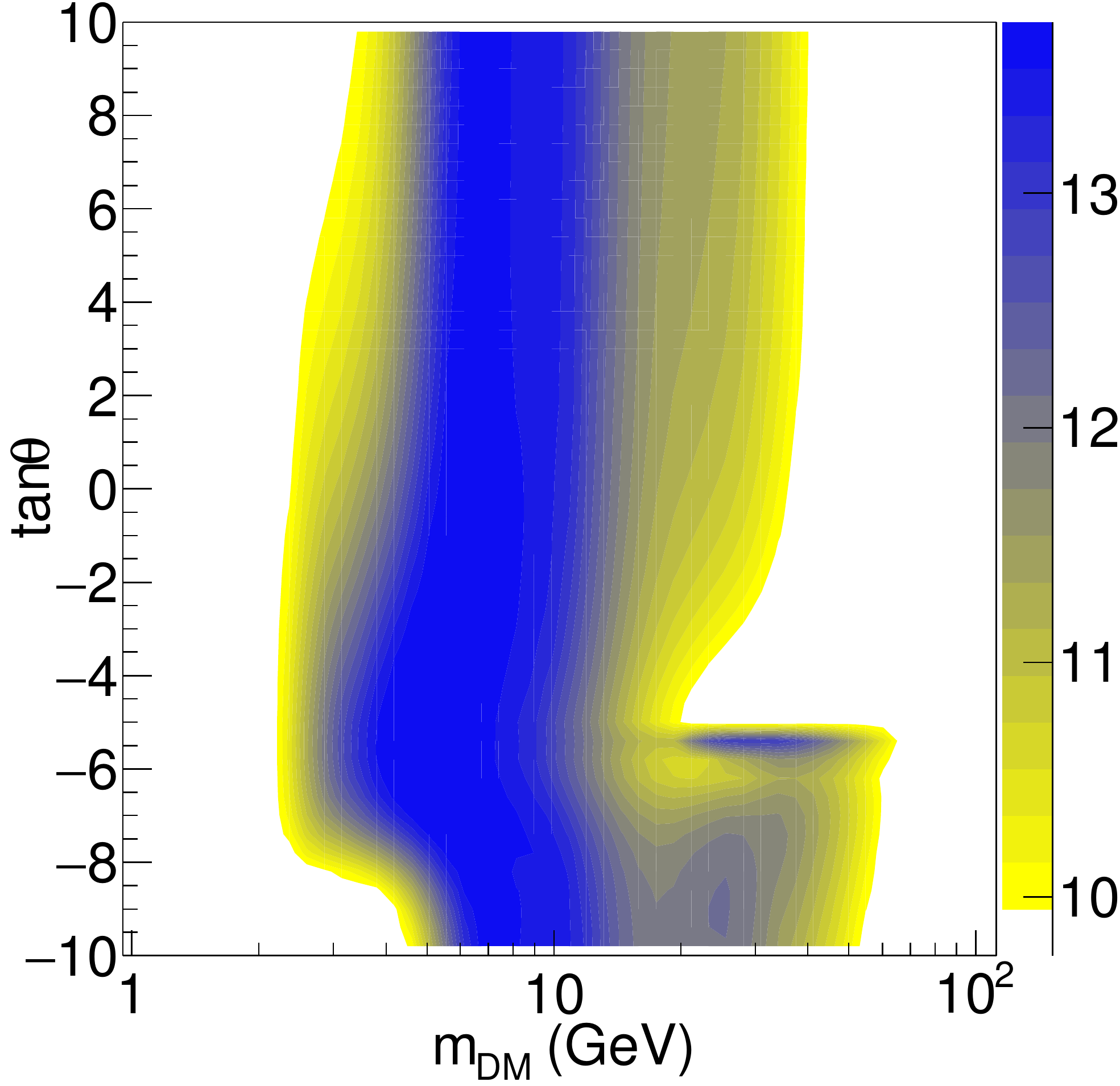}
\end{center}
\vspace{-0.5cm}
\caption{Regions in the $tan \theta$ vs $m_{DM}$ plane allowed by DAMA experiments in the case
of a DM candidate with SD interaction.
The Na and I quenching factors are:
$Q_I$ [$left$ (green on-line)], $Q_{II}$ [$center$ (red on-line)], and with channeling effect [$right$ (blue on-line)].
The considered halos (from top to bottom) are A0 (isothermal sphere), B1, C1, D3 with the $v_0$ and $\rho_0$
in the range of Table III of Ref. \cite{bel02}.
The three possible sets of parameters A, B
and C are considered (see Sect. \ref{data_analysis}); the allowed regions represent the domain where the
likelihood-function values differ more than 10 $\sigma$ from the {\it null hypothesis} (absence of
modulation).
The color scales give the confidence level in units of $\sigma$ from the {\it null hypothesis}.}
\label{fg:sd_theta}
\end{figure}

\vspace{0.6cm}
Finally, Fig. \ref{fg:sd_theta} shows the allowed regions 
in the $tan \theta$ vs $m_{DM}$ plane after marginalizing on $\xi \sigma_{SD}$.
For simplicity four halo models: A0 (isothermal sphere), B1, C1, D3 with the $v_0$ and $\rho_0$ 
in the range of Table III of Ref. \cite{bel02}, and three choices of the Na and I quenching factors:
$Q_I$, $Q_{II}$, and including the channeling effect are considered.

\vspace{0.6cm}
In conclusion, the purely SD scenarios are in good agreement with the DAMA results and can explain the 
different capability of detection among targets with different unpaired nucleon. The large uncertainties 
e.g. in the spin factor also offer additional space for compatibility among different target nuclei.

\vspace{0.6cm}
\subsubsection{Mixed coupling framework}   

The most general case is when both SI and SD couplings are considered. 
Details of related calculations can be found in Ref. \cite{sisd,RNC}. In this scenario, both the uncertainties on the SI and SD frameworks
have to be accounted. The complete result is given by a 4-dimensional allowed volume: ($\xi\sigma_{SI}$, $\xi\sigma_{SD}$,
$m_{DM}$, $\theta$). The isospin violating SI interaction is not included hereafter.
 
\begin{figure} [!p]
\centering
\vspace{-0.4cm}
\includegraphics[width=12.0cm] {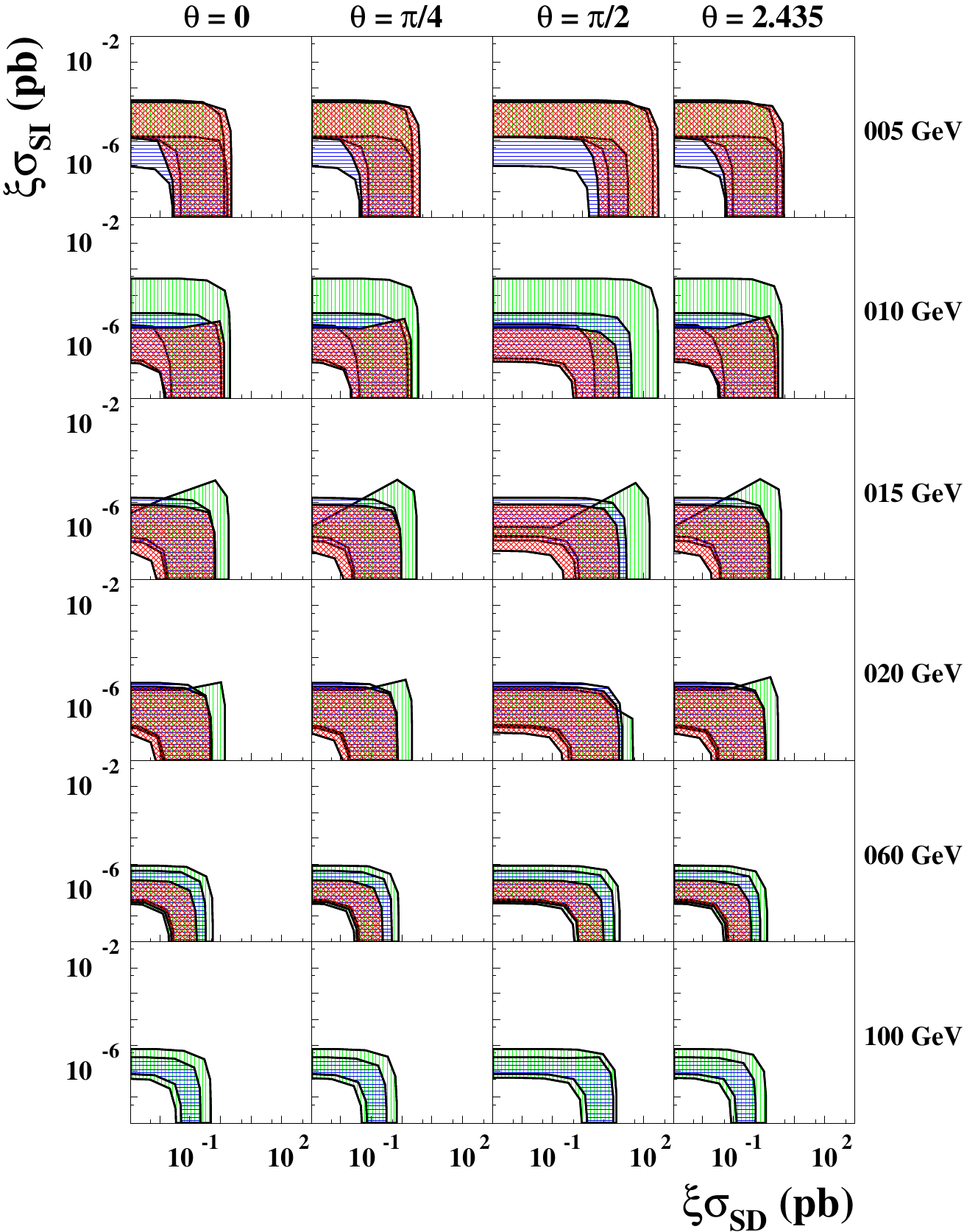}
\vspace{-0.4cm}
\caption{Slices of the 4-dimensional volume ($\xi \sigma_{SI}$, $\xi \sigma_{SD}$, 
$m_{DM}$, $\theta$) allowed by all DAMA experiments in the case of a DM candidate with elastic 
scattering off target nuclei and mixed SI and SD interaction.
Three different instances for the Na and I quenching factors have been considered: 
(i) $Q_I$ case [(green on-line) vertically-hatched region],
(ii) with channeling effect [(blue on-line) horizontally-hatched region] and 
(iii) $Q_{II}$ [(red on-line) cross-hatched region].
The regions have been obtained by marginalizing all the models for each considered scenario 
(see Sect. \ref{data_analysis}) and they represent the domain where the likelihood-function 
values differ more than 10 $\sigma$ from the {\it null hypothesis} (absence of modulation).}
\label{fg:sisd}
\end{figure}

\vspace{0.3cm}
For simplicity examples of slices ($\xi\sigma_{SI}$, $\xi\sigma_{SD}$) 
at 10 $\sigma$ from the {\it null hypothesis} (absence of modulation) 
are shown in Fig. \ref{fg:sisd} for some choices of $m_{DM}$ and $\theta$ values.

\vspace{0.3cm}
Obviously, the proper accounting for the complete 4-dimensional allowed volume and the existing uncertainties 
and complementarity largely extend the results and any comparison.

\begin{figure}[!ht]
\centering
\vspace{-0.4cm}
\includegraphics[width=4.3cm] {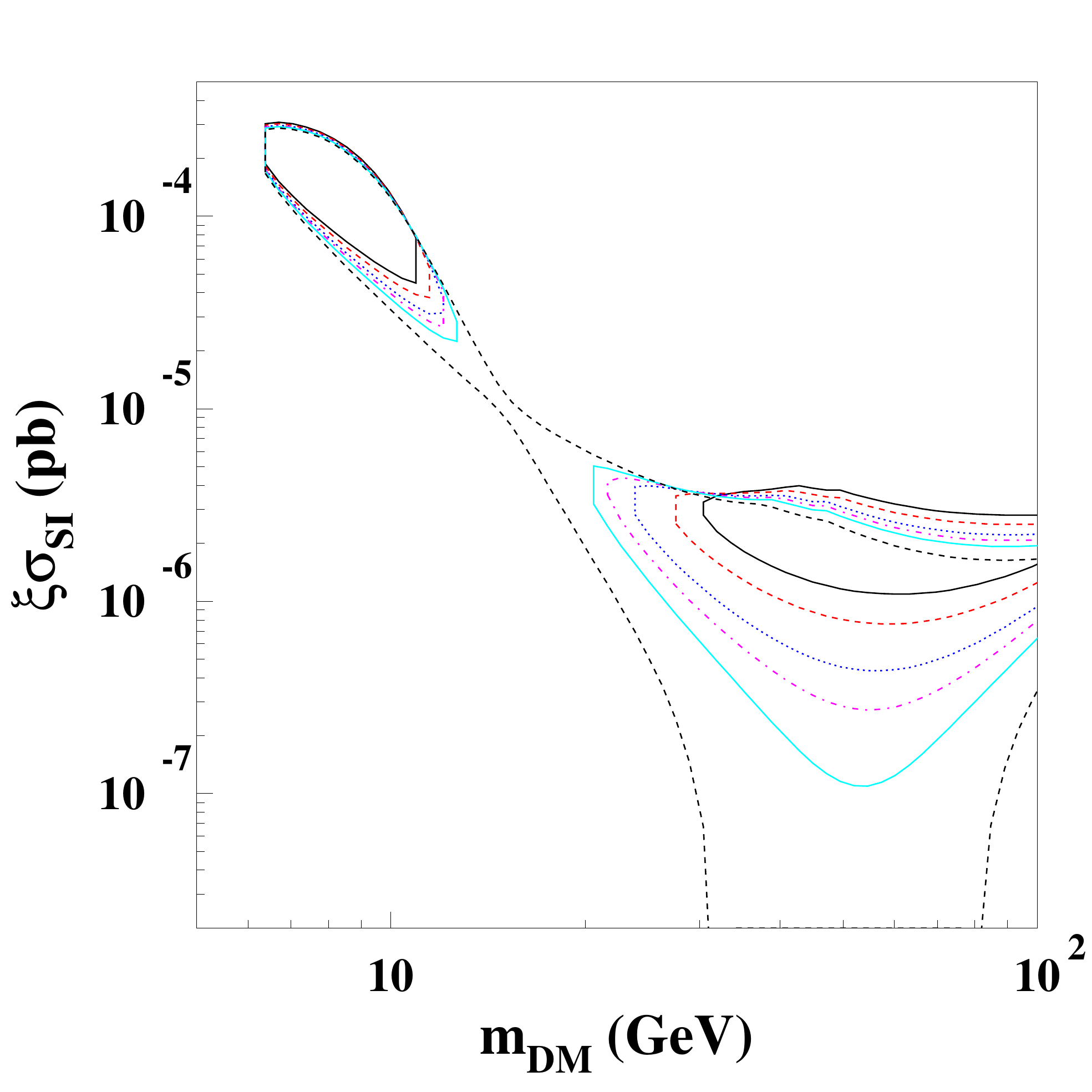}
\includegraphics[width=4.3cm] {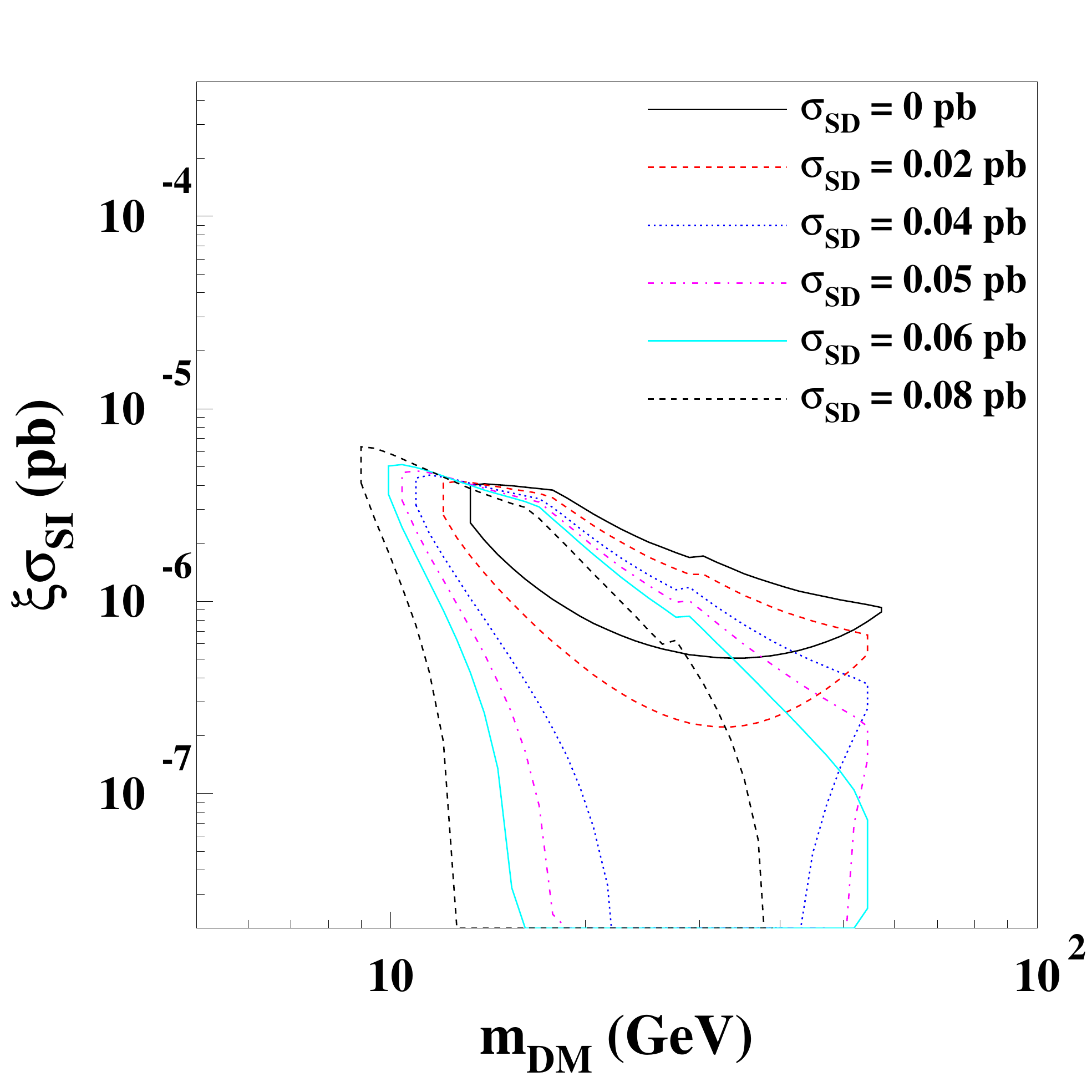}
\includegraphics[width=4.3cm] {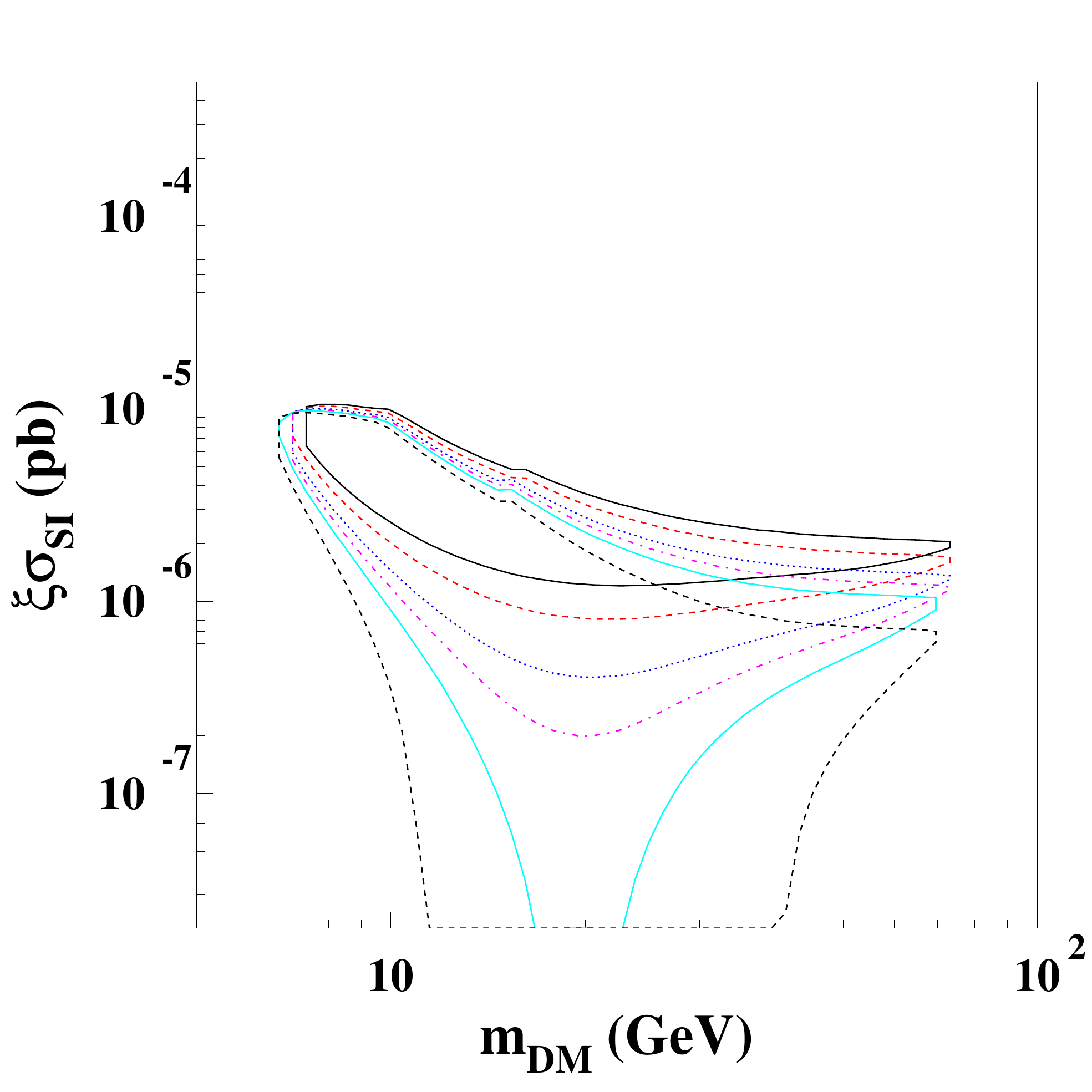}
\vspace{-0.4cm}
\caption{An example of the effect induced by the inclusion of a SD component different from zero
on allowed regions given in the plane $\xi \sigma_{SI}$ vs $m_{DM}$.
In this example the B1 halo model with $v_0 = 170$ km/s and $\rho_0 = 0.42$ GeV/cm$^3$,
the set of parameters A and the particular case of $\theta = 0$ for the SD interaction have been
considered. The used quenching factors are $Q_I$ ($left$), $Q_{II}$ ($center$) and with channeling
effect ($right$).
From top to bottom the contours refer to different SD contributions:
$\sigma_{SD} = 0$ pb (solid black line), 0.02 pb, 0.04 pb, 0.05 pb, 0.06 pb and 0.08 pb.
Analogous situation is found for the other model frameworks.}
\label{fg:si+fsd}
\end{figure}

\begin{figure}[!ht]
\centering
\vspace{-0.4cm}
\includegraphics[width=4.3cm] {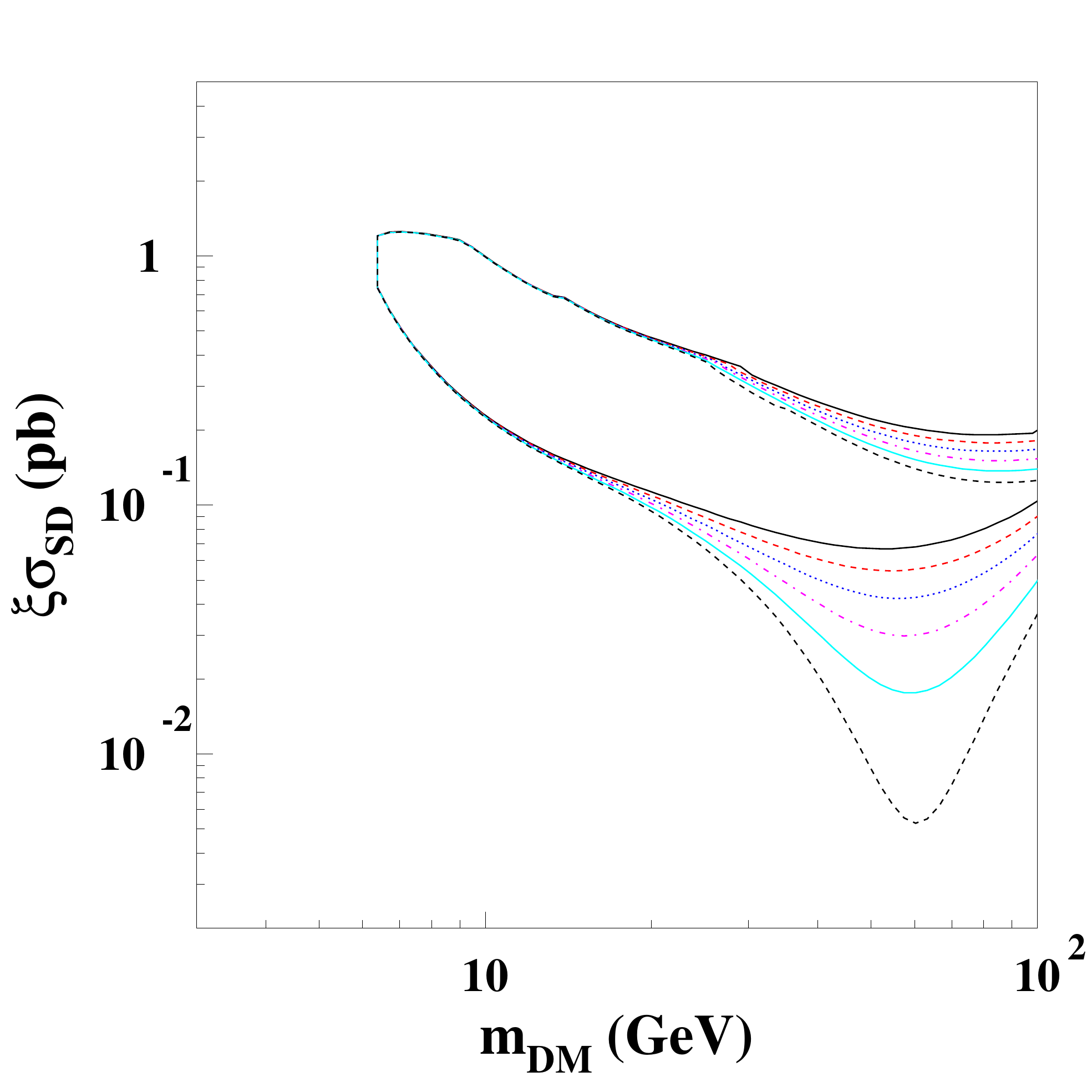}
\includegraphics[width=4.3cm] {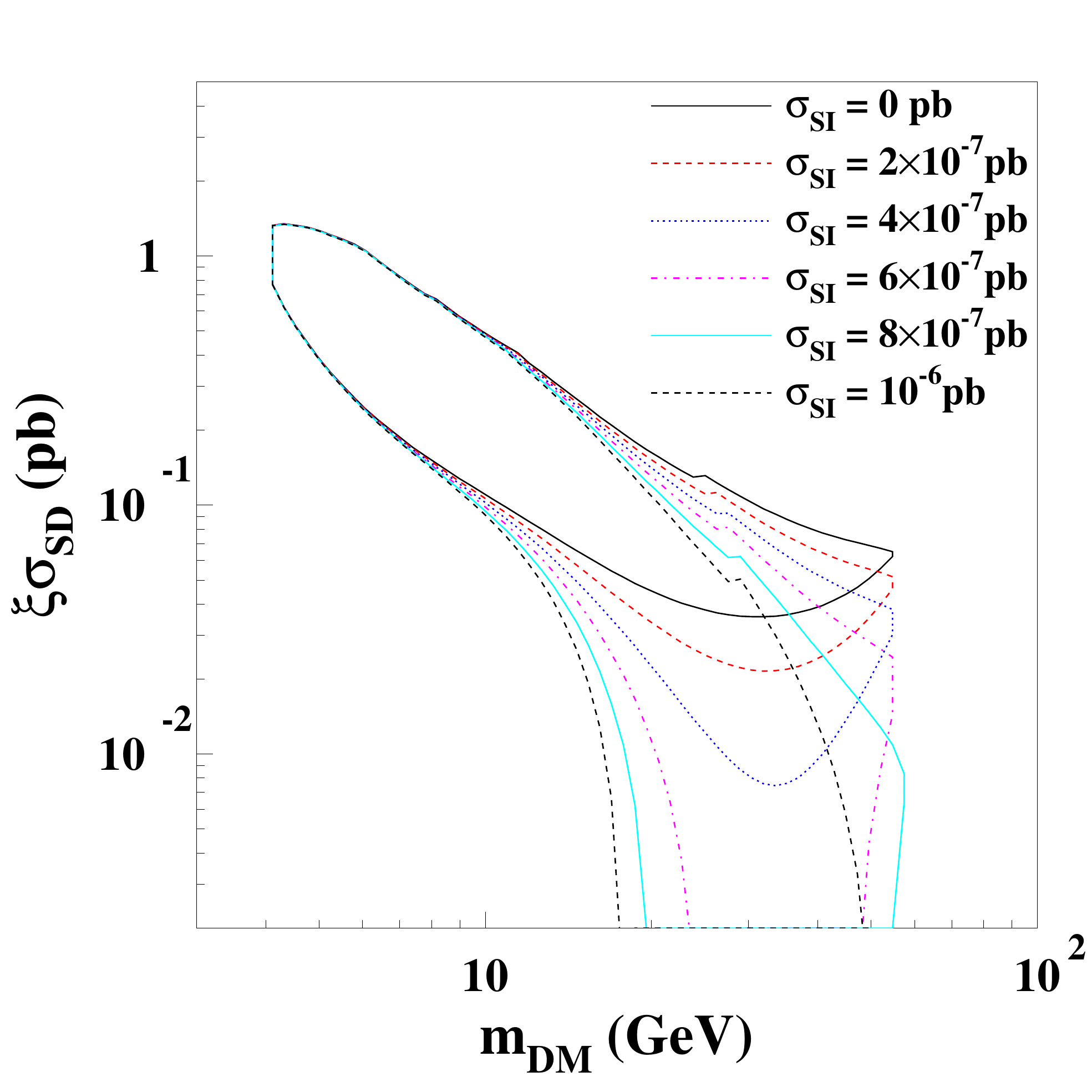}
\includegraphics[width=4.3cm] {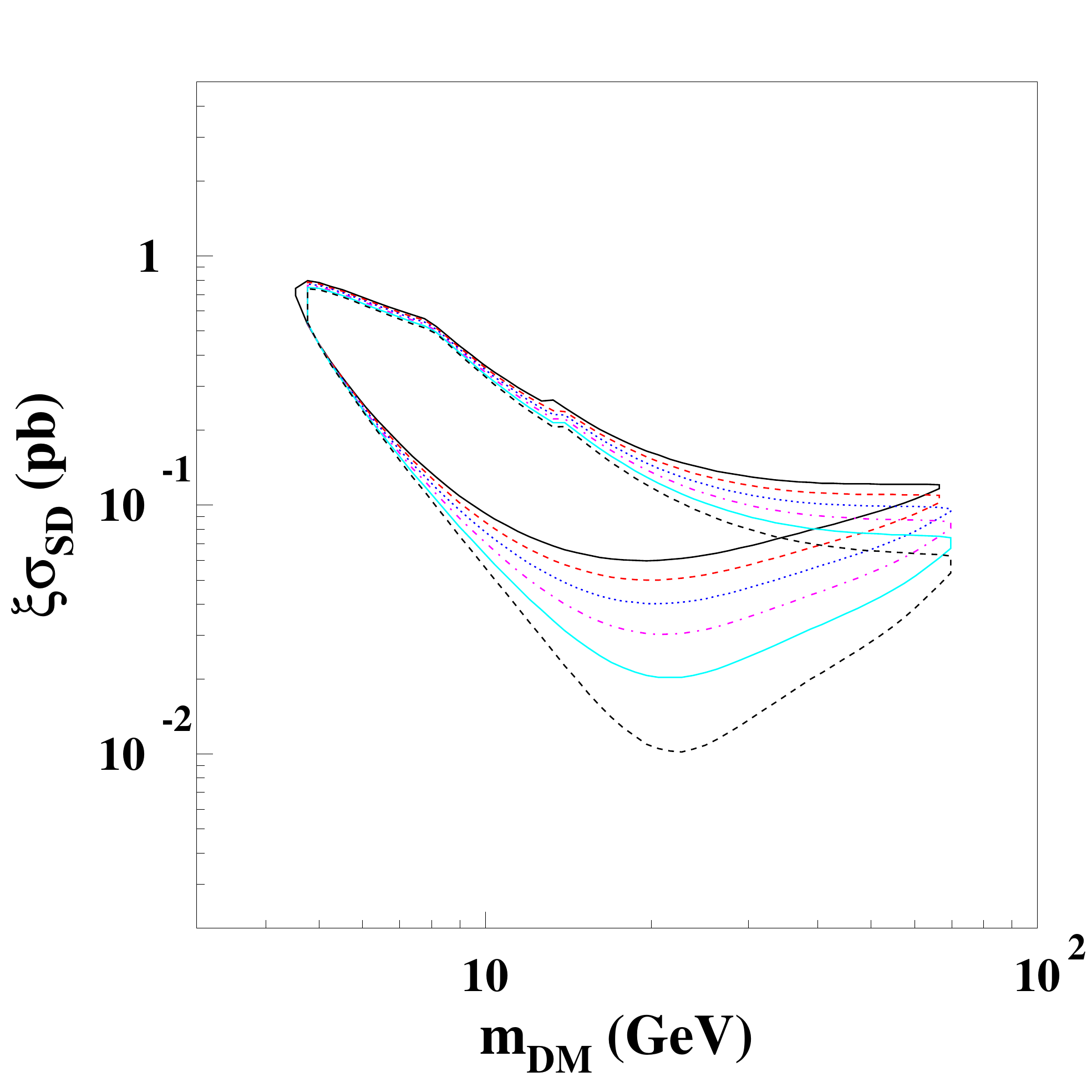}
\vspace{-0.4cm}
\caption{An example of the effect induced by the inclusion of a SI component different from zero
on allowed regions given in the plane $\xi \sigma_{SD}$ vs $m_{DM}$.
The same halo model, parameter set and $\theta$ value of Fig. \ref{fg:si+fsd} have been considered.
The used quenching factors are $Q_I$ ($left$), $Q_{II}$ ($center$) and with channeling
effect ($right$).
From top to bottom the contours refer to different SI contributions:
$\sigma_{SI} = 0$ pb (solid black line), $2 \times 10^{-7}$ pb, $4 \times 10^{-7}$ pb,
$6 \times 10^{-7}$ pb, $8 \times 10^{-7}$ pb, $10^{-6}$ pb.
Analogous situation is found for other model frameworks.}
\label{fg:sd+fsi}
\end{figure}

\vspace{0.3cm}
Finally, let us now point out that configurations with 
$\xi \sigma_{SI}$ ($\xi \sigma_{SD}$) even much lower than those shown in Fig. \ref{fg:si} 
(Fig. \ref{fg:sd}) would be possible if a small SD (SI) contribution would be present in the interaction.
This possibility is clearly pointed out
in Figs.~\ref{fg:si+fsd} and \ref{fg:sd+fsi} where some examples of regions in the plane
cross section vs $m_{DM}$ are reported.
As it can be seen,
these arguments clearly show that even a relatively small SD (SI) contribution
can drastically change the allowed region in the  ($m_{DM}$, $\xi \sigma_{SI(SD)}$)
plane; therefore, the typically shown model-dependent comparison plots
between exclusion limits at a given C.L. and regions of allowed parameter space do not hold e.g. 
for mixed scenarios when comparing experiments with and
without sensitivity to the SD component of the interaction.
The same happens when comparing regions allowed by experiments whose target-nuclei have unpaired proton
with exclusion plots quoted by experiments using target-nuclei with unpaired neutron
when the
SD component of the interaction would correspond
either to $\theta \simeq 0$ or $\theta \simeq \pi.$

\subsection {DM particles with preferred electron interaction}
\label{DM2}

Some extensions of the standard model provide DM candidate 
particles, which can have a dominant coupling with the lepton sector of the 
ordinary matter. Thus, such DM candidate particles can be 
directly detected only through their interaction with electrons in the 
detectors of a suitable experiment, while they cannot be studied 
in those experimental results where subtraction/rejection of the electromagnetic 
component of the experimental counting rate is applied\footnote{If the electron 
is assumed at rest, considering the DM particle velocity, the released energy would be of order 
of few eV, well below the detectable energy in any considered detector in the field. However, the electron is 
bound in the atom and, even if the atom is at rest, the electron can have non-negligible momentum, as shown in 
Ref. \cite{wimpele}.}. These candidates can also offer a 
possible source of the 511 keV photons observed from the galactic bulge. 
This scenario was already investigated by DAMA with lower exposure \cite{wimpele}.

In particular, as shown in Ref. \cite{wimpele}, 
such DM candidate particles with mass $\gsim$ few GeV can interact on bound electrons
with momentum up to $\simeq$ few MeV/c; thus, they can provide signals in the keV 
energy region detectable by low background and low energy threshold detectors, 
such as those of DAMA. The expected differential energy spectrum has been derived in 
Ref. \cite{wimpele}; it depends on a single parameter, $\frac{\xi \sigma_e^0}{m_{DM}}$,
for each halo model. Here, $\sigma_e^0$ is the DM particle cross section on electron at rest \cite{wimpele}.

\begin{figure}[!ht]
\begin{center}
\includegraphics[width=7.0cm] {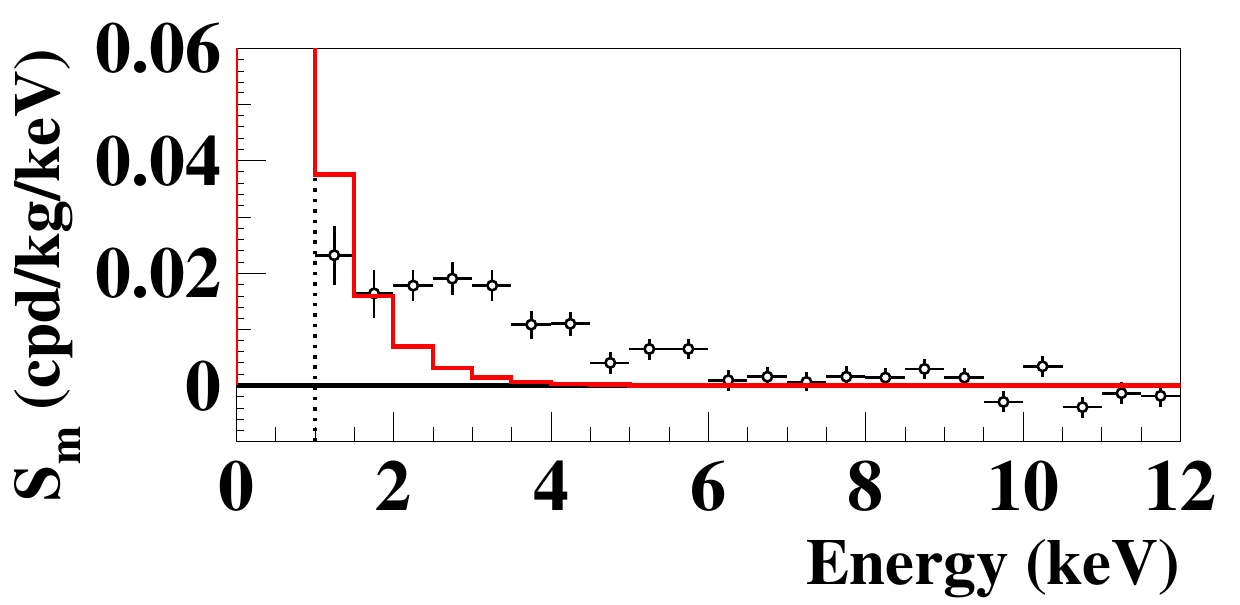}
\end{center}
\vspace{-0.6cm}
\caption{Example of superposition of the measured $\mathcal{S}^{exp}_{m}$ vs energy (points
with error bars) with theoretical expectations (solid histograms) for DM particles with preferred electron interaction.
The case of the C2 (Evans logarithmic) corotating halo model with $\rho_0$ = 0.67 GeV/cm$^3$,
$v_0$ = 170 km/s and $\frac{\xi \sigma_e^0}{m_{DM}} = 9.5 \times 10^{-4}$ pb/GeV
is considered. This is an example how to disentangle among some scenarios, improving the sensitivity of the set-up.
}
\label{fg:smvse_el}
\end{figure}

With the new cumulative exposure we have derived the results 
following the same procedure as in Ref. \cite{wimpele}, and the prescription of Sect. \ref{data_analysis}.
The expected behaviour of the modulation amplitudes rapidly rises at low energy \cite{wimpele};
an example is also reported in Fig. \ref{fg:smvse_el}. On the contrary, the measured modulation amplitudes 
have a smooth trend with energy even below 2 keV. In particular, this has been pointed out by the new 
results of DAMA/LIBRA--phase2. Thus, the lower energy threshold achieved by DAMA/LIBRA--phase2 at 1 keV 
prevents to find configurations for these DM candidates {\it distant} more than 10 $\sigma$ from the {\it null hypothesis}. 
This is an example how to disentangle among some scenarios, improving the sensitivity of the set-up.

\begin{figure}[!ht]
\centering
\includegraphics[width=7.5cm] {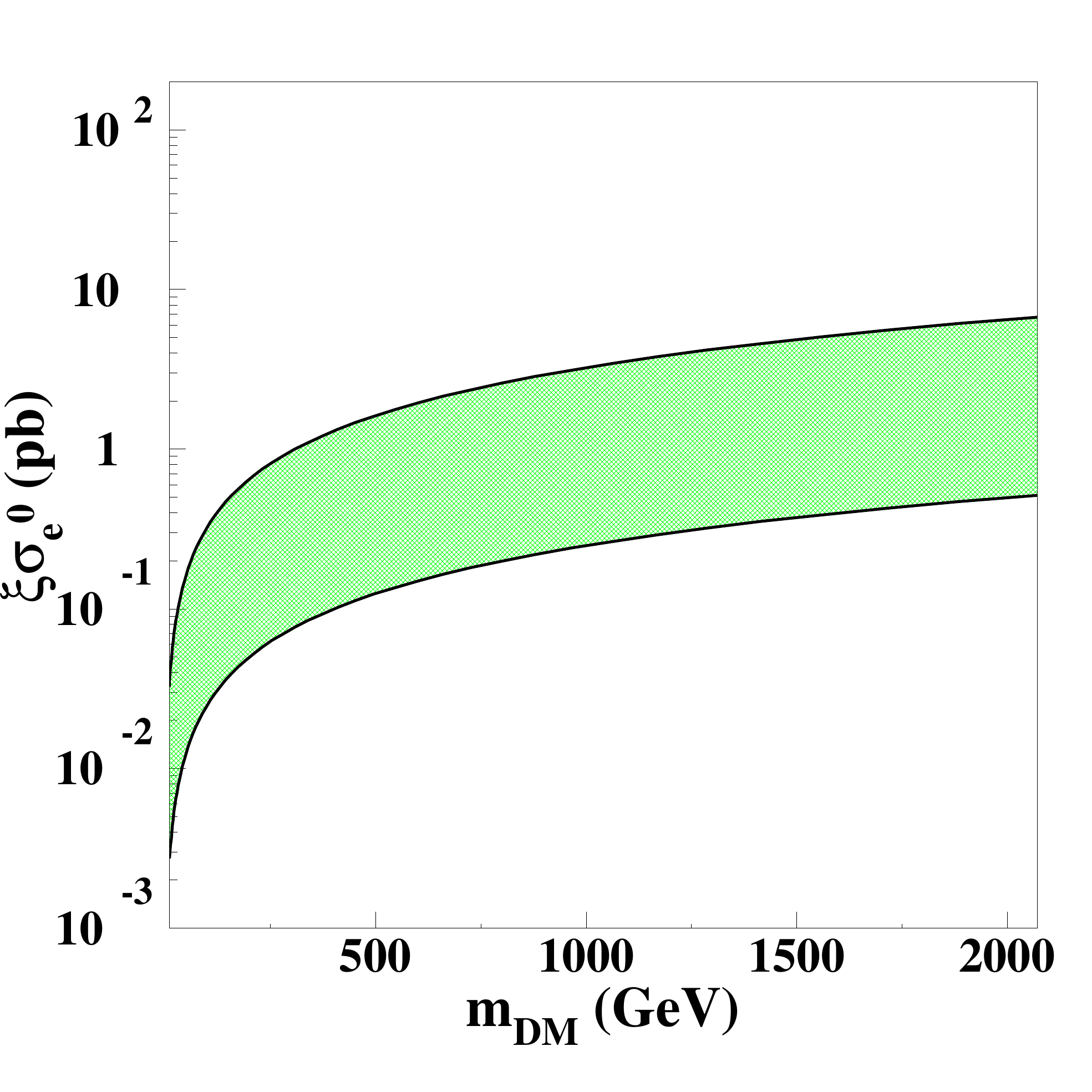}
\vspace{-0.6cm}
\caption{Region allowed in the ($\xi \sigma_e^0$ vs $m_{DM}$) plane
for the same dark halo models and
related parameters described above.
The region encloses configurations {\it distant} more than 8 $\sigma$ from
the {\it null hypothesis}.
The C.L. suggests that this kind of scenario is less favored by the data 
with respect to other ones considered in this paper.
We note that, although the mass region in the plot is up to 2 TeV,
$m_{DM}$ particles with larger masses are also allowed.}
\label{fg:reg}
\end{figure}

However, just for these DM candidates we can apply a less severe confidence level.
Thus, Fig. \ref{fg:reg} shows the allowed region in the ($\xi \sigma_e^0$ vs $m_{DM}$) plane
for the dark halo models and related parameters described above. 
The region encloses configurations {\it distant} more than 8 $\sigma$ from the {\it null hypothesis}.
At such a confidence level about half dark halo models provide allowed interval for the 
$\frac{\xi \sigma_e^0}{m_{DM}}$ parameter.
We note that, although the mass region in the plot is up to 2 TeV,
$m_{DM}$ particles with larger masses are also allowed.

The results given here hold for every kind of DM candidate interacting with electrons
and with cross section $\sigma_e$ having a weak dependence on electron momentum and 
DM particle velocity \cite{wimpele}.

The hypothesis of a 4--fermion point contact interaction
can be described by a possible mediator of the interaction (hereafter $U$ boson)
with mass $M_U$ larger than the transferred momentum ($M_U \gsim 10$ MeV).
In the pure $V\pm A$ and pure scalar scenario,
the effective coupling constant, $G$, depends on the couplings, $c_e$ and $c_{DM}$,
of the $U$ boson with the electron and the DM particle, respectively.
The cross section on electron at rest is:
$$ \sigma_e^0 = \frac{c_e^2 c_{DM}^2 m_e^2}{\pi M_U^4}.$$

\begin{figure} [!ht]
\centering 
\includegraphics[width=7.5cm] {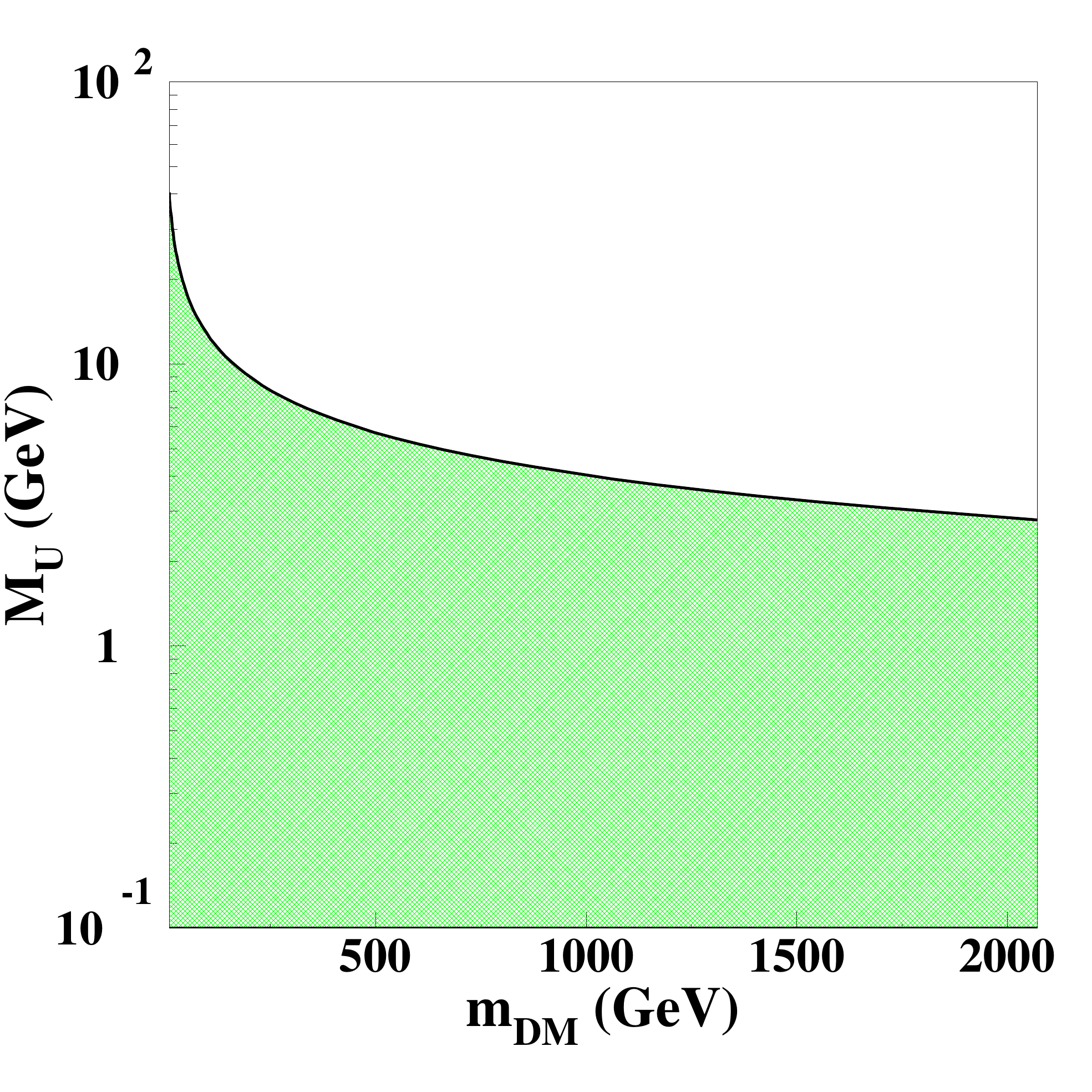}
\vspace{-0.6cm}
\caption{Region of $U$ boson mass allowed by present
analysis considering \cite{wimpele}:
i)   the limit on $c_e$ from $g_e-2$ data; 
ii)  $c_{DM} < \sqrt{4\pi}$, that is the theory is perturbative;
iii) the obtained lower bound $\frac{\xi \sigma_e^0}{m_{DM}} > 2.5 \times 10^{-4}$ pb/GeV
     from the new cumulative data set (as shown in Fig. \ref{fg:reg});
iv)  $\xi \le 1$.
$U$ boson with $M_U$ masses in the sub-GeV range is well allowed
for a large interval of $m_{DM}$.}
\label{fg:uboson}
\end{figure}

Following the procedure of Ref. \cite{wimpele}, that considers: 
i)   the limit on $c_e$ from $g_e-2$ data; 
ii)  $c_{DM} < \sqrt{4\pi}$, that is the theory is perturbative;
iii) the obtained lower bound $\frac{\xi \sigma_e^0}{m_{DM}} > 2.5 \times 10^{-4}$ pb/GeV
     from the cumulative 2.46 ton $\times$ yr data set (as shown in Fig. \ref{fg:reg});
iv)  $\xi \le 1$;
the allowed $U$ boson masses are: $M_U (GeV)  \lsim \sqrt{\frac{16285}{m_{DM}(GeV)}}$, 
for configurations {\it distant} more than 8 $\sigma$ from the {\it null hypothesis}.
They are reported in Fig. \ref{fg:uboson}.

There $U$ boson with $M_U$ masses in the sub-GeV range (see Ref. \cite{wimpele} for details)
is well allowed for a large interval of $m_{DM}$.

In conclusion, the obtained allowed
interval for the mass of the possible mediator of the interaction
is well in agreement with the
typical requirements of the phenomenological analyses available in
literature.

\subsection{Inelastic Dark Matter}
\label{DM3}

Another scenario that will be updated here regards the inelastic Dark Matter: relic particles that cannot 
scatter elastically off nuclei. Following an inelastic scattering off a nucleus, the kinetic energy of the 
recoiling nucleus is quenched and is the detected quantity.
As discussed in Refs. \cite{Wei01,Wei01_2,Wei01_3,dama_inel}, the inelastic Dark Matter
could arise from a massive complex scalar split into two approximately
degenerate real scalars or from a Dirac fermion split into two
approximately degenerate Majorana fermions, namely $\chi_+$ and $\chi_-$,
with a $\delta$ mass splitting. In particular, a specific
model featuring a real component of the sneutrino,
in which the mass splitting naturally arises, has been given in Ref. \cite{Wei01}.

\begin{figure}[!ht]
\centering
\vspace{-0.5cm}
\includegraphics[width=10.cm] {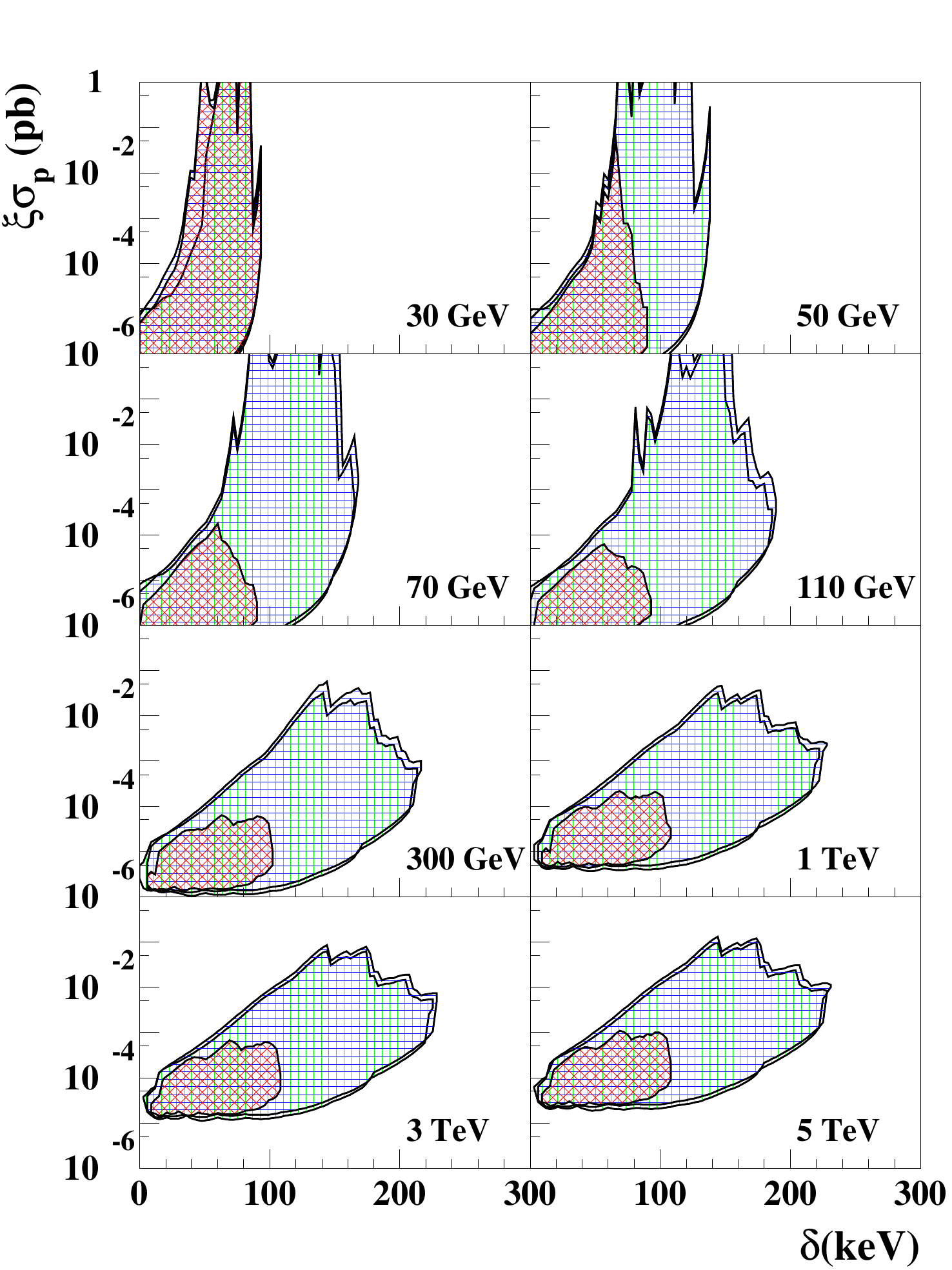}
\vspace{-0.4cm}
\caption{Slices of the 3-dimensional volume ($\xi \sigma_{p}$, $\delta$, $m_{DM}$) allowed by 
DAMA experiments in the case of a Dark Matter candidate with preferred inelastic interaction.
Three different instances for the Na and I quenching factors have been considered: 
(i) $Q_I$ case [(green on-line) vertically-hatched region],
(ii) with channeling effect [(blue on-line) horizontally-hatched region] and 
(iii) $Q_{II}$ [(red on-line) cross-hatched region].
The regions have been obtained by marginalizing all the models for each considered scenario 
(see Sect. \ref{data_analysis}) and they represent the domain where the likelihood-function 
values differ more than 10 $\sigma$ from the {\it null hypothesis} (absence of modulation).}
\label{fg:inel}
\end{figure}

\vspace{0.3cm}
The discussion of the theoretical arguments
on such inelastic Dark Matter can be found e.g. in Ref. \cite{Wei01}, where 
it was shown that for the $\chi_-$ inelastic scattering
off target nuclei a kinematic constraint exists which favors
heavy nuclei (such as $^{127}$I) with respect to
lighter ones (such as e.g. $^{nat}$Ge) as target-detectors media.
In fact, $\chi_{-}$ can only inelastically scatter
by transitioning to $\chi_{+}$ (slightly heavier state than $\chi_{-}$)
and this process can occur
only if the $\chi_{-}$ velocity, $v$, is larger than:
\begin{equation}
v_{thr} = \sqrt{\frac{2\delta}{m_{red}(A,\chi)}},
\label{eq:constraint}
\end{equation}
where $m_{red}(A,\chi)$ is the $\chi-$nucleus reduced mass.
This kinematic constraint becomes increasingly severe
as the nucleus mass, $m_N$, is decreased \cite{Wei01}.
For example, if $\delta \gsim$ 100 keV, a signal rate
measured e.g. in Iodine will be a factor about 10 or more
higher than that measured in Ge \cite{Wei01}.
Moreover, this model scenario implies some characteristic features when
exploiting the DM annual modulation signature
since it gives rise to an enhanced
modulated component, $\mathcal{S}_m$, with respect to the un-modulated one, $\mathcal{S}_0$,
and to largely different behaviors with energy for
both $\mathcal{S}_0$ and $\mathcal{S}_m$ (both show a higher mean value) \cite{Wei01}
with respect to elastic cases.
Details of calculation procedures can be found in Ref. \cite{dama_inel}.

\vspace{0.3cm}
Accounting for the uncertainties mentioned above,
in the inelastic Dark Matter scenario an allowed 3-dimensional volume
in the space ($\xi \sigma_p$, $m_{DM}$, $\delta$)
is obtained. Here, following the notation of Ref. \cite{dama_inel},
$\sigma_p$ is a generalized SI point-like $\chi-$nucleon cross section
and $m_{DM}$ is the $\chi$ mass.

\vspace{0.3cm}
For simplicity, Fig. \ref{fg:inel} shows slices of such an allowed
volume at 10 $\sigma$ from the {\it null hypothesis} for some values of $m_{DM}$;
the different cases of quenching factors are considered as well.
It can be noted that when $m_{DM} \gg m_N$,
the expected differential energy spectrum is trivially dependent on $m_{DM}$
and, in particular, it is proportional to the ratio between $\xi \sigma_p$
and $m_{DM}$. Thus, allowed regions for other $m_{DM} \gg m_N$ can be obtained from
the last panel of Fig. \ref{fg:inel}, straightforward.

\vspace{0.3cm}
Significant enlargement of such regions should be expected when including
complete effects of model (and related experimental and theoretical parameters)
uncertainties.

\begin{figure}[!ht]
\begin{center}
\includegraphics[width=6.0cm] {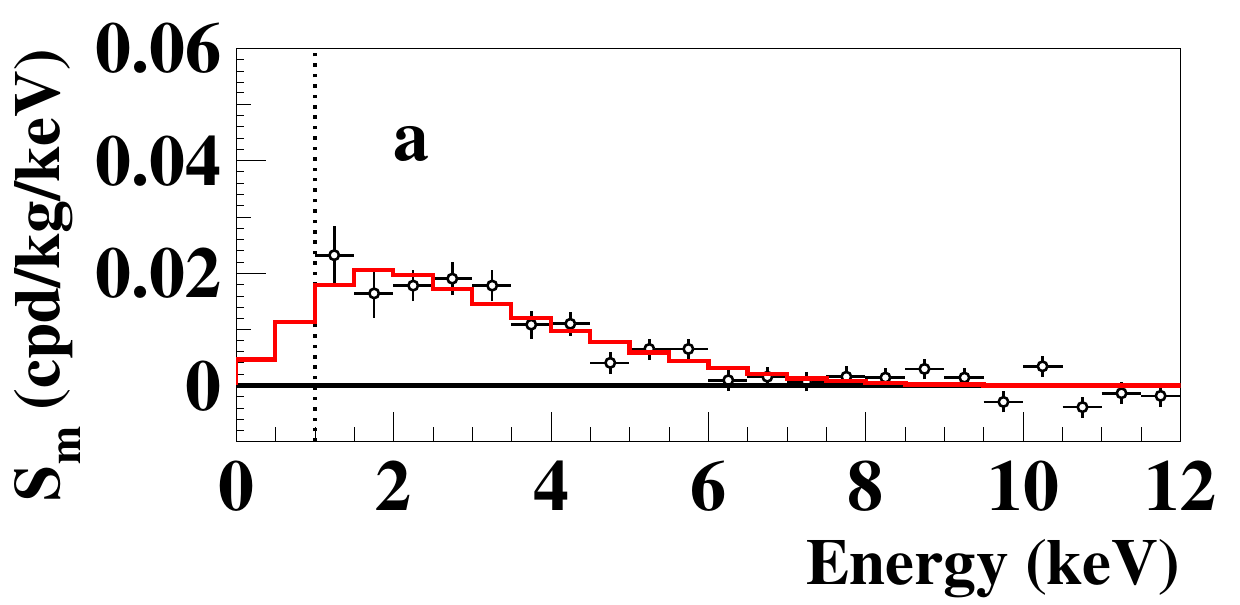}
\includegraphics[width=6.0cm] {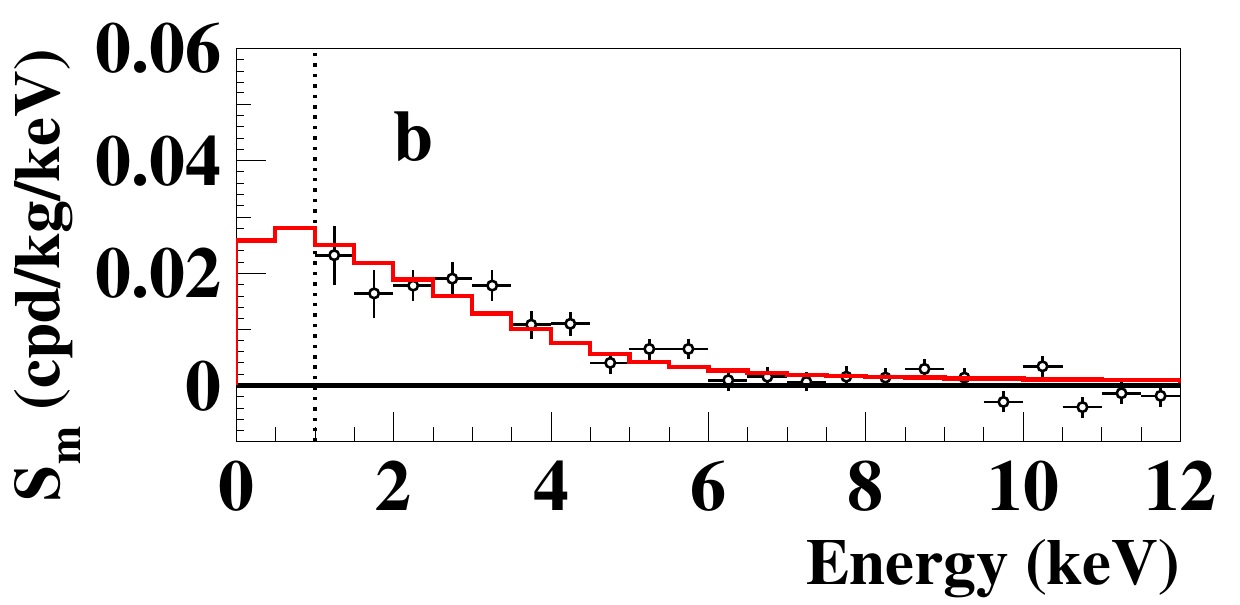}
\end{center}
\vspace{-0.6cm}
\caption{Examples of superposition of the measured $\mathcal{S}^{exp}_{m}$ vs energy (points
with error bars) with theoretical expectations (solid histograms) for Inelastic Dark Matter.
(a) case of the B2 (Evans power-law) halo model with $\rho_0$ = 1.33 GeV/cm$^3$,
    $v_0$ = 270 km/s, set B of parameters values, $m_{DM} = 50$ GeV, $\delta=75$ keV,
    $\xi\sigma_{p} = 1.1 \times 10^{-6}$ pb and quenching $Q_{I}$;
(b) case of the B4 (Jaffe) halo model with $\rho_0$ = 0.44 GeV/cm$^3$,
    $v_0$ = 220 km/s, set C of parameters values, $m_{DM} = 30$ GeV, $\delta=25$ keV,
    $\xi\sigma_{p} = 4.2 \times 10^{-6}$ pb including channeling effect.}
\label{fg:smvsinel}
\end{figure}

\vspace{0.3cm}
In Fig.~\ref{fg:smvsinel} few examples of comparison between $\mathcal{S}_m^{exp}$ and $\mathcal{S}_m^{th}$
are shown.

\begin{figure}[!p]
\begin{center}
\includegraphics[width=4.2cm] {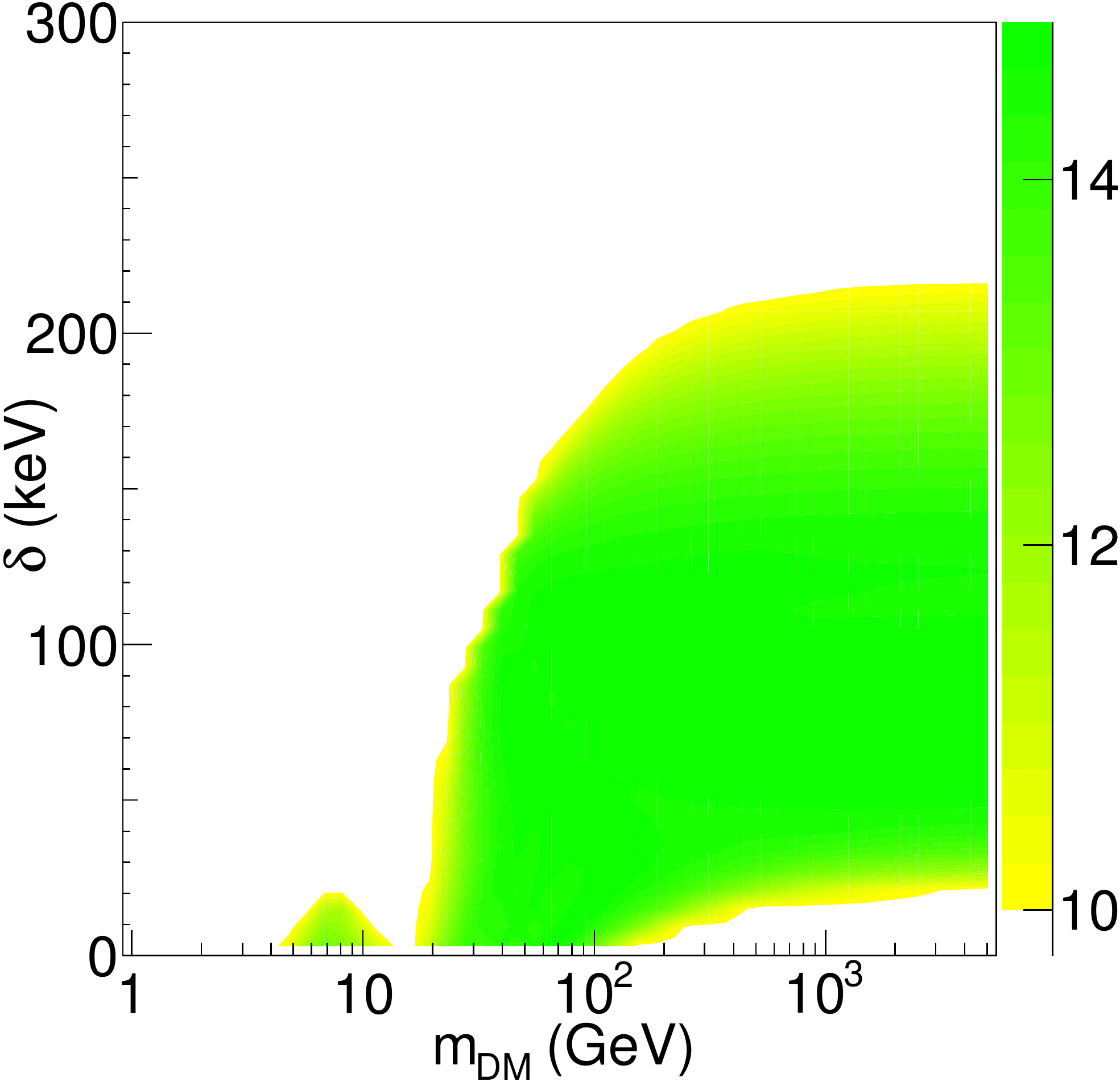}
\includegraphics[width=4.2cm] {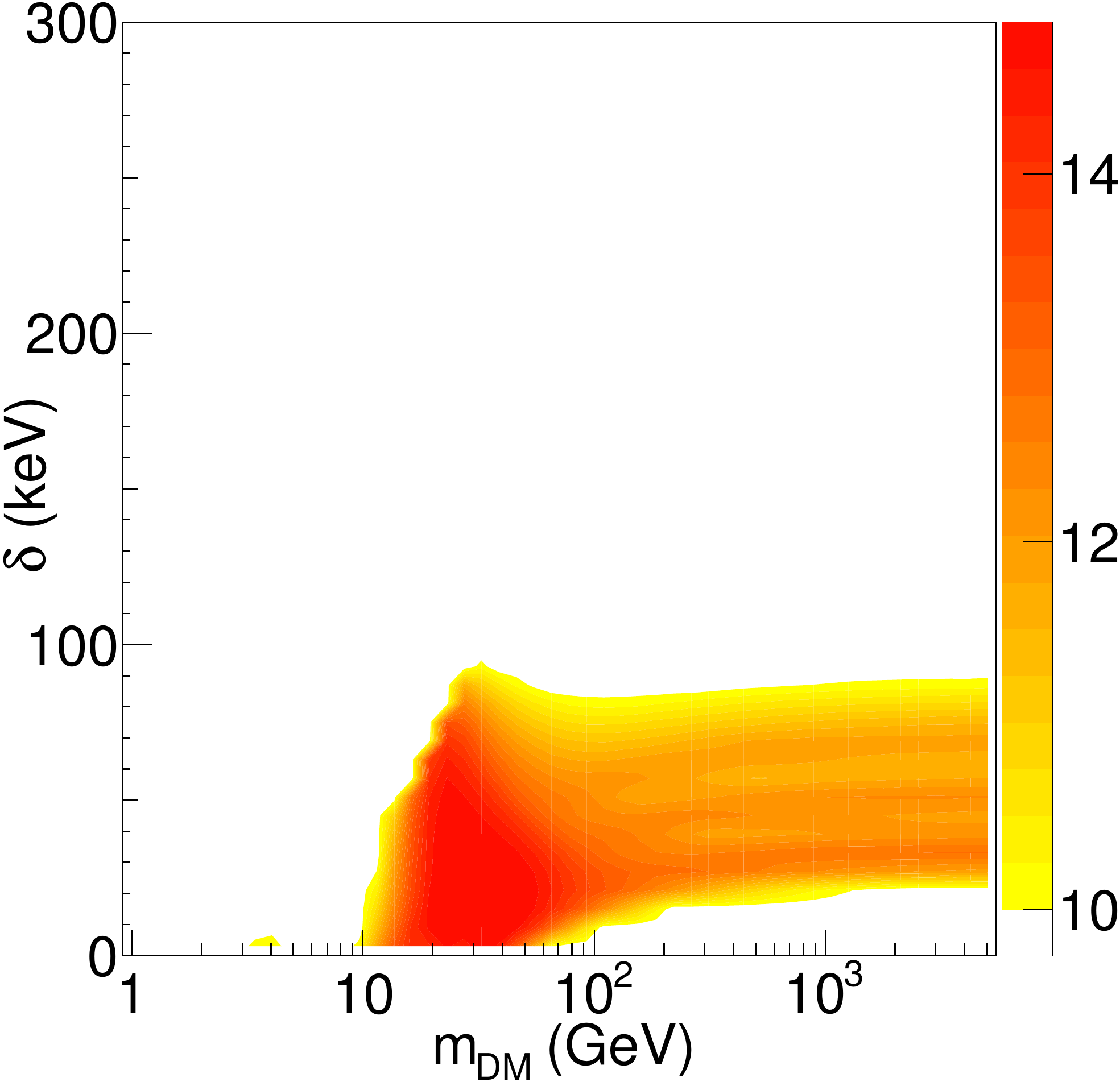}
\includegraphics[width=4.2cm] {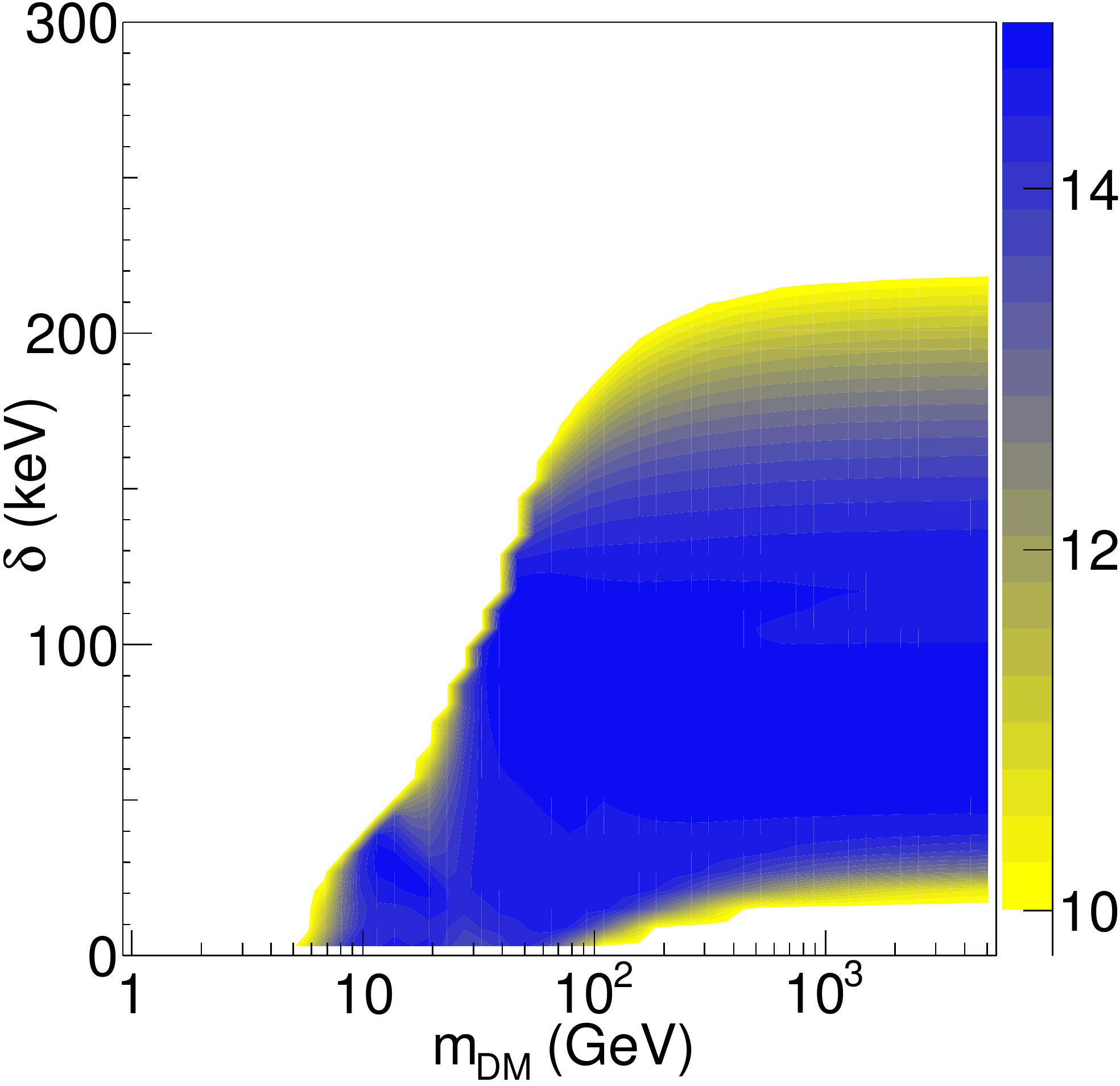}
\includegraphics[width=4.2cm] {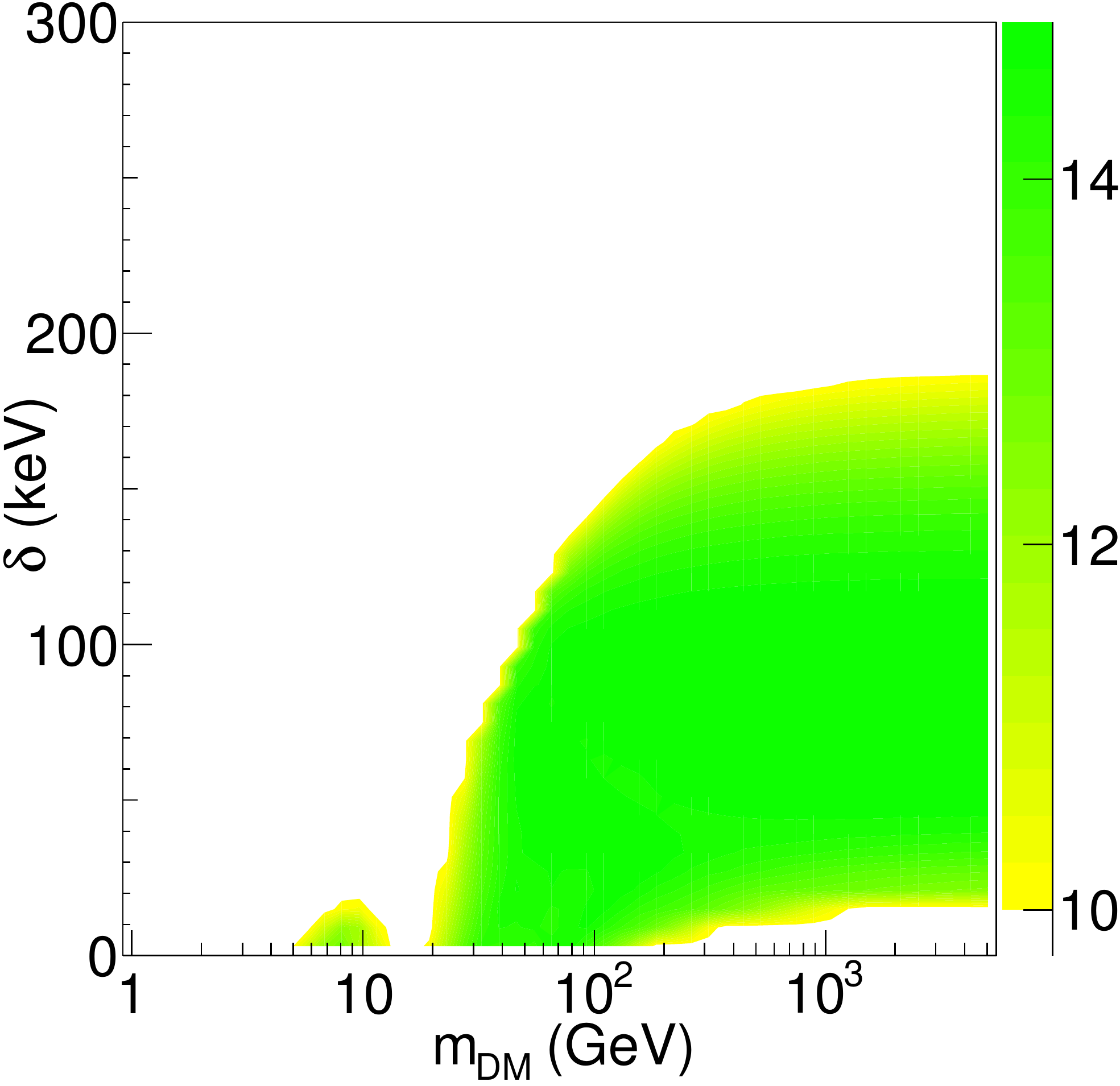}
\includegraphics[width=4.2cm] {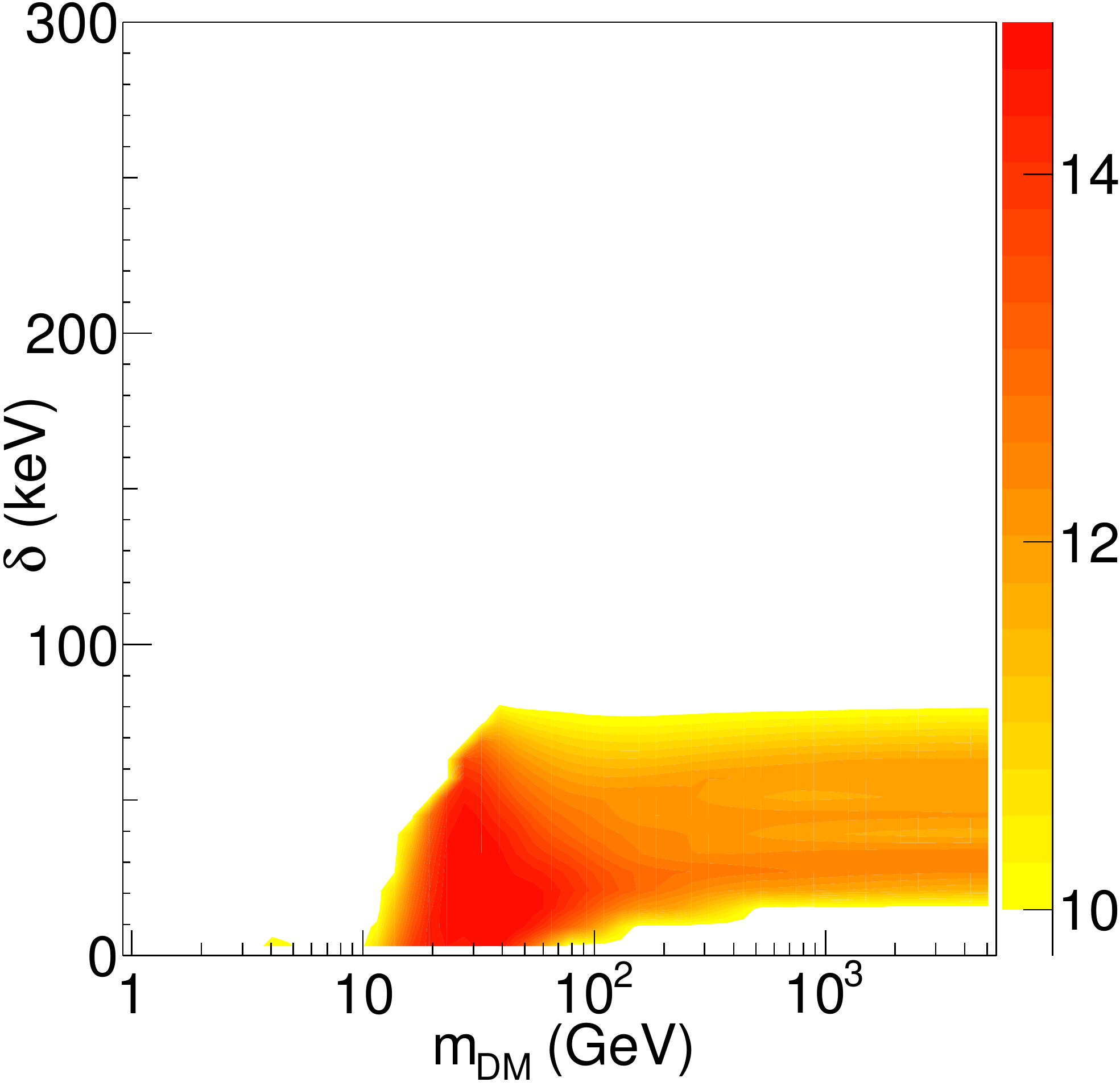}
\includegraphics[width=4.2cm] {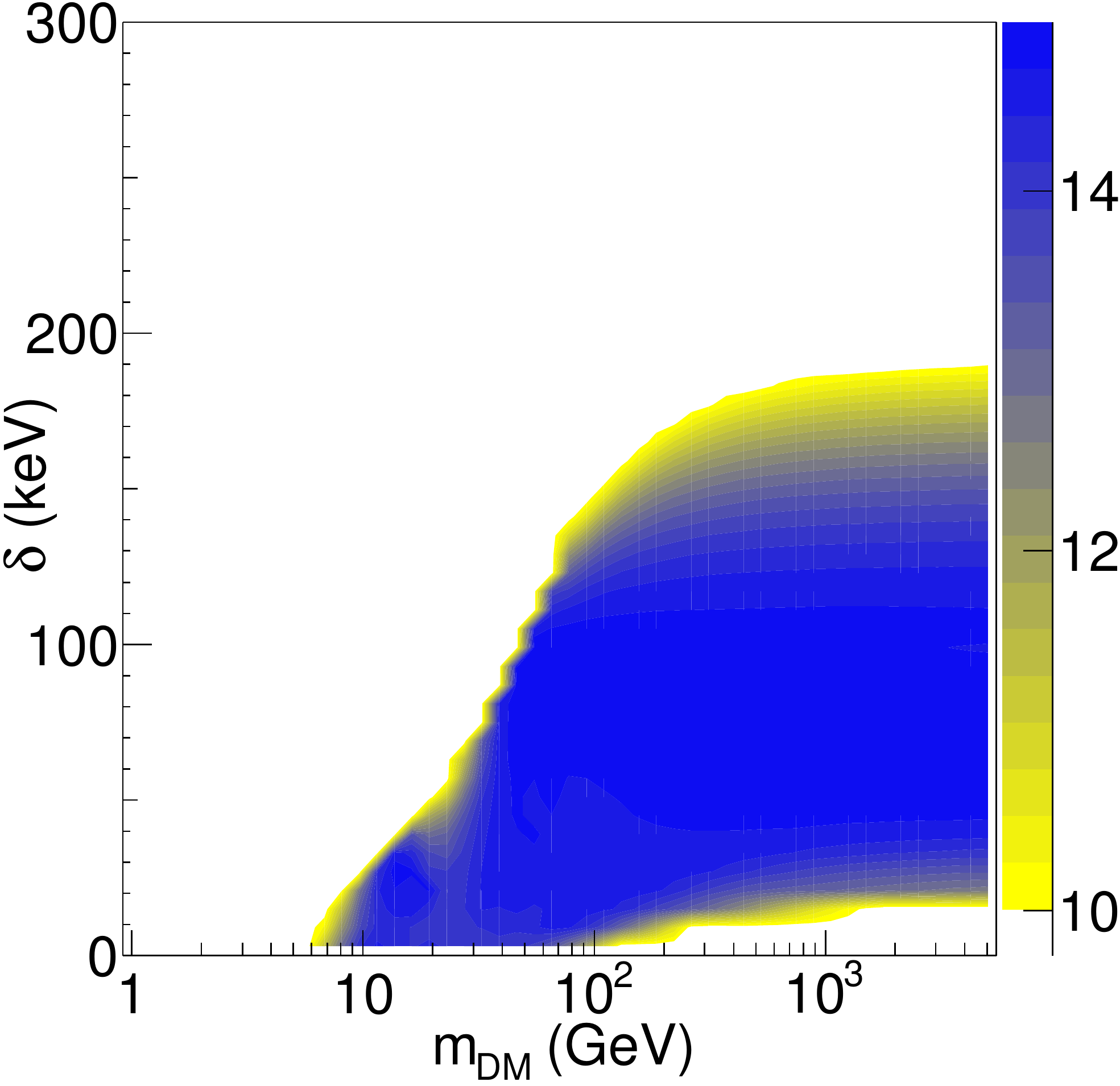}
\includegraphics[width=4.2cm] {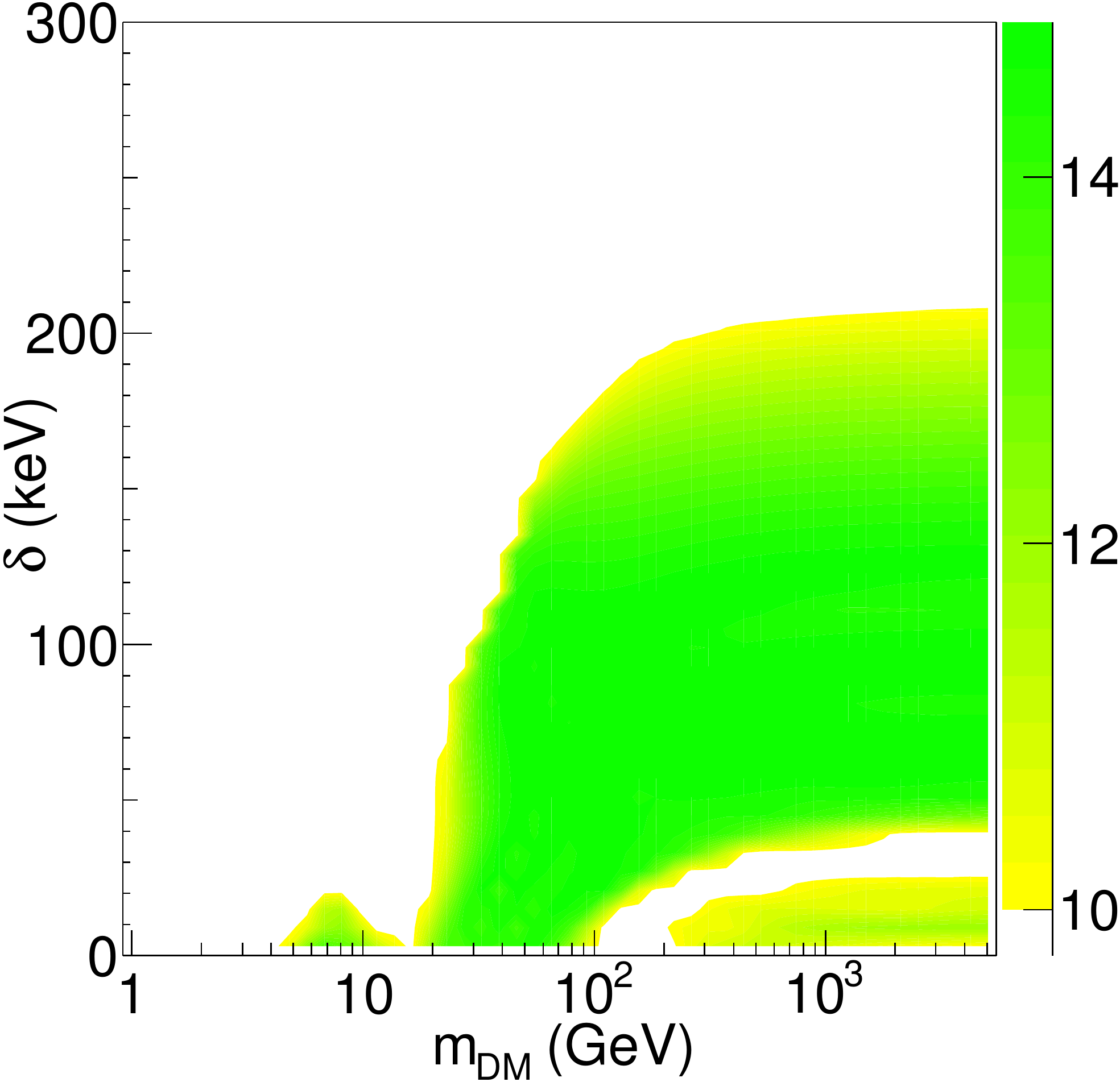}
\includegraphics[width=4.2cm] {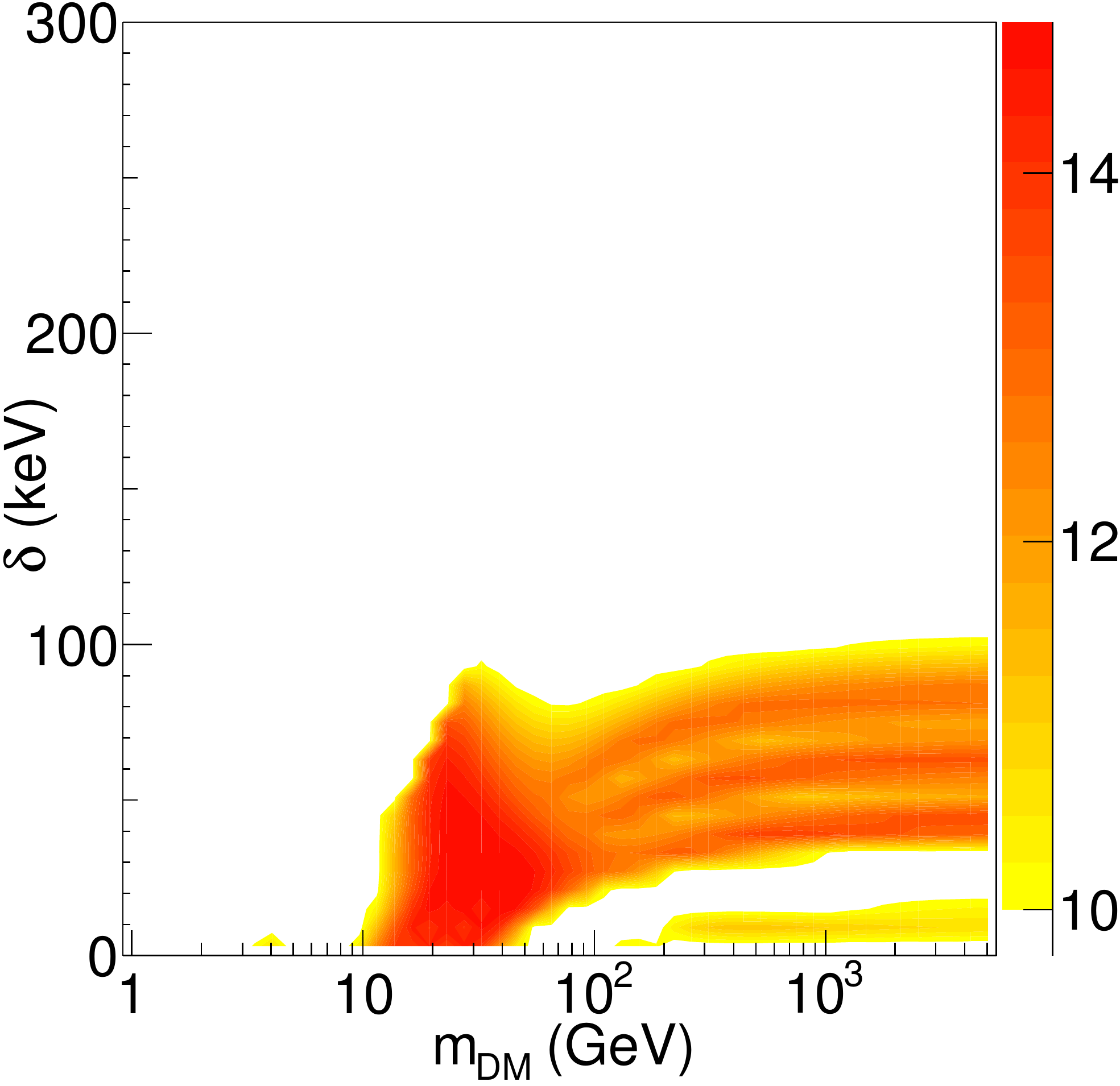}
\includegraphics[width=4.2cm] {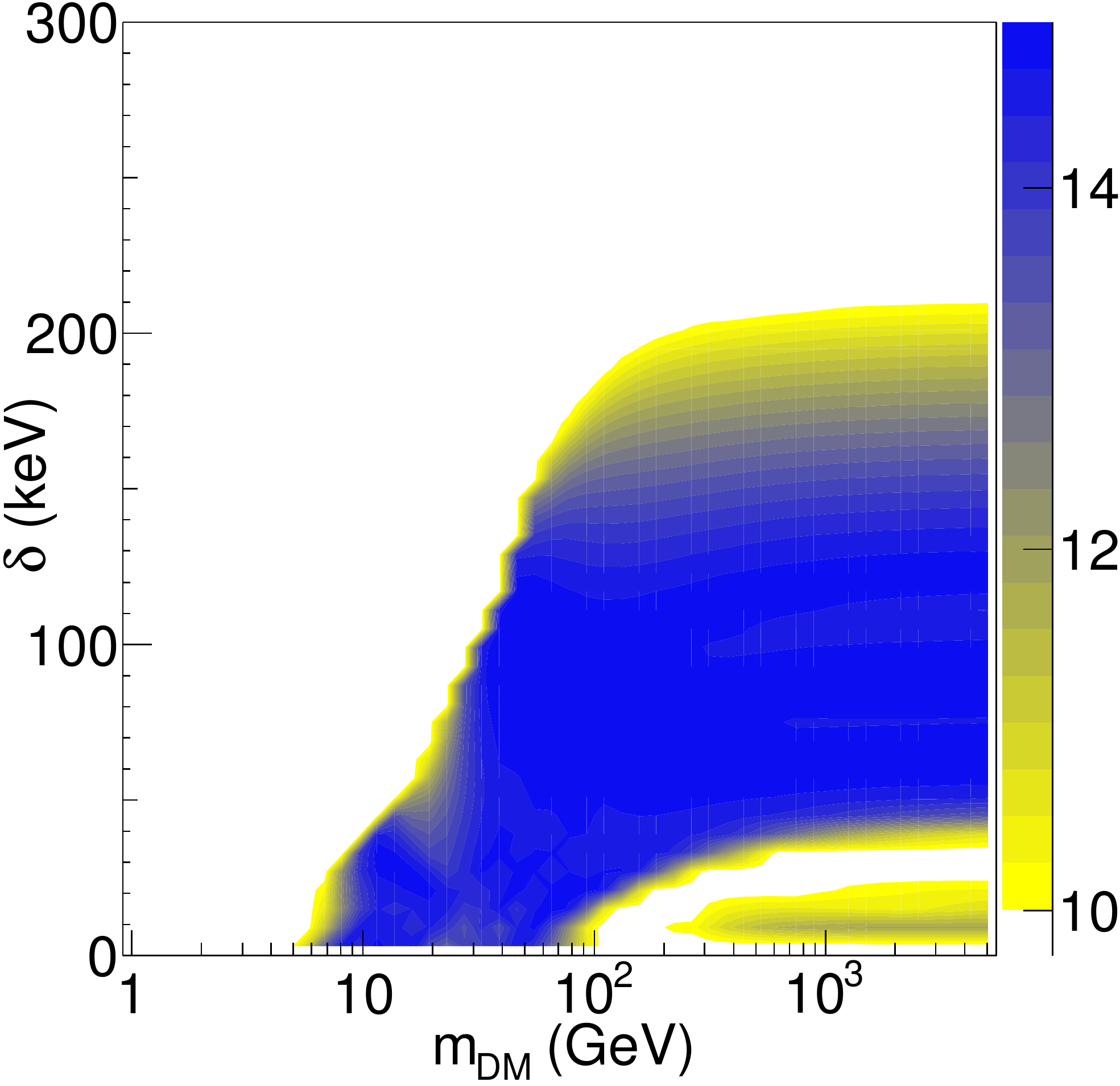}
\includegraphics[width=4.2cm] {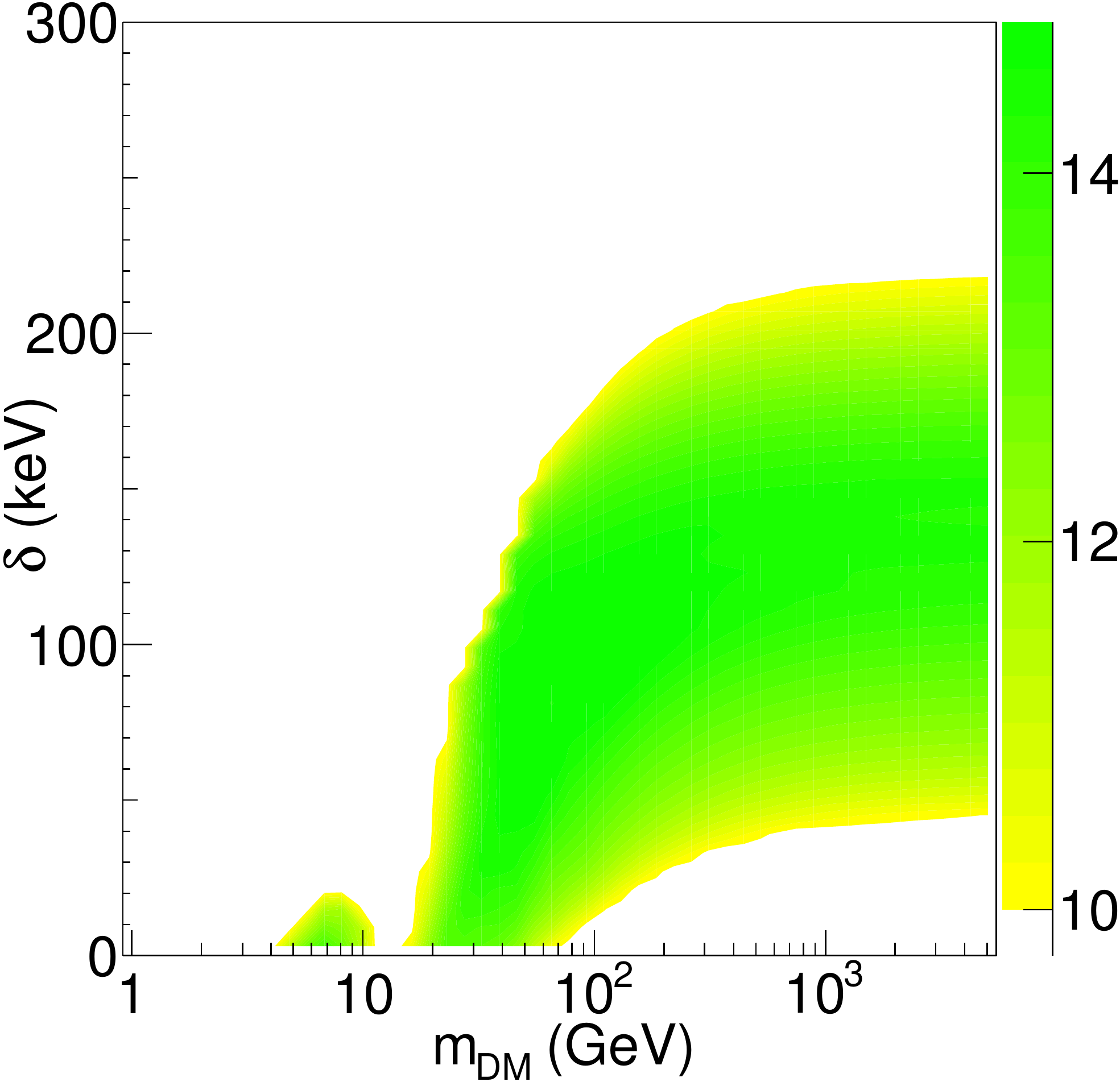}
\includegraphics[width=4.2cm] {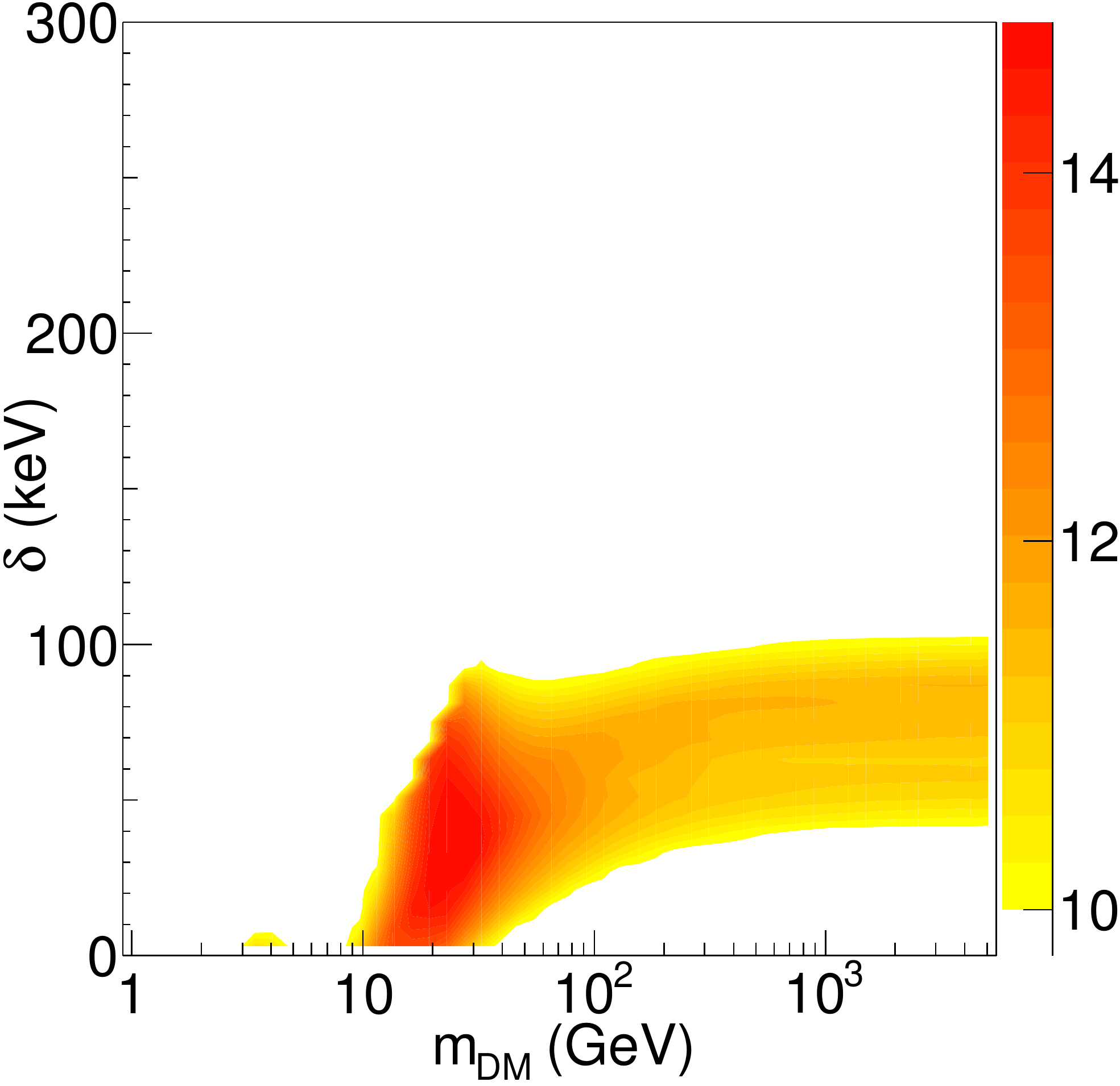}
\includegraphics[width=4.2cm] {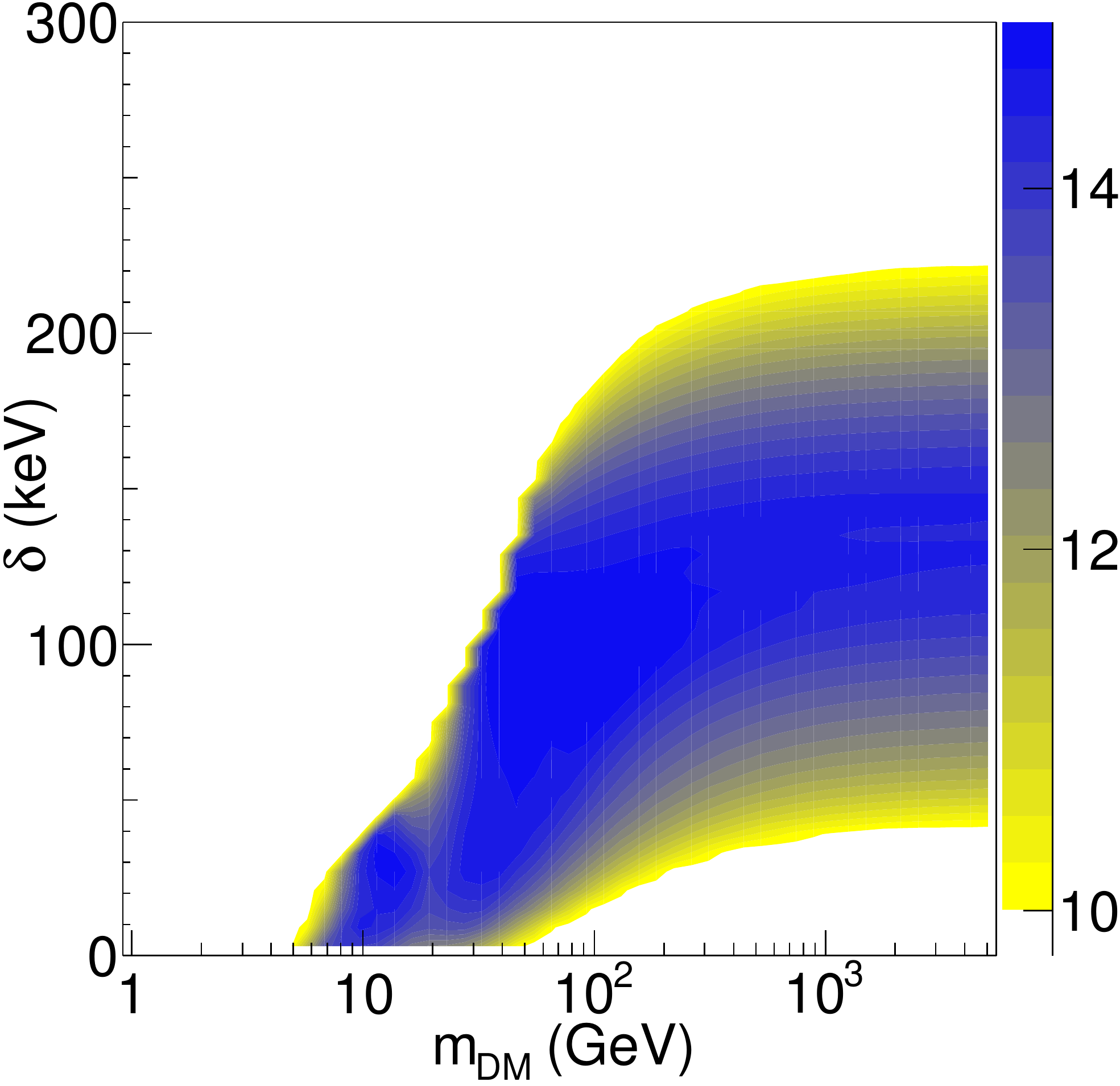}
\end{center}
\vspace{-0.5cm}
\caption{Regions in the $\delta$ vs $m_{DM}$ plane allowed by DAMA experiments in the case
of a Dark Matter candidate with preferred inelastic interaction.
The Na and I quenching factors are:
$Q_I$ [$left$ (green on-line)], $Q_{II}$ [$center$ (red on-line)], and with channeling effect [$right$ (blue on-line)].
The considered halos (from top to bottom) are A0 (isothermal sphere), B1, C1, D3 with the $v_0$ and $\rho_0$
in the range of Table III of Ref. \cite{bel02}.
The three possible sets of parameters A, B and C are considered (see Sect. \ref{data_analysis}); the allowed 
regions represent the domain where the likelihood-function values differ more than 10 $\sigma$ from the 
{\it null hypothesis} (absence of modulation).
The color scales give the confidence level in units of $\sigma$ from the {\it null hypothesis}.}
\label{fg:inel_dvsm}
\end{figure}

\vspace{0.3cm}
Finally, Fig.~\ref{fg:inel_dvsm} shows the allowed regions in the $\delta$ vs $m_{DM}$ plane after 
marginalizing on $\xi \sigma_{p}$.
For simplicity four halo models: A0 (isothermal sphere), B1, C1, D3 with the $v_0$ and $\rho_0$ 
in the range of Table III of Ref. \cite{bel02}, and three choices of the Na and I quenching factors:
$Q_I$, $Q_{II}$, and including the channeling effect are considered.

\vspace{0.5cm}
Let us conclude that here the analysis of the inelastic DM particle has been limited only to
SI coupling. Recently analyses of the inelastic DM candidate with SD coupling
have been reported in Refs. \cite{pSIDM1,pSIDM2}. They show that also
this scenario can be compatible with the DAMA result. This conclusion can be
further confirmed considering e.g. the effects of uncertainties in the models
that in those papers have not been included.

\vspace{0.5cm}
\subsubsection{Including the Thallium}

\begin{figure}[p]
\centering
\includegraphics[width=10cm] {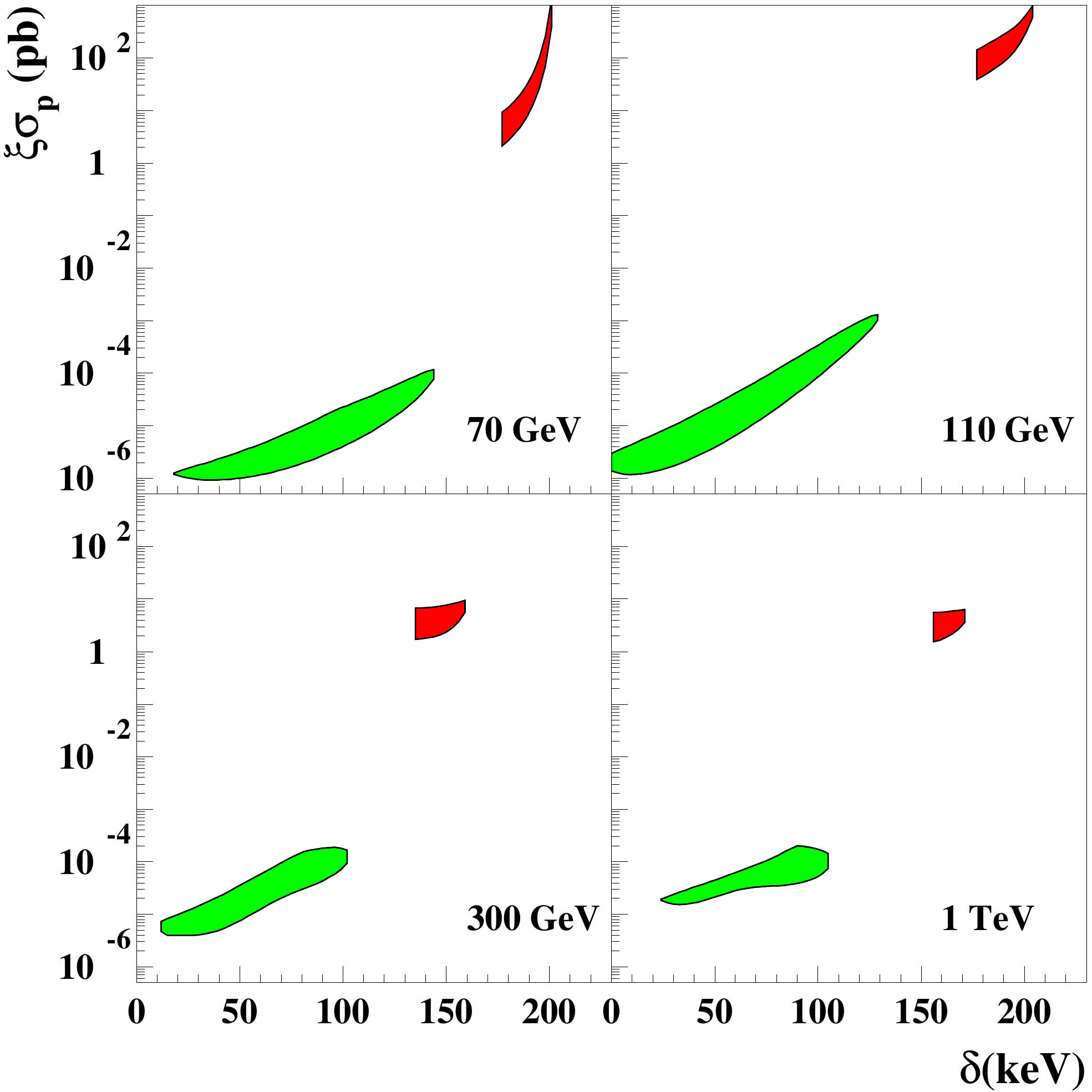}
\vspace{-0.4cm}
\caption{Slices of the 3-dimensional volume ($\xi \sigma_{p}$, $\delta$, $m_{DM}$) allowed by
DAMA experiments for some example of scenarios when inelastic scattering only off Na and I nuclei
is considered (green on-line region) and when the Thallium nuclei is also included (additional red on-line region).
The four examples have been obtained considering the quenching case $Q_I$, the set C of parameters
values and four different DM mass and halo models. In particular:
the A3 (Evans power-law) halo model with $\rho_0$ = 0.52 GeV/cm$^3$ and $v_0$ = 270 km/s for $m_{DM} =  70$ GeV,
the D4 (Triaxial) halo model with $\rho_0$ = 0.30 GeV/cm$^3$ and $v_0$ = 170 km/s for $m_{DM} = 110$ GeV,
the B1 (Evans logarithmic) halo model with $\rho_0$ = 0.20 GeV/cm$^3$ and $v_0$ = 170 km/s for $m_{DM} = 300$ GeV and
the B4 (Jaffe) halo model with $\rho_0$ = 0.26 GeV/cm$^3$ and $v_0$ = 170 km/s for $m_{DM} =   1$ TeV.
The regions represent the domain where the likelihood-function values differ more than 10 $\sigma$
from the {\it null hypothesis} (absence of modulation).}
\label{fg:inel_tl_examp}
\end{figure}

\begin{figure}[p]
\centering
\includegraphics[width=10.cm] {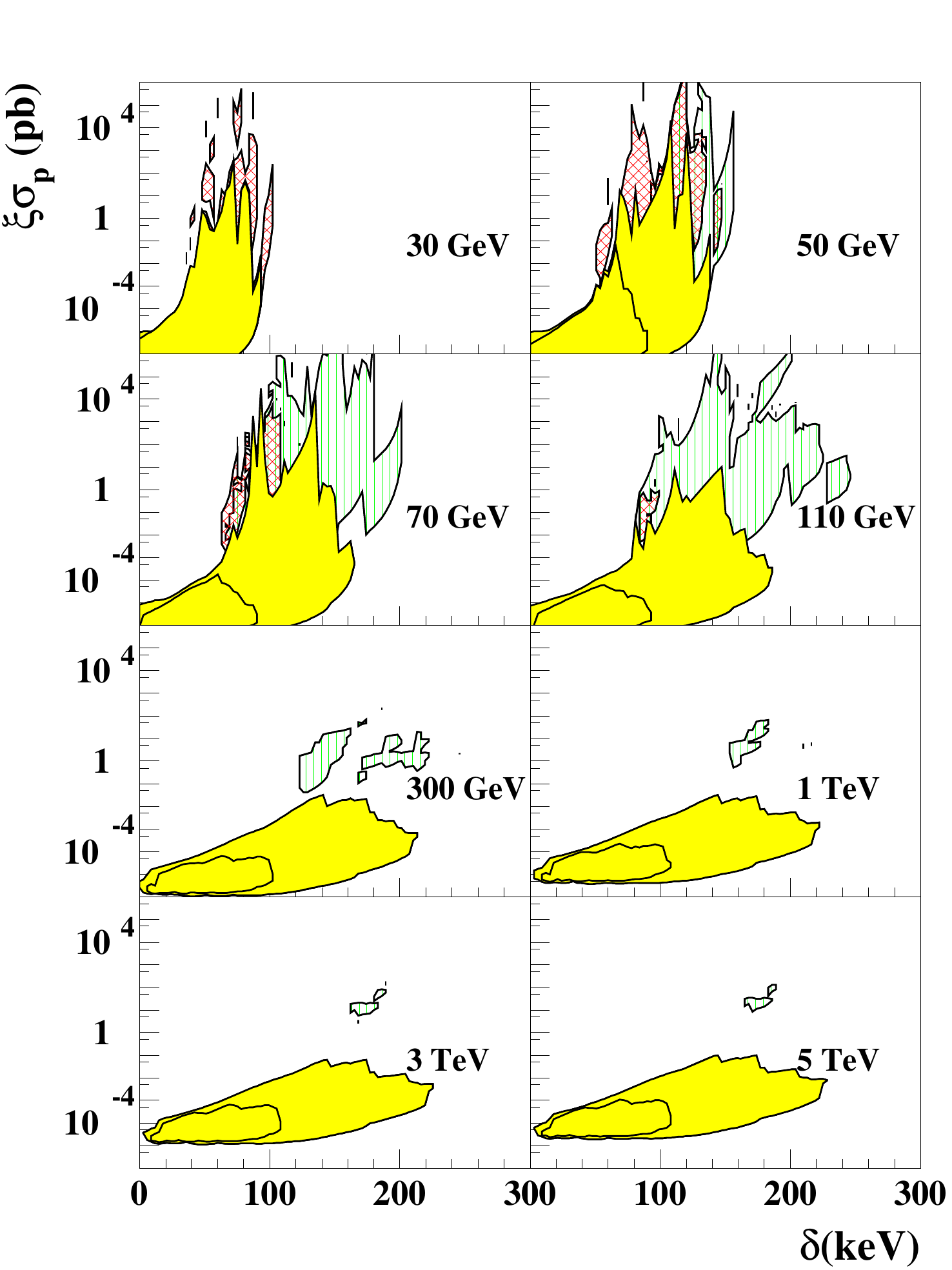}
\vspace{-0.2cm}
\caption{Slices of the 3-dimensional volume ($\xi \sigma_{p}$, $\delta$, $m_{DM}$) allowed by 
DAMA experiments when the inelastic scattering off Thallium nuclei is also included.
The two instances for the Na, I, and Tl quenching factors are considered: 
(i)  $Q_I$ case [(green on-line) vertically-hatched region],
(ii) $Q_{II}$ [(red on-line) cross-hatched region].
The regions due to inelastic scattering only off Na and I nuclei, already shown in Fig. \ref{fg:inel}, are reported in 
(yellow on-line) light-filled.
The regions have been obtained by marginalizing all the models for each considered scenario 
(see Sect. \ref{data_analysis}) and they represent the domain where the likelihood-function 
values differ more than 10 $\sigma$ from the {\it null hypothesis} (absence of modulation).}
\label{fg:inel_thallium}
\end{figure}

Until now, we have considered the NaI(Tl) detectors made of Sodium and Iodine nuclei; 
however, the Thallium dopant (stable isotopes with mass number 203 and 205, and natural abundances 
29.5\% and 70.5\% respectively) can also play a 
role as it has been described in Ref. \cite{cha11}, where 
it has been shown how the DM interaction on Thallium nuclei would give rise to a 
signal which cannot be detected with lower mass target-nuclei. This also can decouple 
theoretical and experimental aspects from different experiments.
The slices of the 3-dimensional volume ($\xi \sigma_{p}$, $\delta$, $m_{DM}$), allowed by 
DAMA experiments when the inelastic scattering off Thallium nuclei is also included,
have been evaluated in Fig. \ref{fg:inel_tl_examp} for some examples of scenarios, and
in Fig. \ref{fg:inel_thallium} marginalizing all the considered models
(see Sect. \ref{data_analysis}).
Two instances for the Tl quenching factor in NaI(Tl) are considered: 
(i)  $Q_I$ case with $q_{Tl} = 0.075$, tentatively obtained by extrapolating the 
     $q_{Na}$ and $q_I$ measured by DAMA with neutrons \cite{psd96};
(ii) $Q_{II}$ quenching factors varying as a function of E$_R$ evaluated as in Ref. \cite{Tretyak}.
Moreover, the Thallium is assumed to be homogeneously distributed in each crystal and among the crystals
at level of 0.1\% in mass (corresponding to $2.95 \times 10^{21}$ Tl atoms/kg).

\vspace{0.6cm}
Thus, we have verified that, as pointed out in Figs.~\ref{fg:inel_tl_examp} and \ref{fg:inel_thallium}, new regions
with $\xi \sigma_{p} \gsim 1$ pb and $\delta \gsim 100$ keV
are allowed by DAMA after the inclusion of the inelastic scattering off Thallium nuclei.  
Such regions are not fully accessible to detectors with target nuclei having mass lower than Thallium.

\begin{figure}[!ht]
\begin{center}
\includegraphics[width=6.0cm] {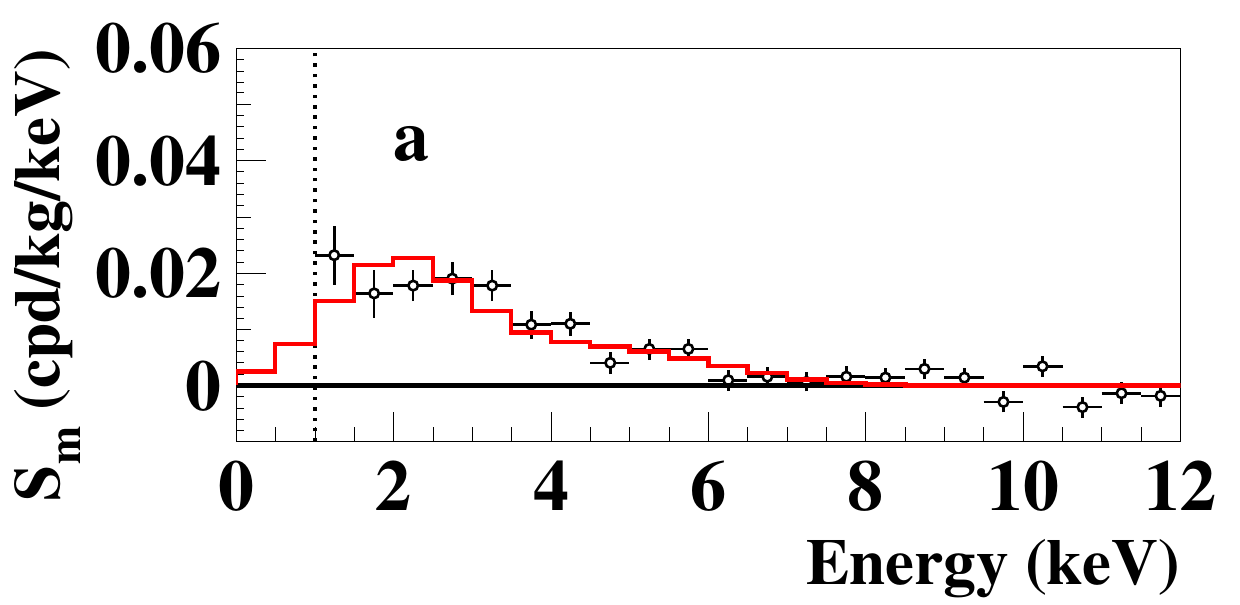}
\includegraphics[width=6.0cm] {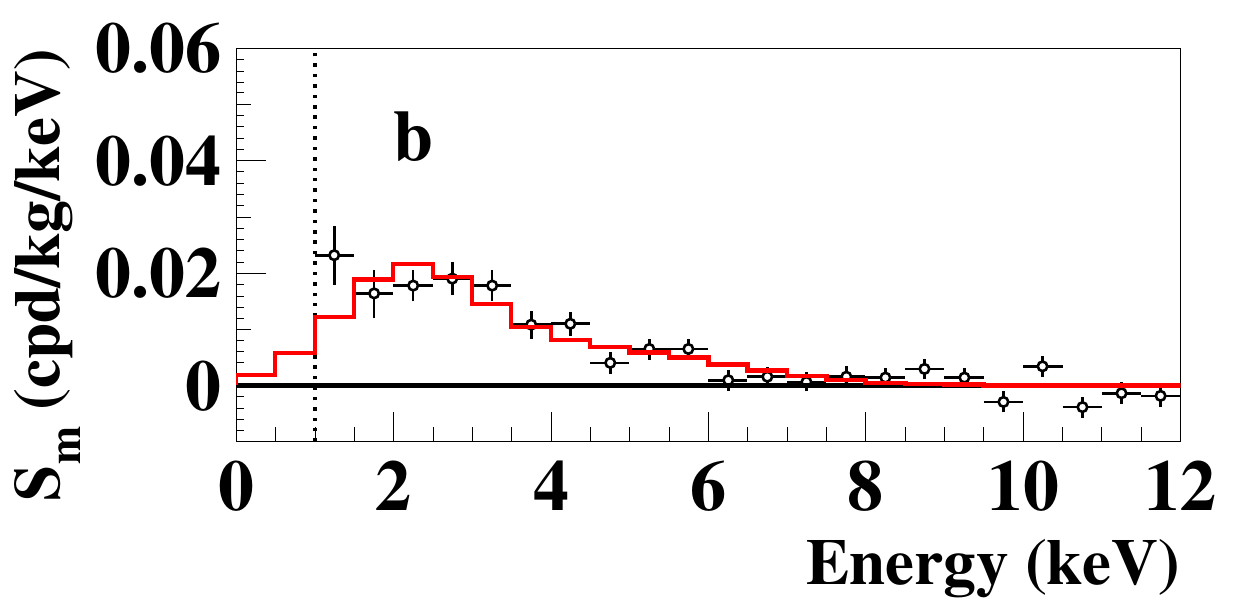}
\end{center}
\vspace{-0.6cm}
\caption{Examples of superposition of the measured $\mathcal{S}^{exp}_{m}$ vs energy (points
with error bars) with theoretical expectations (solid histograms) for Inelastic Dark Matter
when including the Thallium nuclide.
(a) case of the A4 (Jaffe) halo model with $\rho_0$ = 0.44 GeV/cm$^3$,
    $v_0$ = 220 km/s, set A of parameters values, $m_{DM} = 70$ GeV, $\delta=147$ keV,
    $\xi\sigma_{p} = 23$ pb and quenching $Q_{I}$;
(b) case of the A7 (Kravtsov et al.) halo model with $\rho_0$ = 0.32 GeV/cm$^3$,
    $v_0$ = 170 km/s, set A of parameters values, $m_{DM} = 110$ GeV, $\delta=126$ keV,
    $\xi\sigma_{p} = 40$ pb and quenching $Q_{I}$.
Both the configurations belong to the additional allowed regions obtained when taking
into account the presence of the Thallium nuclide in the NaI(Tl) detectors 
(see Figs.~\ref{fg:inel_tl_examp} and \ref{fg:inel_thallium}).}
\label{fg:smvsinel_tl}
\end{figure}

\vspace{0.3cm}
In Fig.~\ref{fg:smvsinel_tl} two examples of comparison between $\mathcal{S}_m^{exp}$ and $\mathcal{S}_m^{th}$
are shown; both the configurations belong to the additional allowed regions obtained when taking
into account the presence of the Thallium nuclide in the NaI(Tl) detectors 
(see Figs.~\ref{fg:inel_tl_examp} and \ref{fg:inel_thallium}).

\subsection{Investigation on light dark matter}
\label{DM4}

Some extensions of the Standard Model provide
DM candidate particles with sub-GeV mass;
in the following these candidates will be indicated as 
Light Dark Matter (LDM).

Several LDM candidates have been proposed in Warm Dark Matter scenarios, as
keV-scale sterile neutrino, axino, gravitino, and MeV-scale particles
(for details see Ref. \cite{ldm}).

In this section the direct detection of LDM candidate
particles is investigated considering the possible inelastic scattering
channels either off the electrons
or off the nuclei of the target; theoretical expectations are compared with the 
recent results obtained by adding DAMA/LIBRA--phase2 \cite{uni18,bled18,npae18,npp19,mg15}. 
Firstly we note that -- since the kinetic
energy for LDM particles in the galactic halo does not exceed hundreds eV --
the elastic scattering of such LDM particles both off electrons and off nuclei
yields energy releases hardly detectable by the detectors used in the field;
this might prevent the exploitation of the elastic scattering as 
detection approach for these candidates.
Thus, the inelastic process could be the only possible viable one 
for the direct detection of LDM; further details were given in Ref. \cite{ldm}.

The following process is, therefore, considered for detection:
the LDM candidate (hereafter named $\nu_H$ with mass $m_H$)
interacts with the ordinary matter target, $T$, with mass $m_T$.
The target $T$ can be either an atomic nucleus or an atomic electron
depending on the nature of the $\nu_H$ particle interaction. 
As result of the interaction a lighter particle is produced
(hereafter $\nu_L$ with mass $m_L < m_H$) 
and the target recoils with an energy $E_R$, which can be detectable
by suitable detectors.

The lighter particle $\nu_L$ is neutral and it is required 
that it interacts very weakly with ordinary matter or not at all;
thus, the $\nu_L$ particle escapes the detector.
In particular, the $\nu_L$ particle can also be another DM halo component 
(dominant or sub-dominant with respect to the $\nu_H$ one), or 
it can simply be a Standard Model particle (e.g. $\nu_L$ can be identified with
an active neutrino). Details can be found in Ref. \cite{ldm}.

Since the sub-GeV LDM wavelength 
($\lambda = \frac{h}{k} > 10^3$ fm) is much larger than the nucleus 
size, the targets can be considered as point-like and the form factors of 
the targets can be approximated by one. 
The cross section of the processes, $\sigma_T$, is generally function 
of the LDM velocity, $v$, and can be written 
by adopting the approximation for the non-relativistic case \cite{ldm}:
\begin{equation}
\sigma_T v \simeq a + b v^2 \; ,
\label{eq:sigmav}
\end{equation}
where $a$ and $b$ are constant depending on the peculiarity of the particle interaction with the 
target $T$. In the analysis, the cross sections 
$\sigma_0^T = \frac{a}{v_0}$ and $\sigma_m^T = b v_0$ are defined \cite{ldm}; 
they are related to the $a$ and $b$ parameters rescaled with the DM local velocity, $v_0$.
In particular, the $\sigma_m^T$
is responsible for the annual modulation of the expected counting rate
for LDM interactions, and in the following it will be used as free parameter,
together with $m_H$ and the mass splitting $\Delta = m_H - m_L$.
Moreover, for the case of LDM interaction on nuclei, following the prescriptions
given in Ref. \cite{ldm}, two different nuclear scaling laws are adopted:
the coherent ($\sigma_m^{coh} \propto \sigma_m^{Na}/A_{Na}^2 \propto \sigma_m^{I}/A_{I}^2$) 
and the incoherent ($\sigma_m^{inc} \propto \sigma_m^{A} \propto \sigma_m^{I} $) ones.

\vspace{0.4cm}
\subsubsection{Interaction with atomic electrons}

After the interaction of $\nu_H$ with an electron in the detector,
the final state can have -- beyond the $\nu_L$ particle -- 
either a prompt electron and an ionized atom or an excited atom plus 
possible X-rays/Auger electrons. 
Therefore, the process produces X-rays and electrons of relatively low energy, 
which are mostly contained with efficiency $\simeq 1$ in a detector of 
a suitable size.

\begin{figure} [!ht]
\centering
\includegraphics[width=7.cm] {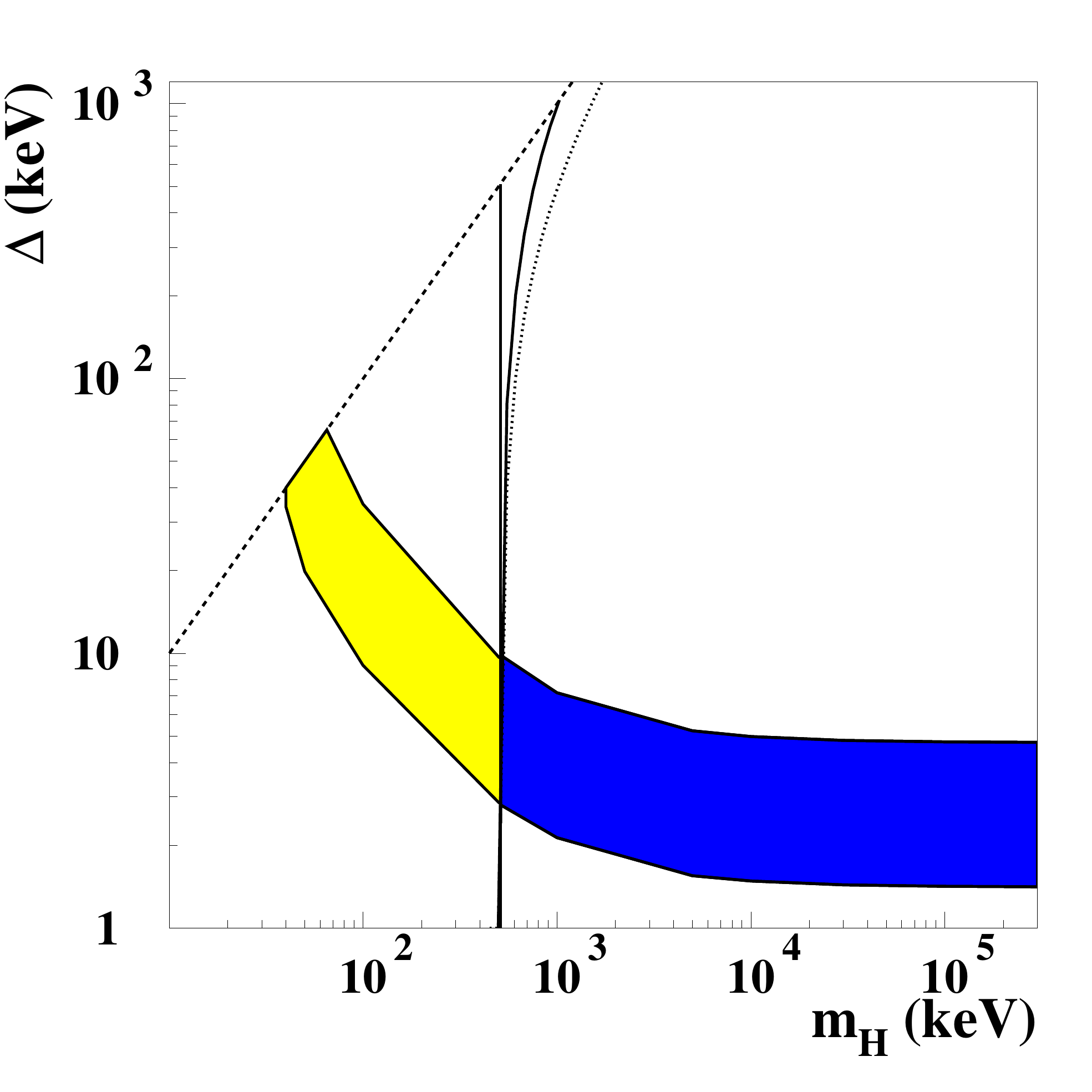}
\caption{Projection of the allowed 3-dimensional volume on the plane ($m_H$, $\Delta$)
for electron interacting LDM. 
The regions have been obtained by marginalizing all the models for each considered scenario 
(see Sect. \ref{data_analysis}) and they represent the domain where the likelihood-function 
values differ more than 10 $\sigma$ from the {\it null hypothesis} (absence of modulation).
The dashed line ($m_H = \Delta$) marks the case where $\nu_L$ is a massless particle.
The decay through the detection channel,
$\nu_H \rightarrow \nu_L e^+ e^-$, is energetically not allowed
for the selected configurations.
The configurations with $m_H \gsim m_e$ (dark area)
are interesting for the possible annihilation processes: 
\mbox{$\nu_H \bar{\nu}_H \rightarrow e^+ e^- $},
$\nu_H \bar{\nu}_L \rightarrow e^+ e^- $,
$\nu_L \bar{\nu}_H \rightarrow e^+ e^- $,
and $\nu_L \bar{\nu}_L \rightarrow e^+ e^- $ 
in the galactic center.
The three nearly vertical curves are the thresholds of these latter processes as
mentioned in Ref. \cite{ldm}.}
\label{fg:elett1}
\end{figure}

\vspace{0.4cm}
Comparing the expected modulated signal for this scenario with the experimental result obtained
cumulatively by DAMA/NaI, DAMA/LIBRA--phase1 and DAMA/LIBRA--phase2 \cite{uni18,bled18,npae18,npp19,mg15},
it is possible to determine a 10 $\sigma$ C.L. allowed volume in the space ($m_H$, $\Delta$, $\xi\sigma_m^e$).
The projection of such a region on the plane ($m_H$, $\Delta$)
for the dark halo models and parameters described before
is reported in Fig. \ref{fg:elett1}. 

\vspace{0.4cm}
The allowed $m_H$ values and the splitting $\Delta$ 
are in the intervals \mbox{40 keV $\lsim m_H \lsim $ O(GeV)}\footnote{\label{fn:1} For values of $m_H$ greater than
O(GeV), the definition of LDM is no longer 
appropriate. Moreover, the kinetic energy of the particle 
would be enough for the detection in DAMA experiments also through 
the elastic scattering process, as demonstrated in Ref. \cite{wimpele}.}
and \mbox{1.5 keV $\lsim \Delta \lsim 70$ keV}, respectively.
It is worth noting that in such a case the decay through the detection channel:
$\nu_H \rightarrow \nu_L e^+ e^-$, is energetically forbidden
for the given $\Delta$ range.
The configurations with $m_H \gsim 511$ keV 
(dark area in Fig. \ref{fg:elett1})
are instead of interest for the possible annihilation processes: 
\mbox{$\nu_H \bar{\nu}_H \rightarrow e^+ e^- $},
$\nu_H \bar{\nu}_L \rightarrow e^+ e^- $,
$\nu_L \bar{\nu}_H \rightarrow e^+ e^- $,
and $\nu_L \bar{\nu}_L \rightarrow e^+ e^- $, 
in the galactic center.

\begin{figure} [!ht]
\centering
\includegraphics[width=7.8cm] {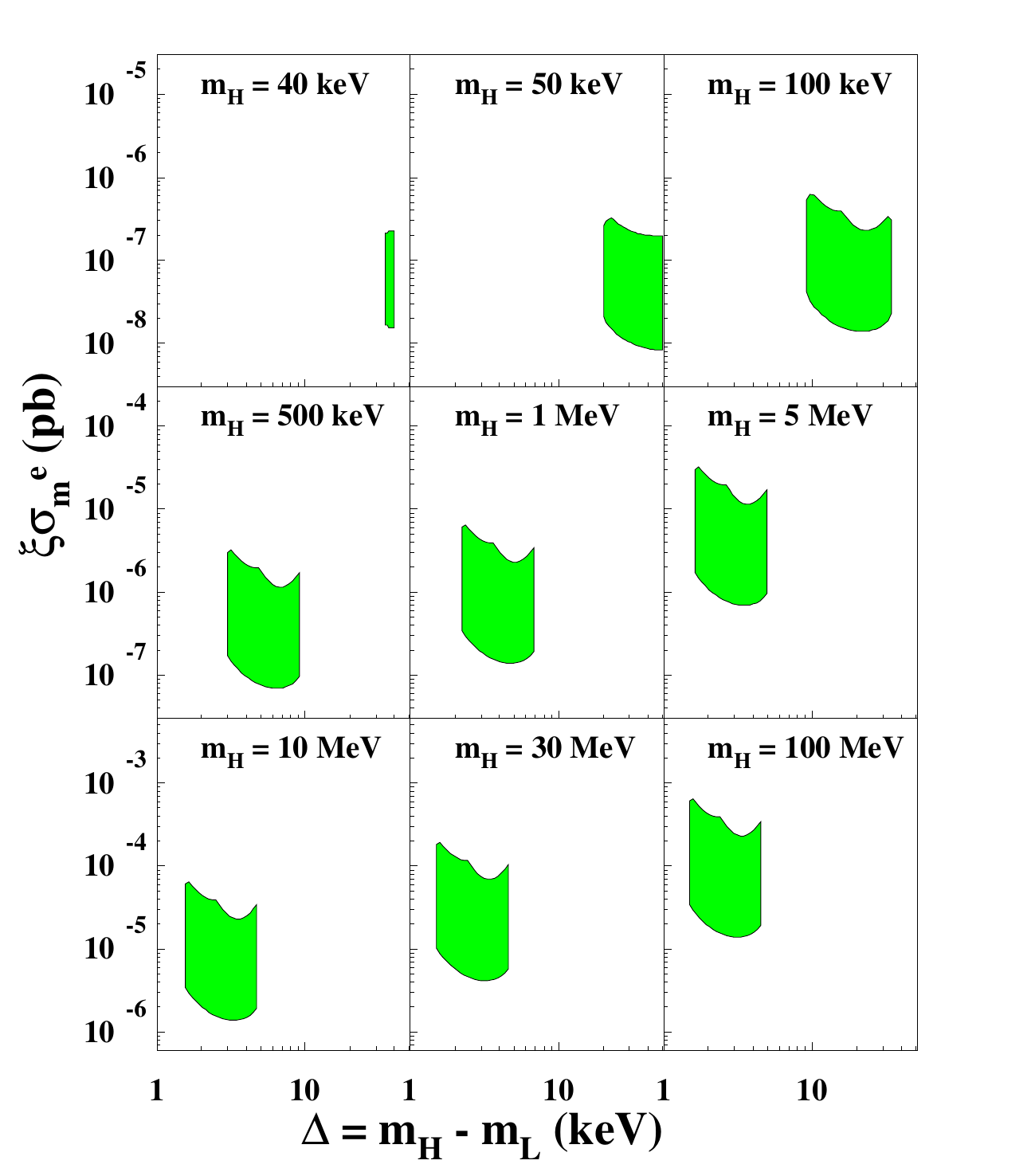}
\raisebox{2cm}{
\includegraphics[width=5.6cm] {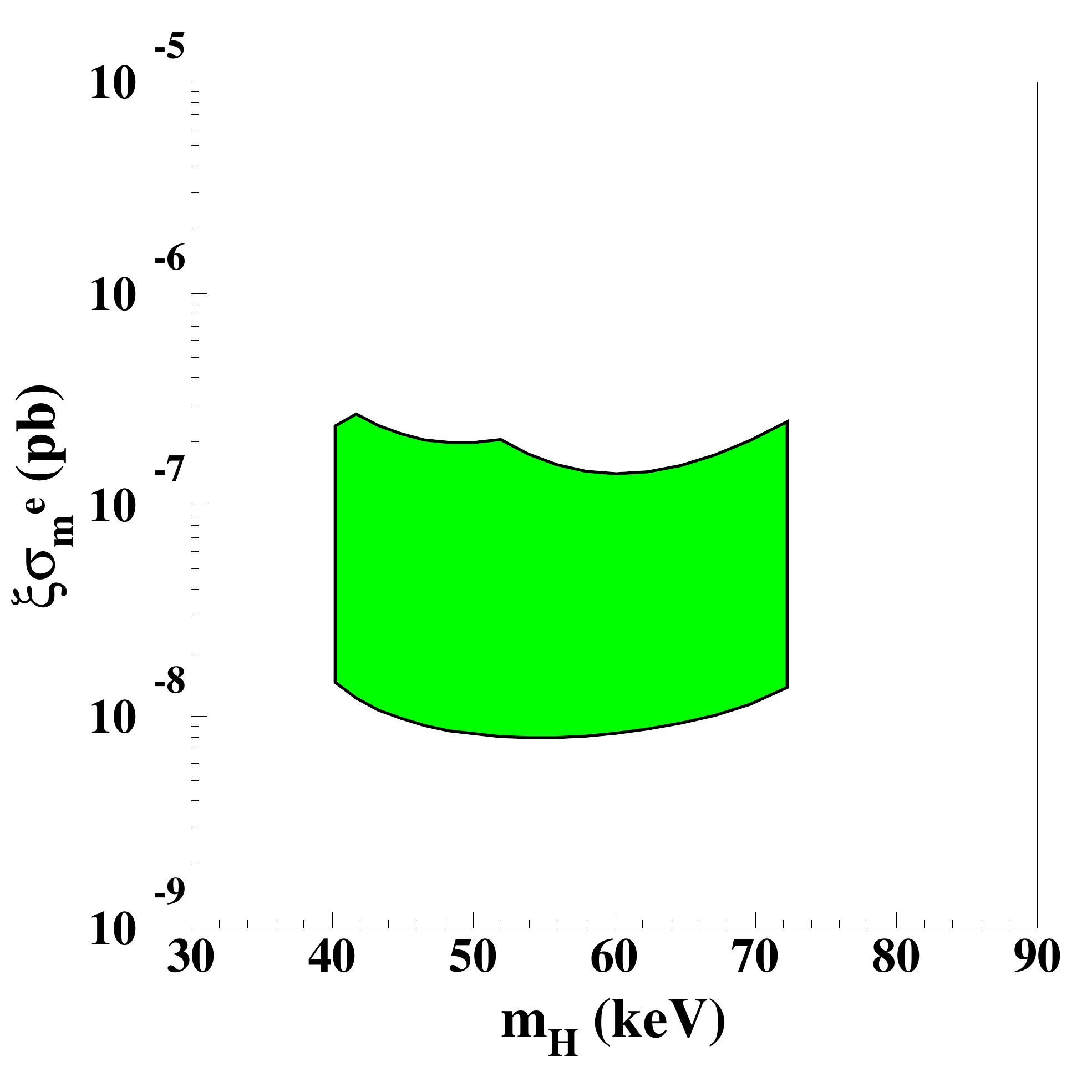}}
\vspace{-0.4cm}
\caption{Case of electron interacting LDM.
{\em Left:} examples of some slices of the allowed 3-dimensional volume 
for various $m_H$ depicted in the ($\xi\sigma^e_m$ vs $\Delta$) plane 
at 10 $\sigma$ from the {\it null hypothesis} (absence of modulation).
{\em Right:} slice of the allowed 3-dimensional volume at 10 $\sigma$ from the {\it null hypothesis}
for $m_H = \Delta$, that is for a massless or a very light $\nu_L$ particle,
as e.g. either an active neutrino or a nearly massless sterile neutrino or the light axion, etc.}
\label{fg:elett}
\end{figure}

\vspace{0.4cm}
As examples, some slices of the 3-dimensional allowed volume for various $m_H$ values 
in the ($\xi\sigma^e_m$ vs $\Delta$) plane are depicted in Fig. \ref{fg:elett}{\it --left}.

\vspace{0.4cm}
The slice of the allowed 3-dimensional volume for $m_H = \Delta$ at 10 $\sigma$ from the {\it null hypothesis}
is shown in Fig. \ref{fg:elett}{\it --right}.
This slice has been taken along the dotted line of Fig. \ref{fg:elett1},
restricting $m_L \simeq 0$, that is for a massless or a very light $\nu_L$ particle,
such as e.g. either an active neutrino or a nearly massless sterile one or the light axion, etc.

\begin{figure}[!ht]
\begin{center}
\includegraphics[width=6.5cm] {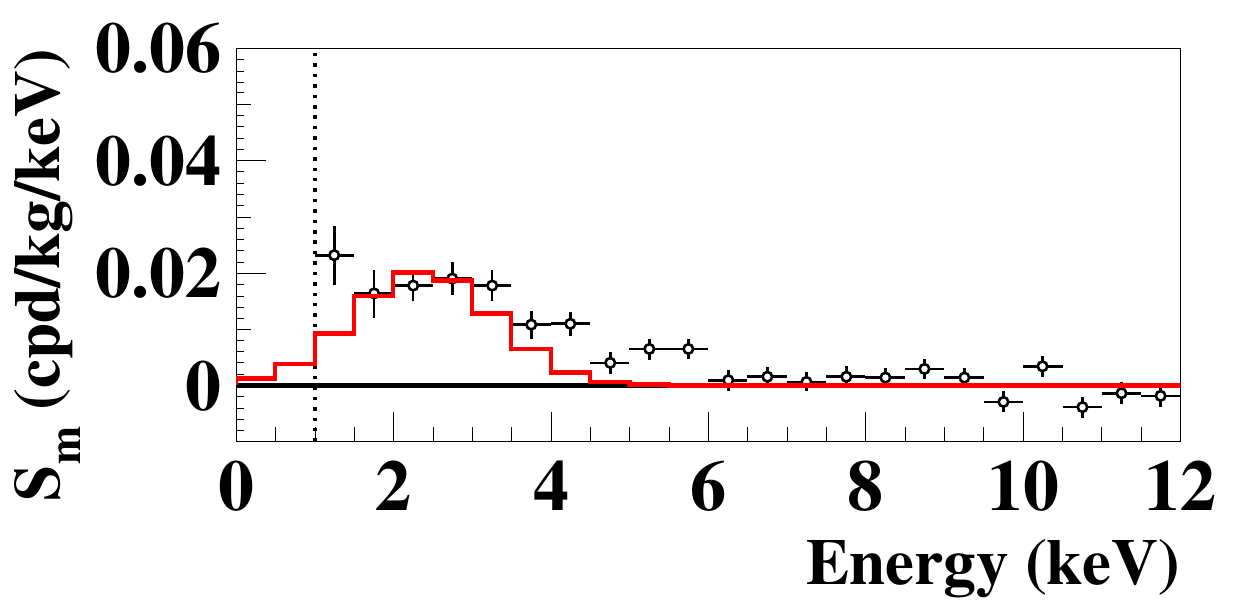}
\end{center}
\vspace{-0.4cm}
\caption{Example of superposition of the measured $\mathcal{S}^{exp}_{m}$ vs energy (points with error bars) 
with theoretical expectations (solid histograms) for electron interacting LDM with $m_H $ = 30 MeV, $\Delta$ = 2.4 keV
and $\xi\sigma^e_m = 1.1 \times 10^{-4}$ pb.
The case of the A3 (Evans power-law) halo model with $\rho_0$ = 0.17 GeV/cm$^3$ and $v_0$ = 170 km/s is considered.}
\label{fg:smvse_LDM_el}
\end{figure}

\vspace{0.4cm}
In Fig. \ref{fg:smvse_LDM_el} an example of superposition of the measured $\mathcal{S}^{exp}_{m}$ vs energy 
(points with error bars) with theoretical expectation (solid histograms) is shown.

\vspace{0.4cm}
In conclusion, it is worthwhile to summarize that electron interacting LDM candidates 
in the few-tens-keV/sub-MeV range are allowed by DAMA experiments (see Figs.~\ref{fg:elett1} and \ref{fg:elett}).
This can be of interest, for example, in the models of Warm Dark Matter particles,
such as e.g. weakly sterile neutrino.
Moreover, configurations with $m_H$ in the MeV/sub-GeV range 
are also allowed; similar LDM candidates 
can also be of interest for the production mechanism of the 511 keV gammas from 
the galactic bulge.

\vspace{0.4cm}
\subsubsection{Interaction with nuclei}

With regard to the interaction of LDM with target nuclei, 
the allowed volume in the space ($m_H$, $\Delta$, $\xi\sigma_m^{nucleus}$)
at 10$\sigma$ from the {\it null hypothesis} can be obtained by
comparing the expected modulated signal with the experimental results obtained
by DAMA/NaI, DAMA/LIBRA--phase1 and DAMA/LIBRA--phase2 \cite{uni18,bled18,npae18,npp19,mg15}. 

\begin{figure} [!ht]
\centering
\includegraphics[width=4.3cm] {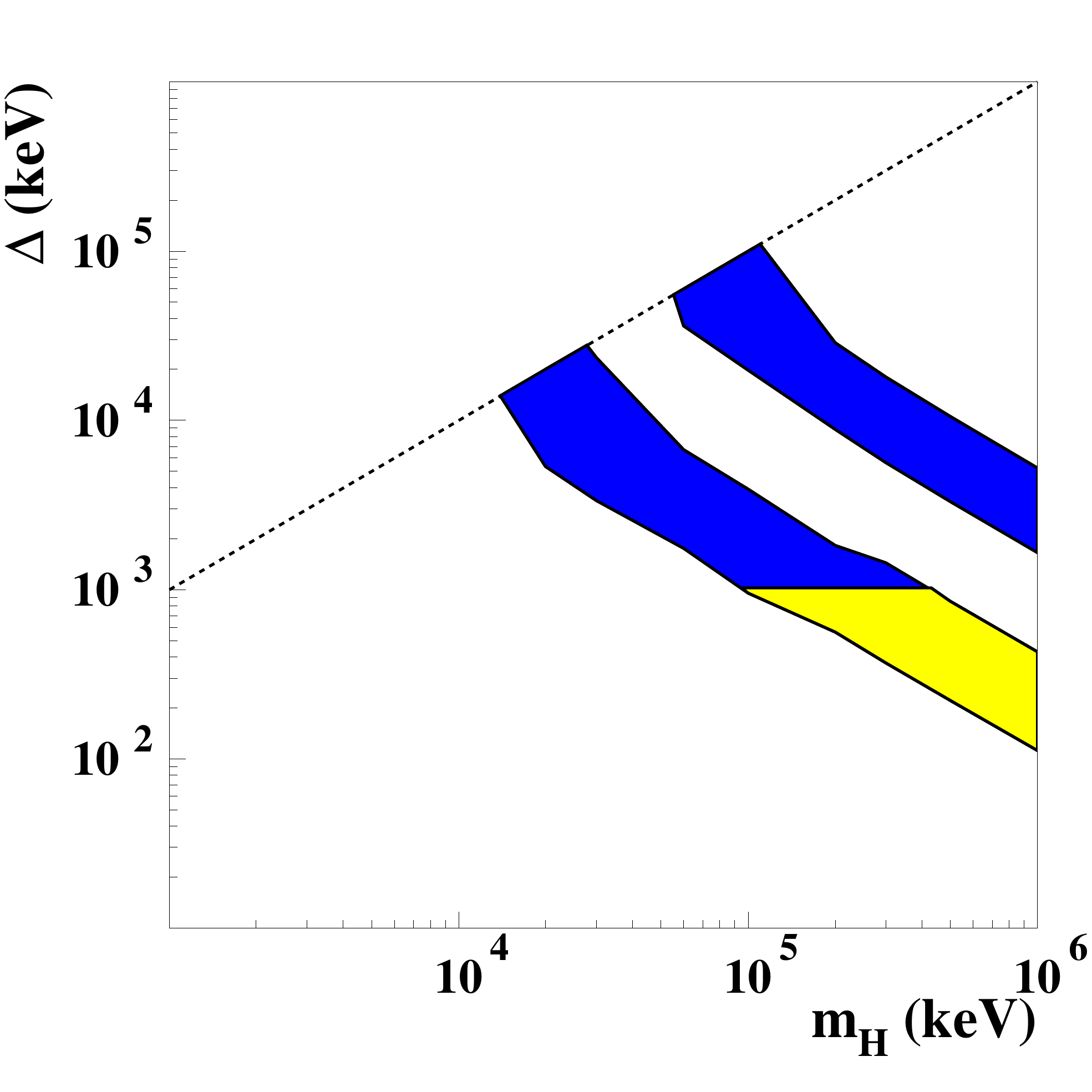}
\includegraphics[width=4.3cm] {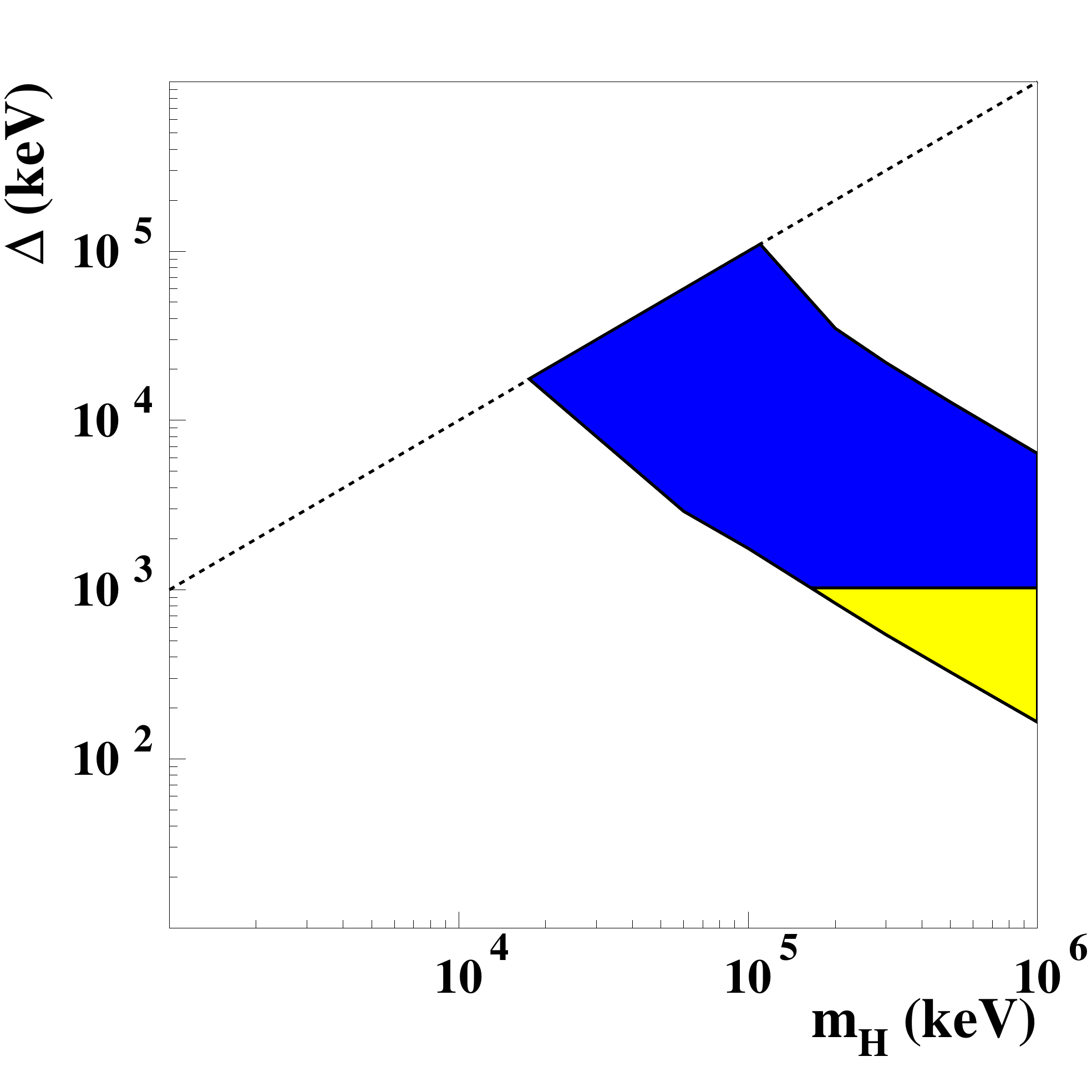}
\includegraphics[width=4.3cm] {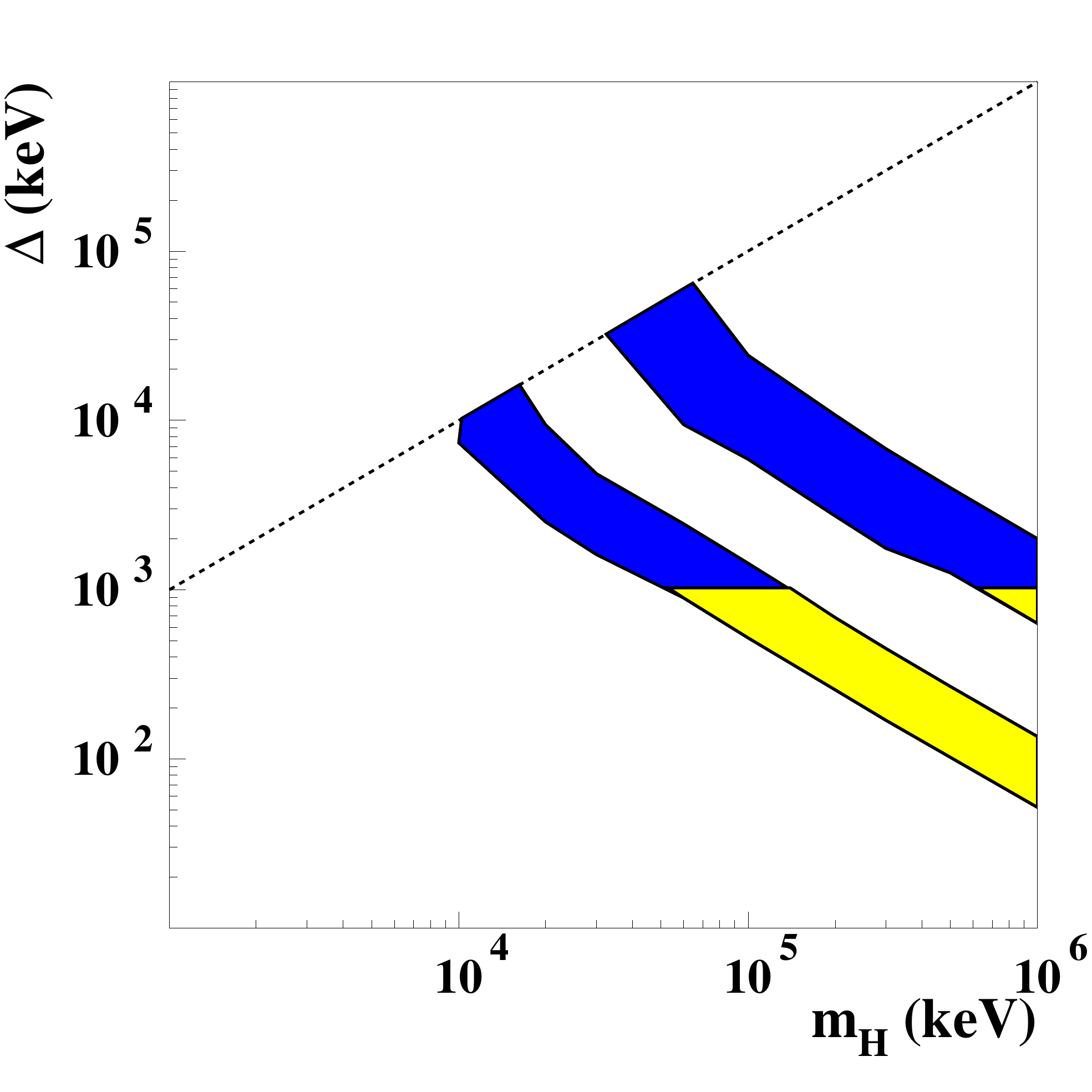}
\caption{Case of nucleus interacting LDM.
Projections of allowed 3-dimensional volumes on the plane ($m_H$, $\Delta$) for
coherent nuclear scaling law, considering for the quenching factors:
(i)   $Q_I$ case ($left$), 
(ii)  with channeling effect ($center$), and
(iii) $Q_{II}$ ($right$). 
The regions have been obtained by marginalizing all the models for each considered scenario 
(see Sect. \ref{data_analysis}) and they represent the domain where the likelihood-function 
values differ more than 10 $\sigma$ from the {\it null hypothesis} (absence of modulation).
The dashed lines ($m_H = \Delta$) mark the case where $\nu_L$ is a massless particle.
The decays through the diagram involved in the detection channel are energetically forbidden.}
\label{fg:regn_coh}
\end{figure}

\begin{figure} [!ht]
\centering
\includegraphics[width=4.3cm] {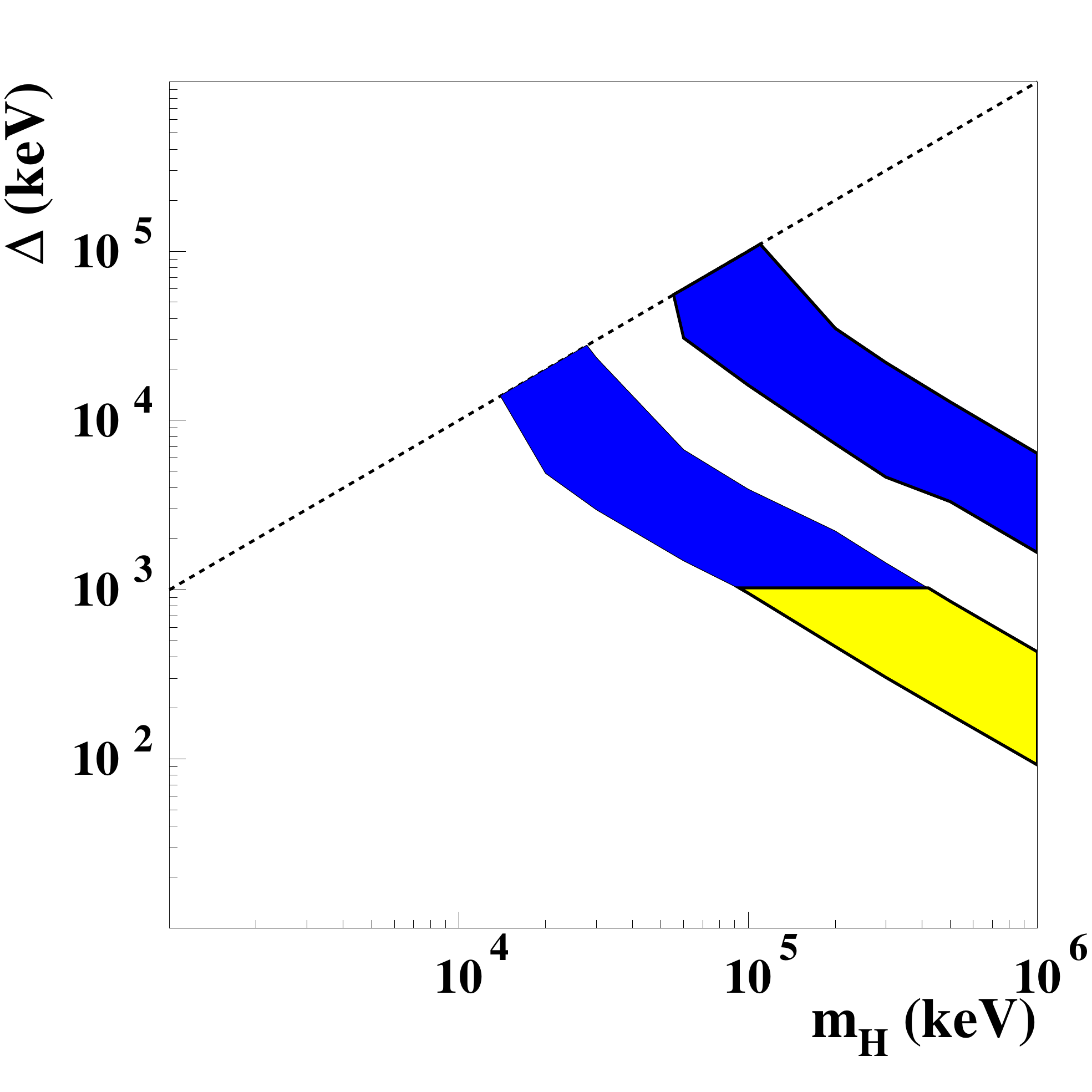}
\includegraphics[width=4.3cm] {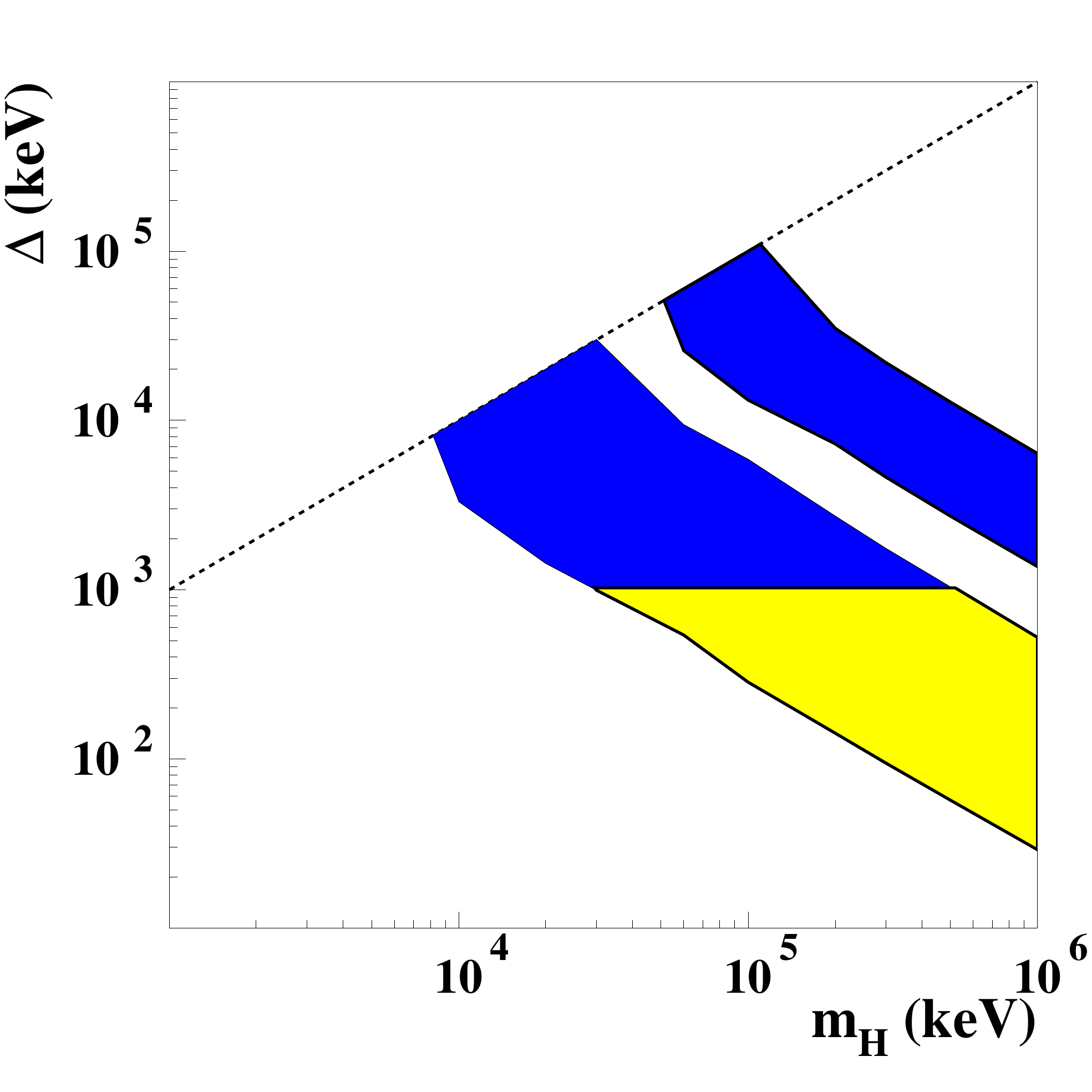}
\includegraphics[width=4.3cm] {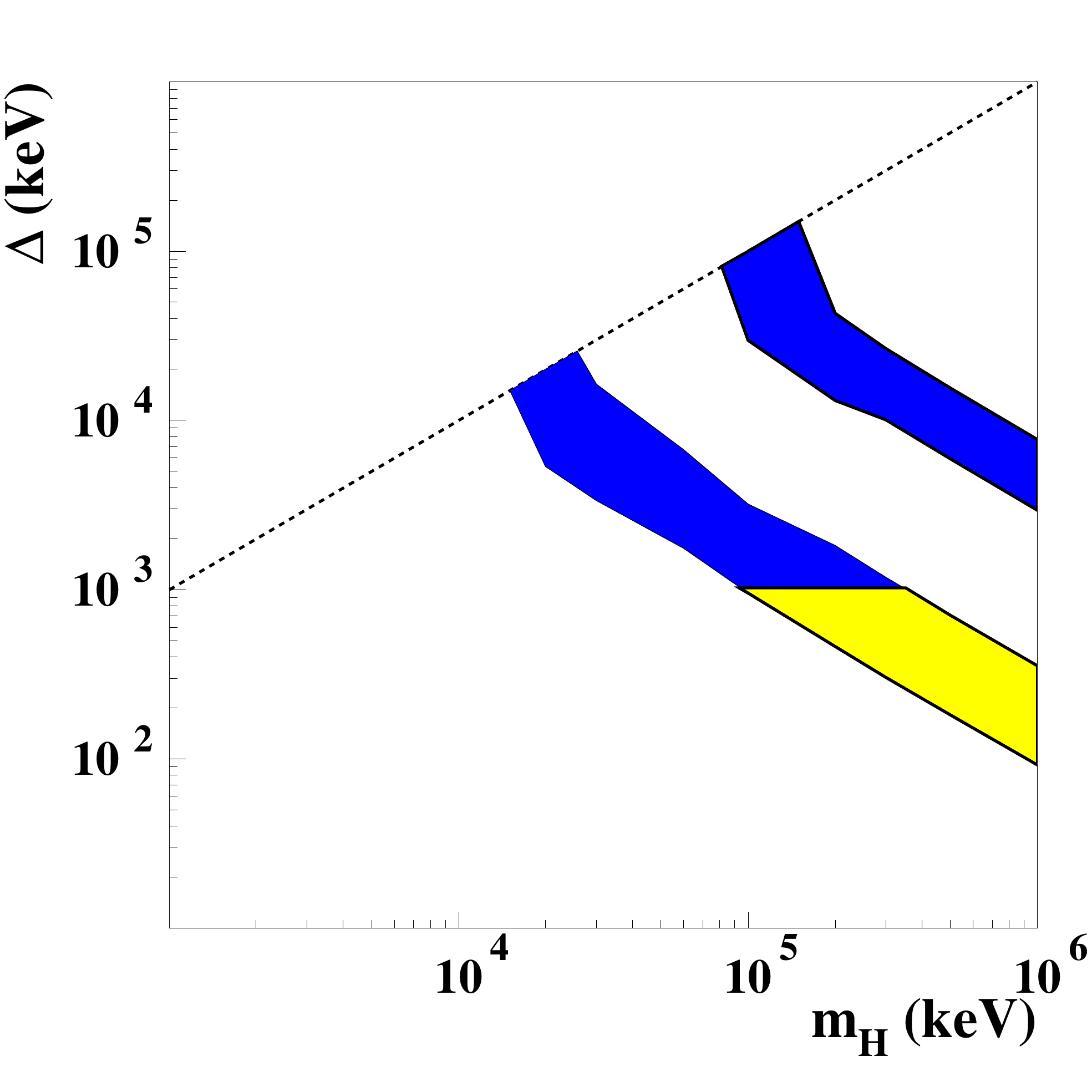}
\caption{Case of nucleus interacting LDM.
Projections of allowed 3-dimensional volumes on the plane ($m_H$, $\Delta$) for
incoherent nuclear scaling law, considering for the quenching factors:
(i)   $Q_I$ case ($left$), 
(ii)  with channeling effect ($center$), and
(iii) $Q_{II}$ ($right$). 
The regions have been obtained by marginalizing all the models for each considered scenario 
(see Sect. \ref{data_analysis}) and they represent the domain where the likelihood-function 
values differ more than 10 $\sigma$ from the {\it null hypothesis} (absence of modulation).
The dashed lines ($m_H = \Delta$) mark the case where $\nu_L$ is a massless particle.
The decays through the diagram involved in the detection channel are energetically forbidden.}
\label{fg:regn_incoh}
\end{figure}

\vspace{0.4cm}
The projections of such a region on the plane ($m_H$, $\Delta$) are reported in Figs.~\ref{fg:regn_coh} and 
\ref{fg:regn_incoh} for the two above-mentioned illustrative cases of coherent and incoherent nuclear scaling laws,
respectively. 
They have been obtained by marginalizing all the models for each considered scenario 
(see Sect. \ref{data_analysis}) and they represent the domain where the likelihood-function 
values differ more than 10 $\sigma$ from the {\it null hypothesis} (absence of modulation).

\begin{figure} [p]
\centering
\vspace{-0.4cm}
\includegraphics[width=0.48\textwidth] {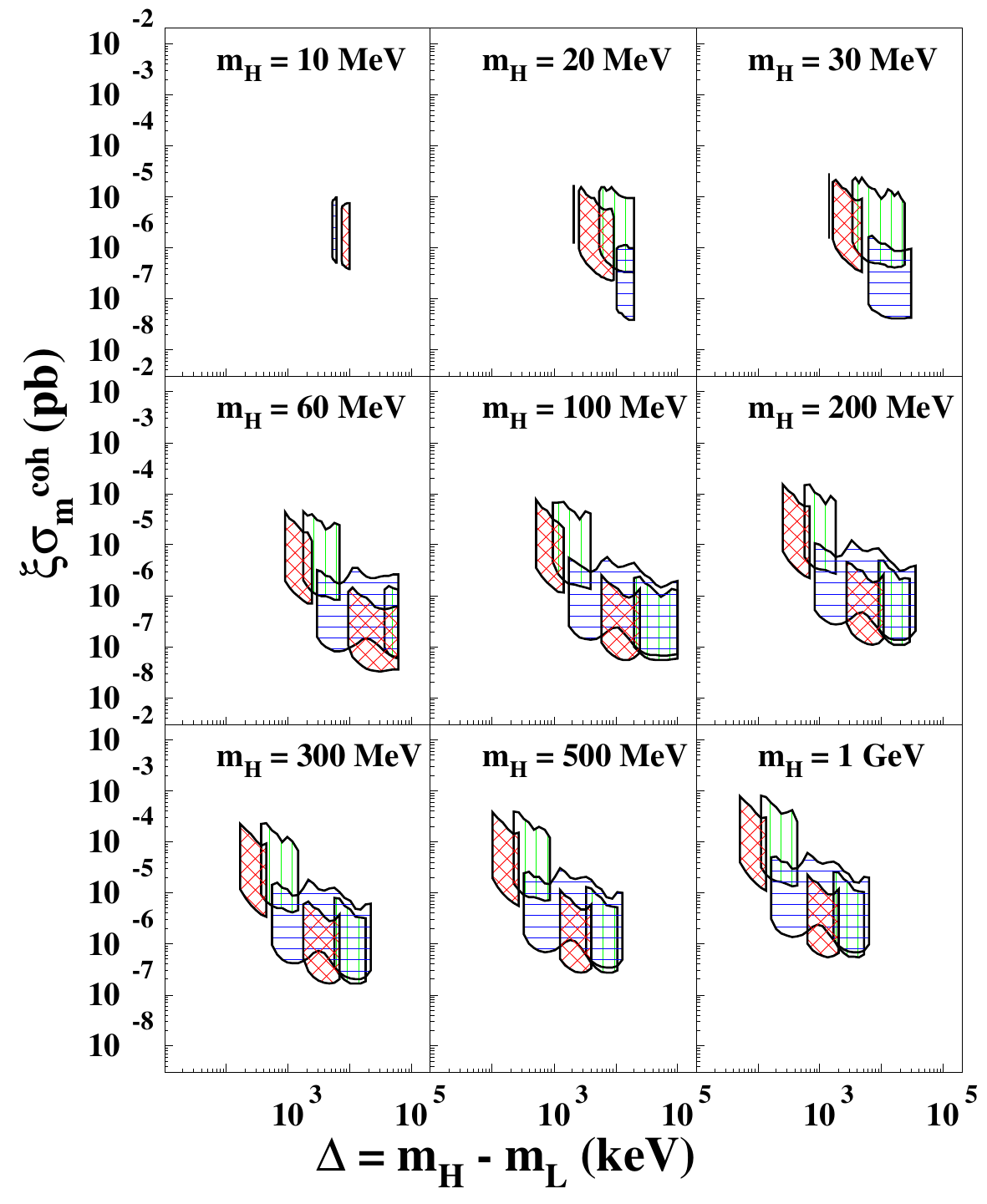}
\includegraphics[width=0.48\textwidth] {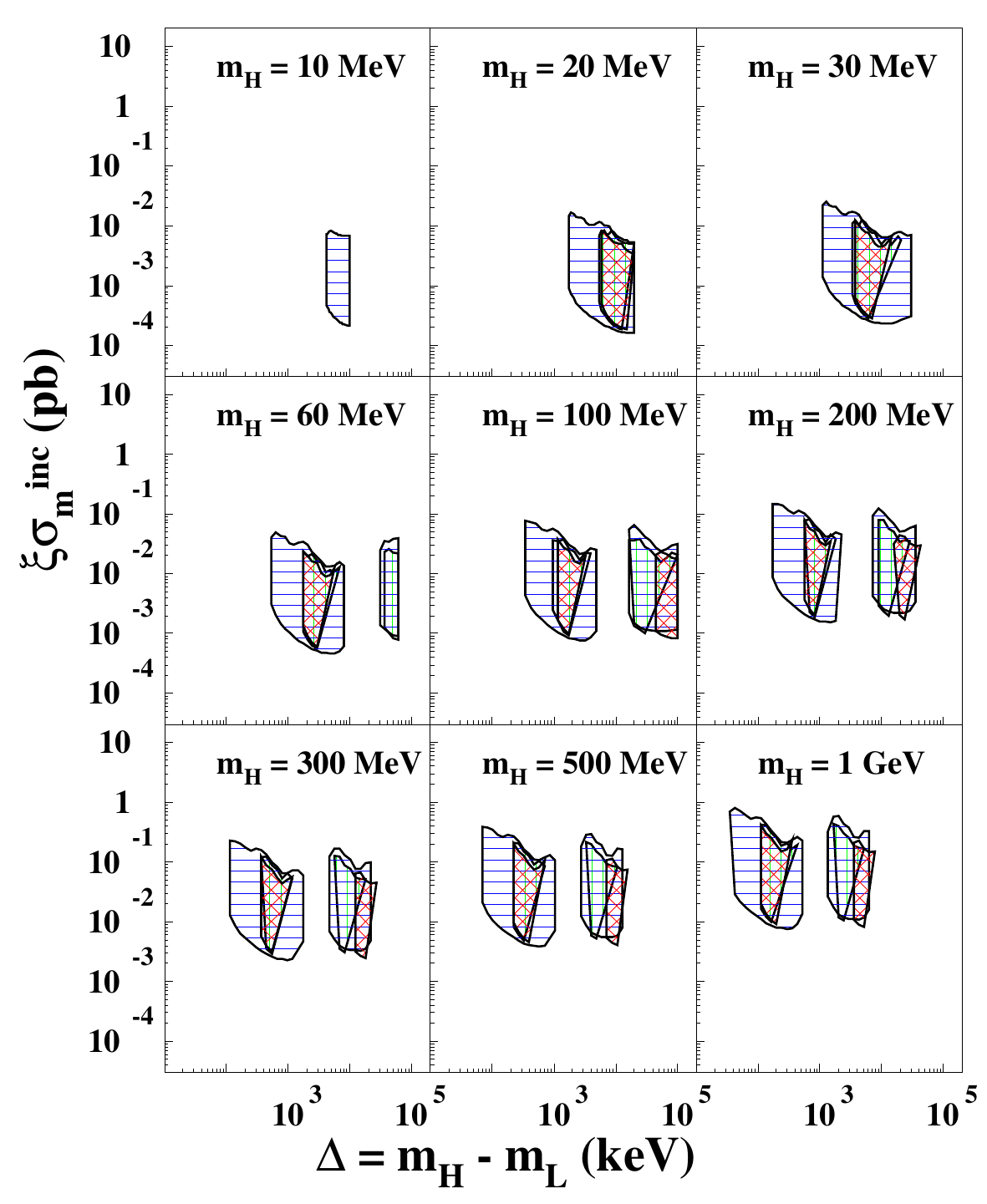}
\vspace{-0.4cm}
\caption{Case of nucleus interacting LDM.
Examples of some slices of the 3-dimensional allowed volumes for various $m_H$ values 
in the ($\xi\sigma_m^{coh,inc}$ vs $\Delta$) plane 
for the two illustrative cases of coherent ({\it left})
and incoherent ({\it right}) nuclear scaling laws.
Three different instances for the Na and I quenching factors have been considered: 
(i) $Q_I$ case [(green on-line) vertically-hatched regions],
(ii) with channeling effect [(blue on-line) horizontally-hatched regions] and 
(iii) $Q_{II}$ [(red on-line) cross-hatched regions].
The 3-dimensional volumes have been obtained by marginalizing all the models for each considered scenario 
(see Sect. \ref{data_analysis}) and they represent the domain where the likelihood-function 
values differ more than 10 $\sigma$ from the {\it null hypothesis} (absence of modulation).}
\label{fg:panel_n}
\end{figure}

The allowed $m_H$ values and the splitting $\Delta$ 
are in the intervals \mbox{8 MeV $\lsim m_H \lsim $ O(GeV)}\footnote{
We remind that for $m_H$ values greater than
O(GeV) the detection in DAMA experiments would also be possible through
the elastic scattering process \cite{RNC,ijmd,epj06,ijma07,chan}.}
and \mbox{29 keV $\lsim \Delta \lsim 150$ MeV}, respectively (see Figs.~\ref{fg:regn_coh} and \ref{fg:regn_incoh}).
It is worth to note that in such a case the decays through the diagram involved in 
the detection channel (e.g. in nucleon anti-nucleon pairs or in meson(s), as $\nu_H \rightarrow \nu_L \pi^0$)
are obviously energetically forbidden.
Moreover, there are allowed configurations that could contribute -- in principle, if suitable
couplings exist -- to the 
positron generation in the galactic center; in fact, 
the decay $\nu_H \rightarrow \nu_L e^+ e^-$ is energetically allowed
for $\Delta > 2m_e$ (dark area in Figs.~\ref{fg:regn_coh} and \ref{fg:regn_incoh}), 
while the annihilation processes into $e^+e^-$ pairs are energetically allowed
for almost all the allowed configurations.

It is worth noting that for nuclear interacting LDM the 3-dimensional allowed configurations 
are contained in two disconnected volumes, as seen e.g. in their projections in Figs.~\ref{fg:regn_coh} and \ref{fg:regn_incoh}.
The one at larger $\Delta$ at $m_H$ fixed is mostly due to interaction on Iodine target, while
the other one is mostly due to interaction on Sodium target.

\vspace{0.4cm}
As examples, some slices of the 3-dimensional allowed volumes for various $m_H$ values 
in the ($\xi\sigma_m^{coh,inc}$ vs $\Delta$) plane are depicted in Fig. \ref{fg:panel_n}
for the two above-mentioned illustrative cases of coherent ({\it left} panel)
and incoherent ({\it right} panel) nuclear scaling laws. 

\begin{figure} [!ht]
\centering
\vspace{-0.4cm}
\includegraphics[width=6.7cm] {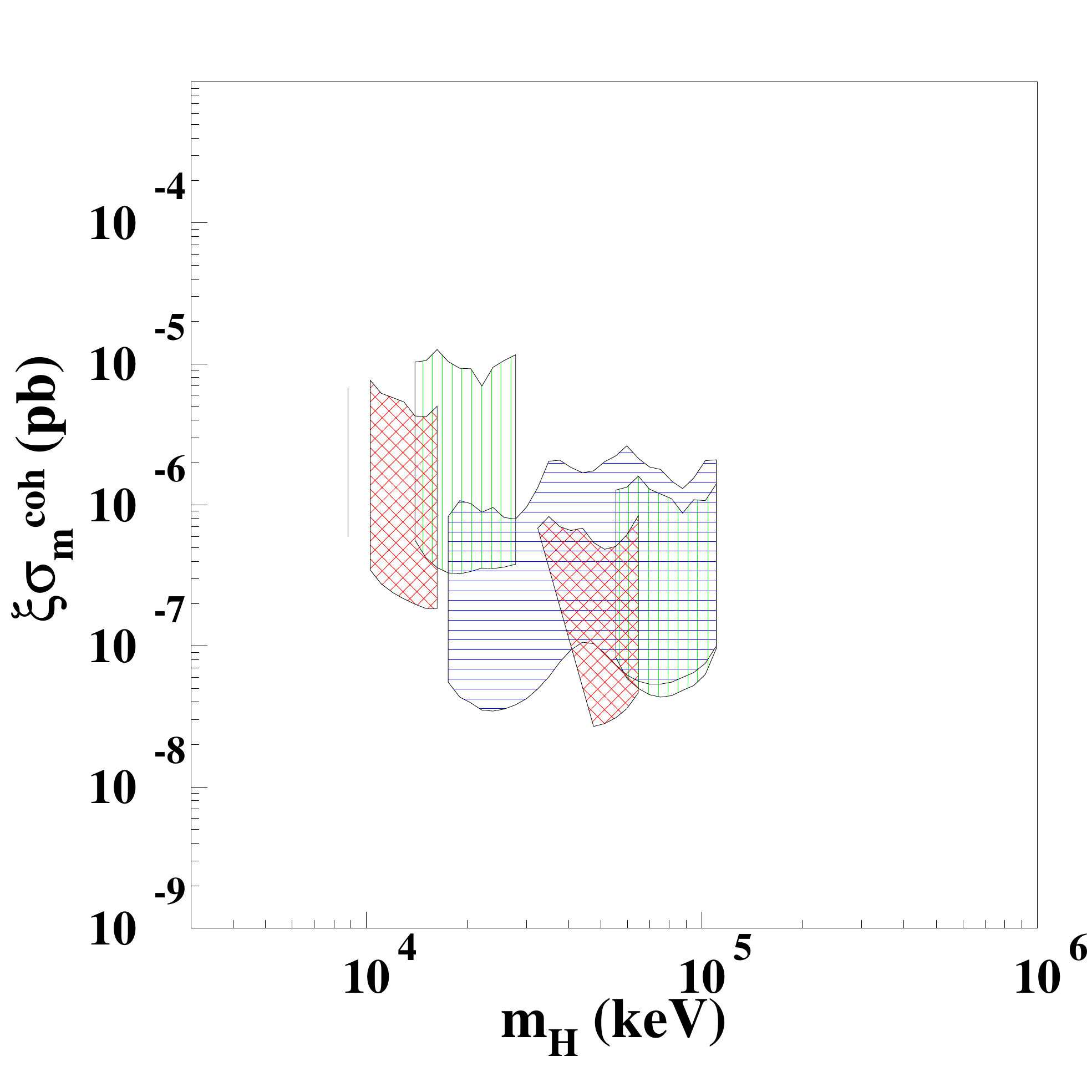}
\includegraphics[width=6.7cm] {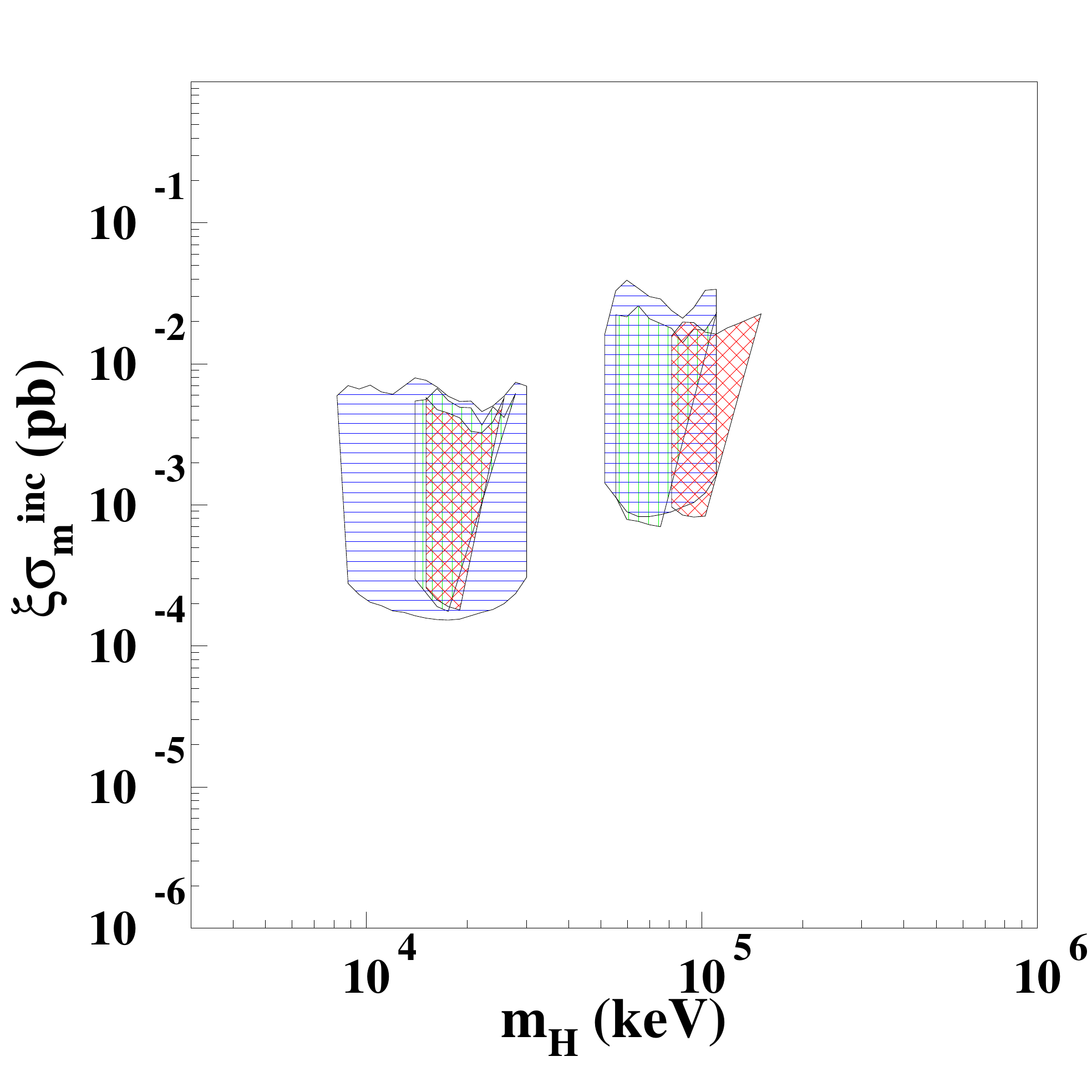}
\vspace{-0.4cm}
\caption{Case of nucleus interacting LDM.
Slices of the 3-dimensional allowed volumes 
for $m_H = \Delta$, that is for a massless or a very light $\nu_L$ particle,
as e.g. either an active neutrino or a nearly massless sterile one or the light axion, etc.
They are evaluated for the two illustrative cases of coherent ({\it left} panel)
and incoherent ({\it right} panel) nuclear scaling laws. 
Three different instances for the Na and I quenching factors have been considered: 
(i) $Q_I$ case [(green on-line) vertically-hatched regions],
(ii) with channeling effect [(blue on-line) horizontally-hatched regions] and 
(iii) $Q_{II}$ [(red on-line) cross-hatched regions].
For the coherent case with channeling effect the low mass region is very narrow 
so that appears as a vertical segment. 
The 3-dimensional volumes have been obtained by marginalizing all the models for each considered scenario 
(see Sect. \ref{data_analysis}) and they represent the domain where the likelihood-function 
values differ more than 10 $\sigma$ from the {\it null hypothesis} (absence of modulation).}
\label{fg:ster_n}
\end{figure}

\vspace{0.4cm}
The slices of the 3-dimensional allowed volumes 
for $m_H = \Delta$ are shown in Fig. \ref{fg:ster_n}
for the two illustrative cases of coherent ({\it left} panel)
and incoherent ({\it right} panel) nuclear scaling laws. 
These slices have been taken along the dotted lines of Figs.~\ref{fg:regn_coh} and \ref{fg:regn_incoh},
restricting $m_L \simeq 0$, that is for a massless or a very light $\nu_L$ particle,
as e.g. either an active neutrino or a nearly massless sterile one or light axion, etc.

\begin{figure}[!ht]
\begin{center}
\includegraphics[width=6.5cm] {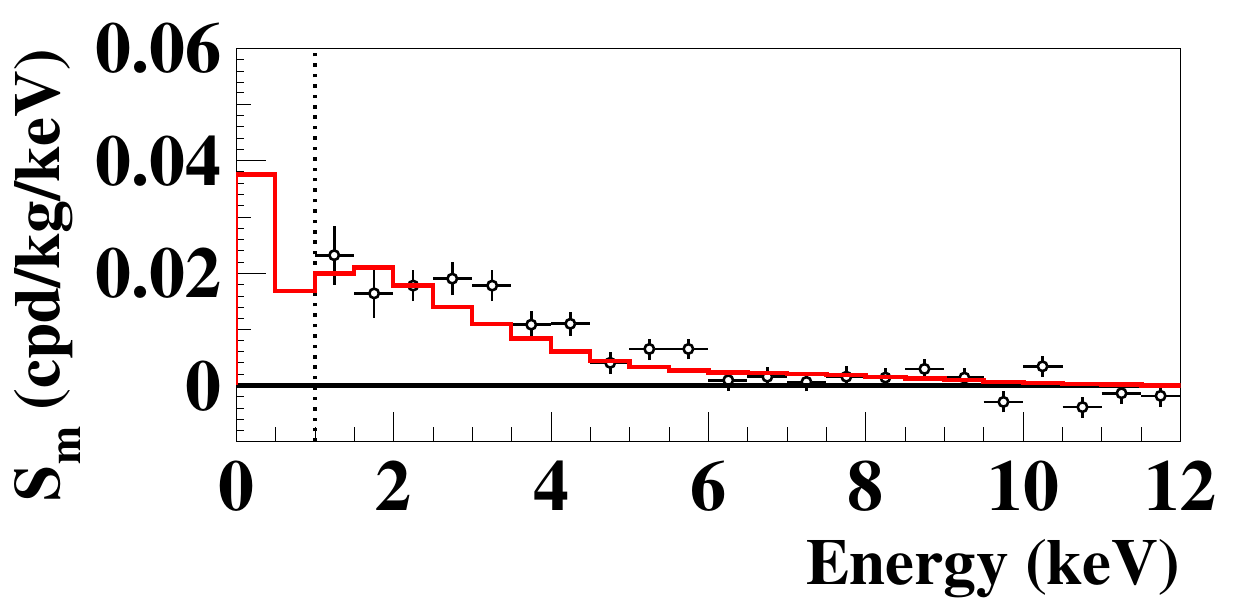}
\end{center}
\vspace{-0.4cm}
\caption{Example of superposition of the measured $\mathcal{S}^{exp}_{m}$ vs energy (points with error bars) 
with theoretical expectations (solid histograms) for nucleus interacting LDM with incoherent nuclear scaling 
law, $m_H $ = 60 MeV, $\Delta$ = 2.9 MeV and $\xi\sigma_m^{inc} = 3.8 \times 10^{-3}$ pb.
The case of the C2 (Evans logarithmic) corotating halo model with $\rho_0$ = 0.67 GeV/cm$^3$ and $v_0$ = 170 km/s
is considered. Moreover, the channeling effect is included and the set B of parameters values is used.}
\label{fg:smvse_nuc}
\end{figure}

\vspace{0.4cm}
In Fig. \ref{fg:smvse_nuc} an example of superposition of the measured $\mathcal{S}^{exp}_{m}$ vs energy 
(points with error bars) with a theoretical expectation (solid histograms) is shown.

\vspace{0.4cm}
Finally, it is worthwhile to summarize that LDM candidates in the MeV/sub-GeV range 
are allowed by DAMA experiments (see Figs.~\ref{fg:regn_coh}, \ref{fg:regn_incoh}, \ref{fg:panel_n} and \ref{fg:ster_n}).
Also these candidates, such as e.g. axino, sterile neutrino,
can be of interest for the positron production in the galactic bulge.

\vspace{0.4cm}
\subsection{Mirror Matter}
\label{DM5}

Well-motivated Dark Matter candidates are represented by
the so called Mirror particles. The Mirror scenario can be introduced by considering 
a parallel gauge sector with particle
physics exactly identical to that of ordinary particles, coined
as mirror world. In this theory the Mirror particles 
belong to the hidden or shadow gauge sector and
can constitute the DM particles of the Universe. A comprehensive discussion about
Mirror Matter as DM component can be found in Refs. \cite{mirasim,mirsim}.
In these two papers, in addition, the annual modulation effect measured by DAMA experiments -- 
with lower exposure than presently -- has been analyzed
in the framework of Asymmetric and Symmetric Mirror Matter scenarios.
In the following these analyses are updated by including the new data of the first six annual cycles of 
DAMA/LIBRA--phase2 with lower software energy threshold. 
This new analysis restricts a significant part of the parameters' space of the Mirror DM scenarios.

\subsubsection{Asymmetric Mirror Matter}

In the Asymmetric Mirror scenario  
the mirror world is a heavier and deformed copy of our world, 
with mirror particle masses scaled in different ways with respect to the masses of the ordinary particles.   
Taking the mirror weak scale e.g. of the order of 10 TeV,   
the mirror electron would become two orders of magnitude heavier than our electron       
while the mirror nucleons $p'$ and $n'$  only about 5 times heavier than the ordinary nucleons. 
The dark matter would exist in the form of mirror Hydrogen 
with mass of about 5 GeV (which is a rather interesting mass range for DM particles), 
composed of mirror proton and electron.
The interaction of mirror atomic-type DM 
candidates with the detector nuclei occurs via the interaction portal which is 
the kinetic mixing $\frac{\epsilon}{2} F^{\mu\nu} F'_{\mu\nu}$  of two massless states, 
ordinary photon and mirror photon. To fulfill the phenomenology 
the mixing parameter is $\epsilon \ll 1$;
this mixing mediates the mirror atom scattering off the ordinary target nuclei in 
the NaI(Tl) detectors at DAMA/LIBRA set-ups with the Rutherford-like cross sections.

\begin{figure}[!ht]
\centering
\vspace{-0.5cm}
\includegraphics[width=10.cm] {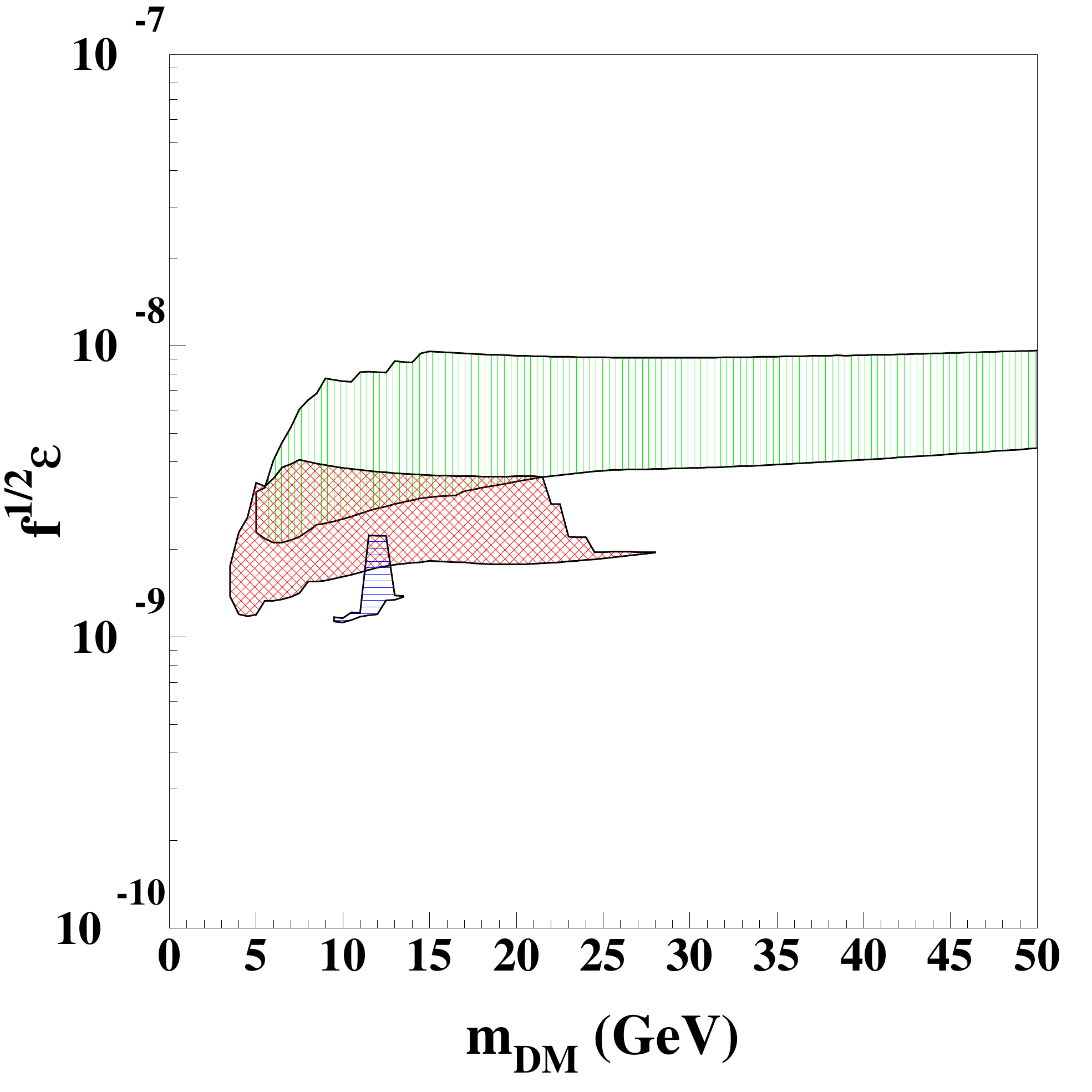}
\caption{Regions in the plane $\sqrt{f}\epsilon$ vs $M_{A'}$ allowed by DAMA experiments in the case 
of Asymmetric Mirror Matter, when the assumption $M_{A'}\simeq 5 m_{p}$ is released.
Three different instances for the Na and I quenching factors have been considered: 
(i) $Q_I$ case [(green on-line) vertically-hatched region],
(ii) with channeling effect [(blue on-line) horizontally-hatched region)] and 
(iii) $Q_{II}$ [(red on-line) cross-hatched region].
The regions have been obtained by marginalizing all the models for each considered scenario 
(see Sect. \ref{data_analysis}) and they represent the domain where the likelihood-function 
values differ more than 10 $\sigma$ from the {\it null hypothesis} (absence of modulation).}
\label{fg:asim_mirr}
\end{figure}

The low-energy differential cross-section of the interaction between mirror ($A'$)
and ordinary ($A$) nuclei has the Rutherford-like form:
\be{}
\frac{d\sigma_{A,A'}
}{dE_{R}}=\frac{\mathcal{C}_{A,A'}}{E_{R}^{2}v^{2}}
\ee
where $E_{R}$ is the energy of the ordinary nucleus recoil, $v=|\vect{v}|$ is the relative velocity between the mirror nucleus and the ordinary one,
and:
\begin{equation}
\mathcal{C}_{A,A'}=\frac{2\pi \epsilon^{2}\alpha^{2}Z^{2}Z'^{2}}{M_{A}}\mathcal{F}^2_{A}\mathcal{F}^2_{A'}
\end{equation}
where $\alpha$ is the fine structure constant, $Z$ and $Z'$ are the charge numbers of the ordinary and mirror nuclei,
$M_{A}$ is the mass of the ordinary nucleus, and  
$\mathcal{F}_{X}$$(qr_{X})$ $(X=A,A')$ are the Form-factors of ordinary and mirror nuclei, which depend on the momentum transfer, $q$,
and on the radius of $X$ nucleus.

In the case of asymmetric mirror model we consider just one species of mirror nuclei. 
Our benchmark model is the mirror Hydrogen ($A'=Z'=1$), with mass $M_{A'} \simeq m_p \simeq 5$~GeV ($m_p$ is the proton mass). 
Alternatively, one can consider the Helium like $\Delta$-atom, 
with $A'=1, Z'=2$ and with mass again $M_{A'}\simeq 5$~GeV.
For backward compatibility with our previous papers on this subject, 
the fractional amount of local density in terms of mirror matter 
is named $f$ only in this Section, instead of the already defined $\xi$.
Hence, the signal rate is proportional to $Z'^2f \epsilon^2$. 
All the numerical results are presented in the case of mirror Hydrogen in terms of $\sqrt{f}\epsilon$.
They would be equivalent to $Z' \sqrt{f}\epsilon$ in the case of mirror nuclei with $Z'>1$  with 
the same mass. So, for $\Delta$-atom one just puts  $Z'=2$.  

\begin{figure}[!ht]
\begin{center}
\includegraphics[width=4.3cm] {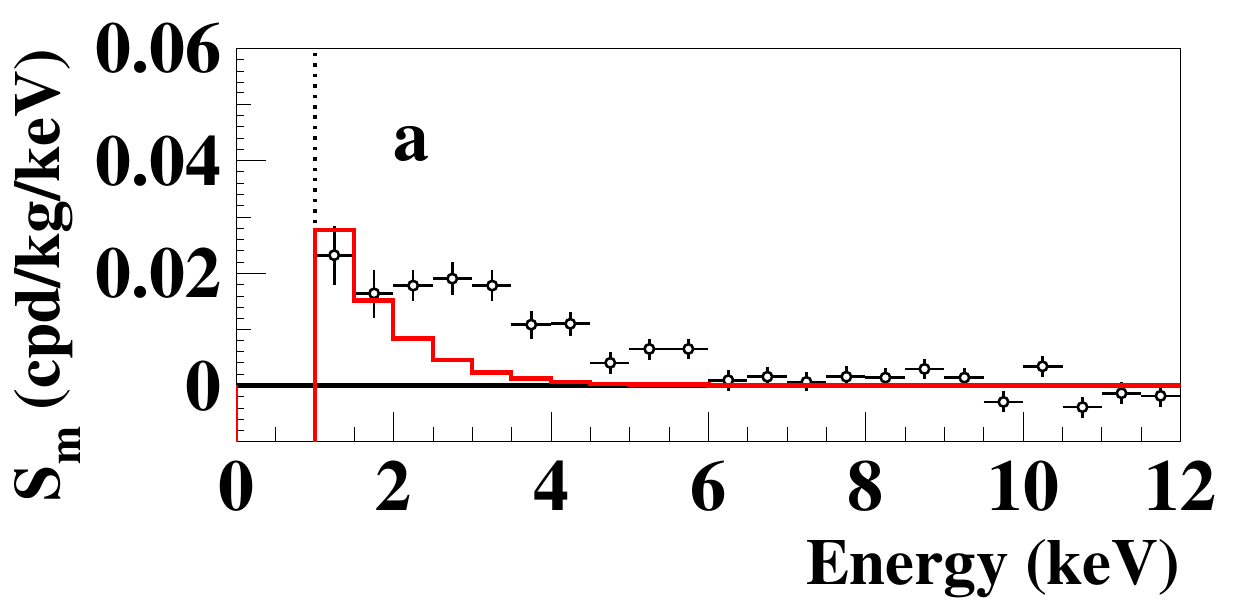}
\includegraphics[width=4.3cm] {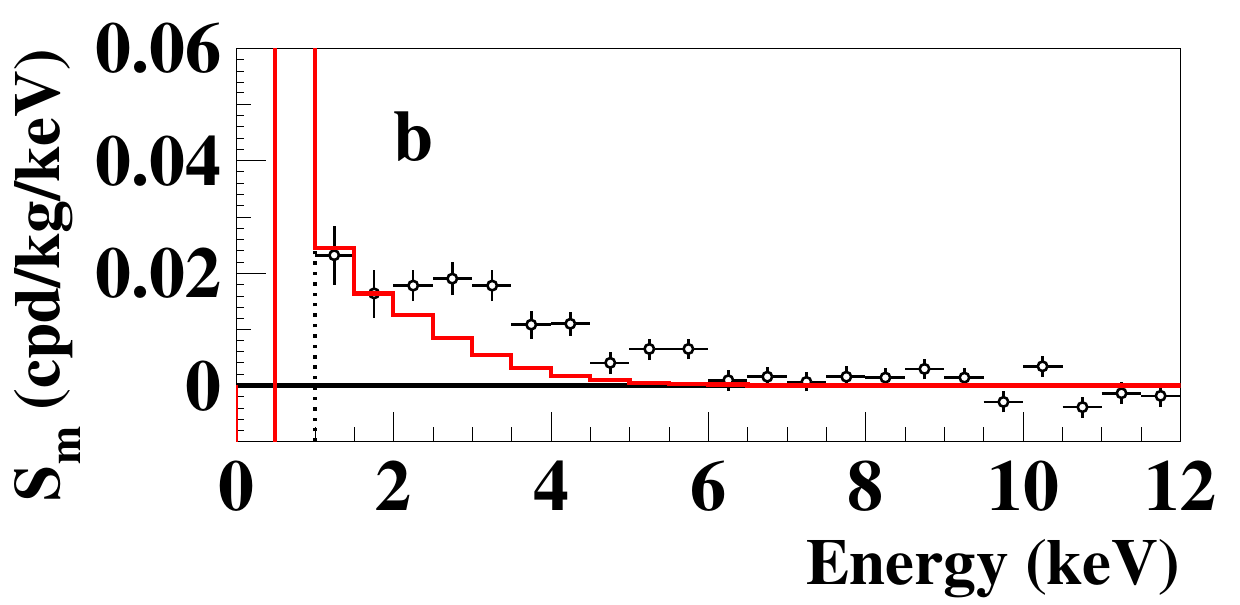}
\includegraphics[width=4.3cm] {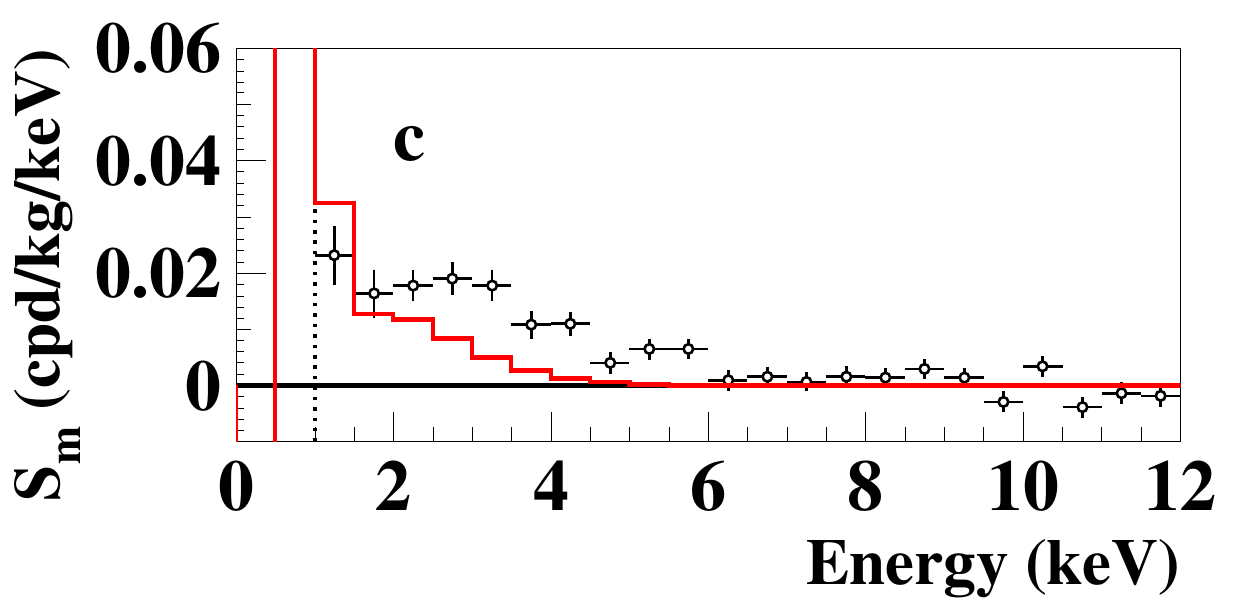}
\end{center}
\vspace{-0.4cm}
\caption{Examples of superposition of the measured $\mathcal{S}^{exp}_{m}$ vs energy
(points with error bars) with theoretical expectations (solid histograms) for
Asymmetric Mirror Matter.
(a) case of the C2 (Evans logarithmic) halo model with $\rho_0$ = 0.67 GeV/cm$^3$,
    $v_0$ = 170 km/s, set A of parameters values, $m_{DM} = 11$ GeV,
    $\sqrt{f}\epsilon = 1.2 \times 10^{-9}$ including channeling effect;
(b) case of the C2 (Evans logarithmic) halo model with $\rho_0$ = 0.67 GeV/cm$^3$,
    $v_0$ = 170 km/s, set B of parameters values, $m_{DM} = 14$ GeV,
    $\sqrt{f}\epsilon = 3.9 \times 10^{-9}$ and quenching $Q_{I}$;
(c) case of the C1 (Evans logarithmic) halo model with $\rho_0$ = 0.56 GeV/cm$^3$,
    $v_0$ = 170 km/s, set B of parameters values, $m_{DM} =  7$ GeV,
    $\sqrt{f}\epsilon = 2.5 \times 10^{-9}$ and quenching $Q_{II}$. As it can be seen, the data allow discriminating the various scenarios.}
\label{fg:asym_theo}
\end{figure}

The data analysis in the Mirror DM model framework considered here allows the determination of the
$\sqrt{f}\epsilon$ parameter. It has been taken into account the uncertainties 
by marginalizing all the models for each considered scenario 
(see Sect. \ref{data_analysis}) for the three instances of the quenching factors:
(i)   $Q_I$ case, 
(ii)  with channeling effect, and
(iii) $Q_{II}$. 
The obtained allowed intervals of the $\sqrt{f}\epsilon$ parameter 
identify the $\sqrt{f}\epsilon$ values corresponding to C.L. 
larger than 10 $\sigma$ from the {\it null hypothesis}, that is $\sqrt{f}\epsilon=0$. 

These allowed intervals are:
$ \sqrt{f}\epsilon \in [2.28 - 3.13] \times 10^{-9}$, and 
$ \sqrt{f}\epsilon \in [1.19 - 3.38] \times 10^{-9}$
for the $Q_{I}$ and $Q_{II}$ cases, respectively,
while no interval is selected for the case when the channeling effect is included.
The obtained values of the $\sqrt{f}\epsilon$ parameter
are well compatible with cosmological bounds (see Refs. \cite{mirasim,mirsim} and references therein).

It is worth noting that in all the considered scenarios for mirror DM the DAMA signal in the 
1-6 keV energy interval arises mainly from interactions with Sodium nuclei.
This effect is due to the 
fact that the considered Mirror DM particle is quite light: $M_{A'}\simeq 5 m_{p}$.
            
If the assumption $M_{A'}\simeq 5 m_{p}$ is released, the allowed regions for the $\sqrt{f}\epsilon$ parameter 
as function of $M_{A'}$ ($= m_{DM}$) can be obtained by marginalizing all the models 
for the three instances for the Na and I quenching factors.
This is shown in Fig. \ref{fg:asim_mirr} where the $m_{DM}$ interval from few GeV up to 50 GeV is explored.
These allowed intervals identify the $\sqrt{f}\epsilon$ values corresponding to C.L. 
larger than 10 $\sigma$ from the {\it null hypothesis}, that is $\sqrt{f}\epsilon=0$.
The regions obtained for the three  instances for quenching factors 
can be recognized on the basis of different hatching of the allowed regions.

In Fig. \ref{fg:asym_theo} comparisons between the DAMA experimental modulation amplitudes and some expectations for
Mirror DM are shown.  

Thus, as shown in Figs.~\ref{fg:asim_mirr} and \ref{fg:asym_theo}, the restrictions on the mirror DM candidate become more severe 
thanks to the new DAMA/LIBRA--phase2 data (see Ref. \cite{mirasim} for comparison).

\subsubsection{Symmetric Mirror Matter}

In Symmetric Mirror Matter scenario, the mirror parity exchanging mirror to ordinary particles 
is an exact symmetry; thus for all ordinary particles: 
the electron $e$, proton $p$, neutron $n$, photon $\gamma$, neutrinos $\nu$ etc., 
with interactions described by the Standard Model $SU(3)\times SU(2)\times U(1)$,  
there should exist their mirror twins: $e'$, $p'$, $n'$, $\gamma'$, $\nu'$ etc.
which are sterile to our strong, weak and electromagnetic interactions  
but have instead their own gauge interactions $SU(3)'\times SU(2)'\times U(1)'$ 
with exactly the same  coupling constants. 
Ordinary and mirror particles are degenerate in mass, 
and the ordinary and mirror sectors have identical microphysics
at all levels from particle to atomic physics. 

In this context, to analyze the annual modulation observed by DAMA experiments in the framework of 
Symmetric Mirror matter, it has been exploited the interaction portal related to  the 
{\it photon-mirror photon kinetic mixing} term 
$ \frac{\epsilon}{2} \, F^{\mu\nu} F'_{\mu\nu} $
with a small parameter $\epsilon \ll 1$. 
This mixing renders the mirror nuclei mini-charged with respect to ordinary electromagnetic force,  
and thus mediates the scattering of  mirror nuclei off ordinary ones 
with the Rutherford-like cross sections. 
The low-energy differential cross-section of the interaction between mirror 
and ordinary nuclei has the same form as reported in the previous section.
In the Symmetric Scenario there is a different chemical compositions of mirror sector. 
The dominant components should be mirror Hydrogen and mirror Helium-4 but 
a contribution up to few per cent can be due to heavier mirror atoms as Oxygen, Carbon, etc..

In this framework (whose details can be found in Ref. \cite{mirsim}), the Dark Matter particles are expected
to form, in the Galaxy, clouds and bubbles with diameter which could be 
as large as the solar system. In this model a dark halo, at the present epoch, is 
crossing a region close to the Sun with a velocity
in the Galactic frame that could be, in principle, arbitrary.
Hereafter we will refer to such local bubbles simply as halo. 
The halo can be composed by different species of mirror DM particles (different
mirror atoms) that have been thermalized 
and in a frame at rest with the halo. They have a velocity distribution
that can be considered Maxwellian with the characteristic velocity 
related to the temperature of the halo and to the mass
of the mirror atoms. We assume that the halo has its own local equilibrium temperature, 
$T$, and that the velocity parameter of the $A'$ mirror atoms is given by
$ \sqrt{2 k_{\scriptscriptstyle B} T / M_{\scriptscriptstyle A'} }$.
In this scenario lighter mirror atoms have larger velocities than 
the heavier ones, on the contrary of the CDM model where the velocity distribution is
mass independent.  

\begin{figure}[!ht]
\begin{center}
\includegraphics[width=4.3cm] {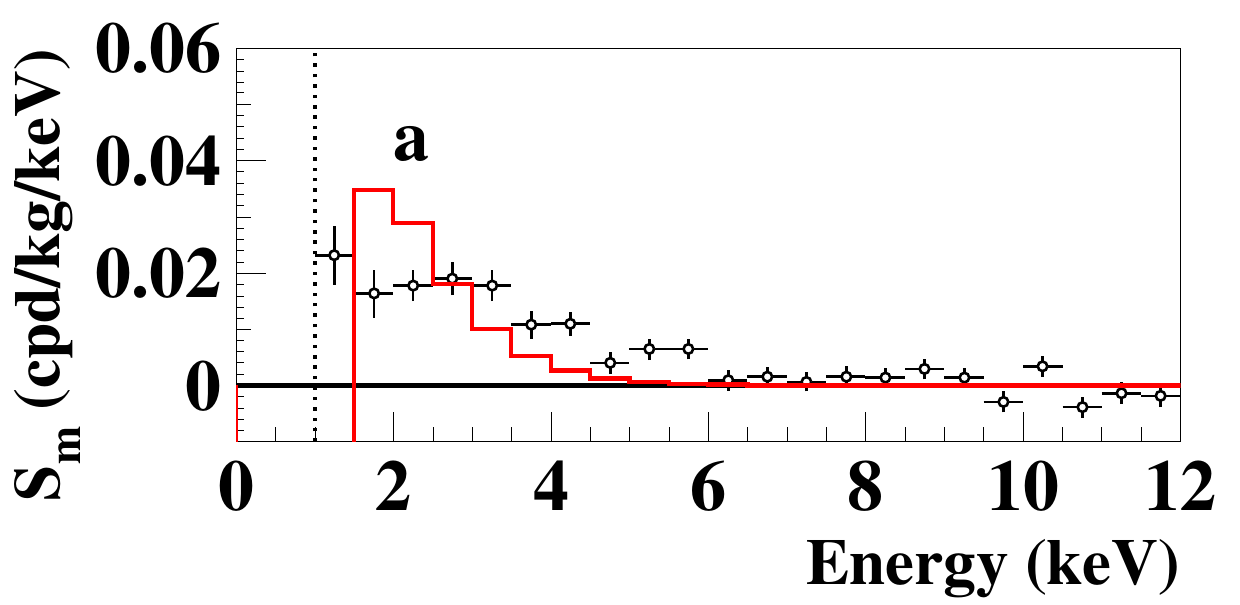}
\includegraphics[width=4.3cm] {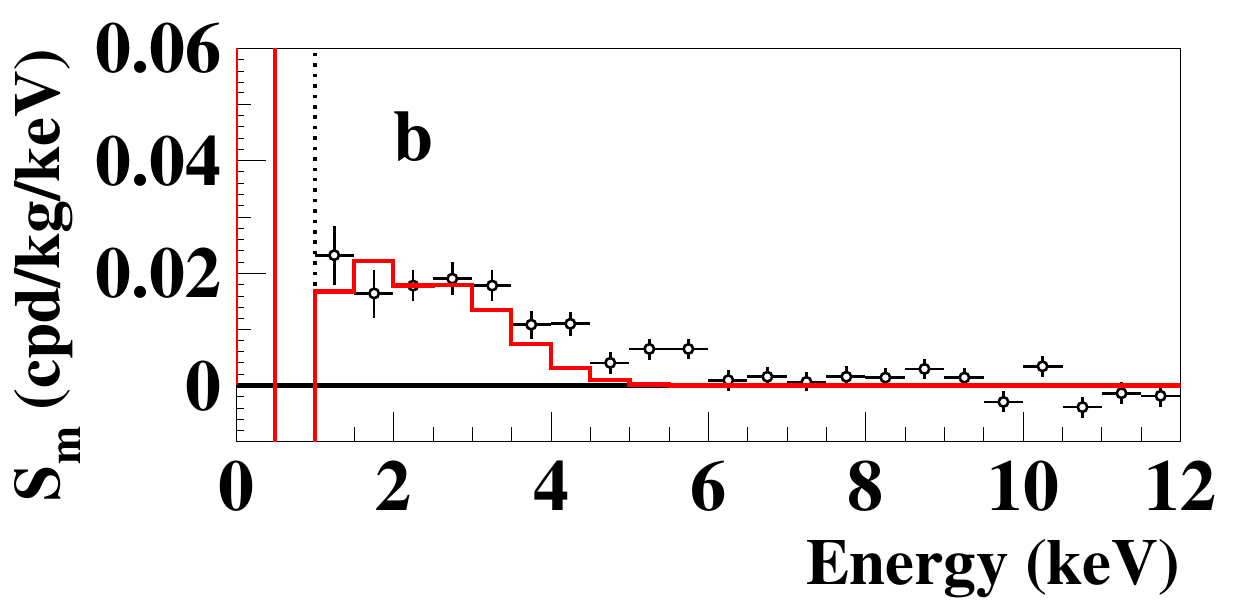}
\includegraphics[width=4.3cm] {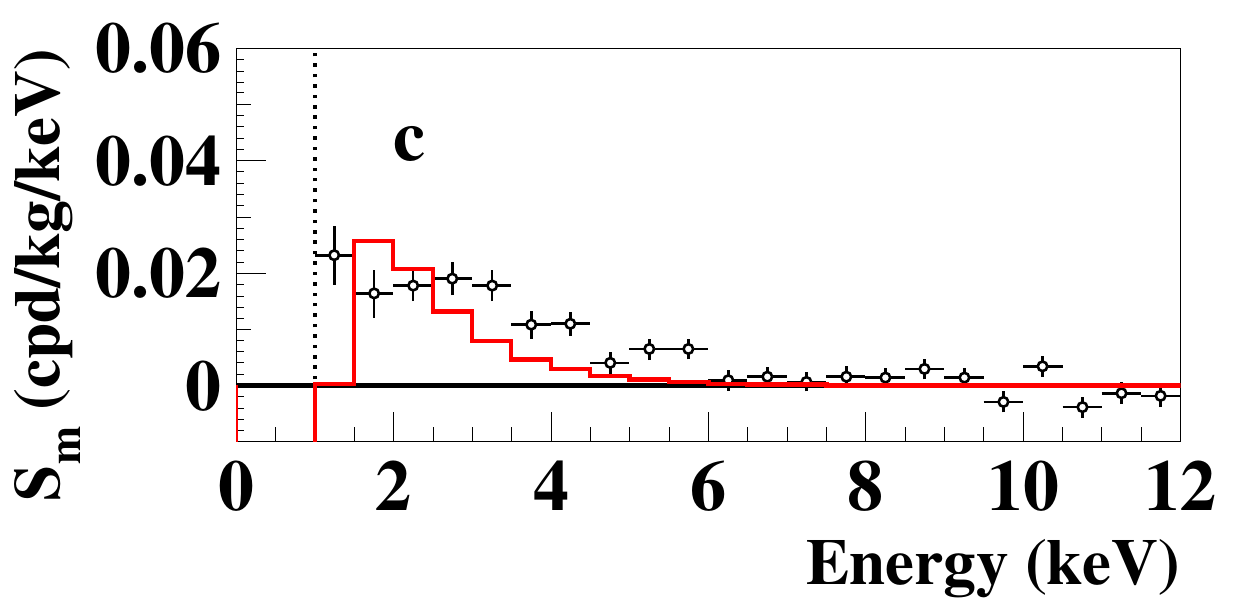}
\end{center}
\vspace{-0.4cm}
\caption{Examples of superposition of the measured $\mathcal{S}^{exp}_{m}$ vs energy
(points with error bars) with theoretical expectations (solid histograms) for
Symmetric Mirror Matter for a composite dark halo H$'$(24\%), He$'$(75\%),
Fe$'$(1\%) and $\rho_0$ = 0.3 GeV/cm$^{3}$.
(a) case with $v_{0} = 270$ km/s, $v_{halo} = 200$ km/s,
    $T=3.1 \times 10^{6} \;{\textrm K}$, parameters in the set A,
    $\sqrt{f}\epsilon = 7.1 \times 10^{-10}$ including channeling effect;
(b) case with $v_{0} = 220$ km/s, $v_{halo} =  60$ km/s,
    $T=10^{5} \;{\textrm K}$, parameters in the set B,
    $\sqrt{f}\epsilon = 2.4 \times 10^{-9}$ and quenching $Q_{I}$;
(c) case with $v_{0} = 220$ km/s, $v_{halo} = 100$ km/s,
    $6.2 \times T=10^{6} \;{\textrm K}$, parameters in the set B,
    $\sqrt{f}\epsilon = 1.1 \times 10^{-9}$ and quenching $Q_{II}$.
    As it can be seen, the data allow discriminating the various scenarios.}
\label{fg:sym_theo}
\end{figure}

It is worthwhile to remind that the expected phase of the annual modulation signal induced by
the mirror particles depends on the halo velocity (module and direction) with respect to 
the laboratory in the Galactic frame. The detailed study of the behavior of the phase in the
Symmetric Mirror Model is reported in Ref. \cite{mirsim}. 
In the data analysis we have taken into account all the uncertainties discussed in the previous sections.
The scenarios and halo composition described in Ref. \cite{mirsim} have been considered.
As example, in Fig. \ref{fg:sym_theo} the expectations of the modulation amplitude 
calculated for some Symmetric Mirror models favored by DAMA experiments, superimposed to
the DAMA experimental modulation amplitudes, are shown. 

\begin{figure}[!ht]
\centering
\includegraphics[width=4.3cm] {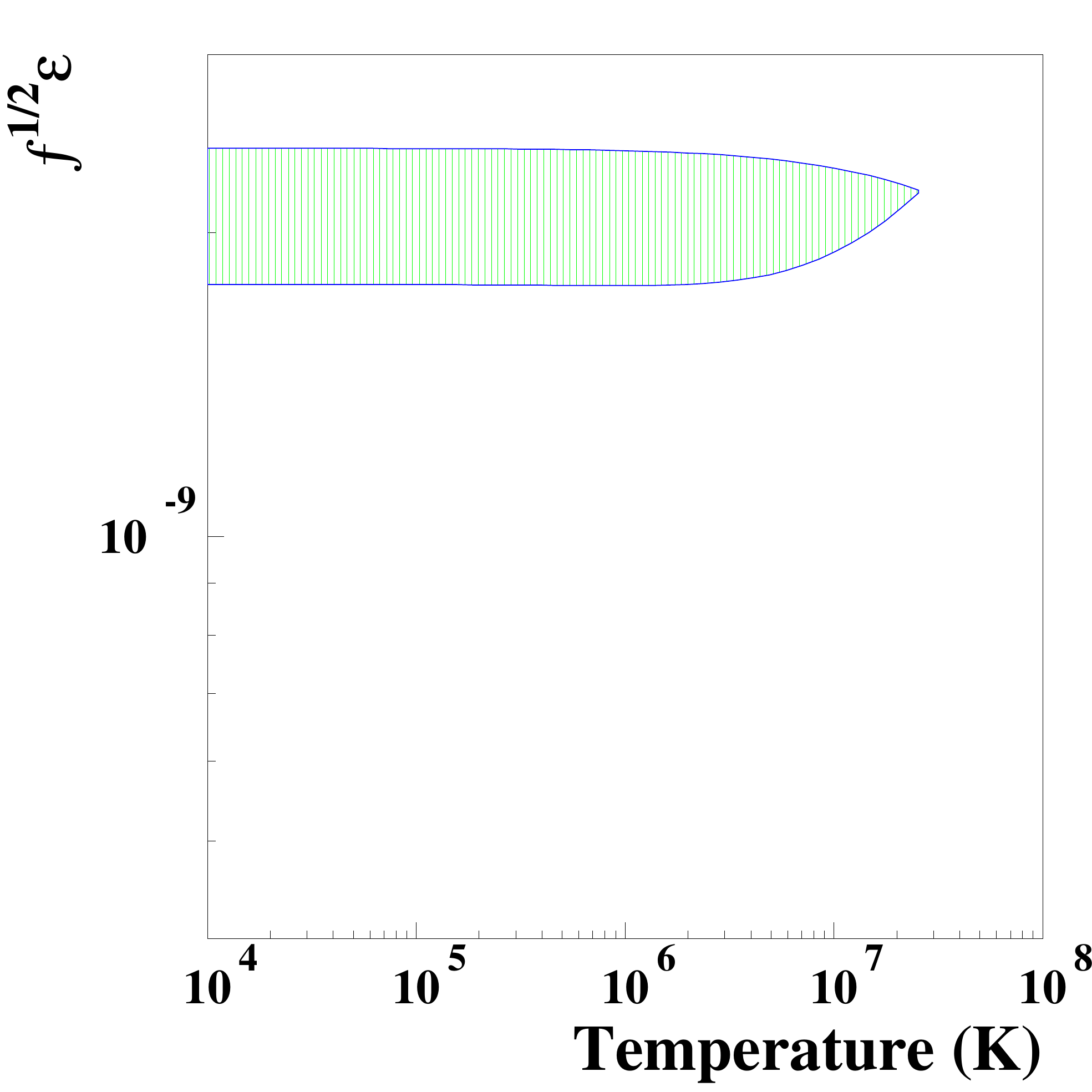}
\includegraphics[width=4.3cm] {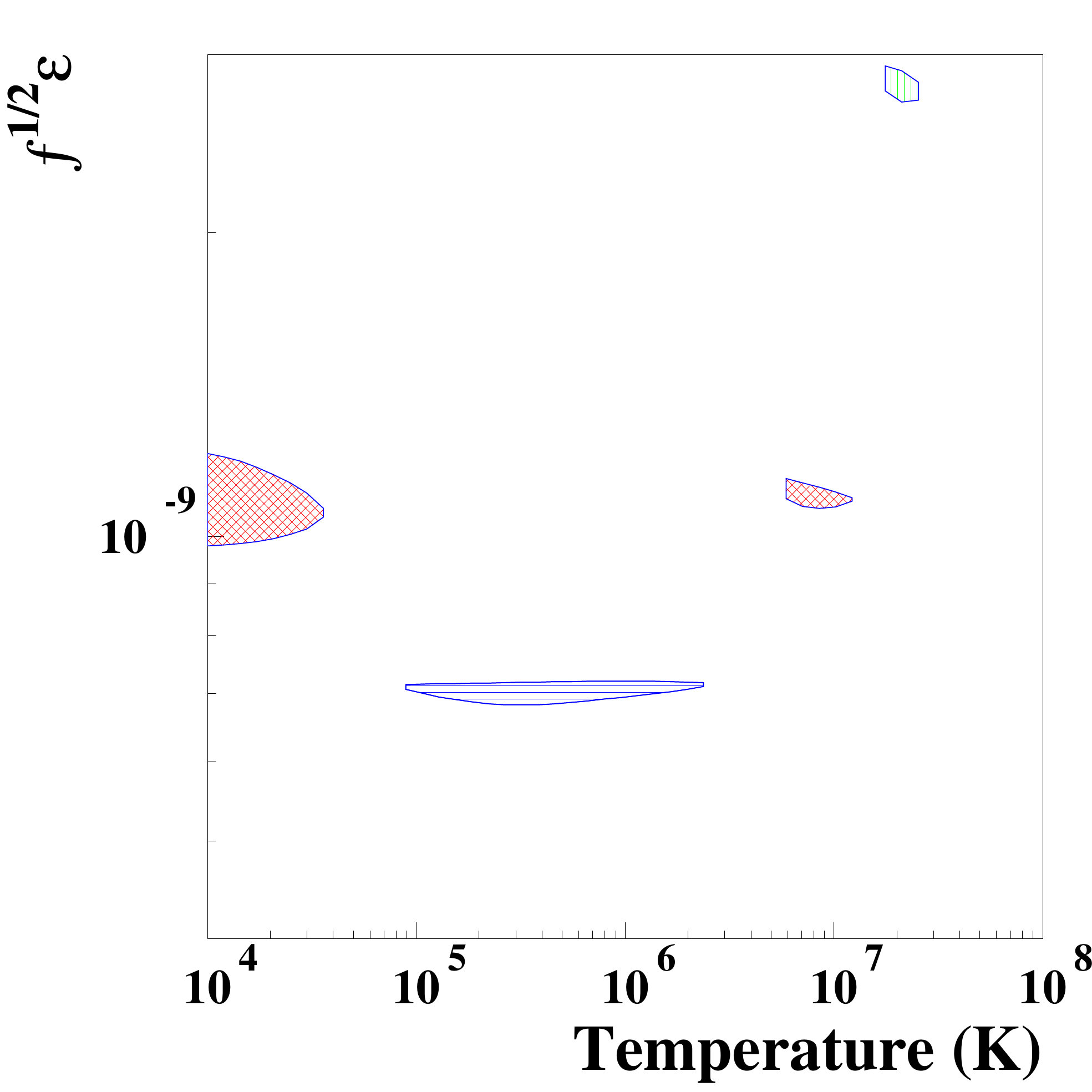}
\includegraphics[width=4.3cm] {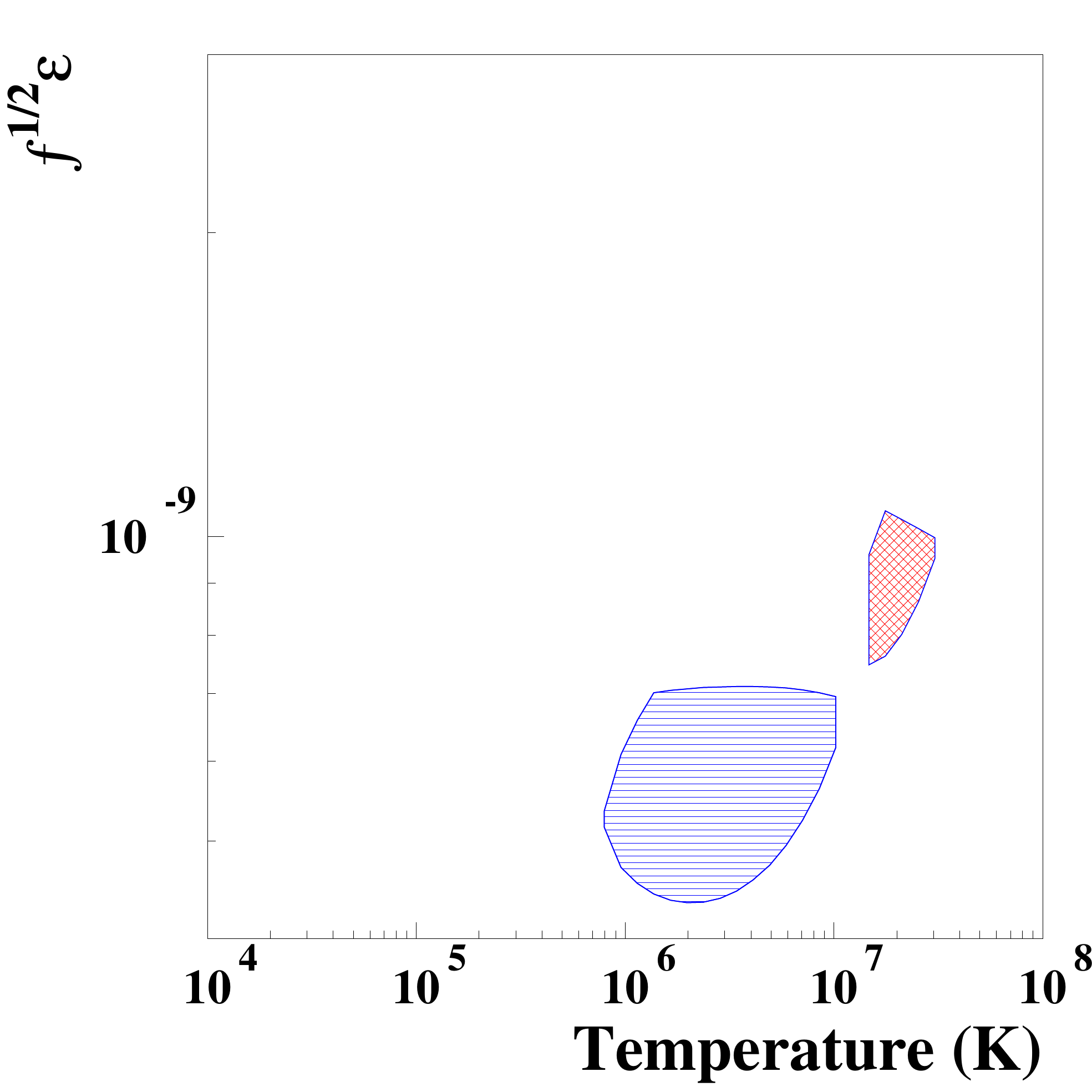}
\caption{Some examples of regions in the plane $\sqrt{f}\epsilon$ vs halo temperature allowed by DAMA experiments
in the case of Symmetric Mirror Matter.
The regions represent the domain where the likelihood-function values differ more than 10 
$\sigma$ from the {\it null hypothesis} (absence of modulation).
The three graphs refer to different dark halo composition:
$Left:$ composite dark halo H$'$(12.5\%), He$'$(75\%), C$'$(7\%), O$'$(5.5\%), with 
$v_{0} = 220$ km/s, $v_{halo} = -100$ km/s and parameters in the set C.
$Center:$ composite dark halo H$'$(20\%), He$'$(74\%), C$'$(0.9\%), O$'$(5\%), 
Fe$'$(0.1\%), with $v_{0} = 220$ km/s, $v_{halo} = 0$ km/s and parameters in the set C.
$Right:$  composite dark halo H$'$(24\%), He$'$(75\%), Fe$'$(1\%), with $v_{0} = 220$ km/s, 
$v_{halo} = 150$ km/s and parameters in the set C.
Three different instances for the Na and I quenching factors have been considered: 
(i) $Q_I$ case [(green on-line) vertically-hatched regions],
(ii) with channeling effect [(blue on-line) horizontally-hatched regions] and 
(iii) $Q_{II}$ [(red on-line) cross-hatched regions].}
\label{fg:simm_mirr13_5}
\end{figure}

\begin{figure}[!ht]
\centering
\includegraphics[width=4.3cm] {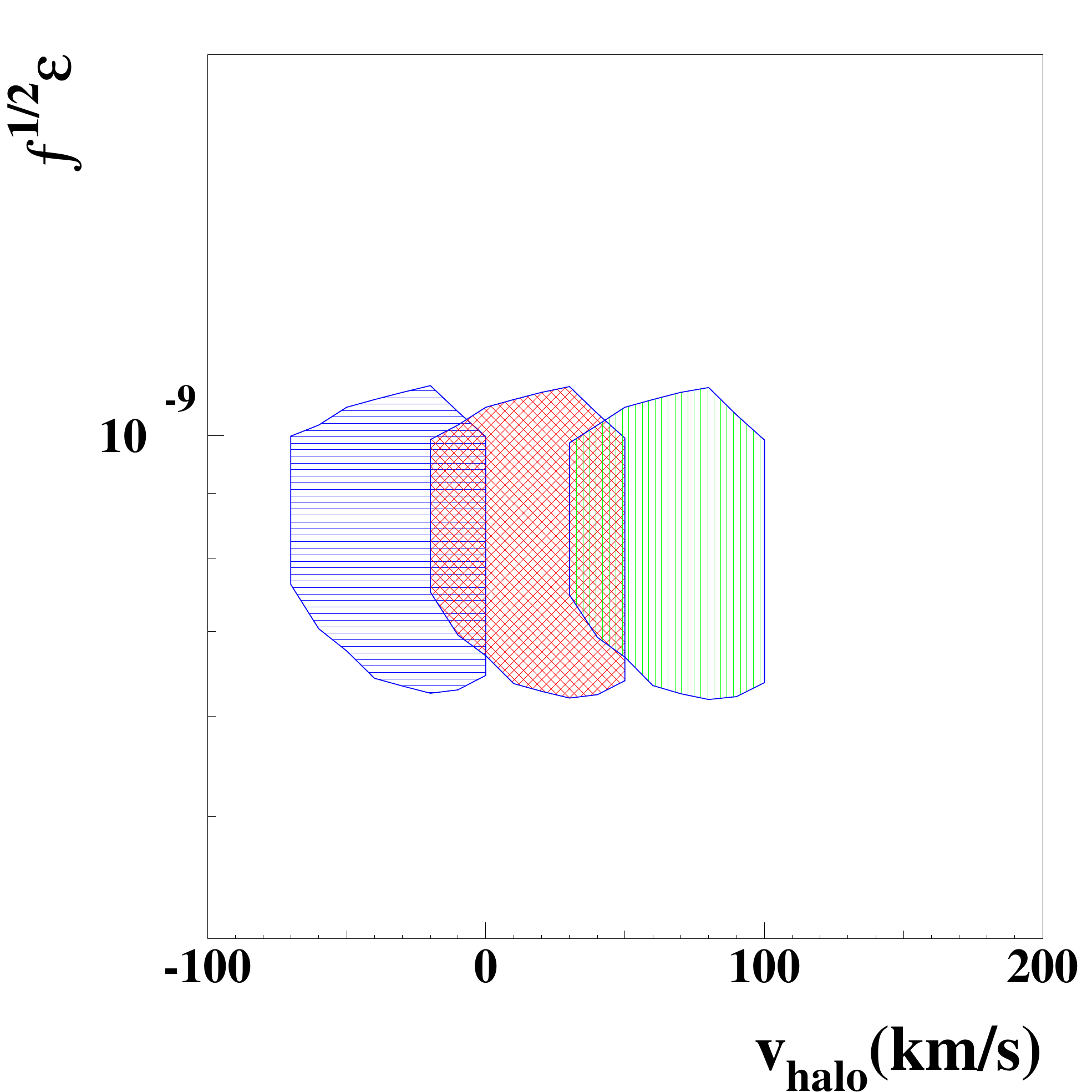}
\includegraphics[width=4.3cm] {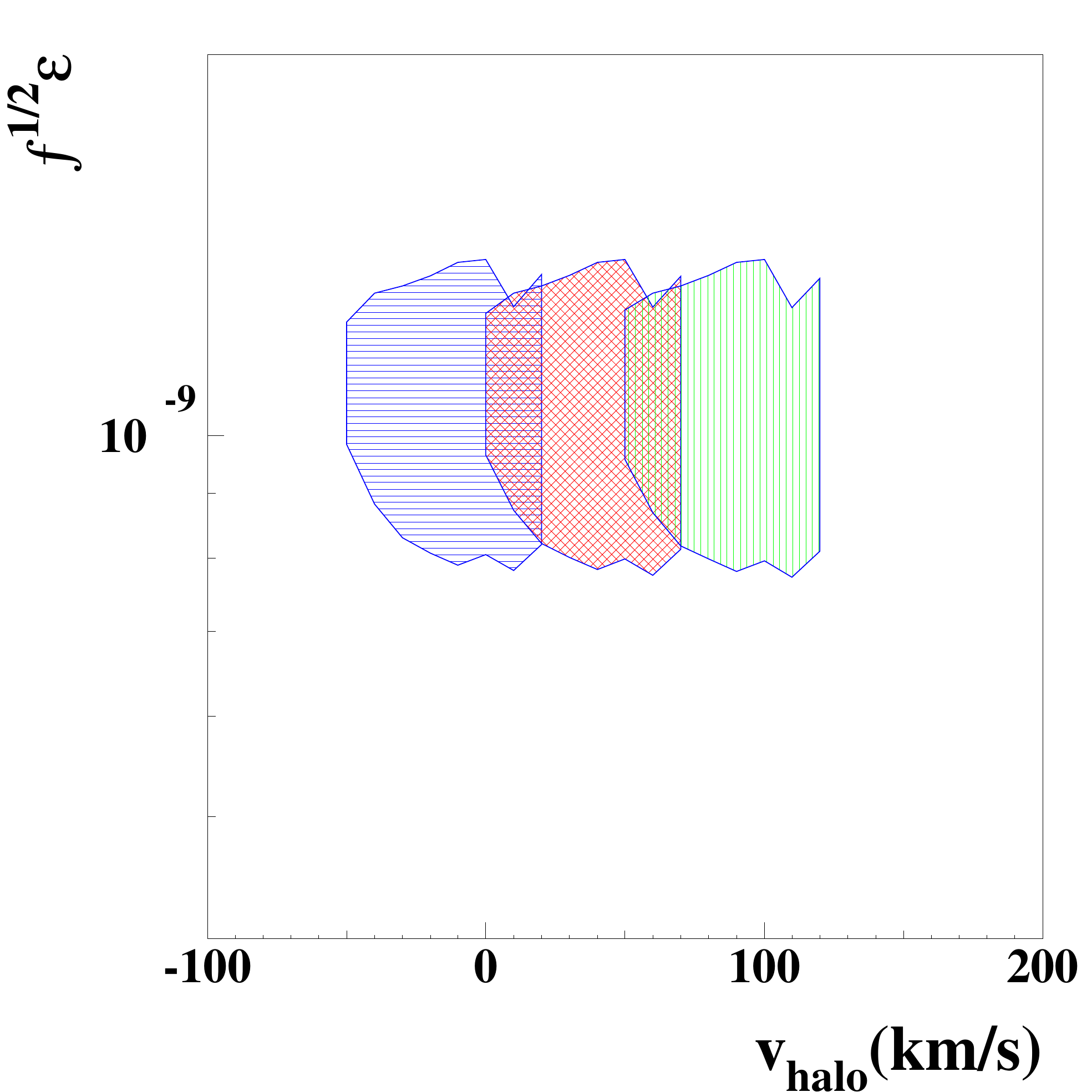}
\includegraphics[width=4.3cm] {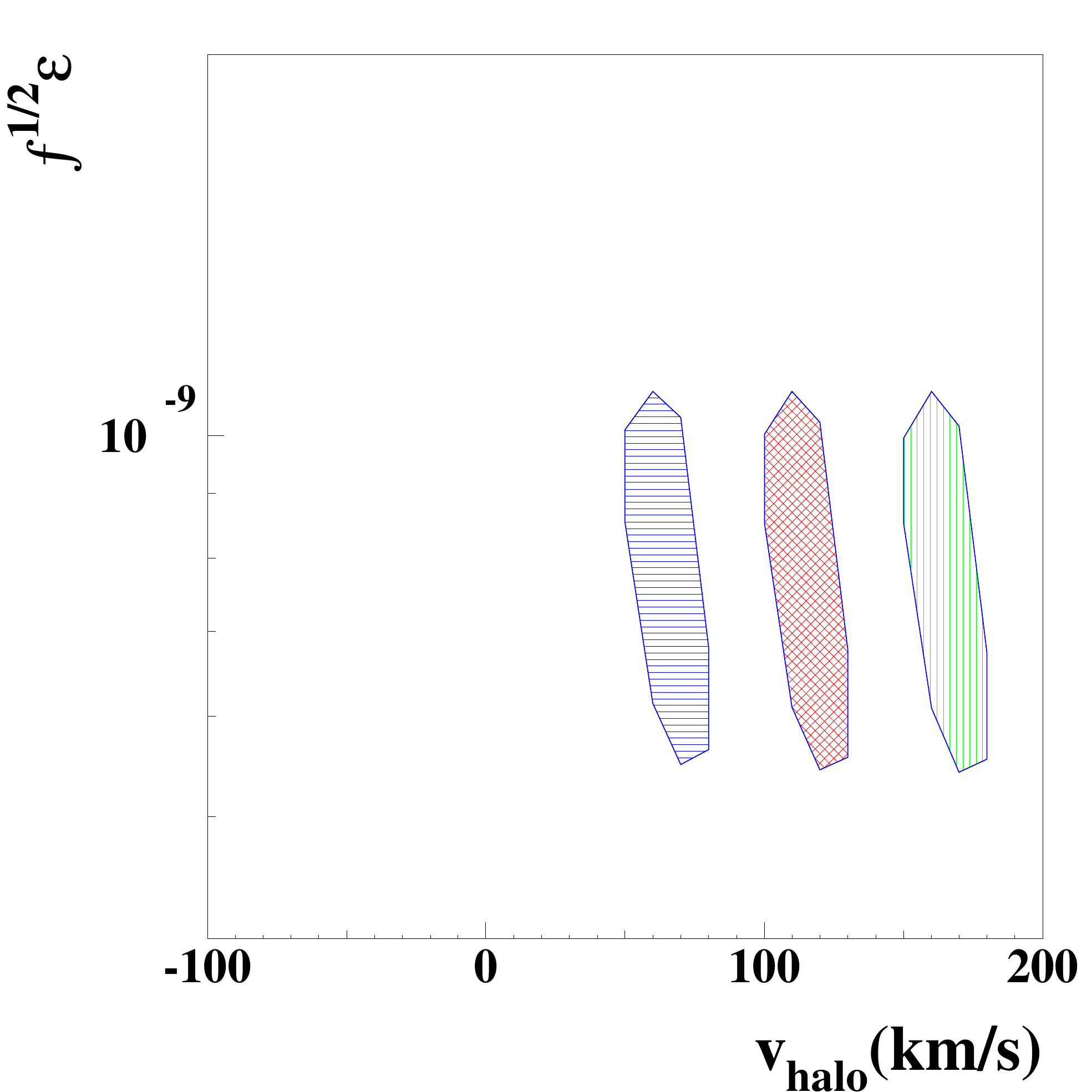}
\caption{Regions in the plane $\sqrt{f}\epsilon$ vs $v_{halo}$ allowed by DAMA experiments in the 
case of Symmetric Mirror Matter.
The regions represent the domain where the likelihood-function values differ more than 10 
$\sigma$ from the {\it null hypothesis} (absence of modulation).
The three graphs refer to different dark halo compositions with the same temperature 
$T=10^{4} \;{\textrm K}$, the same set A and the $Q_{II}$ scenario:
$Left:$   composite dark halo H$'$(12.5\%), He$'$(75\%), C$'$(7\%), O$'$(5.5\%).
$Center:$ composite dark halo H$'$(20\%), He$'$(74\%), C$'$(0.9\%), O$'$(5\%), Fe$'$(0.1\%).
$Right:$  composite dark halo H$'$(24\%), He$'$(75\%), Fe$'$(1\%).
The three contours in each plot correspond to 
$v_{0} = 170$ km/s [(blue on-line) horizontally-hatched region],
$v_{0} = 220$ km/s [(red on-line) cross-hatched region] and
$v_{0} = 270$ km/s [(green on-line) vertically-hatched region].}
\label{fg:simm_mirr14_5}
\end{figure}

In the following, the $\sqrt{f}\epsilon$ values allowed by DAMA experiments 
in different halo models and some scenarios are reported
in order to study how the inclusion of the new data from DAMA/LIBRA--phase2
with 1 keV energy threshold helps to restrict a significant part of the parameters' space.
In particular, two different plots for each halo composition are shown:  
i) allowed regions for the $\sqrt{f}\epsilon$ parameter
as a function of the halo temperature for different values of the halo velocity in the Galactic frame,
Fig. \ref{fg:simm_mirr13_5};
ii) allowed regions for the $\sqrt{f}\epsilon$ parameter
as a function of the halo velocity in the Galactic frame 
for the same temperature $T = 10^{4}$ K, and for different $v_0$ values, Fig. \ref{fg:simm_mirr14_5}.
The figures refer to two different compositions of halo models. 
All the reported allowed intervals identify the $\sqrt{f}\epsilon$ values corresponding to C.L.
larger than 10 $\sigma$ from the {\it null hypothesis}, that is $\sqrt{f}\epsilon=0$.

In conclusion, seemingly the symmetric mirror DM has a better agreement with respect to the 
asymmetric mirror DM within the considered scenarios. 
Finally, the mirror DM scenarios are still of interest at the same most stringent C.L. considered
for some other scenarios above, and the allowed parameters' space is largely reduced when including the data of the
first six annual cycles by DAMA/LIBRA--phase2.

\section{Conclusions}
\label{concl}

The investigation on the nature of the DM particles is an open problem; it always requires a large number of assumptions.
In this paper several possible scenarios for DM candidates are analyzed on the basis of the long-standing DAMA results 
exploiting the DM annual modulation signature.

In particular, the DAMA/LIBRA--phase2 data, collected over the first six full annual cycles (1.13 ton $\times$ yr)  
with a software energy threshold down to 1 keV, are analyzed with the DAMA/NaI and DAMA/LIBRA--phase1 data
for several scenarios, improving the confidence levels and restricting the allowed parameters' space of the considered DM candidate particles
with respect to previous analyses. For example, in the case of DM particles inducing nuclear recoils through SI elastic 
scattering low mass candidates are allowed in particular when the channeling effect is included. In the case of 
a DM candidate with SI isospin violating interaction (that is the effective DM particle couplings to
protons and neutrons are not equal), very good agreements are obtained for most of the considered scenarios.
Moreover, the cases of  
a DM candidate with isospin violating interaction and the case of a DM candidate with preferred inelastic interaction
including the Thallium contribution are analyzed here by DAMA for the first time.

As shown, in this paper several scenarios are compatible with the observed signal; other possibilities are open as well. In 
particular, we remind the interest in including tidal stream effects from dwarf satellite galaxies of the Milky Way as e.g. the 
Sagittarius one (other data from GAIA are expected to add significant information on the topic in near future) 
or other possible non-thermalized components in the galactic halo as e.g. presence of caustics, as suggested in Ref. \cite{caus}.
The presence of similar effects could play an important role in the corollary model-dependent results. It is also worth noting that 
even a suitable particle not yet foreseen by theories may be the- or one-of-the- solution for DM particles. Let us also highlight
that in the DM field the case of a single candidate accounting for all DM is generally adopted, as done in the 
present paper. However, other possibilities exist, as the case of two DM candidates recently proposed to explain the
DAMA results \cite{her18}. In addition, considering the richness in different particles of the visible matter which is $\ll 1\%$  
of the Universe density, one could expect that the DM component (about 27\% of the Universe density) may also be multicomponent. 
This latter possibility is natural in some cases, as e.g. in the mirror DM, as stressed in the present paper.

Similar considerations and the improved results presented in this paper show how important is to improve the capability of the 
experiment to effectively disentangle among the many possible different scenarios.
For such a purpose an increase of exposure in the new lowest energy bins and the lowering of the software energy threshold below 1 keV 
are important. Thus, DAMA/LIBRA--phase2 has continued its data taking.
Moreover, related R\&D's towards the so-called phase3 have been funded and are in progress. In particular, new PMTs with
high quantum efficiency have been especially developed by HAMAMATSU: R11065-20MOD, which satisfy all the needed requirements,
and a new voltage divider allocating the preamplifier on the same basis has been designed and already tested.

In conclusion, the new data have allowed significantly improving the confidence levels and restricting the allowed parameters' space 
for the various considered scenarios with respect to previous DAMA analyses; efforts towards further improvements are in progress.

\end{document}